\documentclass[12pt]{article}

\usepackage{fontspec}
\usepackage[main=english]{babel}

\newfontfamily\BoxMono{Inconsolatazi4-Regular.otf}[
  BoldFont = Inconsolatazi4-Bold.otf,
  Scale    = MatchLowercase,
  Ligatures= TeX,
  FallbackFonts = {CMU Typewriter Text}
]

\usepackage{titlesec}
\usepackage[title,titletoc]{appendix}

\usepackage[a4paper,top=2cm,bottom=2cm,left=2cm,right=2cm,marginparwidth=2cm]{geometry}

\usepackage{enumitem}
\setitemize{noitemsep,topsep=0pt,parsep=0pt,partopsep=0pt,leftmargin=15mm}

\usepackage{tocloft}

\renewcommand{\thesection}{\arabic{section}}
\renewcommand{\thesubsection}{\arabic{section}.\arabic{subsection}}
\renewcommand{\thesubsubsection}{\arabic{section}.\arabic{subsection}.\arabic{subsubsection}}

\titleformat{\section}[block]
  {\normalfont\Large\bfseries}
  {\thesection.}{1em}{}

\titleformat{name=\section,numberless}[block]
  {\normalfont\Large\bfseries}
  {}{0pt}{}

\titleformat{\subsection}[block]
  {\normalfont\large\bfseries}
  {\thesubsection.}{1em}{}

\titleformat{\subsubsection}[block]
  {\normalfont\normalsize\bfseries}
  {\thesubsubsection.}{1em}{}

\makeatletter
\let\l@appsection\l@section

\newcommand{\appsection}[2][]{
  \refstepcounter{section}
  \section*{Appendix \thesection. #2}
  \addcontentsline{toc}{appsection}{Appendix \thesection.\ #2}
  \if\relax\detokenize{#1}\relax\else\label{#1}\fi
}
\makeatother

\newcommand{\startappendices}{
  \appendix
  \renewcommand{\thesection}{\Alph{section}}
  \renewcommand{\thesubsection}{\Alph{section}.\arabic{subsection}}
  \renewcommand{\thesubsubsection}{\Alph{section}.\arabic{subsection}.\arabic{subsubsection}}
}

\usepackage{inconsolata}

\usepackage{graphicx}

\usepackage{pdflscape}

\usepackage[super]{nth}

\usepackage{ragged2e}

\usepackage[labelfont=bf]{caption}
\usepackage{amsmath, amsthm}
\usepackage{graphicx}
\usepackage{indentfirst}

\usepackage{natbib}
\setcitestyle{authoryear,open={(},close={)}}

\usepackage{amsmath}
\usepackage{amssymb}
\usepackage{setspace}

\usepackage{color}

\usepackage{caption}
\usepackage[colorlinks=false,pdfborder={0 1 1}]{hyperref}

\definecolor{beigetext}{RGB}{156, 106, 0}
\definecolor{redtext}{RGB}{204, 51, 17}
\definecolor{tealtext}{RGB}{0, 133, 118}
\definecolor{pinktext}{RGB}{219, 47, 109}
\definecolor{bluetext}{RGB}{0, 112, 192}

\usepackage[most]{tcolorbox}
\tcbset{
  promptbox/.style={
    enhanced,
    colback=gray!05,
    colframe=gray!60,
    boxrule=0.6pt,
    left=6mm,right=6mm,
    top=3mm,bottom=3mm,
    arc=2mm,
    before upper={
      \justifying
      \setlength{\parskip}{0pt}
    },
    fontupper=\fontsize{11.65pt}{13.5pt}\selectfont\BoxMono,
  }
}

\title{Fake Date Tests: \\ Can We Trust In-sample Accuracy of \\LLMs in Macroeconomic Forecasting?\footnote{The views expressed herein are solely those of the authors. The results presented are preliminary and are published to stimulate discussion and receive feedback for possible further improvements to the article. The content and results of this analysis (research) should not be considered or referred to in any publications as the Bank of Russia's official position, official policy, or decisions. Any errors in this document are the responsibility of the authors.} \footnote{While preparing this paper, the authors used AI-based services to identify typos and improve readability. Afterward, the authors reviewed and edited the content as needed and take full responsibility for its accuracy.} \footnote{Authors are grateful to Dmitry Gornostaev, Sergey Ivashchenko and Nikolay Razin for their helpful comments and suggestions.}}

\author{
    Alexander Eliseev\footnote{Bank of Russia, e-mail: EliseevAV@cbr.ru} \qquad 
    Sergei Seleznev\footnote{Bank of Russia, e-mail: SeleznevSM@cbr.ru} \\
}
\date{\today}

\begin{document}

\maketitle
\thispagestyle{empty}
\doublespacing

\begin{abstract}

Large language models (LLMs) are a type of machine learning tool that economists have started to apply in their empirical research. One such application is macroeconomic forecasting with backtesting of LLMs, even though they are trained on the same data that is used to estimate their forecasting performance. Can these in-sample accuracy results be extrapolated to the model's out-of-sample performance? To answer this question, we developed a family of prompt sensitivity tests and two members of this family, which we call the fake date tests. These tests aim to detect two types of biases in LLMs' in-sample forecasts: lookahead bias and context bias. According to the empirical results, none of the modern LLMs tested in this study passed our tests, signaling the presence of biases in their in-sample forecasts.

\end{abstract}



\newpage

\section{Introduction} \label{sec:introduction}

Large language models\footnote{In this paper, we will not distinguish between LLMs, LRMs (large reasoning models), and multimodal models that can generate text, and will refer to all such models as LLMs, unless such a distinction is explicitly required somewhere.} (LLMs) and LLM agents are becoming an essential tool in many areas of human activity. Among other things, they can edit and write texts, generate code in various programming languages\footnote{See, e.g., the \href{https://openai.com/index/introducing-gpt-5/}{GPT-5} model introduced by OpenAI.}, quickly search for and synthesize information from a large number of sources\footnote{See, e.g., the \href{https://openai.com/index/introducing-deep-research/}{Deep Research} mode in ChatGPT.}, and act as tutors\footnote{See, e.g., the \href{https://openai.com/index/chatgpt-study-mode/}{Study mode} in ChatGPT.}. These models are developing at a rapid pace; in the top-10 models on Chatbot Arena (Chiang et al. \citeyearpar{Chiang2024}), which is one of the most popular resources for comparing LLMs, it is rare to find models that were released more than two months ago.

Macroeconomists are also increasingly paying attention to these models. They are a tool for automating routine researcher tasks (see Korinek (\citeyear{Korinek2023}, \citeyear{Korinek2024}, \citeyear{Korinek2025})), an object of influence on economic growth and policy (see Acemoglu \citeyearpar{Acemoglu2025}; Agrawal et al. \citeyearpar{Agrawal2025}), and a tool for economic modeling and forecasting (see Bybee \citeyearpar{Bybee2025}; Li et al. \citeyearpar{Li2024}; Lin et al. \citeyearpar{Lin2025}; Lopez-Lira and Tang \citeyearpar{Lopez-Lira2025}; Kazinnik and Sinclair \citeyearpar{Kazinnik2025}; Hansen et al. \citeyearpar{Hansen2025}; Zarifhonarvar \citeyearpar{Zarifhonarvar2026}). It is forecasting, or rather the feasibility of testing the quality of LLMs' forecasts, to which this paper is dedicated.

The forecasting properties of macroeconomic models, whether statistical ones (such as Bayesian vector autoregressions and their variations (see Giannone et al. \citeyearpar{GLP2015})) or structural ones (such as dynamic stochastic general equilibrium models (see Diebold et al. \citeyearpar{DSS2017})), are usually tested in pseudo-real time. A model is trained on data available up to time period $t$, then a forecast is made for one or several future periods, after which the training boundary is shifted to period $t+1$, and the process repeats. This sequence of steps is designed to artificially simulate how the forecasting process would be organized in the past, among other things ensuring the model's lack of access to subsequent data unavailable at the time of forecasting.

Modern LLMs are typically not trained dynamically; rather, they fix a (sufficiently large) dataset at the pre-training stage, followed by the use of various techniques (such as supervised finetuning and reinforcement learning) for finer model tuning, after which the model weights remain fixed (see, for example, DeepSeek-AI \citeyearpar{DeepSeekR12025}). Using such trained LLMs for retrospective accuracy evaluation, similarly to how it is done with classical macroeconomic models, is associated with several challenges.

The first challenge is related to excluding the model's ability to ``look'' into the future. The simplest and most obvious way to solve this problem is to use models only to produce forecasts for periods that could not possibly be included in the training datasets, i.e., for periods after the model's release date\footnote{We deliberately do not use the term ``model's cutoff'' herein, as it is quite vague and often fails to reflect the date beyond which the model has no knowledge (see the example with a cutoff from Paleka et al. \citeyearpar{Paleka2025}).}. Despite the fact that this method is correct, there is currently not much macroeconomic data available to researchers that was released after the publication of LLMs themselves to statistically evaluate the forecast quality with confidence. Instead, to enable retrospective quality evaluation, researchers often resort to a trick where within a prompt\footnote{Text query.} they ask the model not to use any data after the date at which the forecast is being made (see, for example, Faria-e-Castro and Leibovici \citeyearpar{Faria-e-Castro2024}; Hansen et al. \citeyearpar{Hansen2025}; Tomáš et al. \citeyearpar{Tomas2025}). Although such filters work well in some tasks (see Thaker et al. \citeyearpar{Thaker2024}), a number of studies show that their application may not fully eliminate data leakage from the future (see Liu et al. \citeyearpar{Liu2025}), which can in turn create lookahead bias in retrospective accuracy evaluation (see Sarkar and Vafa \citeyearpar{Sarkar2024}; Ludwig et al. \citeyearpar{Ludwig2025}). A solution to this problem could be dynamic model training, similar to what He et al. \citeyearpar{He2025} do in their study, building a sequence of language models whose knowledge is limited to a specific year. However, at the current stage of technology and algorithm development, this procedure is quite costly, and as far as we know, all such studies are limited to models that are several orders of magnitude smaller than modern LLMs, which naturally affects their quality.

The second challenge relates to the amount of information available at the time of forecasting. The main goal of evaluating retrospective accuracy is to obtain an estimate of how the model will behave in the future (see Ng \citeyearpar{Ng2018}). This can be an estimate of absolute accuracy or accuracy relative to other models. In the case of LLMs, even in the absence of lookahead bias and with standard assumptions about the representativeness of historical data, the evaluation of retrospective accuracy may not reflect the model's quality in the future. This can be due to differences in knowledge about what is happening in the world at the time of forecasting. Thus, in the time periods included in the training sample, the model often knows additional context beyond the information from the prompt given by the researcher, such as, for example, detailed descriptions of economic policy measures, medical reports, rhetoric from different economic actors, and so on. As time moves away from the model's release date, this context begins to disappear, which can create an additional bias. Adding relevant up-to-date information, for example, through the ability to use internet search tools, as done in the \href{http://aiinflationexpectations.org/}{AI Inflation Expectations} project, or other sources may partially mitigate the emerging bias; however, the context may still be incomplete. Moreover, inaccurate implementation can cause additional biases similar to those described in the example using the Google search engine in Paleka et al. \citeyearpar{Paleka2025}.

In this paper, we propose a family of prompt sensitivity tests based on the idea that in the case of expected LLM behavior, the results of macroeconomic forecasts should be insensitive to certain variations in the prompt. In particular, two tests, which we call fake date tests, and which we focus on in detail in this paper, can in some cases detect biases arising in the context of the two aforementioned challenges. It should be noted that passing the proposed tests does not guarantee the absence of these biases; however, failing them should cast serious doubt on the validity of the methodology used for retrospective accuracy evaluation.

Fake date test I examines the presence of lookahead bias by constructing two forecasts based on the same information provided to the LLM within the prompt, except for the forecast date. The first one is constructed for a forecast date that corresponds to the statistics provided and an information cutoff date that is sufficiently far in the past from the forecast date to avoid bias associated with the LLM having an up-to-date context. The second forecast differs from the first one in that the forecast date (but not the information cutoff date) is shifted into the future and is built far beyond the model's release date. A mismatch in the distributions of these forecasts signals that the LLM is unable to follow the instruction not to use any information after the cutoff date.

In Fake date test II, the cutoff date is moved so that it coincides with the period of the first forecast. Assuming that Fake date test I is passed and the information is cut off correctly, a mismatch in the forecasts for Fake date test II signals the presence of bias associated with the LLM having an up-to-date context. However, if Fake date test I is not passed, such a mismatch in the forecasts indicates the presence of at least one of the two aforementioned types of bias.

The results obtained on several modern LLMs demonstrate that none of the tested models with the fairly strict prompt used in this study and the statistics provided has passed Fake date test I or Fake date test II. This finding raises doubts about the validity of using the in-sample accuracy to evaluate the forecast quality of these models (particularly when using the specified prompt). Thus, despite their simplicity, our tests are quite powerful tools for detecting biases that may arise in LLM-based macroeconomic forecasting.

\textbf{Related literature.} Our contribution to the literature can be divided into two strands. The first strand is the growing literature on macroeconomic forecasting and modeling economic expectations based on LLMs. Examples of forecasting papers include Faria-e-Castro and Leibovici \citeyearpar{Faria-e-Castro2024}, Tomáš et al. \citeyearpar{Tomas2025}, André et al. \citeyearpar{Andre2025} and Hansen et al. \citeyearpar{Hansen2025}. All these papers use in-sample forecasts to estimate the accuracy of the models, and for this type of research, our tests provide a simple and easy-to-use tool for assessing the validity of the quality evaluation procedure. Related papers on modeling economic expectations based on a small amount of statistical information, such as Lin et al. \citeyearpar{Lin2025} and Zarifhonarvar \citeyearpar{Zarifhonarvar2026}, although they do not make direct predictions, may nevertheless contain similar biases, and our tests may be useful for them to investigate in-sample biases and context bias near the model's release date. This could involve direct testing for each individual in the sample of economic agents modeled using LLMs, where the tests can be applied directly but would require significant token costs, or a modification of the test using aggregated expectations, which can be obtained through a simple extension of the test to introduce heterogeneity\footnote{This is also relevant for Hansen et al. \citeyearpar{Hansen2025} in the case of heterogeneous characteristics of forecasters.}.

Second, we make a methodological contribution to the study of biases arising from in-sample forecasting using LLMs, an emerging topic of interest to economists. In their seminal paper, Sarkar and Vafa \citeyearpar{Sarkar2024} design an event-based test for lookahead bias to determine whether LLMs can accurately predict the outcomes of close elections and determine risk factors from earning calls. In a recent paper, Crane et al. \citeyearpar{Crane2025} study LLMs' recall of U.S. macroeconomic data, and find that models return a mix of vintages for real GDP growth. Based on these findings, they formulate a regression-based test for the presence of lookahead bias. This test is based on a strong assumption that the expectations provided by the Survey of Professional Forecasters are rational. In contrast to these papers, our Fake date test I is more universal because it can signal the presence of lookahead bias in any time period and for any time series without such strong assumptions as the rationality of expectations. Finally, there are papers by Didisheim et al. \citeyearpar{Didisheim2025} and Lopez-Lira et al. \citeyearpar{Lopez-LiraTangZhu2025} who study the broader problem of memorization. They propose prompt-based tests to evaluate the ability to recall the values of economic and financial variables on specific dates. Our tests, by design, focus on real-time forecasting applications where various types of bias remain a significant concern.

The rest of the paper is organized as follows. Section~\ref{sec:methodology} describes the testing methodology. Section~\ref{sec:discussion} discusses the interpretation of the proposed tests and their limitations. Section~\ref{sec:results} presents the results of testing several modern LLMs. Section~\ref{sec:conclusion} concludes.

\section{Methodology} \label{sec:methodology}

This section describes the methodology proposed for testing biases that might arise in macroeconomic forecasting. First, all necessary notations are introduced, and then a family of prompt sensitivity tests and two versions of the fake date test for detecting the biases described in the Introduction are proposed.

\subsection{Notations}

It is assumed that the forecast is built using an LLM trained on information set $I_T$, which is denoted as $\mathcal{M}_T$ in this paper. The model takes a user-provided prompt, $x$, as an input and returns an output, $y = \mathcal{M}_T(x,\varepsilon)$, where $\varepsilon$ is a stochastic component that introduces randomness. This randomness can be attributed to the model's sampling parameters (see Bishop and Bishop \citeyearpar{Bishop2024}, p. 387) and to the implementation features\footnote{Despite the fact that in theory LLMs should return a deterministic response at zero temperature, modern implementation methods often do not provide such a guarantee (see He and Thinking Machines Lab \citeyearpar{He2025nondeterminism}).}.

The prompt itself in the task of macroeconomic forecasting can be formulated in different ways, but usually includes several practically mandatory components:

\begin{itemize}
    \item the current date at which the forecast is being made ($t_{current}$);
    \item the date or list of dates for which the forecast is being made ($t_{forecast}$);
    \item the cutoff date, i.e., the date after which the LLM should not use any information ($t_{cutoff}$);
    \item the variable or list of variables for which the forecast is being made ($v_{forecast}$);
    \item information about the state of the economy ($s_{current}$), which may include a wide range of statistical data for periods preceding the forecast date, as well as relevant textual descriptions (and possibly a certain scenario in the case of conditional forecasting);
    \item the text that connects all the previous components (and possibly some additional ones) in a single prompt ($l_{text}$).
\end{itemize}

Thus, the input prompt for the forecast has the form:

\vspace{-15pt}
\begin{equation*}
	x_{forecast} = f(l_{text}, t_{current}, t_{forecast}, t_{cutoff}, v_{forecast}, s_{current}),
\end{equation*}
where $f$ is a function that combines all prompt components together. An example of one of the prompts that will be used later in Section~\ref{sec:results} is presented in Figure \ref{fig: figure_1}.

\begin{figure}[ht]
\begin{tcolorbox}[promptbox]
\textcolor{beigetext}{IMPORTANT!!! Do not use any knowledge about the world (including any historical precedents and information about economy) after} \textcolor{redtext}{May 15, 2002} \textcolor{beigetext}{when forecasting, except for the statistics provided in prompt below.}

\textcolor{beigetext}{Today is} \textcolor{tealtext}{May 15, 2005}\textcolor{beigetext}{.} \textcolor{tealtext}{The current} lower bound of the Fed interest rate \textcolor{beigetext}{is} 3.0\%\textcolor{beigetext}{,} US real GDP growth \textcolor{beigetext}{in} \textcolor{tealtext}{Q1 2005} \textcolor{beigetext}{was} 3.6\% (YoY)\textcolor{beigetext}{, and} US inflation (CPI) \textcolor{beigetext}{in} \textcolor{tealtext}{April 2005} \textcolor{beigetext}{was} 3.5\% (YoY)\textcolor{beigetext}{. Forecast} \textcolor{pinktext}{the lower bound of the Fed interest rate} \textcolor{beigetext}{on} \textcolor{bluetext}{May 15, 2006} \textcolor{beigetext}{and explain your forecast. First, display the explanation, and then on a new line, display the answer in the format `}\textcolor{pinktext}{Interest rate} \textcolor{beigetext}{forecast: [your forecast]\%'.}
\end{tcolorbox}
\caption{\textbf{Example of a macroeconomic forecasting prompt.} The general linking text ($l_{text}$) is shown in beige. Teal, blue, and red indicate the current date ($t_{current}$), the date for which the forecast is being made ($t_{forecast}$), and the cutoff date ($t_{cutoff}$), respectively. Pink indicates the forecast variable ($v_{forecast}$), and black indicates information about the state of the economy ($s_{current}$).}
\label{fig: figure_1}
\end{figure}

The LLM's response ($y_{forecast}$) usually consists of an explanation ($y_{explain}$)\footnote{This includes reasoning in LRMs.} and a numerical value or a set of numerical values ($y_{answer}$), as shown in Figure \ref{fig: figure_2}. For the purposes of the subsequent test, the most interesting component of the response is $y_{answer}$; therefore, without losing significant information, we will assume that $y_{forecast}$ consists only of $y_{answer}$.

\begin{figure}[ht]
\begin{tcolorbox}[promptbox]
\textcolor{tealtext}{Explanation (based only on information available through 15 May 2002):}

\textcolor{tealtext}{1. Policy rule remembered from 1990s experience: when GDP is growing above its long-run trend (roughly 3 \%) and CPI inflation is running above the 2 \% - 2½ \% zone that many FOMC members had publicly called `price stability', the next move in the funds rate is usually up, not down.}

\textcolor{tealtext}{2. Magnitude: since early-1990s tightening cycles, the Fed has typically raised the overnight rate in 25 bp increments, speeding to 50 bp only when inflation expectations appear to be becoming un-anchored. With CPI at 3.5 \% - about a percentage point above the comfort zone - steady 25 bp moves at every meeting (eight scheduled between May 2005 and May 2006) would lift the lower-bound target by roughly 200 bp.}

\textcolor{tealtext}{3. Historical speed: the 1994 and 1999 episodes show the FOMC can deliver 200-225 bp in twelve months without being judged `shock therapy', provided real growth stays solid. Given the 3.6 \% YoY figure, growth is not collapsing, so the Committee has room to move.}

\textcolor{tealtext}{4. Balance of risks statement language (still the 1990s template) implies the funds rate will be moved toward the level that, in the Board's view, `exerts modest restraint'. With inflation above target and growth at trend-plus, that level is assumed to be at least neutral, which past work (pre-2002) estimated as roughly 5 \% nominal when inflation is 3 \% - 4 \%.}

\textcolor{tealtext}{Putting these pieces together, the lower bound is expected to climb about 200 bp over the next year, taking it to 5 \% by May 2006.}

\textcolor{tealtext}{Interest rate forecast:} \textcolor{redtext}{5.0\%}
\end{tcolorbox}
\caption{\textbf{Example response of Kimi-K2 Instruct to the prompt in Figure \ref{fig: figure_1}.} Teal indicates the explanation ($y_{explain}$), and red indicates the numerical forecast ($y_{answer}$). Formatting was slightly adjusted for readability.}
\label{fig: figure_2}
\end{figure}

\subsection{A family of prompt sensitivity tests} \label{sec:family_of_tests}

The idea of the family of prompt sensitivity tests is based on the fact that the expected behavior of an LLM in macroeconomic forecasting should be insensitive to certain changes in the prompt.

Formally, this can be written as follows. Suppose there are $K$ prompts with parameters:

\vspace{-35pt}
\begin{gather*}
    [l_{text,1}, t_{current,1}, t_{forecast,1}, t_{cutoff,1}, v_{forecast,1}, s_{current,1}], \\
    ... \\
    [l_{text,K}, t_{current,K}, t_{forecast,K}, t_{cutoff,K}, v_{forecast,K}, s_{current,K}].
\end{gather*}
\vspace{-35pt}

For each of these prompts, $N$ forecasts are sampled:

\vspace{-35pt}
\begin{gather*}
    y^1_{forecast,1}, ..., y^1_{forecast,N}, \\
    ... \\
    y^K_{forecast,1}, ..., y^K_{forecast,N},
\end{gather*}
\vspace{-35pt}

\noindent where each forecast

\vspace{-40pt}
\begin{multline}
    y^k_{forecast,n} = \mathcal{M}_T(f(l_{text}=l_{text,k}, t_{current}=t_{current,k}, t_{forecast}=t_{forecast,k}, \\ t_{cutoff}=t_{cutoff,k}, v_{forecast}=v_{forecast,k}, s_{current}=s_{current,k}), \varepsilon^k_n)
\end{multline}
\vspace{-40pt}

The null hypothesis involves testing a set of assumptions $\mathcal{H}$, in the presence of a number of other assumptions $\mathcal{A}$, which are considered true. If the set of assumptions $\mathcal{H}$ is satisfied, then for functions $s$ and $h$, which are responsible for aggregating the characteristics of forecast distributions and comparing them respectively:

\vspace{-15pt}
\begin{equation*}
    \displaystyle{\lim_{N \to \infty}} h(s(y^1_{forecast,1}, ..., y^1_{forecast,N}),...,s(y^K_{forecast,1}, ..., y^K_{forecast,N})) \to 0,
\end{equation*}

To detect lookahead bias and context bias, we propose to sequentially conduct two tests (which we further refer to as Fake date test I and Fake date test II), comparing the forecasts with real and fake forecast dates.

\subsubsection{Fake date test I}

Fake date test I evaluates the model's ability to follow the instruction not to use data after the cutoff boundary and is considered a proxy test for detecting lookahead bias. To check this, for each date $t_{retro}$ that is planned to be used when calculating retrospective accuracy, we calculate $N$ forecasts $y^1_{forecast}$, which only differ in the values of the realization of random variables $\varepsilon^1_n$, and approximate the forecast distribution:

\vspace{-40pt}
\begin{multline} \label{eq: eq_2}
    y^1_{forecast,n} = \mathcal{M}_T( f(l_{text}=l, \textcolor{redtext}{t_{current}=t_{retro}}, \textcolor{redtext}{t_{forecast}=t_{retro}+h}, t_{cutoff}=t_{retro}-d, \\ v_{forecast}=v, s_{current}=s_{retro}), \varepsilon^1_n), \qquad n=1,...,N
\end{multline}
\vspace{-40pt}

\noindent where the date for which the forecast is made ($t_{forecast}$) is $h$ time periods ahead of the forecast date ($t_{current}$)\footnote{In the case of forecasting for several horizons within one prompt, $h$ is a set of horizons.}, and the information cutoff date ($t_{cutoff}$) is $d$ periods in the past from the forecast date ($t_{current}$). The forecast is built using information about the economic situation ($s_{retro}$), which corresponds to time $t_{retro}$, fixed textual description ($l$), and variables for the forecast ($v$).

The only parameter that differs from typical prompts used in macroeconomic research on forecasting with LLMs is the information cutoff date, which is shifted to the past. This is done to maximally exclude the influence of additional information beyond $s_{retro}$ (additional economic context) affecting the forecast and to focus solely on lookahead bias.

In addition to forecast (\ref{eq: eq_2}), a forecast $y^2_{forecast}$ is generated for a fake date $t_{fake}$:

\vspace{-40pt}
\begin{multline} \label{eq: eq_3}
    y^2_{forecast,n} = \mathcal{M}_T( f(l_{text}=l, \textcolor{bluetext}{t_{current}=t_{fake}}, \textcolor{bluetext}{t_{forecast}=t_{fake}+h}, t_{cutoff}=t_{retro}-d, \\ v_{forecast}=v, s_{current}=s_{retro}), \varepsilon^2_n), \qquad n=1,...,N
\end{multline}
\vspace{-40pt}

In fact, this forecast differs from the previous one only in changing the forecast date ($t_{current}$) and the date for which the forecast is made ($t_{forecast}$). These dates are shifted in parallel beyond the model's release date. Examples of two prompts used later in Section~\ref{sec:results} are shown in Figure \ref{fig: figure_3}.

\vspace{5pt}

\begin{figure}[ht]
\centering

\begin{minipage}[t]{0.50\textwidth}
  \begin{tcolorbox}[promptbox, equal height group=prompts]
  IMPORTANT!!! Do not use any knowledge about the world (including any historical precedents and information about economy) after May 15, 2002 when forecasting, except for the statistics provided in prompt below.

  Today is \textcolor{redtext}{May 15, 2005. The current} lower bound of the Fed interest rate is 3.0\%, US real GDP growth in \textcolor{redtext}{Q1 2005} was 3.6\% (YoY), and US inflation (CPI) in \textcolor{redtext}{April 2005} was 3.5\% (YoY). Forecast the lower bound of the Fed interest rate on \textcolor{redtext}{May 15, 2006} and explain your forecast. First, display the explanation, and then on a new line, display the answer in the format `Interest rate forecast: [your forecast]\%'.
  \end{tcolorbox}
\end{minipage}\hfill
\begin{minipage}[t]{0.48\textwidth}
  \begin{tcolorbox}[promptbox, equal height group=prompts]
  IMPORTANT!!! Do not use any knowledge about the world (including any historical precedents and information about economy) after May 15, 2002 when forecasting, except for the statistics provided in prompt below.

  Today is \textcolor{bluetext}{February 15, 2030. The current} lower bound of the Fed interest rate is 3.0\%, US real GDP growth in \textcolor{bluetext}{Q4 2029} was 3.6\% (YoY), and US inflation (CPI) in \textcolor{bluetext}{January 2030} was 3.5\% (YoY). Forecast the lower bound of the Fed interest rate on
  \textcolor{bluetext}{February 15, 2031} and explain your forecast. First, display the explanation, and then on a new line, display the answer in the format `Interest rate forecast: [your forecast]\%'.
  \end{tcolorbox}
\end{minipage}
\caption{\textbf{Examples of prompts for Fake date test I.} The left prompt corresponds to forecast (\ref{eq: eq_2}), and the right one to forecast (\ref{eq: eq_3}).}
\label{fig: figure_3}
\end{figure}

To formalize the test, the following two assumptions are made:

\noindent \textbf{(A1) Assumption 1.} The distribution of the forecast with cutoff date $t_{cutoff}$ is measurable\footnote{The measurability of the forecast here and further is understood in the sense that for a fixed $\varepsilon$, the value $\mathcal{M_T}(f(\cdot), \varepsilon)$ is measurable.} with respect to the $\sigma$-algebra generated by the union of $s_{current}$ and $I_{t_{cutoff}}$ (the intersection of $I_T$ and all information available up to time $t_{cutoff}$).

\noindent \textbf{(A2) Assumption 2.} For two forecasts with cutoff boundary $t_{cutoff}$, measurable with respect to the $\sigma$-algebra generated by the union of $s_{retro}$ and $I_{t_{cutoff}}$, and differing only in the forecast date ($t_{current,1} \geq t_{cutoff}+d$ and $t_{current,2} \geq t_{cutoff}+d$, where $t_{current,1} \neq t_{current,2}$) and the date for which the forecast is made ($t_{forecast,1}=t_{current,1}+h$ and $t_{forecast,2}=t_{current,2}+h$, where $t_{forecast,1} \neq t_{forecast,2}$), the distributions coincide.

Assumption 1 formalizes the absence of lookahead bias and requires that only information up to the cutoff boundary and the information about the state of the economy specified in the prompt be used in the forecast. Assumption 2 requires that any two forecasts, sufficiently distant in time from the information set available to the LLM (except for the information about the state of the economy specified in the prompt), do not differ. This assumption can be interpreted as follows: the LLM should be insensitive to a specific date outside its information set (see more discussion in Sections~\ref{sec:discussion} and~\ref{sec:results}), and this information set should only be used to build general economic principles relevant to the forecasting task, not to form context regarding the forecast date. However, strictly speaking, this is merely what we tried to embed in Assumption 2 and these conditions may not necessarily be met, compensating each other to satisfy the assumption (see more discussion in Section~\ref{sec:discussion}).

Given these two assumptions, we can formulate Fake date test I.

\noindent \textbf{Proposition 1 (Fake date test I).} When $\mathcal{A} = \{A2\}$ and $\mathcal{H} = \{A1\}$ are satisfied, the distributions of forecasts $y^1_{forecast}$ and $y^2_{forecast}$ must coincide.

\noindent \textbf{Proof.} See Appendix~\ref{app:appendix_a}.

Proposition 1 means that if the distributions of $y^1_{forecast}$ and $y^2_{forecast}$ do not coincide, the model fails to correctly follow the instruction to cut off information. Empirically, for each date this can be tested using any test for comparing distributions, for example, a permutation test with the Kolmogorov-Smirnov statistic (choosing the cumulative distribution as $s$, and the test statistic as $h$), or tests for comparing the sample characteristics of distributions, for example, a test for equality of means ($s$ is the mean of the distribution, and $h$ is the test statistic). Multiple comparisons for all periods can also be made using classical techniques, for example, Bonferroni correction.

Both assumptions can be selected as $\mathcal{H}$ in the test (for example, $\mathcal{A}=\{\}$ and $\mathcal{H}=\{A1,A2\}$), which will weaken its interpretation in terms of lookahead bias; however, as is further discussed in Sections~\ref{sec:discussion} and~\ref{sec:results}, a number of supporting experiments show that the influence of not satisfying Assumption 2 for the tested LLMs is potentially not as great as the influence of Assumption 1.

\subsubsection{Fake date test II}

Fake date test II is similar to the first one and differs only in that depth $d$ in formulas (\ref{eq: eq_2}) and (\ref{eq: eq_3}) is taken to be zero (see example in Figure \ref{fig: figure_4}). Assumption 1 is preserved, while the depth in Assumption 2 is taken to be zero, i.e., it becomes:

\noindent \textbf{(A2$'$) Assumption 2$'$.} For two forecasts with cutoff boundary $t_{cutoff}$, measurable with respect to the $\sigma$-algebra generated by the union of $s_{retro}$ and $I_{t_{cutoff}}$, and differing only in the forecast date ($t_{current,1} \geq t_{cutoff}$ and $t_{current,2} \geq t_{cutoff}$, where $t_{current,1} \neq t_{current,2}$) and the date for which the forecast is made ($t_{forecast,1}=t_{current,1}+h$ and $t_{forecast,2}=t_{current,2}+h$, where $t_{forecast,1} \neq t_{forecast,2}$), the distributions coincide.

\vspace{5pt}

\begin{figure}[ht]
\centering

\begin{minipage}[t]{0.50\textwidth}
  \begin{tcolorbox}[promptbox, equal height group=prompts]
   IMPORTANT!!! Do not use any knowledge about the world (including any historical precedents and information about economy) after \textcolor{tealtext}{May 15, 2005} when forecasting, except for the statistics provided in prompt below.

   Today is \textcolor{redtext}{May 15, 2005. The current} lower bound of the Fed interest rate is 3.0\%, US real GDP growth in \textcolor{redtext}{Q1 2005} was 3.6\% (YoY), and US inflation (CPI) in \textcolor{redtext}{April 2005} was 3.5\% (YoY). Forecast the lower bound of the Fed interest rate on \textcolor{redtext}{May 15, 2006} and explain your forecast. First, display the explanation, and then on a new line, display the answer in the format `Interest rate forecast: [your forecast]\%'. 
  \end{tcolorbox}
\end{minipage}\hfill
\begin{minipage}[t]{0.48\textwidth}
  \begin{tcolorbox}[promptbox, equal height group=prompts]
   IMPORTANT!!! Do not use any knowledge about the world (including any historical precedents and information about economy) after \textcolor{tealtext}{May 15, 2005} when forecasting, except for the statistics provided in prompt below.

   Today is \textcolor{bluetext}{February 15, 2030. The current} lower bound of the Fed interest rate is 3.0\%, US real GDP growth in \textcolor{bluetext}{Q4 2029} was 3.6\% (YoY), and US inflation (CPI) in \textcolor{bluetext}{January 2030} was 3.5\% (YoY). Forecast the lower bound of the Fed interest rate on \textcolor{bluetext}{February 15, 2031} and explain your forecast. First, display the explanation, and then on a new line, display the answer in the format `Interest rate forecast: [your forecast]\%'.
  \end{tcolorbox}
\end{minipage}

\caption{\textbf{Examples of prompts for Fake date test II.} The left prompt corresponds to forecast (\ref{eq: eq_2}) with $d=0$, and the right one to forecast (\ref{eq: eq_3}) with $d=0$.}
\label{fig: figure_4}
\end{figure}

If $\mathcal{H}$ contains only A2$'$, and A1 relates to $\mathcal{A}$, i.e., it is assumed that the LLM correctly cuts off information beyond the cutoff date, this test is a proxy for context bias. If $\mathcal{H}=\{A1, A2'\}$, then the test is a proxy for the presence of at least one of the two biases. If Fake date test I is passed, we choose the first partition, since the hypothesis that A1 holds is not rejected; the test is then formulated as follows:

\noindent \textbf{Proposition 2 (Fake date test II).} When $\mathcal{A} = \{A1\}$ and  $\mathcal{H} = \{A2'\}$ are satisfied, the distributions of forecasts $y^1_{forecast}$ and $y^2_{forecast}$ with $d=0$ must coincide.

\noindent \textbf{Proof.} See Appendix~\ref{app:appendix_a}.

However, if Fake date test I fails, A1 cannot be assumed to hold; therefore, in this case Fake date test II is formulated as follows:

\noindent \textbf{Proposition 2$'$ (Fake date test II).} When $\mathcal{A} = \{\}$ and  $\mathcal{H} = \{A1,A2'\}$ are satisfied, the distributions of forecasts $y^1_{forecast}$ and $y^2_{forecast}$ with $d=0$ must coincide.

\section{Interpretation of the tests and their limitations} \label{sec:discussion}

The fake date tests proposed above are formal, but they can be interpreted in different ways. We interpret Fake date test I as a check for lookahead bias, and Fake date test II as a check of how the model uses knowledge near the cutoff boundary. This section discusses why we interpret these tests in this way, and what pitfalls may arise in this.

Passing these tests serves precisely as another advantage of the LLM not containing the corresponding biases, but is not the ground truth. At the same time, failing the tests does not necessarily mean the presence of the corresponding biases, but serves as a serious signal to question the validity of the methodology used to evaluate the model's retrospective accuracy and rank it relative to other models.

If the hypothesis of the equality of distributions is not rejected in Fake date test I, then even with A2 satisfied, this does not mean that the model does not use information outside the cutoff boundary. It may happen that in both forecasts (\ref{eq: eq_2}) and (\ref{eq: eq_3}), the LLM uses data outside the cutoff boundary while leading to the same forecasts. One example of such behavior could be a situation where the LLM cuts off data not after the cutoff boundary, but slightly later, and A2 is satisfied with the availability of this information set. In this case, the forecast distributions will also coincide, but information from the future will leak into both forecasts. One can come up with a number of more complex potential situations where the hypothesis that the distributions coincide will not be rejected despite information leakage. For example, if the LLM with a forecast date preceding the model release date cuts off information correctly, while failing to do so when the forecast date is beyond the release date (or vice versa). At the same time, the LLM uses this information differently to build a forecast, obtaining identical distributions. We do not claim to provide a complete and comprehensive description of all such cases, but only want to additionally emphasize that the test results should not be interpreted too optimistically and unequivocally.

A variation of the latter example, when the LLM correctly cuts off information with a forecast date within the training sample and incorrectly with a forecast date outside it, can also be used in the reverse situation when Fake date test I is not passed. If the LLM uses the available information equally, but there is more information for a fake date, the forecasts must not coincide. At the same time, there is no lookahead bias for in-sample forecasts. A second example could be a situation where, for both forecast dates, the LLM does not use information outside the cutoff boundary, but for different dates inside this boundary the model looks at different information, i.e., it is sensitive to the forecast date. In this case, the test may also fail despite correctly following the instruction to ignore information outside the cutoff boundary. The model may also have a bias toward specific numbers in the prompt. For example, the model may be sensitive to dates under identical conditions and generate answers that differ for even or odd months/quarters/years, or years that most frequently appear in science fiction. Although the last example with science fiction seems highly artificial, it is meant to emphasize that LLMs are black boxes, and therefore, unexpected effects can arise inside them that are very difficult to predict in advance without facing them in a specific case. As in the situation with passing Fake date test I, the list of examples is naturally not exhaustive. Moreover, in both the case of passing the test and the reverse situation, various effects can overlap each other. By giving the examples outlined above, we only want to say that the test results should be interpreted with a degree of caution and try to verify the assumptions on which the test is based whenever possible.

We tried to partially verify Assumption 2 for Fake date test I by choosing a cutoff date that definitely satisfies Assumption 1. Such a date was chosen close to the models' release dates (August 15, 2025). For all models used in Section~\ref{sec:results}, we selected statistics for five dates\footnote{November 15, 2008 (a date during the Great Recession when forecasts with real and fake dates differ significantly), August 15, 2013 (a random date from the sample), May 15, 2020 (the beginning of the pandemic), August 15, 2021 (a date when the YoY GDP growth indicator has a low base effect), and February 15, 2025 (the last date in our sample).} and looked at how the dynamics of the forecast distributions behave until 2035. Despite the fact that the p-values for the Kolmogorov-Smirnov permutation test (Tables \ref{tab:table_C.2.1.1}-\ref{tab: table_C.2.3.2} in Appendix~\ref{app:appendix_c}), which is used when analyzing the results in Section~\ref{sec:results}, are slightly shifted toward zero, which should not be the case with ideal fulfillment of Assumption 2, we did not see such significant differences in forecasts, except for a number of situations, as when comparing forecasts with real and fake dates. Although these results do not allow us to conclude that Assumption 2 is fulfilled ideally, its approximate fulfillment seems quite sufficient for Fake date test I in practice (see more detailed discussion in Subsection~\ref{sec:case_studies}). For other prompts and models, the results may differ, so we recommend running such tests anew for each experiment.

It should also be noted that Assumption 2 is quite strict in the sense that it requires not just the convergence of forecasts as the forecast date increases, but their exact matching starting from a certain period. This assumption can be weakened by introducing a certain neighborhood in which forecasts must lie starting from $t_{current} \geq t_{cutoff} + d$. Adjusting statistical tests for the selected neighborhood is not so complex; however, selecting this neighborhood in practice is associated with a number of technical and computational difficulties, and to avoid going into details, we introduce a less realistic but technically simpler assumption\footnote{Assumption 2 can also be adjusted to account for seasonality, which, as shown in Subsection~\ref{sec:case_studies}, may be relevant for some models and forecast indicators. This adjustment in practice requires that when forecasting with real and fake dates, only the year differs, but not the day or month of the forecast within a year. As a consequence, adjustment of the statistical test is not required in this case.}

As for Fake date test II, its main assumption is A1, which is checked within Fake date test I. If it is satisfied, arguments similar to those in Fake date test I can also be applied to Fake date test II regarding interpretation. Thus, if the test fails, situations may arise where the mismatch of distributions is the result of the model not using information near the forecast date, but interpreting the mechanisms of how the economy works differently depending on which forecast date is used. In the case of passing, one can come up with a number of examples where several effects overlap in such a way as to compensate for each other. For example, due to the compensation of the latter effect with bias associated with context. Alternatively, the LLM may find a historically similar precedent in the data given in the prompt, which exactly corresponds to the historical forecast date, and use this information to fill in the missing context, which will thus lead to identical forecast distributions.

It is also worth emphasizing that Fake date test I uses different forecast and cutoff dates, which, strictly speaking, is not equivalent to a typical LLM-based macroeconomic forecasting setup, in which these dates coincide. Consequently, beyond the arguments outlined earlier, following (or not following) the instruction to cut off information in Fake date test I may not fully translate into following (or not following) this instruction in a real-world problem and should be considered only as an additional argument supporting the absence (or presence) of lookahead bias. A simple example of this interpretational gap occurs when the model misinterprets the temporal boundaries and implicitly treats the forecast date as the cutoff date. In such cases, a failure to pass Fake date test I stems from misinterpreting the task rather than disregarding the cutoff instruction. In Fake date test II, the forecast and cutoff dates coincide; therefore, this problem is absent. Although Fake date test II cannot disentangle lookahead bias from context bias, it provides an additional argument for or against the validity of using in-sample forecasts to assess LLM accuracy when Fake date test I fails.

Despite the potentially possible effects described above, we tend to interpret the proposed tests as a test for lookahead bias and a test for bias associated with additional knowledge in the vicinity of the forecast date (if Fake date test I does not fail). These explanations seem to us the most likely from the point of view of how models that mimic expert reasoning with broad knowledge about how the world is organized would behave. Nevertheless, we understand that LLMs are just complex statistical models, and such interpretations may contain anthropomorphic bias (see Ludwig et al. \citeyearpar{Ludwig2025}), and we urge caution in respect of their conclusions, using them only as one of the arguments for or against the presence of the corresponding biases.

It is also worth noting that, assuming that our interpretation is correct and Fake date test I is passed, the forecast for a fake date and zero depth (forecast (\ref{eq: eq_3}) with $d = 0$) can be considered as a candidate for evaluating in-sample accuracy. This forecast, provided that the interpretation is correct and Fake date test I is passed, should not contain lookahead bias and may not contain context bias. However, we leave the study of such possibilities for future research for two reasons. First, none of the models in Section~\ref{sec:results} passes Fake date test I, which does not allow us to obtain such a candidate in practice. Second, a number of additional tests should be developed to test the assumptions that must be satisfied so as to avoid various additional biases, such as bias associated with memorizing precedents in the description of the limitations of Fake date test II.

\section{Results} \label{sec:results}

This section describes the results of our experiments that illustrate the application of the methodology outlined in Section~\ref{sec:methodology}. First, we describe the experimental setup, including the models, data, and prompts used. Then we demonstrate general results based on histograms of forecast distributions. At the end of the section, we show a number of cases that help better understand the results obtained.

\subsection{Setup}

For the experiments, we selected three open-weight models: \href{https://huggingface.co/moonshotai/Kimi-K2-Instruct}{Kimi-K2 Instruct} (see Kimi Team \citeyearpar{Kimi2025}), \href{https://huggingface.co/Qwen/Qwen3-235B-A22B-Instruct-2507}{Qwen3 Instruct} (see Qwen Team \citeyearpar{Qwen2025}) and \href{https://huggingface.co/deepseek-ai/DeepSeek-V3.1}{DeepSeek-V3.1} (see DeepSeek-AI (\citeyear{DeepSeek2025})). At the start of conducting the experiments, all the three models were among the top-10 of Chatbot Arena (among text models) and leading open-weight non-reasoning models. We choose non-reasoning models because, compared to reasoning models, they allow relatively quick responses, which significantly saves time in experiments. Nevertheless, without any changes, the proposed methodology can be applied to reasoning models as well.

The prompts used in the experiments\footnote{All requests to the models were made through \href{https://bothub.chat}{BotHub}, a provider that allows access to different LLMs via one application programming interface (API).} are similar to those shown in Figure \ref{fig: figure_3} in Section~\ref{sec:methodology}. As variables for forecasting, we selected the U.S. data for the lower bound of the Fed interest rate, YoY CPI growth, and YoY real GDP growth one year ahead from the last known statistical publication date at the time of forecasting. As statistics that the forecast is based on, we took the latest available values of the said variables (vintages\footnote{The first release of seasonally adjusted data for the CPI and real GDP were taken from \href{https://www.philadelphiafed.org/surveys-and-data/real-time-data-research/real-time-data-set-for-macroeconomists}{The Philadelphia Fed's Real-Time Data Set}.} for the CPI and GDP). Forecasts were built for the period from May 2005 to February 2025 with quarterly frequency (80 forecasts in total). As the forecast date, we took the fifteenth day of the second month of a quarter, as a date close to the time of release of inflation data for the previous month and GDP data for the previous quarter. The fake forecast date was chosen as February 15, 2030, as a date that is sufficiently far from the model release dates. The cutoff boundary for Fake date test I was shifted three years back from the real forecast date. Such a period corresponds to the standard medium-term horizon of macroeconomic forecasts, and with this time horizon, forecasts should be almost independent of the current economic context. Furthermore, the selected period does not contradict the results regarding the stabilization horizon of forecasts in the test of Assumption 2. Figure \ref{fig: figure_5} shows the prompts used to forecast all three variables.

It should be noted that the set of macroeconomic context variables selected for provision to the LLM is far from exhaustive. However, our goal is not to build the best forecast, but only to illustrate the proposed methodology, for which a specific set of indicators does not play the key role, so we focused solely on the main macroeconomic variables.

For each date, $N=100$ forecasts were sampled with real and fake dates with a temperature equal to 0.7. To compare the sample distributions of forecasts, a permutation test (see Ritzwoller et al. \citeyearpar{Ritzwoller2025}) with the Kolmogorov-Smirnov statistic with 10,000 permutations was used.

\begin{figure}[ht]
\begin{tcolorbox}[promptbox]
\textcolor{redtext}{Fed interest rate}

IMPORTANT!!! Do not use any knowledge about the world (including any historical precedents and information about economy) after \{date\_cutoff\} when forecasting, except for the statistics provided in prompt below.

Today is \{date\}. The current lower bound of the Fed interest rate is \{rate\}\%, US real GDP growth in Q\{GDP\_value\_period\} \{GDP\_value\_year\} was \{GDP\}\% (YoY), and US inflation (CPI) in \{CPI\_value\_period\} \{CPI\_value\_year\} was \{CPI\}\% (YoY). Forecast the lower bound of the Fed interest rate on \{date\_forecast\} and explain your forecast. First, display the explanation, and then on a new line, display the answer in the format `Interest rate forecast: [your forecast]\%'.

\textcolor{redtext}{CPI growth (YoY)}

IMPORTANT!!! Do not use any knowledge about the world (including any historical precedents and information about economy) after \{date\_cutoff\} when forecasting, except for the statistics provided in prompt below.

Today is \{date\}. The current lower bound of the Fed interest rate is \{rate\}\%, US real GDP growth in Q\{GDP\_value\_period\} \{GDP\_value\_year\} was \{GDP\}\% (YoY), and US inflation (CPI) in \{CPI\_value\_period\} \{CPI\_value\_year\} was \{CPI\}\% (YoY). Forecast US YoY inflation for \{date\_forecast\} and explain your forecast. First, display the explanation, and then on a new line, display the answer in the format `Inflation forecast: [your forecast]\%'.

\textcolor{redtext}{Real GDP growth (YoY)}

IMPORTANT!!! Do not use any knowledge about the world (including any historical precedents and information about economy) after \{date\_cutoff\} when forecasting, except for the statistics provided in prompt below.

Today is \{date\}. The current lower bound of the Fed interest rate is \{rate\}\%, US real GDP growth in Q\{GDP\_value\_period\} \{GDP\_value\_year\} was \{GDP\}\% (YoY), and US inflation (CPI) in \{CPI\_value\_period\} \{CPI\_value\_year\} was \{CPI\}\% (YoY). Forecast US YoY real GDP growth for \{date\_forecast\} and explain your forecast. First, display the explanation, and then on a new line, display the answer in the format `GDP growth forecast: [your forecast]\%'.
\end{tcolorbox}
\caption{\textbf{Prompts for forecasting macroeconomic variables.} `\{\}' are placeholders for substituted values of variables and dates.}
\label{fig: figure_5}
\end{figure}

\subsection{Results of comparing forecasts} \label{sec:forecasts_comparison}

\begin{figure}[h!]
	\hspace{-0.5cm}
	\includegraphics[scale=0.57]{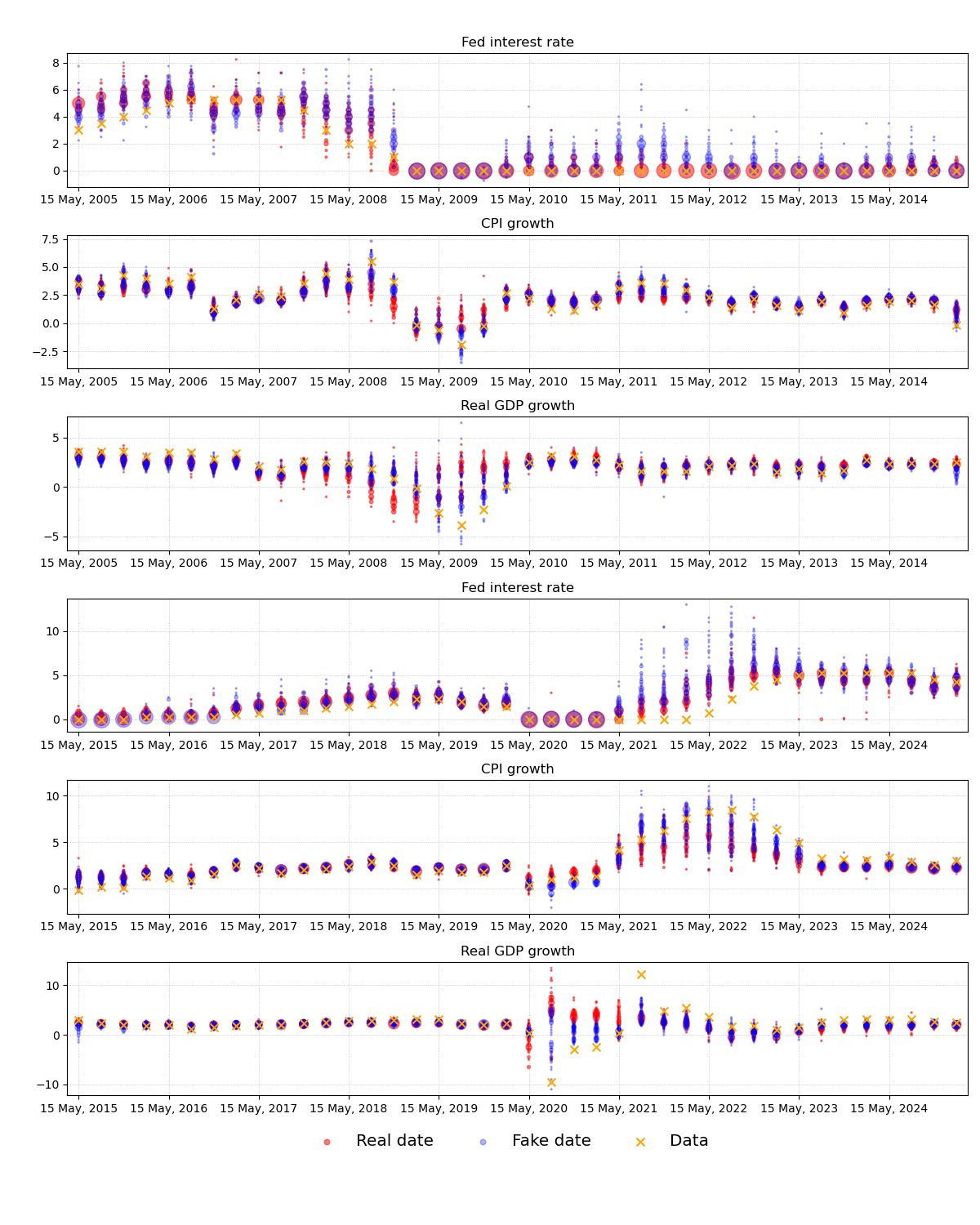}
	\caption{\textbf{Distributions of forecasts in Fake date test I (with the cutoff boundary 3 years back from the forecast date for real data) for 1 year produced by the Kimi-K2 Instruct model.} The marker size is proportional to the density of the forecast distribution.}
    \label{fig: figure_6}
\end{figure}

\begin{figure}[h!]
	\hspace{-0.5cm}
	\includegraphics[scale=0.57]{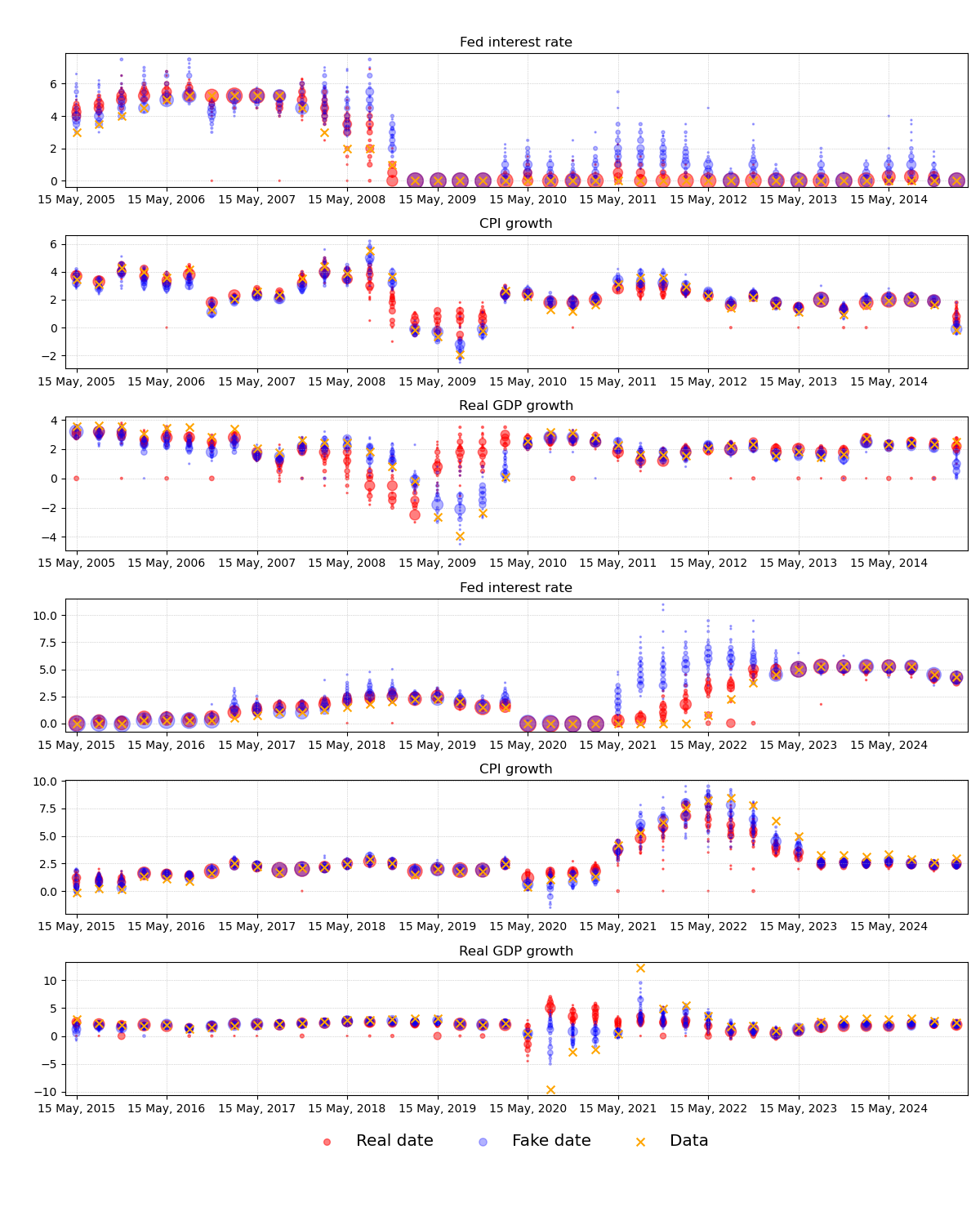}
	\caption{\textbf{Distributions of forecasts in Fake date test I (with the cutoff boundary 3 years back from the forecast date for real data) for 1 year produced by the Qwen3 Instruct model.} The marker size is proportional to the density of the forecast distribution.}
    \label{fig: figure_7}
\end{figure}

\begin{figure}[h!]
	\hspace{-0.5cm}
	\includegraphics[scale=0.57]{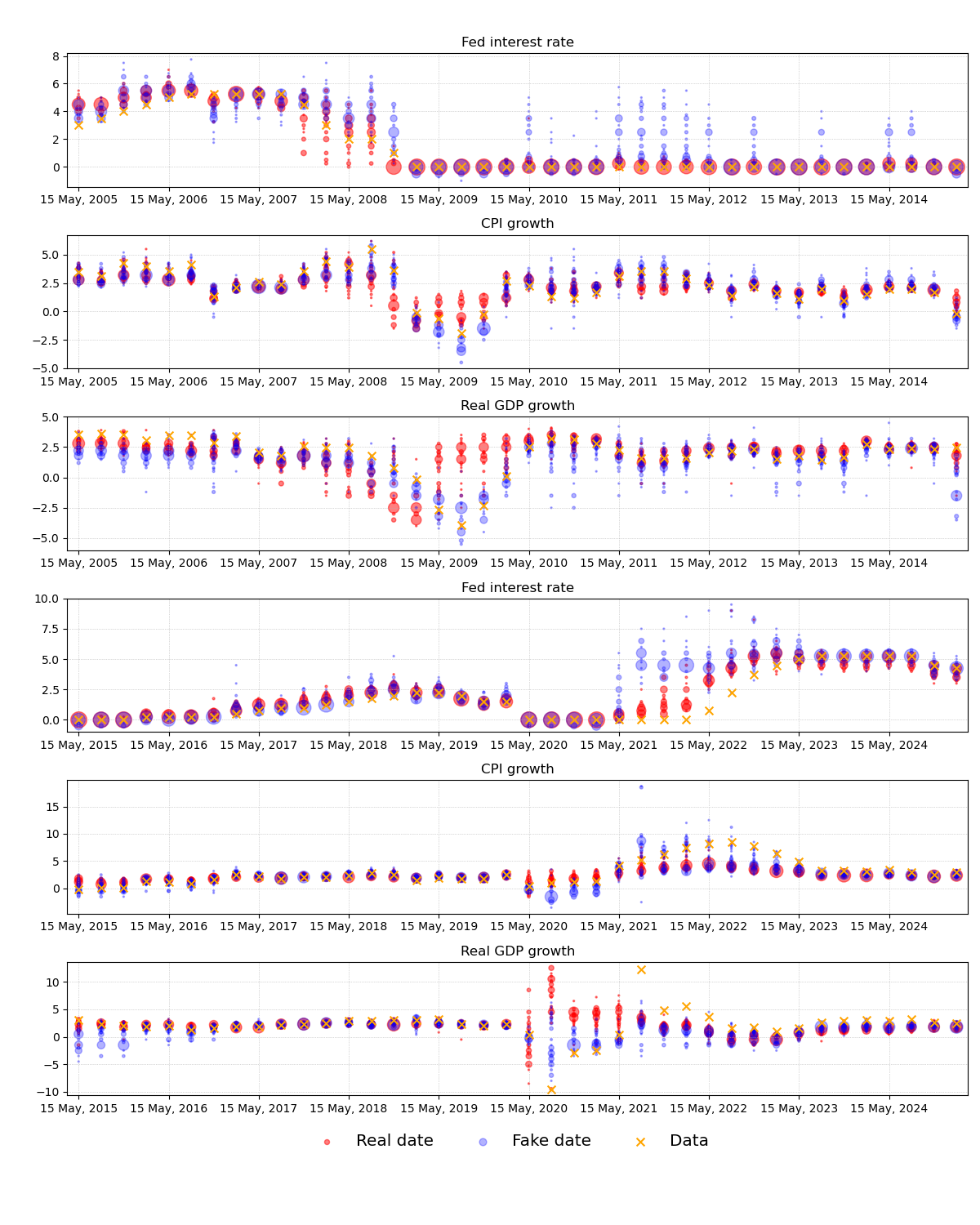}
	\caption{\textbf{Distributions of forecasts in Fake date test I (with the cutoff boundary 3 years back from the forecast date for real data) for 1 year produced by the DeepSeek-V3.1 model.} The marker size is proportional to the density of the forecast distribution.}
    \label{fig: figure_8}
\end{figure}

Figures \ref{fig: figure_6}-\ref{fig: figure_8} show general results of Fake date test I for all three models. The results are presented in the format of bubble histograms with the circle size proportional to the density of the forecast distribution. Red indicates forecasts for a real date, blue shows forecasts for a fake date, yellow denotes the real data corresponding to the vintage available on the forecast date. More illustrative forecast distributions together with the p-values of the permutation test with the Kolmogorov-Smirnov statistic for each model and for each date are presented in Figures \ref{fig: figure_B.1.1.1}-\ref{fig: figure_B.3.3.5} in Appendix~\ref{app:appendix_b}\footnote{Results are presented in histogram format with the bin size equal to 0.25 for the lower bound of the Fed interest rate and 0.1 for CPI and real GDP growth, corresponding to the natural forecast step of these indicators. The headers show the p-values of the permutation test with the Kolmogorov-Smirnov statistic.}.

The visual analysis of Figures \ref{fig: figure_6}-\ref{fig: figure_8} suggests that the forecast distributions\footnote{For some queries, the models, and especially Qwen3 Instruct, state insufficient data with our prompt and output a forecast of 0.0 or `N/A'. We did not exclude such values from the sample, as their share, as can be seen from the figures, is small.} for some dates significantly differ, which is confirmed by the p-values for the Kolmogorov-Smirnov permutation test\footnote{We also conducted tests with the Wasserstein distance. They do not differ qualitatively from those presented in Appendix~\ref{app:appendix_b}, so to save space we do not present them herein.} presented in Appendix~\ref{app:appendix_b}. When applying the Bonferroni correction for conservative multiple hypothesis testing, the hypothesis of equal distributions is also rejected for any conventional significance level for all models and all variables\footnote{The maximum p-value among all models and variables was 0.008 for 10,000 permutations, due to the p-value discretization level of 0.0001 for individual hypothesis testing. To obtain less discretized p-values ​​in a computationally efficient manner, we estimated the p-value for 1,000,000 permutations for the forecast based on statistics as of November 15, 2008. Since the Bonferroni correction for multiple hypothesis testing depends only on the smallest p-value, this value can be used as the estimate of the upper bound for the p-value of the multiple hypothesis. The maximum value in this case was 0.00008.}. This is a signal that the Kimi-K2 Instruct, Qwen3 Instruct, and DeepSeek-V3.1 models might potentially fail to effectively follow the instruction to cut off information, which is a reason to doubt the correctness of the procedure for evaluating the quality of these LLMs by calculating in-sample metrics in the proposed macroeconomic forecasting task.

The differences in forecasts vary from date to date and from model to model. However, it can be noted that these differences become especially noticeable around the 2007-2009 financial crisis and the 2020 pandemic. Also, forecasts significantly differ on some dates of significant changes in macroeconomic variables outside these periods. This can be seen, for example, in forecasts of the lower bound of the Fed interest rate in 2010-2012 when CPI growth (YoY) and real GDP growth (YoY) left negative territory. The forecasts for a real date in this period are almost always in the vicinity of zero, while the forecasts for a fake date are often higher. Another example could be a decrease in forecasts of real GDP growth (YoY) for a fake date when CPI growth (YoY) decreased in 2015.

\begin{figure}[h!]
	\hspace{-0.5cm}
	\vspace{0.5cm}
	\includegraphics[scale=0.57]{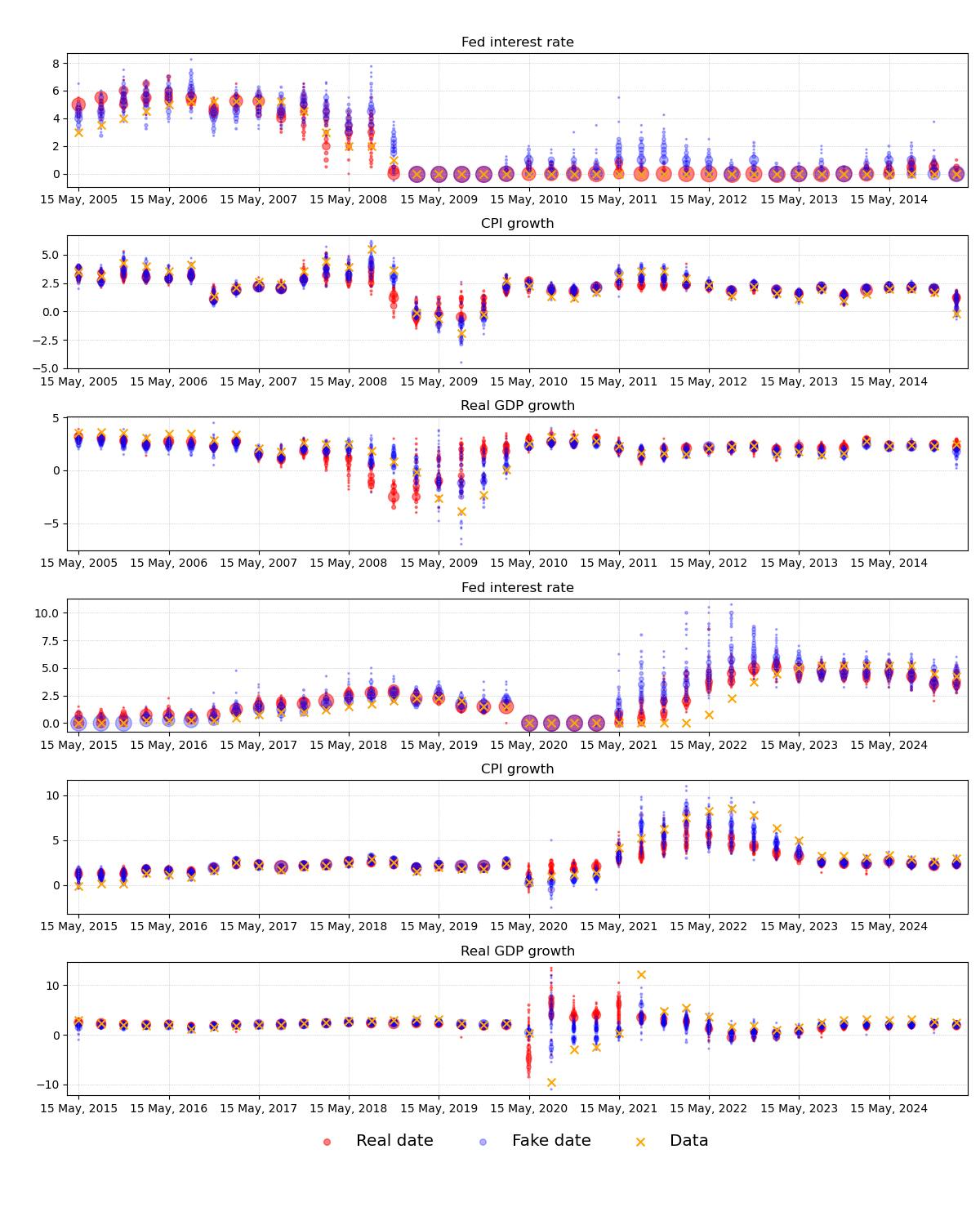}
	\caption{\textbf{Distributions of forecasts in Fake date test II for 1 year produced by the Kimi-K2 Instruct model.} The marker size is proportional to the density of the forecast distribution.}
    \label{fig: figure_9}
\end{figure}

\begin{figure}[h!]
	\hspace{-0.5cm}
	\vspace{0.5cm}
	\includegraphics[scale=0.57]{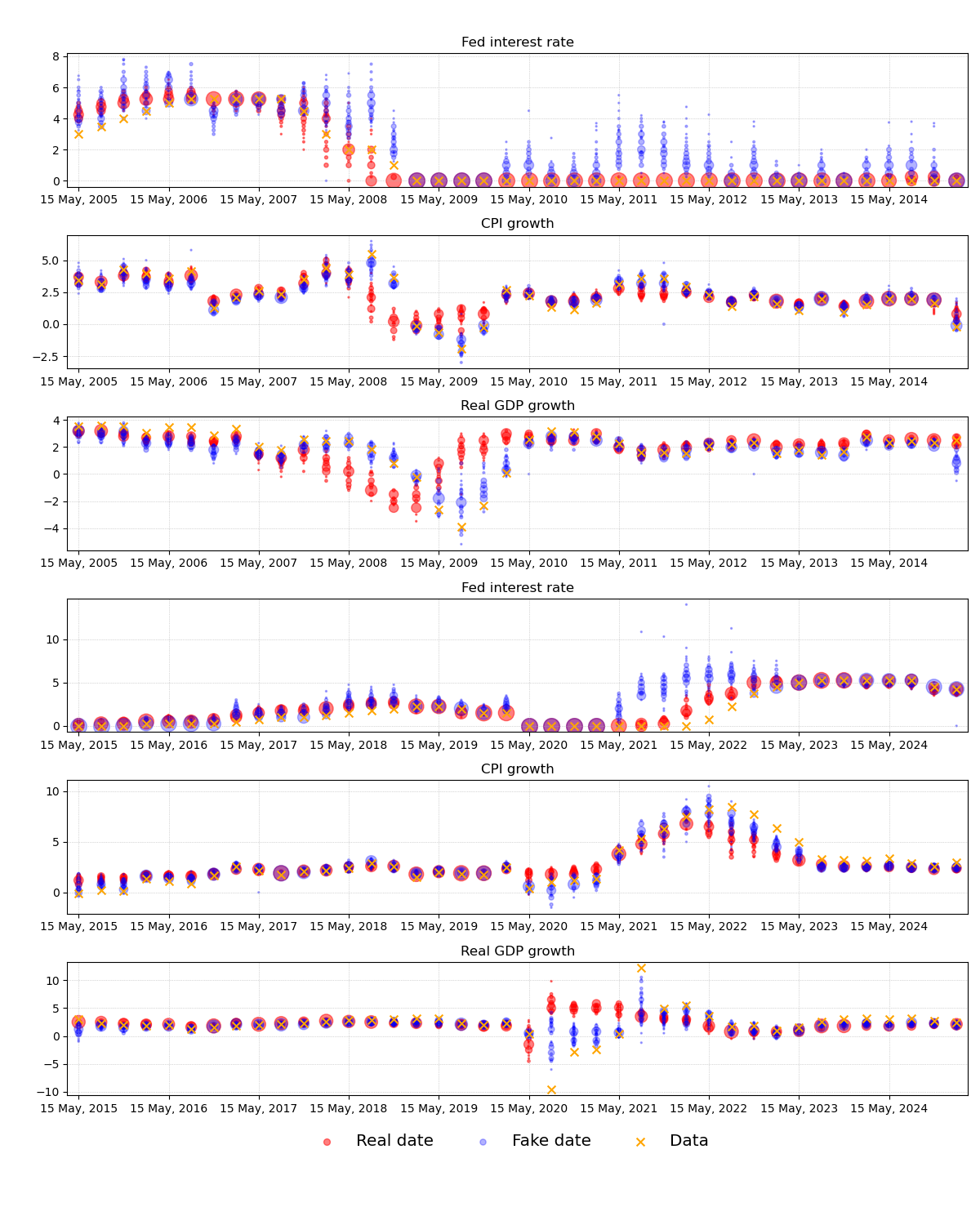}
	\caption{\textbf{Distributions of forecasts in Fake date test II for 1 year produced by the Qwen3 Instruct model.} The marker size is proportional to the density of the forecast distribution.}
    \label{fig: figure_10}
\end{figure}

\begin{figure}[h!]
	\hspace{-0.5cm}
	\vspace{0.5cm}
	\includegraphics[scale=0.57]{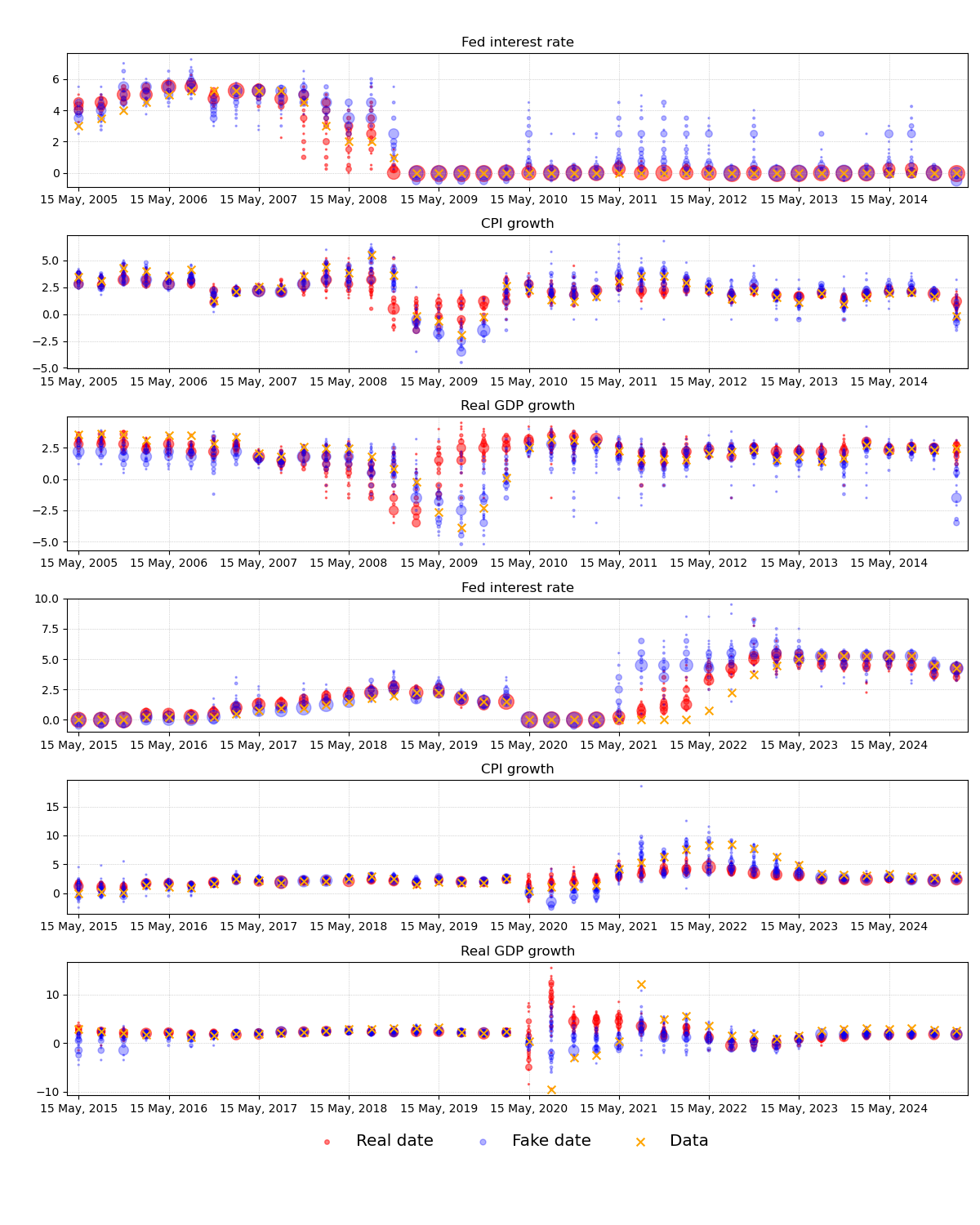}
	\caption{\textbf{Distributions of forecasts in Fake date test II for 1 year produced by the DeepSeek-V3.1 model.} The marker size is proportional to the density of the forecast distribution.}
    \label{fig: figure_11}
\end{figure}

Since none of the LLMs considered has passed Fake date test I, we conduct Fake date test II as a proxy test for the presence of at least one of the two biases, not only context bias. Figures \ref{fig: figure_9}-\ref{fig: figure_11} show bubble histograms similar to those in Figures \ref{fig: figure_6}-\ref{fig: figure_8}. To save space, we omit figures similar to those given in Appendix~\ref{app:appendix_b}. The p-values of the tests are provided in Tables \ref{tab: table_E.1}-\ref{tab: table_E.4} in Appendix~\ref{app:appendix_e}. As with Fake date test I, the distributions of forecasts for real and fake dates differ markedly for many time periods, exhibiting patterns similar to those observed in Fake Date Test I. The test with the Bonferroni correction also rejects the hypothesis of equality of distributions at all conventional significance levels. As noted in Section~\ref{sec:discussion}, this result, in addition to Fake date test I, serves as another signal of the unsuitability of in-sample metrics for evaluating the forecast quality of the Kimi-K2 Instruct, Qwen3 Instruct, and DeepSeek-V3.1 models in the forecasting task considered herein.

\clearpage

\clearpage

\clearpage

\clearpage

\clearpage

\clearpage

\clearpage

\subsection{Case studies} \label{sec:case_studies}

To better understand the results, we will consider several examples that help shed some light on the nature of the demonstrated forecast distributions.

The first thing we would like to focus on is a more detailed analysis of forecast distributions within Assumption 2. As noted in Section~\ref{sec:discussion}, this assumption is not perfectly satisfied. Tables \ref{tab:table_C.2.1.1}-\ref{tab: table_C.2.3.2} in Appendix~\ref{app:appendix_c} present the p-values for the Kolmogorov-Smirnov permutation test. In the case of ideal fulfillment of Assumption 2, conservative p-values (see Ritzwoller et al. \citeyearpar{Ritzwoller2025}) for the permutation test should be either distributed uniformly or partially shifted toward one. The values in Tables \ref{tab:table_C.2.1.1}-\ref{tab: table_C.2.3.2} have a higher density in the neighborhood of zero\footnote{It should be noted that individual p-values in one column are not independent due to the common forecast as of February 15, 2030, and therefore the analysis was conducted for all dates and tables.}. Nevertheless, for most forecast dates, individual p-values are generally adequate (far from zero or not extremely close to zero), which allows us to speak of approximate fulfillment of Assumption 2.

To illustrate how sensitive the results obtained in the previous subsection are to the non-fulfillment of Assumption 2 in the exact form, we selected several triples --- a set of statistical data, a model, and a forecasting indicator --- for which the deviations are the largest. For this, we excluded the first three years of forecasting from the analysis to avoid potential biases associated with the presence of context, i.e., only dates starting from August 15, 2028 were taken. Then, we selected the triples for which the p-values from Tables \ref{tab:table_C.2.1.1}-\ref{tab: table_C.2.3.2} do not exceed the minimum threshold of 0.001 in at least 10 percent of cases.

\begin{figure}[h!]
	\hspace{-0.5cm}
	\includegraphics[scale=0.57]{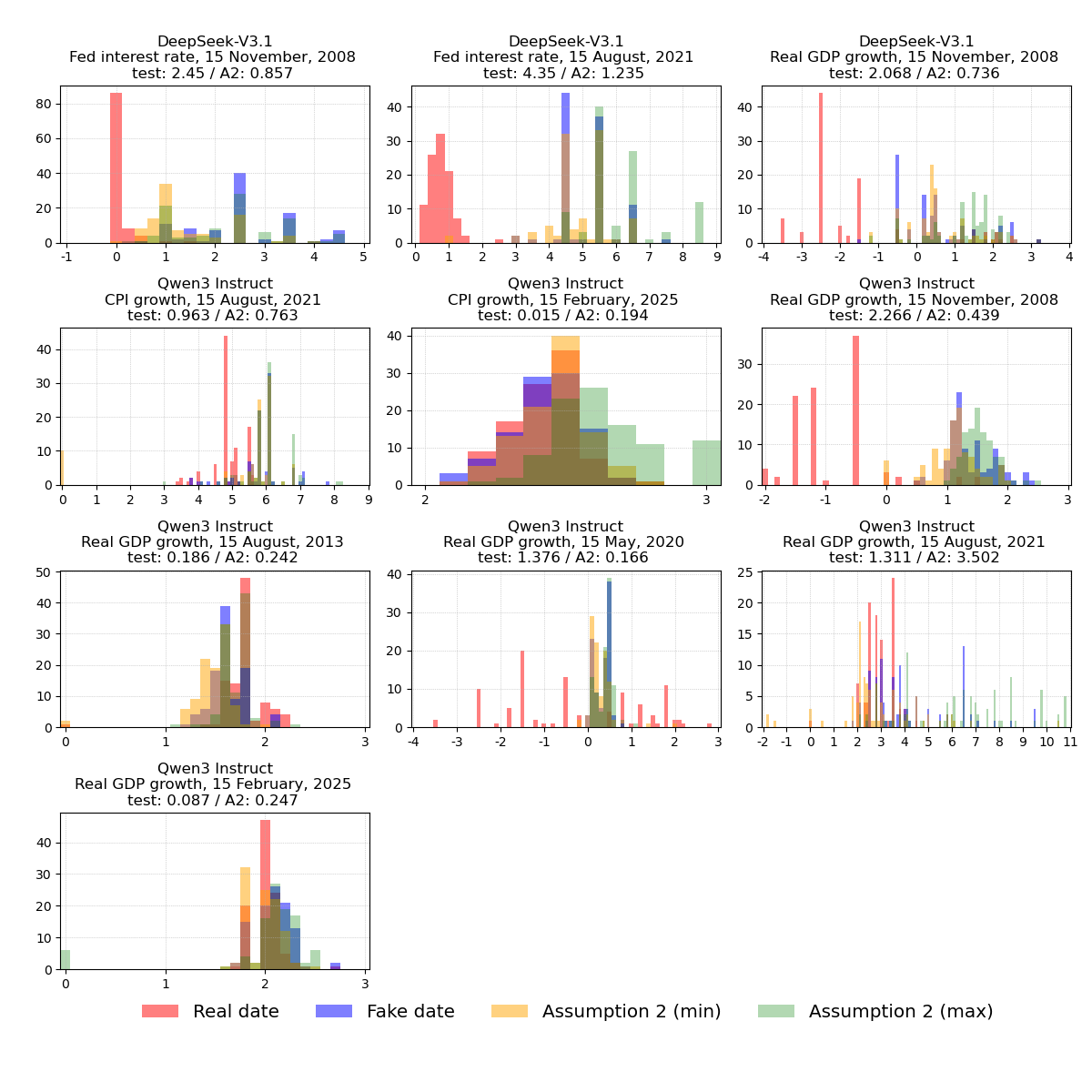}
	\caption{\textbf{Comparison of forecast distributions with real and fake dates (identical to forecasts in Appendix~\ref{app:appendix_b}), and forecasts with the maximum Wasserstein distance for testing Assumption 2.} The header shows the model, forecast indicator, date for which statistics were taken, and the Wasserstein distance between forecasts with real and fake dates (the first number), as well as forecasts with the maximum Wasserstein distance (the second number).}
    \label{fig: figure_12}
\end{figure}

For the Kimi-K2 Instruct model, no such pairs were found. In the case of the DeepSeek-V3.1 model, there are three such situations: the forecast of the lower bound of the Fed interest rate and the forecast of real GDP growth based on statistics as of November 15, 2008, the forecast of the lower bound of the Fed interest rate based on statistics as of August 15, 2021. As can be seen from Figure \ref{fig: figure_12}, which shows forecasts for real and fake dates, as well as two forecasts within the testing of Assumption 2 with the maximum Wasserstein distance (as a distance having interpretation in terms of the transportation problem and being an upper bound to the difference of means of distributions), the sensitivity of the model to the forecast date when unable to look ahead is significantly lower than the difference of forecasts with real and fake dates, which indirectly confirms the weak influence of pure sensitivity to the forecast date on the results of Subsection~\ref{sec:forecasts_comparison} in the DeepSeek-V3.1 model.

For the Qwen3 Instruct model, seven such pairs were found. These are all forecasts of real GDP growth, as well as forecasts of CPI growth based on statistics as of August 15, 2021 and February 15, 2025. Figure \ref{fig: figure_12} shows that the cases of forecasts of real GDP growth based on statistics as of November 15, 2008 and May 15, 2020 and of CPI growth based on statistics as of August 15, 2021 are similar to those obtained by the DeepSeek-V.3.1 model, and the differences are not as great as those for forecasts with real and fake dates. Moreover, the main divergence in CPI growth forecasts is due to a large number of model refusals to provide a forecast. Excluding such points significantly reduces the Wasserstein distance between the forecast distributions within the testing of Assumption 2. Forecasts with statistics as of August 15, 2013 and August 15, 2021, as we saw from the additional analysis, have pronounced seasonality. When comparing the forecast distributions not against February 15, 2030, but against the corresponding month of 2030, the p-values of the tests similar to those presented in Tables \ref{tab:table_C.2.1.1}-\ref{tab: table_C.2.3.2} satisfy the selected threshold. This result suggests that choosing a single date when conducting Fake date test I may not be the best solution in some cases and casts doubt on the test results for real GDP growth for Qwen3 Instruct in Subsection~\ref{sec:forecasts_comparison} for May, August, and November. However, as shown in Figure \ref{fig: figure_13} and Figures \ref{fig: figure_C.1.1}-\ref{fig: figure_C.1.5} in Appendix~\ref{app:appendix_c}, a significant difference in forecasts when applying Fake date test I remains even when comparing forecasts for a real date against the corresponding month of 2030 as a fake date.

\begin{figure}[h!]
	\hspace{-0.25cm}
	\includegraphics[scale=0.55]{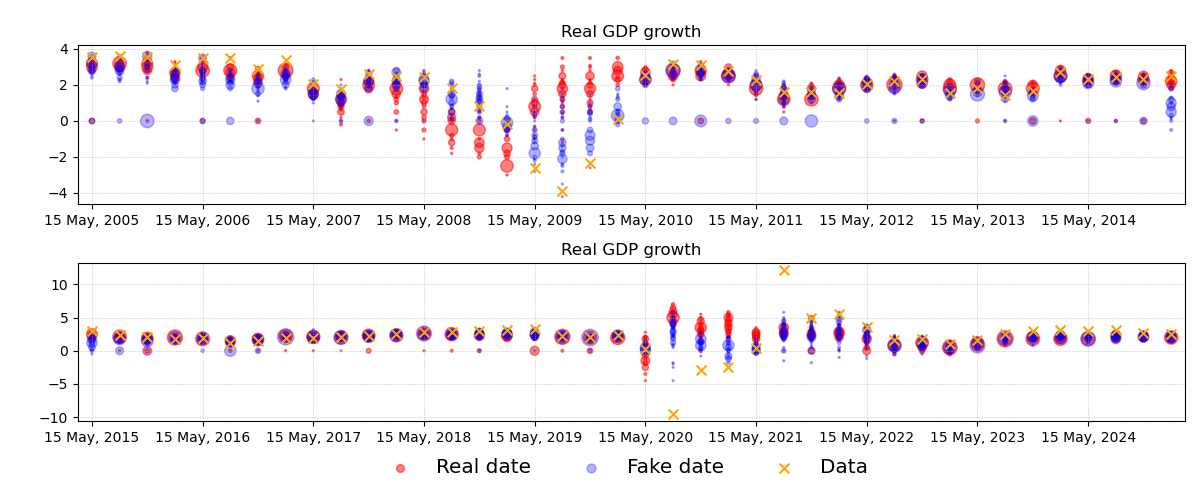}
	\caption{\textbf{Distributions of real GDP growth forecasts in Fake date test I for a seasonal fake date (with the cutoff boundary 3 years back from the forecast date for real data) for 1 year produced by the Qwen3 Instruct model.} The marker size is proportional to the density of the forecast distribution.}
    \label{fig: figure_13}
\end{figure}

Forecasts of CPI growth and real GDP growth based on statistics as of February 15, 2025 also show sensitivity to the date, which is comparable in magnitude to the difference in forecasts with real and fake dates. For real GDP growth, this happens with a small but statistically significant difference in the metric for comparing the forecast distributions with real and fake dates. In such cases, sensitivity to the date can indeed become a reason for additional false rejection of the hypothesis of the equality of distributions, but as can be seen from the figures presented in the previous subsection and figures in Appendix~\ref{app:appendix_b}, the forecast distributions differ quite strongly for a large number of dates and, although potentially influencing part of the results, this effect cannot fully explain the difference in the results of Fake date test I.

To better understand the reasons for large discrepancies in forecasts with real and fake dates in Fake date test I for some periods, we selected two dates: November 15, 2008 and May 15, 2021. For each of the previously used LLMs, each forecast indicator, and both dates, we randomly selected 20 responses with real and fake dates and analyzed them manually.

For the forecast based on statistics available as of November 15, 2008, all models for all indicators give forecasts that for a real date are lower than for a fake date. The main reason for this is how the models interpret the statistics on the lower bound of the Fed interest rate, CPI growth, and real GDP growth. When forecasting for a real date, these data are interpreted as the beginning of a recession in most cases. Kimi-K2 Instruct in the most obvious way does not follow the instruction not to use data after November 15, 2005, often referring to indicators and events between the cutoff date and the forecast date, such as the rate-cutting cycle, financial system stress, and trends in commodity market price dynamics (examples of model responses are provided in Appendix~\ref{app:appendix_d}). DeepSeek-V3.1, on the other hand, does not always directly reference real events of this period, but nevertheless concludes that the economy is either entering or already is in a recession phase. One of the arguments that recurred in the responses in favor of the fact that the risks of a decline in GDP (real GDP growth of 0.8\% (YoY)) dominate over inflation risks (CPI growth of 3.7\% (YoY)) is a historically low level of the lower bound of the Fed interest rate (1 percentage point). This seems to be a very logical conclusion, and the forecast based on the model's explanation is often very difficult to suspect of ``looking'' beyond the cutoff boundary. However, the comparison with identical forecasts for a fake date, where DeepSeek-V3.1 along with other LLMs, even with low rates, evaluate the risks of rising inflation as more important for the economy, makes one doubt that the forecast for a real date does not use additional information contained in the LLM weights, even if this is in no way reflected in the explanation of the obtained forecasts.

The analysis of forecasts based on statistics as of May 15, 2021 shows similar patterns. The data provided are largely similar to the information available in November 2008: high inflation (CPI growth of 4.2\% (YoY)), low growth (real GDP growth of 0.4\% (YoY)) and interest rate (the lower bound of the Fed interest rate equal to 0 percentage points). The indicator we would like to focus on is real GDP growth. Forecasts of this variable for a real date turn out to be higher in all models than forecasts for a fake date. The reason for this is that when forecasting for a real date, all LLMs understand that the economy is in the recovery phase after the COVID-19 shock. All models in one way or another signal this in their explanations of the forecasts given. In some cases, this happens explicitly, as, for example, in many responses of Kimi-K2 Instruct, while in others, as with most responses of Qwen3 Instruct, in the form of information that the model cannot use when producing a forecast due to prompt restrictions related to the unavailability of information after May 15, 2018.

In the analyzed LLM responses for November 15, 2008 and May 15, 2021, we did not detect, except for a few cases, any direct references to data and events beyond the forecast date; however, we are not inclined to interpret this as the models not using this information in their responses in any way. This can happen implicitly (see Arcuschin et al. \citeyearpar{Arcuschin2025}, Chen et al. \citeyearpar{Chen2025Anthropic}), through the use of model weights where information about these periods is stored, without explicit indication in the tokens generated for explanation due to the direct restriction in the prompt. This question requires additional study and potentially more advanced and computationally complex tools\footnote{As far as we know, the question of systematically detecting a discrepancy between model explanations and its ``true motives" is open in the literature. However, new techniques, such as, for example, interpretation based on sparse autoencoders or transcoders (see, for example, research by Ameisen et al. \citeyearpar{Ameisen2025} and Lindsey et al. \citeyearpar{Lindsey2025}), allow increasingly deep penetration into the ``decision-making'' processes of LLMs (see Marks et al. \citeyearpar{Marks2025}).}, which is beyond the scope of this study. Nevertheless, the analysis results show that all models cannot accurately follow the instruction not to use information beyond the cutoff boundary, which, together with the results of Fake date test II, is a signal regarding the potential incorrectness of using the procedures for evaluating the accuracy of macroeconomic forecasts of the considered LLMs based on in-sample metrics (at least for the prompt used in this paper and the set of macroeconomic statistics provided to the models).

Neither did we see in the analyzed responses, except for a few cases, any references to historical data or precedents beyond the information cutoff date in forecasts with the fake date\footnote{In addition to the analysis described above, we conducted a focused analysis of the occurrence of the character strings `pandem' and `covid' in responses generated with fake dates for both tests. We found that such mentions were nearly absent in responses dated prior to May 15, 2020. Moreover, almost all earlier occurrences were associated with explicit statements regarding the impossibility of utilizing pandemic-related information for subsequent forecasting.} \footnote{It should be noted that in a number of preliminary experiments with a softer prompt regarding data cutoff, we noticed mentions of historical precedents beyond the cutoff boundary in forecasts for a fake date.}, which, in addition to the candidate with $d=0$ discussed earlier in Section~\ref{sec:discussion}, makes these forecasts a potential candidate for using in in-sample testing; however, as in the case of using data beyond the forecast date with the real date, this can happen implicitly. A more detailed study of the properties of such candidates requires separate research.

The final point we would like to address in this subsection is a brief analysis of LLM responses within Fake date test II. This test checks for the presence of at least one of two biases: lookahead bias and context bias. Due to the absence of a restriction on the use of knowledge for the past three years, models explicitly incorporate information about ongoing economic processes when forecasting for a real date. For example, when forecasting based on statistics as of May 15, 2021, all models --- in contrast to the forecast generated for a fake date --- refer to the COVID-19 shock of 2020 (see Appendix~\ref{app:appendix_f}), which is a sign of context bias in the in-sample forecasts. Furthermore, the use of information beyond the forecast date is particularly difficult to detect from model explanations; however, as noted earlier, such leakage may occur implicitly. A comprehensive analysis of the explanations given by the three models tested in this study to identify all such cases is beyond the scope of this paper. Nevertheless, a selective manual review reveals that such examples are present in the dataset, signaling not only context bias but also a model's failure to follow the instruction to truncate information beyond the cutoff date. As an example of such behavior, Figure \ref{fig: figure_14} shows the Qwen3 Instruct model's response for the inflation forecast based on statistics as of August 15, 2008.  This date falls at the start of the oil price decline from the July 2008 peaks. Although the decline had only just begun and the subsequent trajectory was not obvious from futures prices or analytical reports of the time, the model's explanations refer to the sharp drop in oil prices in late 2008 and early 2009, demonstrating that the model has explicitly used of future information when generating the forecast.

\begin{figure}[ht]
\begin{tcolorbox}[promptbox]
To forecast U.S. year-over-year (YoY) inflation for July 2009, we consider the available data as of August 15, 2008: the Federal Reserve's lower bound on the interest rate is 2.0\%, real GDP growth in Q2 2008 was 1.8\% (YoY), and inflation as measured by CPI in July 2008 was 5.5\% (YoY).

Inflation at 5.5\% in July 2008 reflects significant price pressures, likely driven in part by elevated energy and commodity prices earlier in the year. However, real GDP growth of 1.8\% in Q2 2008, while positive, shows signs of slowing momentum compared to earlier expansions. Additionally, the Fed's interest rate lower bound of 2.0\% - though relatively high compared to later crisis-era levels - may still be below the neutral rate given 5.5\% inflation, implying a relatively accommodative monetary stance if nominal rates do not sufficiently counteract inflation expectations.

Nevertheless, by mid-2008, financial conditions were tightening due to the deepening housing and credit market turmoil. Labor markets were beginning to weaken, and consumer demand was showing signs of softening due to declining wealth and tighter credit. These factors would likely dampen aggregate demand over the coming year, exerting downward pressure on inflation.

Moreover, commodity prices - particularly oil - had peaked around mid-2008 and were beginning to decline sharply in the second half of the year due to weakening global demand, though this decline would fully register in 2009 prices. Given that July 2009 inflation will reflect price changes over the 12 months ending in July 2009, \textcolor{redtext}{the sharp drop in oil and commodity prices in late 2008 and early 2009} would significantly reduce CPI inflation on a YoY basis, even if core inflation remains somewhat sticky.

Thus, while inflation was high in July 2008, the combination of slowing economic growth, tightening financial conditions, falling commodity prices, and weakening demand suggests a substantial decline in YoY inflation by July 2009. A reversal into deflation is unlikely given still-positive (though slowing) growth in mid-2008 and the stickiness of core prices, but disinflation is expected.

Inflation forecast: 2.1\%
\end{tcolorbox}
\caption{\textbf{Example of a response for 1-year forecast of CPI growth (YoY) for a real date by the Qwen3 Instruct model with the forecast date of August 15, 2008.} Formatting was slightly adjusted for readability.}
\label{fig: figure_14}
\end{figure}

\section{Conclusion} \label{sec:conclusion}

In this paper, we propose tests designed to shed at least some light on the possibility of using in-sample forecasts of LLMs to evaluate the quality of their forecasts. The fake date tests are positioned as an applied tool that gives signals regarding the presence of lookahead bias and bias associated with an additional context beyond that specified in the prompt, and consequently are a reason to question the validity of using the procedure for evaluating the accuracy based on in-sample forecasts if the tests are not passed. Although this study focuses on macroeconomic forecasts, the proposed tests can be applied to a wider range of tasks, which includes, at a minimum, tasks of time series forecasting.

The biases discussed in this study may potentially be relevant in a number of other tasks that use LLMs. Such tasks include, for example, studies related to sentiment-based forecasting, modeling of survey data or human decision-making. Evaluation of the predictive ability of news, inflation expectations based on LLM responses, and agent-based models, where agents are modeled not based on simple predetermined rules or reinforcement learning, but based on LLMs, can serve as examples of such tasks in economics. How significant the biases studied herein are in these tasks is a separate and rather complex question that we do not investigate, but we believe that with some modifications and, possibly, optimization of the number of requests, the proposed tests can help in detecting biases in these directions as well.

\newpage

\bibliography{fake-date-tests.bib}

\clearpage

\startappendices

\appsection[app:appendix_a]{Proofs}

\subsection{Proof of Proposition 1}

Let's consider two forecast distributions:

\vspace{-40pt}
\begin{multline*}
    y^1_{forecast} = \mathcal{M}_T( f(l_{text}=l, t_{current}=t_{retro}, t_{forecast}=t_{retro}+h, t_{cutoff}=t_{retro}-d, \\ v_{forecast}=v, s_{current}=s_{retro}), \varepsilon)
\end{multline*}
\vspace{-80pt}

\begin{multline*}    
    y^2_{forecast} = \mathcal{M}_T( f(l_{text}=l, t_{current}=t_{fake}, t_{forecast}=t_{fake}+h, t_{cutoff}=t_{retro}-d, \\ v_{forecast}=v, s_{current}=s_{retro}), \varepsilon)
\end{multline*}
\vspace{-40pt}

According to A1, with fixed $\varepsilon$, both of these forecasts are measurable with respect to the $\sigma$-algebra generated by the union of $s_{retro}$ and $I_{t_{retro}-d}$. Since both forecasts are measurable with respect to the $\sigma$-algebra generated by the union of $s_{retro}$ and $I_{t_{retro}-d}$ and differ only in the current date ($t_{current,1} = t_{retro}$ and $t_{current,2} = t_{fake}$) and the date for which the forecast is being made ($t_{forecast,1} = t_{retro} + h$ and $t_{forecast,2} = t_{fake} + h$), then by A2, for any $t_{retro} \geq t_{cutoff} + d$ and $t_{retro} \geq t_{fake} + d$, the distributions $y^1_{forecast}$ and $y^2_{forecast}$ coincide.

By construction, $t_{retro} = t_{cutoff} + d \geq t_{cutoff} + d$ and $t_{fake} > t_{retro} \geq t_{cutoff} + d$, which completes the proof.

\subsection{Proof of Proposition 2}

Analogous to the proof of Proposition 1 for $d = 0$.

\clearpage

\newgeometry{top=0.25cm,bottom=1.75cm,left=2cm,right=2cm,marginparwidth=2cm}

\appsection[app:appendix_b]{Results of Fake date test I}
\subsection
  [Visualization of the distributions of Kimi-K2 Instruct forecasts]
  {\texorpdfstring
    {\makebox[5cm][l]{Visualization of the distributions of Kimi-K2 Instruct forecasts}}
    {Visualization of the distributions of Kimi-K2 Instruct forecasts}}

\subsubsection{Fed interest rate}

\renewcommand{\thefigure}{B.1.1.\arabic{figure}}
\setcounter{figure}{0}

\begin{figure}[h!!]
	\centering \hspace*{-0.5cm}
	\includegraphics[scale=0.55]{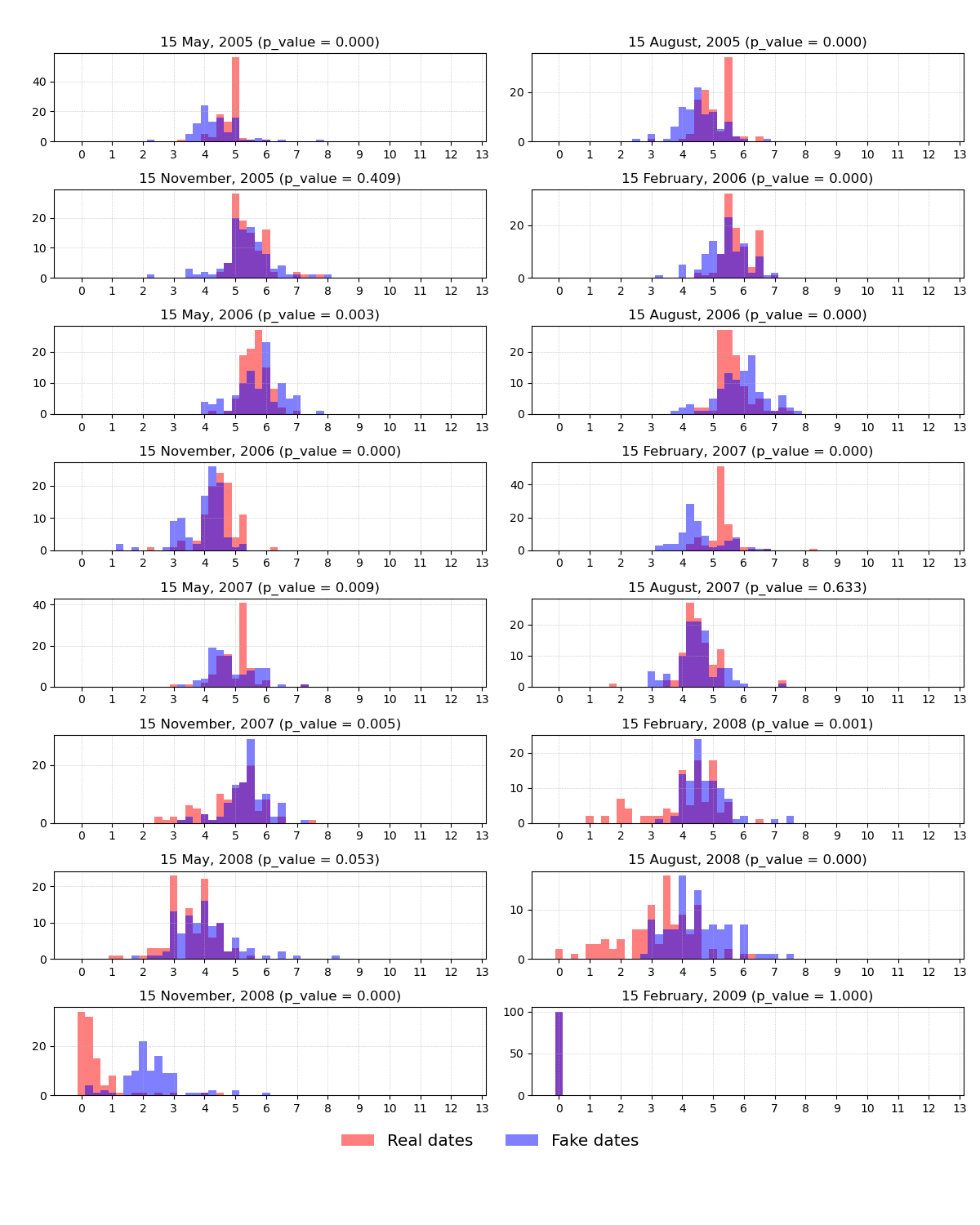}
	\caption{\textbf{Distributions of 1-year forecasts of the lower bound of the Fed interest rate produced by the Kimi-K2 Instruct model.} The header shows the forecast date and the p-value for the Kolmogorov-Smirnov permutation test.}
    \label{fig: figure_B.1.1.1}
\end{figure}

\clearpage

\restoregeometry

\begin{figure}[ht]
	\centering \hspace*{-0.5cm}
	\includegraphics[scale=0.55]{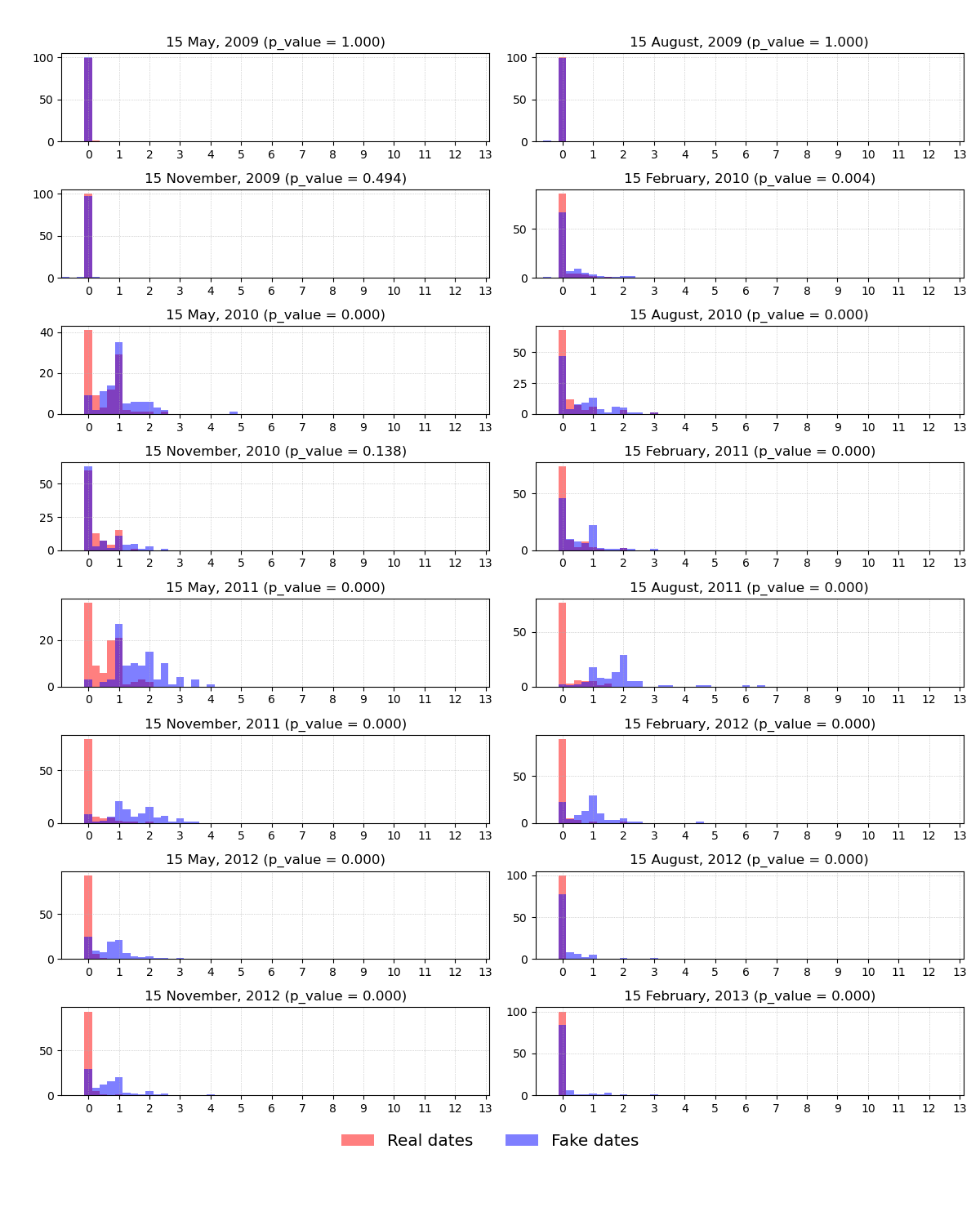}
	\caption{\textbf{Distributions of 1-year forecasts of the lower bound of the Fed interest rate produced by the Kimi-K2 Instruct model.} The header shows the forecast date and the p-value for the Kolmogorov-Smirnov permutation test.}
    \label{fig: figure_B.1.1.2}
\end{figure}

\begin{figure}[h!]
	\centering \hspace*{-0.5cm}
	\includegraphics[scale=0.55]{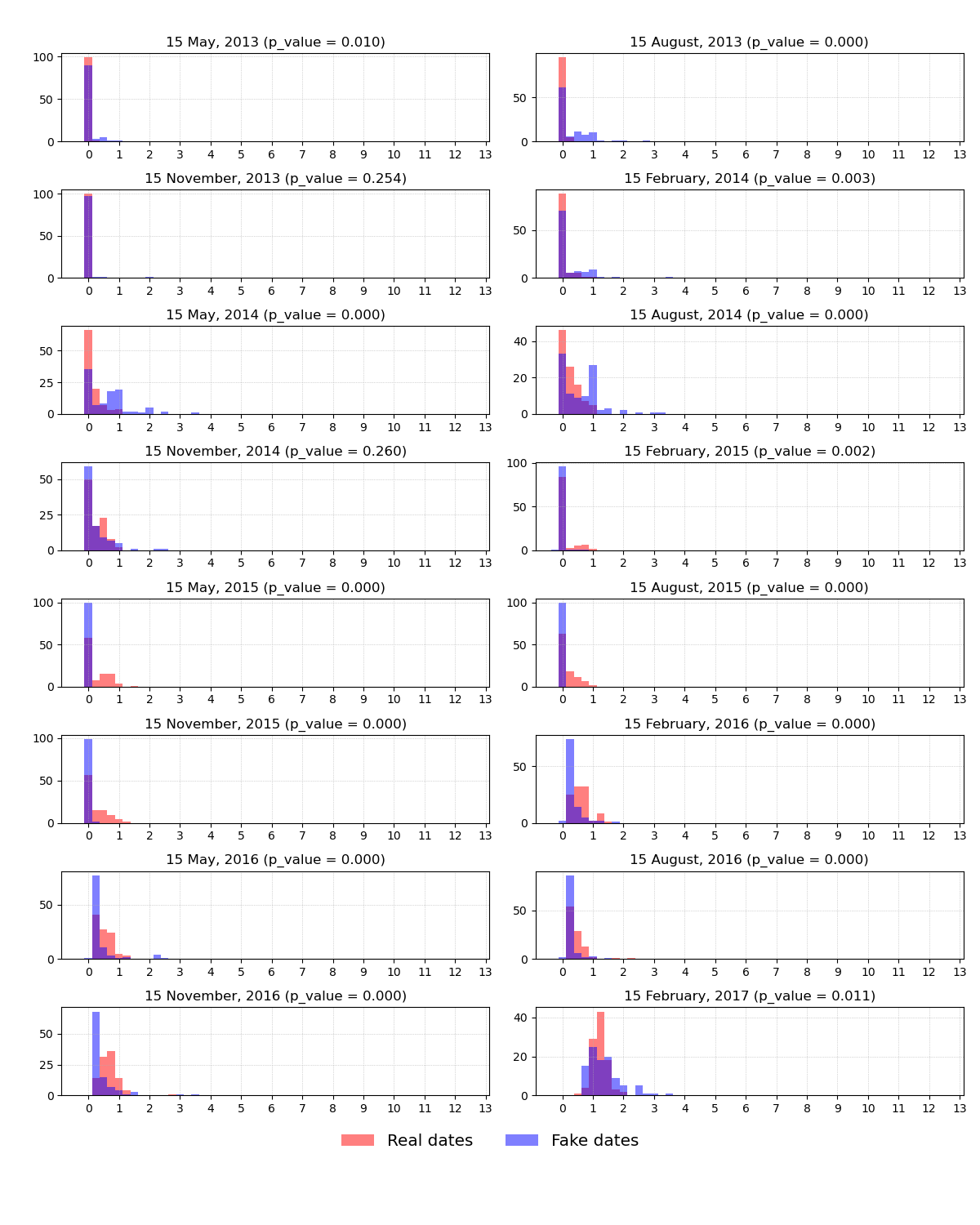}
	\caption{\textbf{Distributions of 1-year forecasts of the lower bound of the Fed interest rate produced by the Kimi-K2 Instruct model.} The header shows the forecast date and the p-value for the Kolmogorov-Smirnov permutation test.}
    \label{fig: figure_B.1.1.3}
\end{figure}

\begin{figure}[h!]
	\centering \hspace*{-0.5cm}
	\includegraphics[scale=0.55]{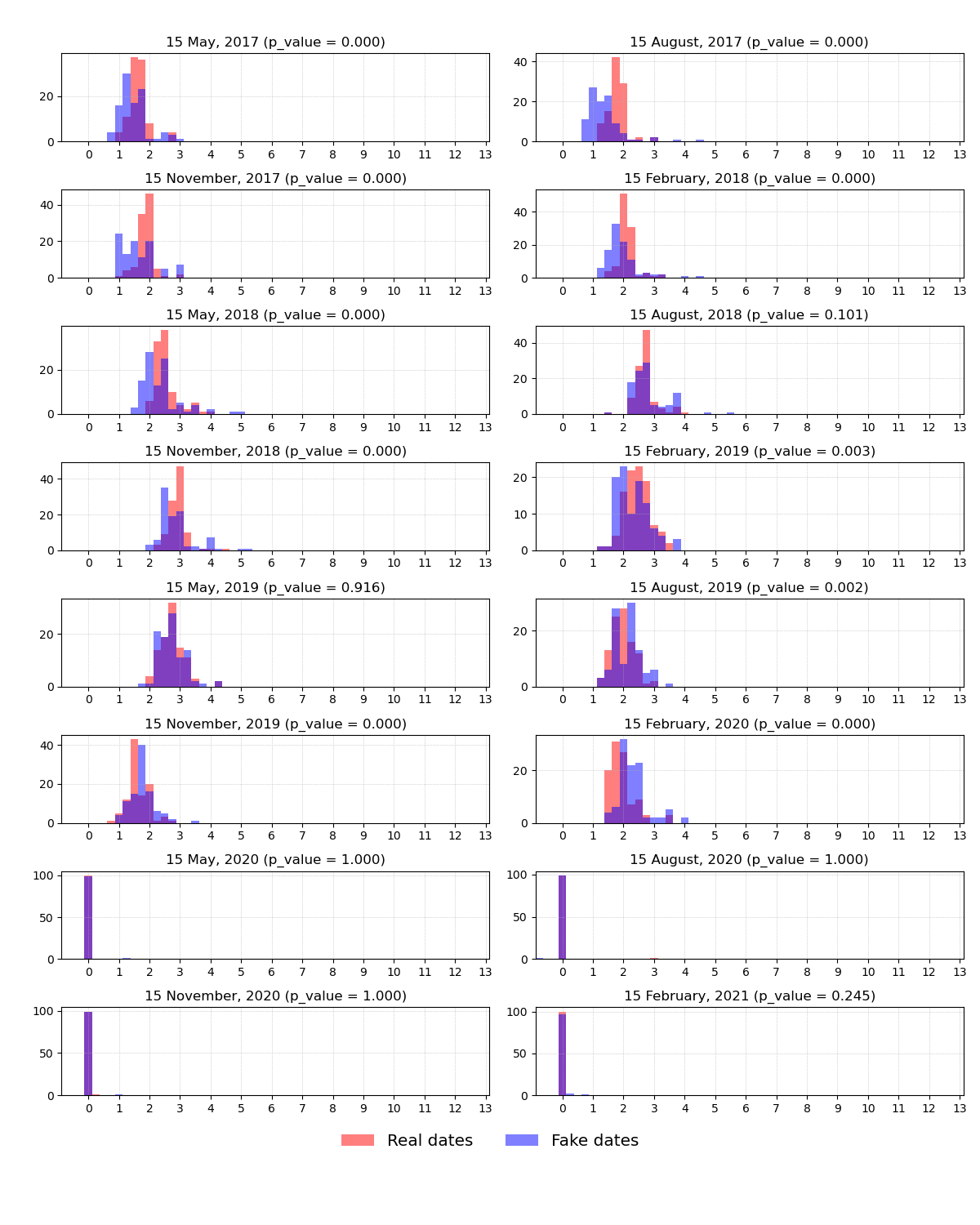}
	\caption{\textbf{Distributions of 1-year forecasts of the lower bound of the Fed interest rate produced by the Kimi-K2 Instruct model.} The header shows the forecast date and the p-value for the Kolmogorov-Smirnov permutation test.}
    \label{fig: figure_B.1.1.4}
\end{figure}

\begin{figure}[h!]
	\centering \hspace*{-0.5cm}
	\includegraphics[scale=0.55]{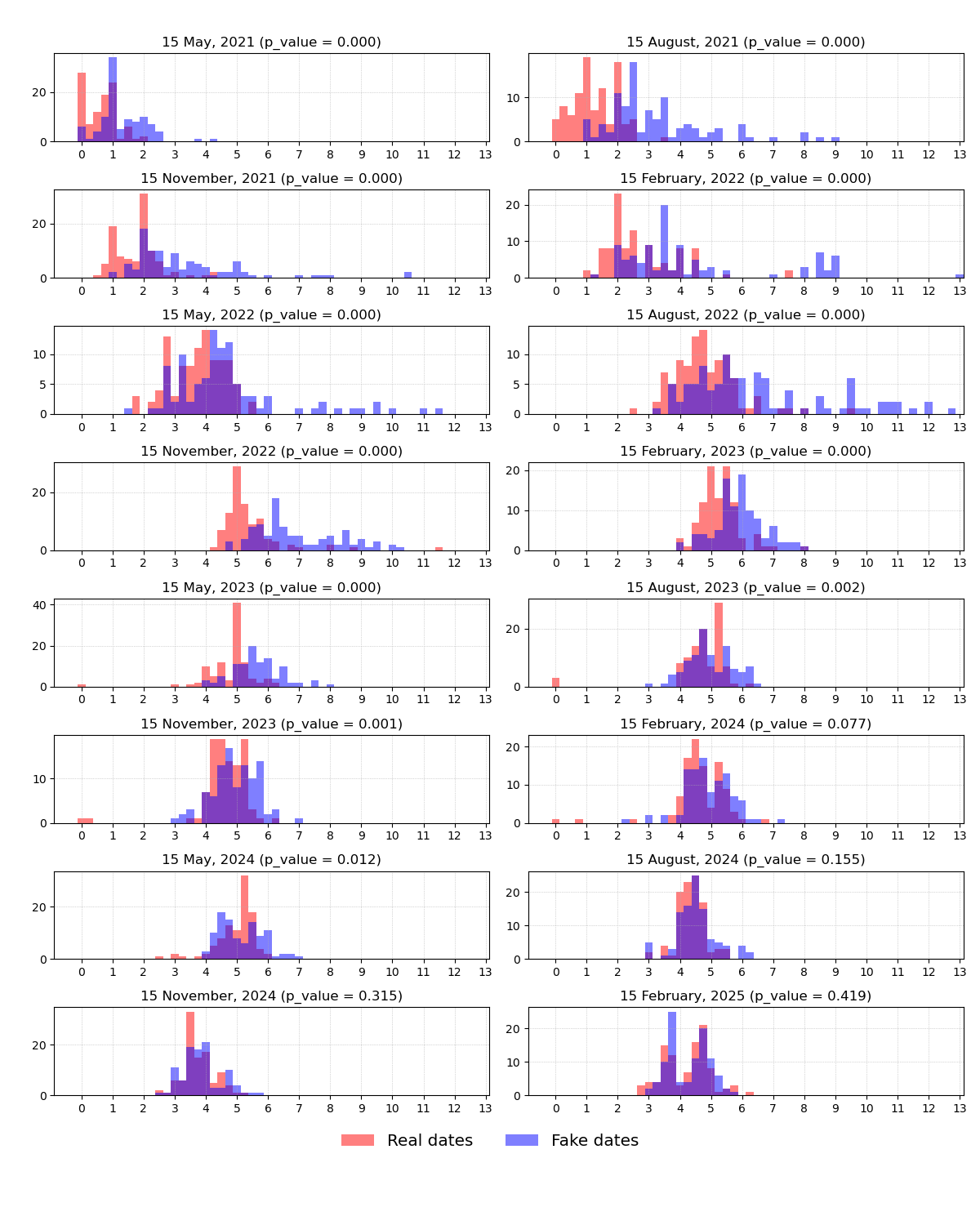}
	\caption{\textbf{Distributions of 1-year forecasts of the lower bound of the Fed interest rate produced by the Kimi-K2 Instruct model.} The header shows the forecast date and the p-value for the Kolmogorov-Smirnov permutation test.}
    \label{fig: figure_B.1.1.5}
\end{figure}

\clearpage

\newgeometry{top=1.85cm,bottom=2cm,left=2cm,right=2cm,marginparwidth=2cm}

\subsubsection{CPI growth (YoY)}

\renewcommand{\thefigure}{B.1.2.\arabic{figure}}
\setcounter{figure}{0}

\begin{figure}[h!]
	\centering \hspace*{-0.5cm}
	\includegraphics[scale=0.55]{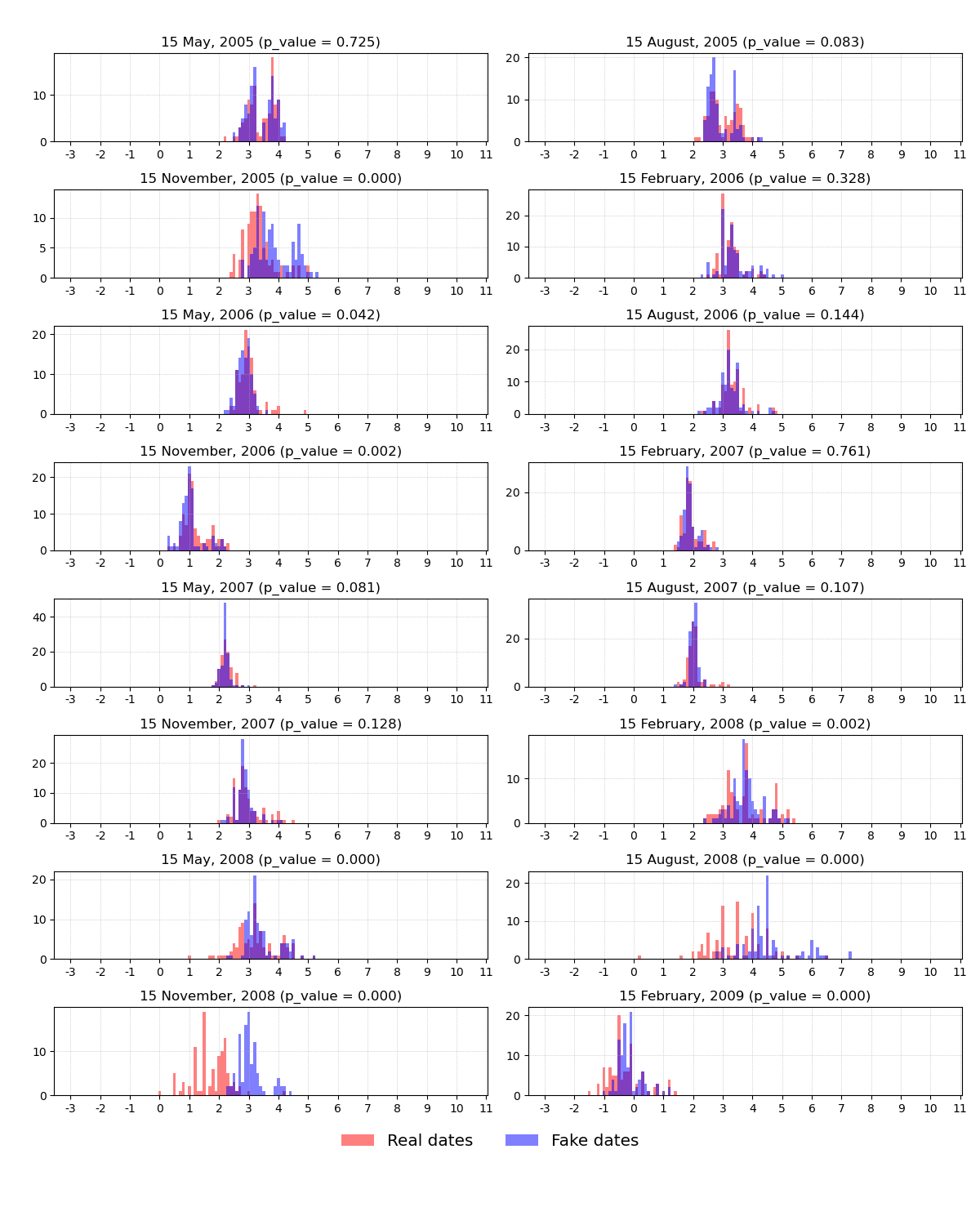}
	\caption{\textbf{Distributions of 1-year forecasts of CPI growth (YoY) produced by the Kimi-K2 Instruct model.} The header shows the forecast date and the p-value for the Kolmogorov-Smirnov permutation test.}
    \label{fig: figure_B.1.2.1}
\end{figure}

\clearpage

\restoregeometry

\begin{figure}[ht]
	\centering \hspace*{-0.5cm}
	\includegraphics[scale=0.55]{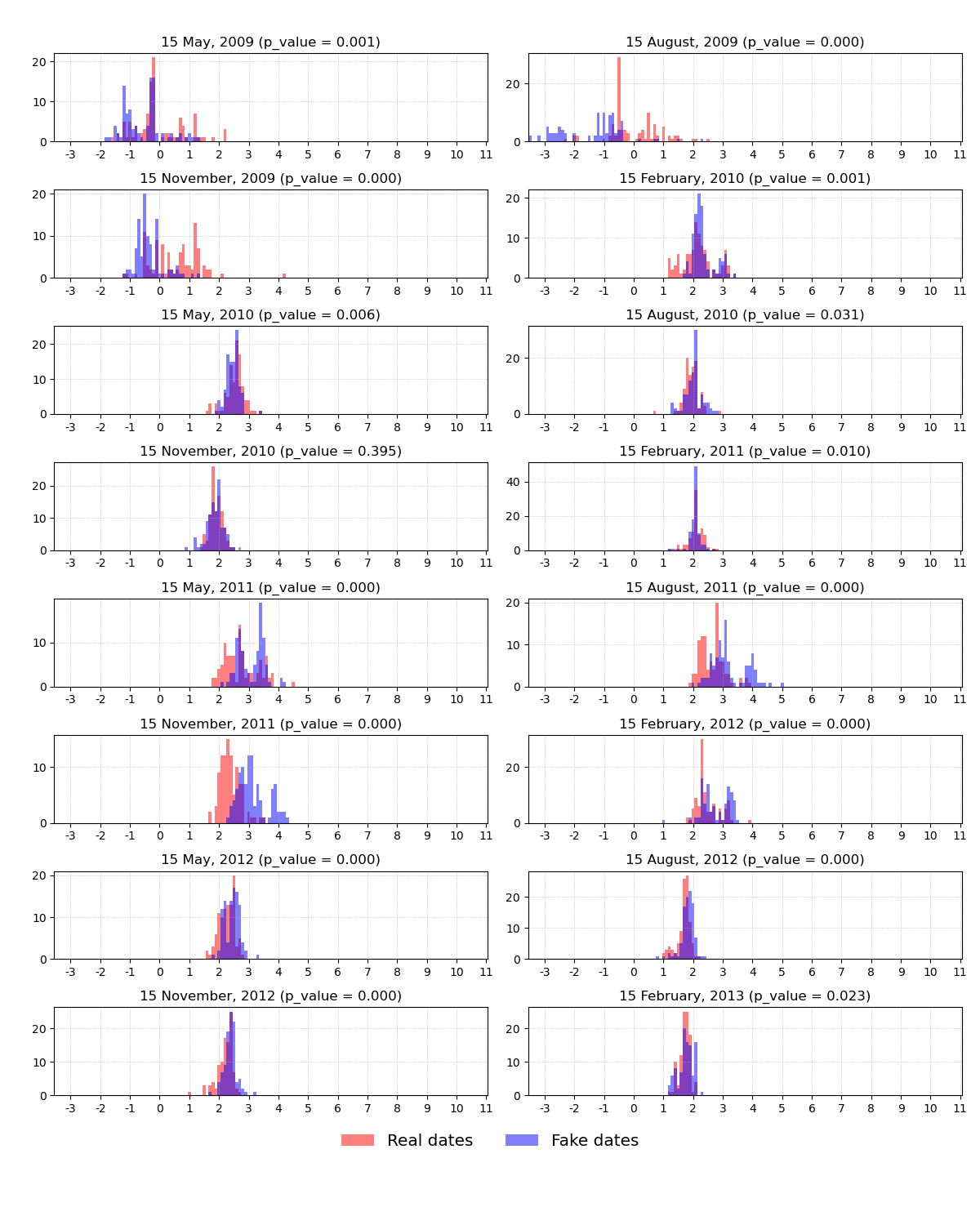}
	\caption{\textbf{Distributions of 1-year forecasts of CPI growth (YoY) produced by the Kimi-K2 Instruct model.} The header shows the forecast date and the p-value for the Kolmogorov-Smirnov permutation test.}
    \label{fig: figure_B.1.2.2}
\end{figure}

\begin{figure}[h!]
	\centering \hspace*{-0.5cm}
	\includegraphics[scale=0.55]{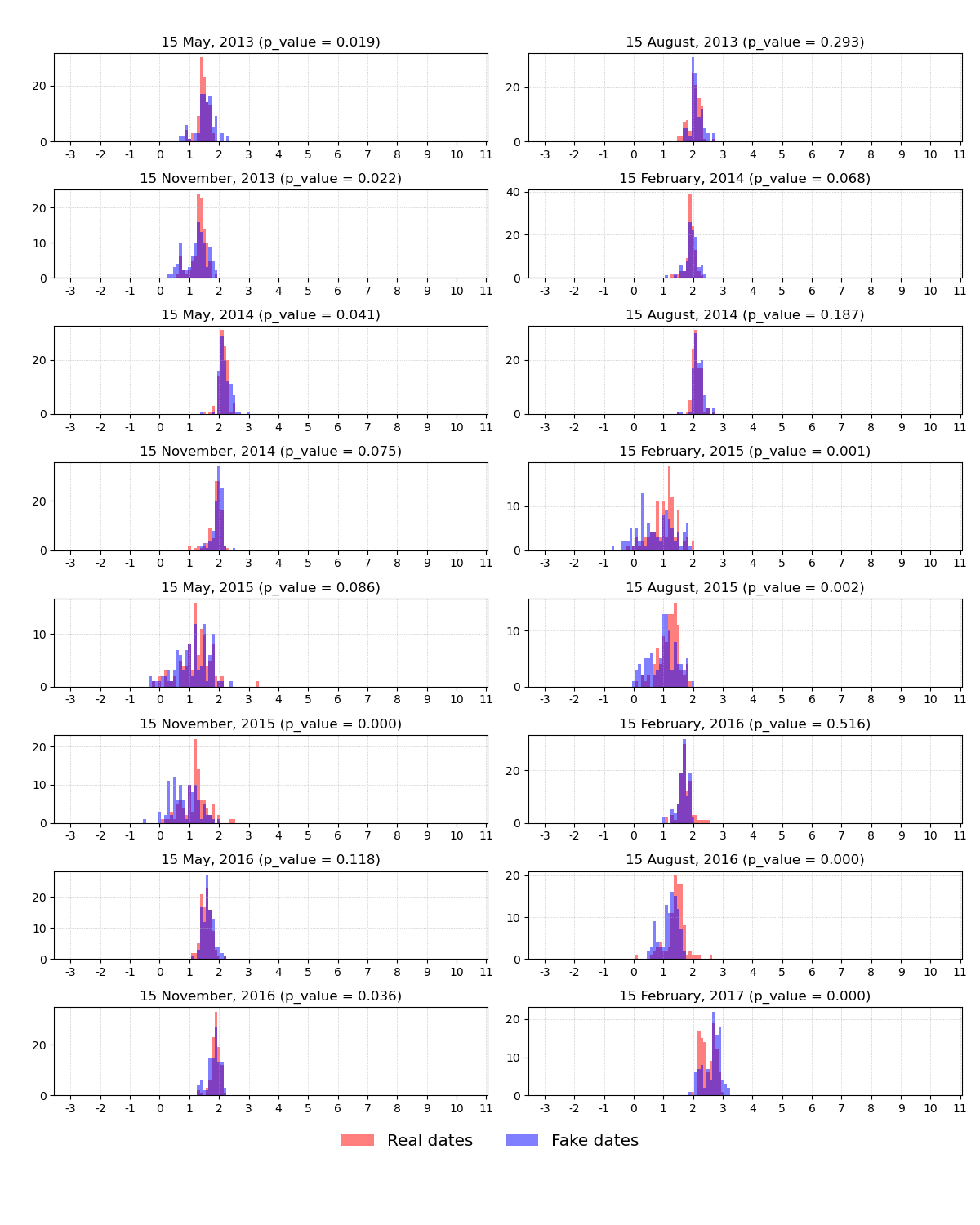}
	\caption{\textbf{Distributions of 1-year forecasts of CPI growth (YoY) produced by the Kimi-K2 Instruct model.} The header shows the forecast date and the p-value for the Kolmogorov-Smirnov permutation test.}
    \label{fig: figure_B.1.2.3}
\end{figure}

\begin{figure}[h!]
	\centering \hspace*{-0.5cm}
	\includegraphics[scale=0.55]{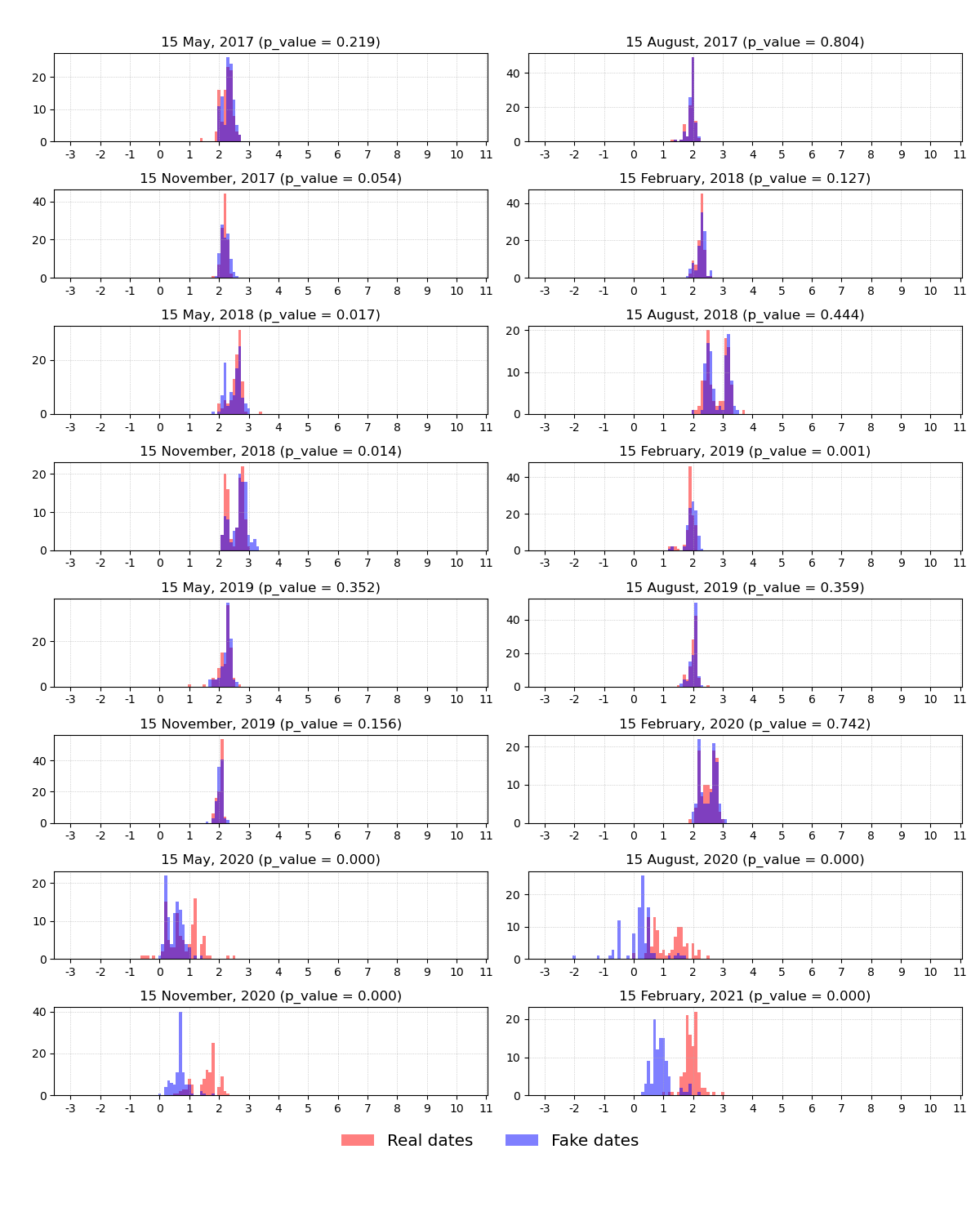}
	\caption{\textbf{Distributions of 1-year forecasts of CPI growth (YoY) produced by the Kimi-K2 Instruct model.} The header shows the forecast date and the p-value for the Kolmogorov-Smirnov permutation test.}
    \label{fig: figure_B.1.2.4}
\end{figure}

\begin{figure}[h!]
	\centering \hspace*{-0.5cm}
	\includegraphics[scale=0.55]{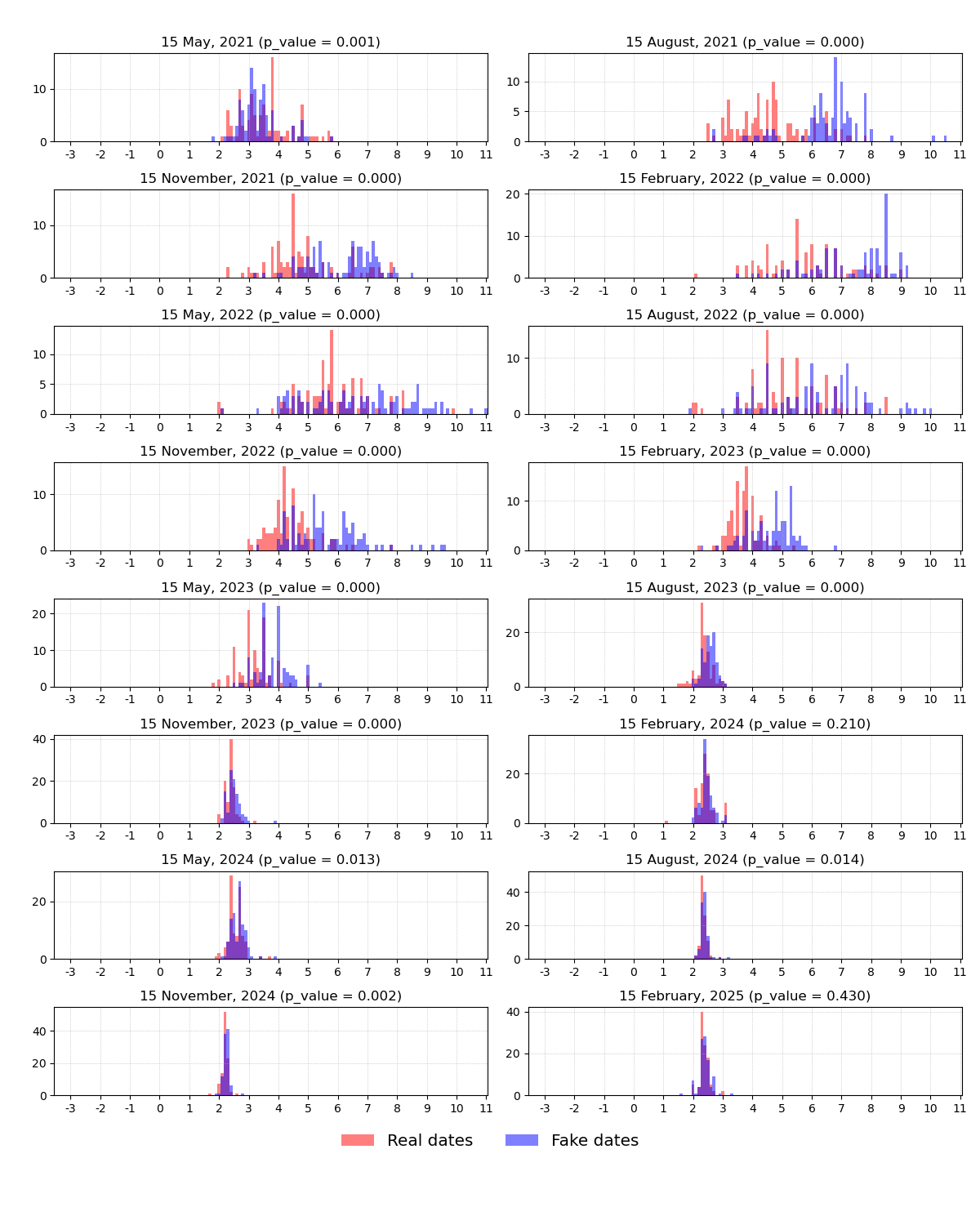}
	\caption{\textbf{Distributions of 1-year forecasts of CPI growth (YoY) produced by the Kimi-K2 Instruct model.} The header shows the forecast date and the p-value for the Kolmogorov-Smirnov permutation test.}
    \label{fig: figure_B.1.2.5}
\end{figure}

\clearpage

\newgeometry{top=1.85cm,bottom=2cm,left=2cm,right=2cm,marginparwidth=2cm}

\subsubsection{Real GDP growth (YoY)}

\renewcommand{\thefigure}{B.1.3.\arabic{figure}}
\setcounter{figure}{0}

\begin{figure}[h!]
	\centering \hspace*{-0.5cm}
	\includegraphics[scale=0.55]{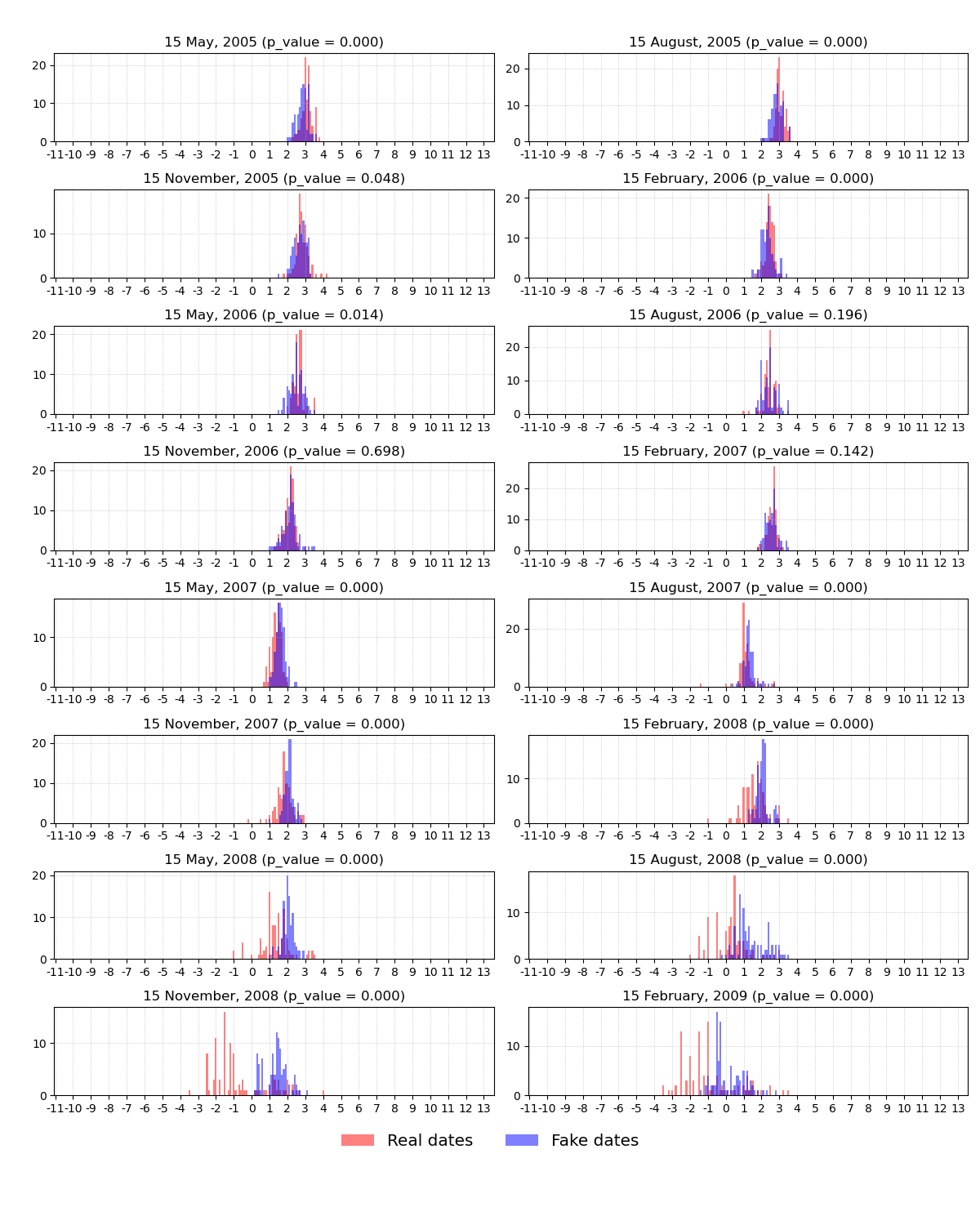}
	\caption{\textbf{Distributions of 1-year forecasts of real GDP growth (YoY) produced by the Kimi-K2 Instruct model.} The header shows the forecast date and the p-value for the Kolmogorov-Smirnov permutation test.}
    \label{fig: figure_B.1.3.1}
\end{figure}

\clearpage

\restoregeometry

\begin{figure}[ht]
	\centering \hspace*{-0.5cm}
	\includegraphics[scale=0.55]{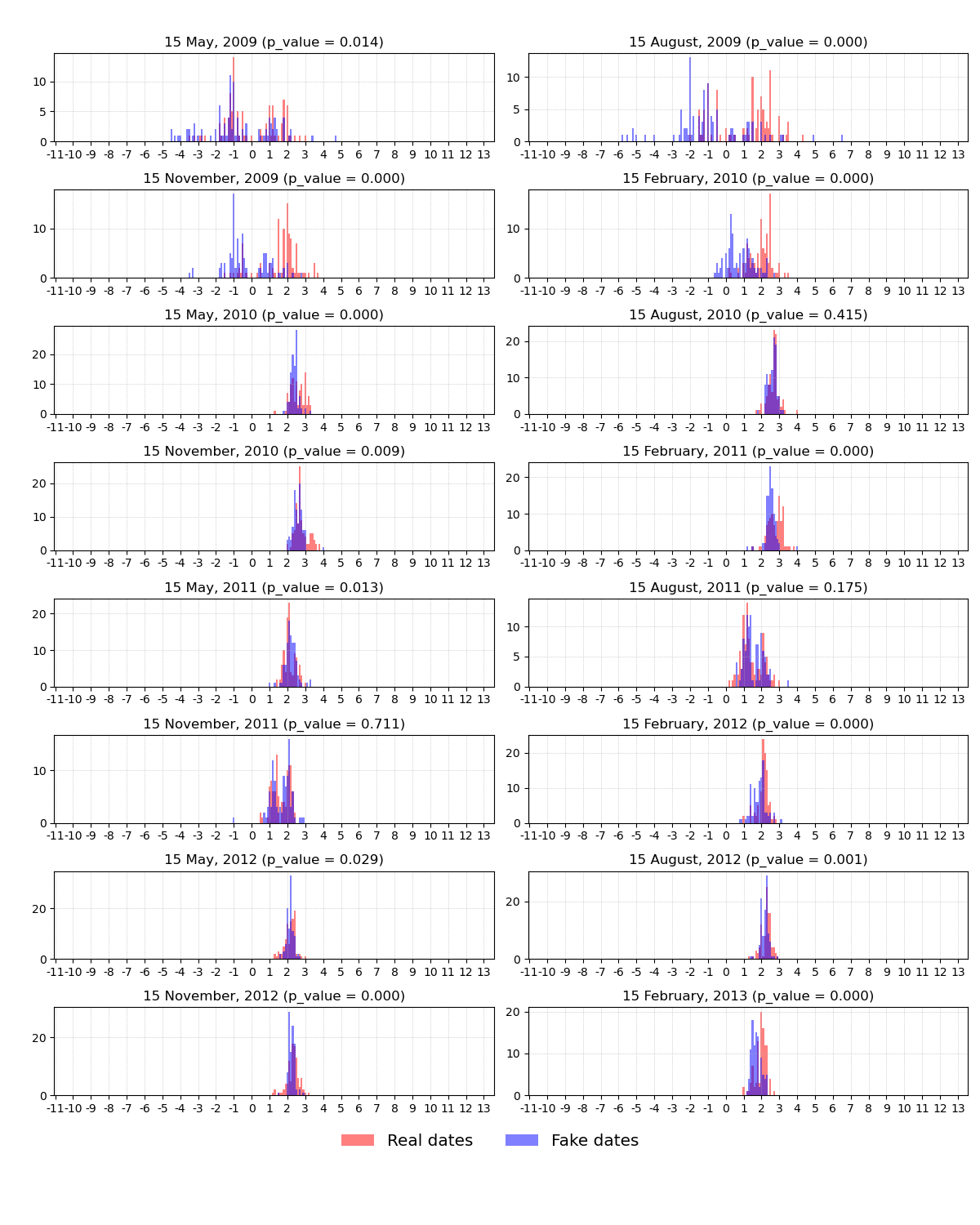}
	\caption{\textbf{Distributions of 1-year forecasts of real GDP growth (YoY) produced by the Kimi-K2 Instruct model.} The header shows the forecast date and the p-value for the Kolmogorov-Smirnov permutation test.}
    \label{fig: figure_B.1.3.2}
\end{figure}

\begin{figure}[h!]
	\centering \hspace*{-0.5cm}
	\includegraphics[scale=0.55]{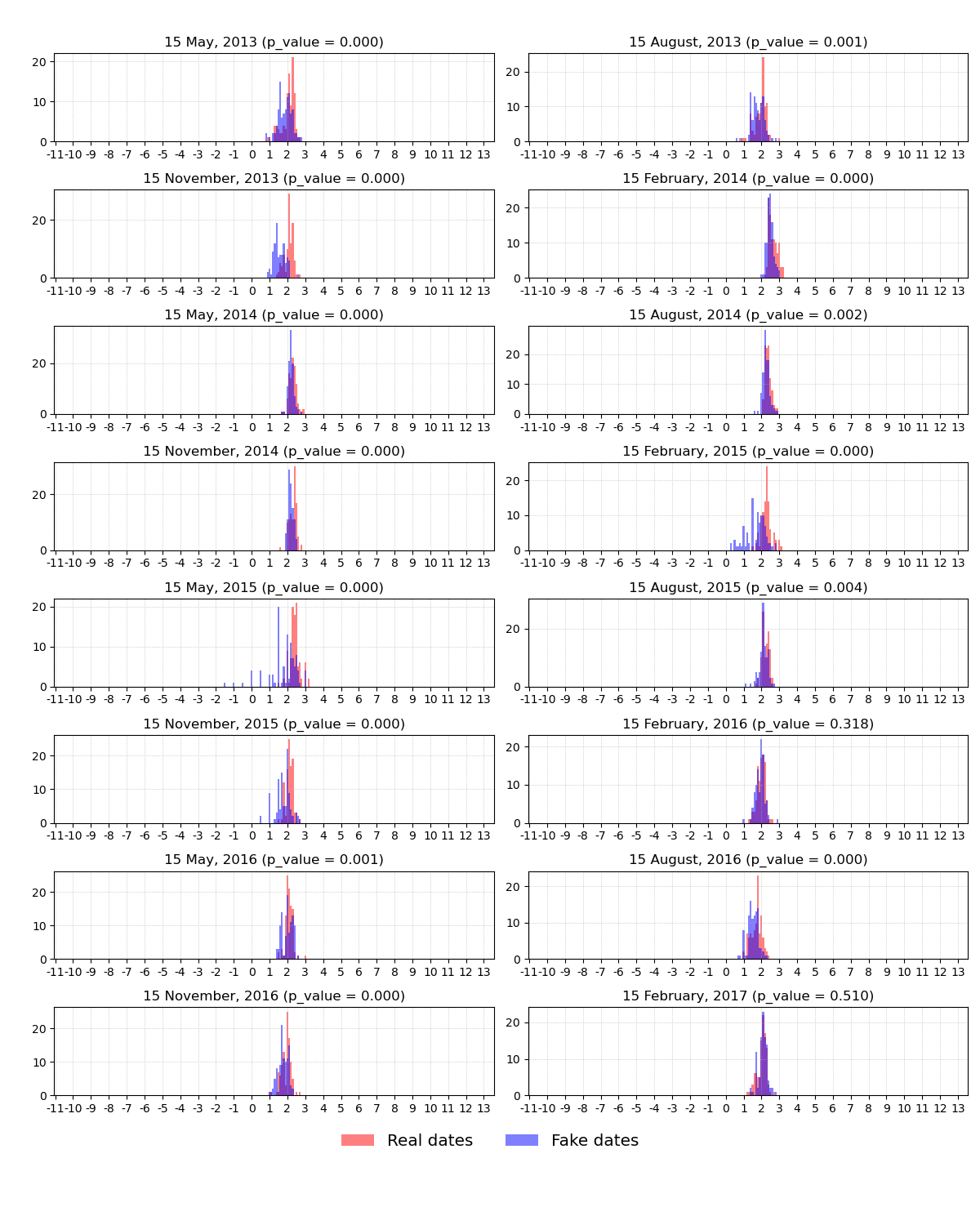}
	\caption{\textbf{Distributions of 1-year forecasts of real GDP growth (YoY) produced by the Kimi-K2 Instruct model.} The header shows the forecast date and the p-value for the Kolmogorov-Smirnov permutation test.}
    \label{fig: figure_B.1.3.3}
\end{figure}

\begin{figure}[h!]
	\centering \hspace*{-0.5cm}
	\includegraphics[scale=0.55]{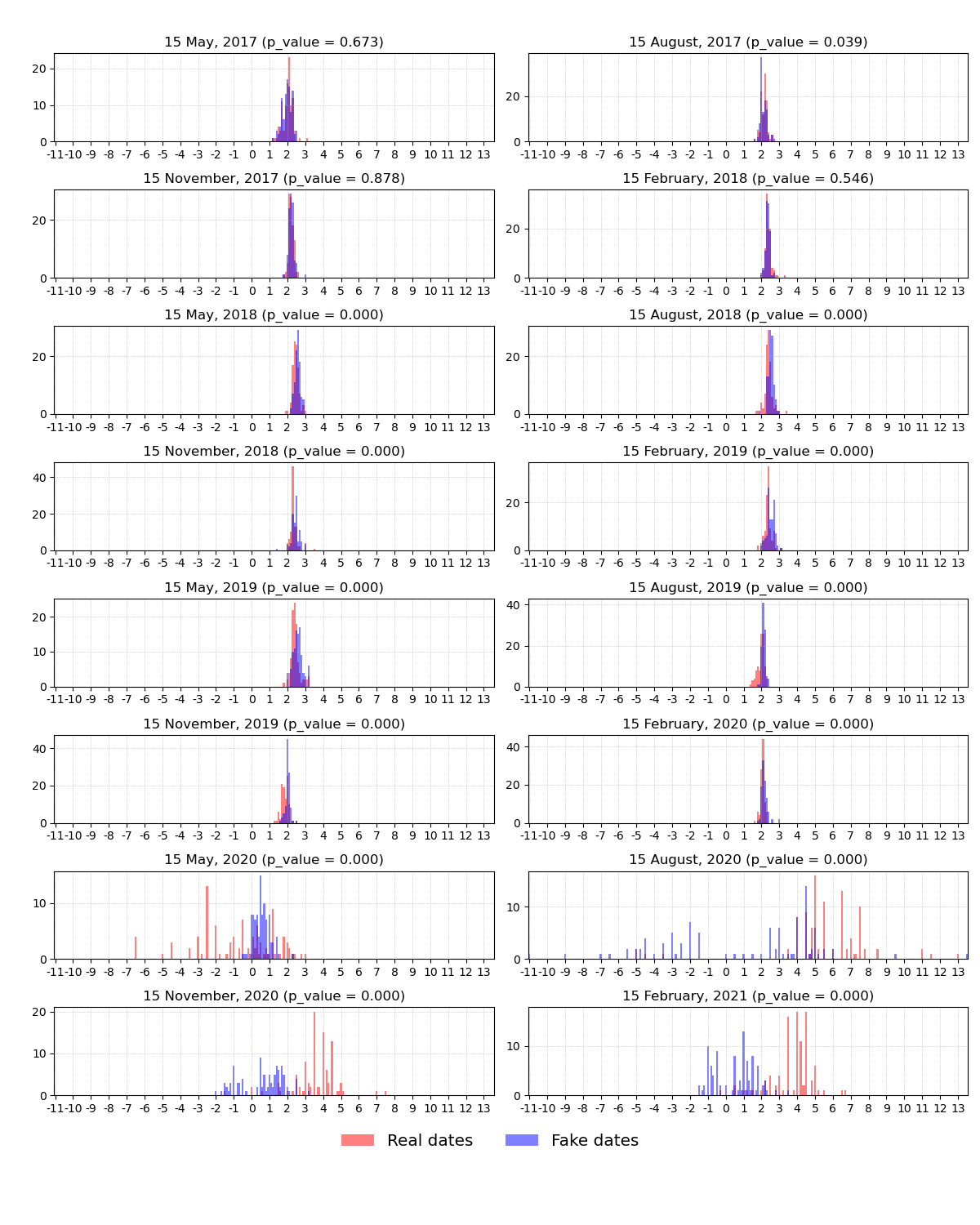}
	\caption{\textbf{Distributions of 1-year forecasts of real GDP growth (YoY) produced by the Kimi-K2 Instruct model.} The header shows the forecast date and the p-value for the Kolmogorov-Smirnov permutation test.}
    \label{fig: figure_B.1.3.4}
\end{figure}

\begin{figure}[h!]
	\centering \hspace*{-0.5cm}
	\includegraphics[scale=0.55]{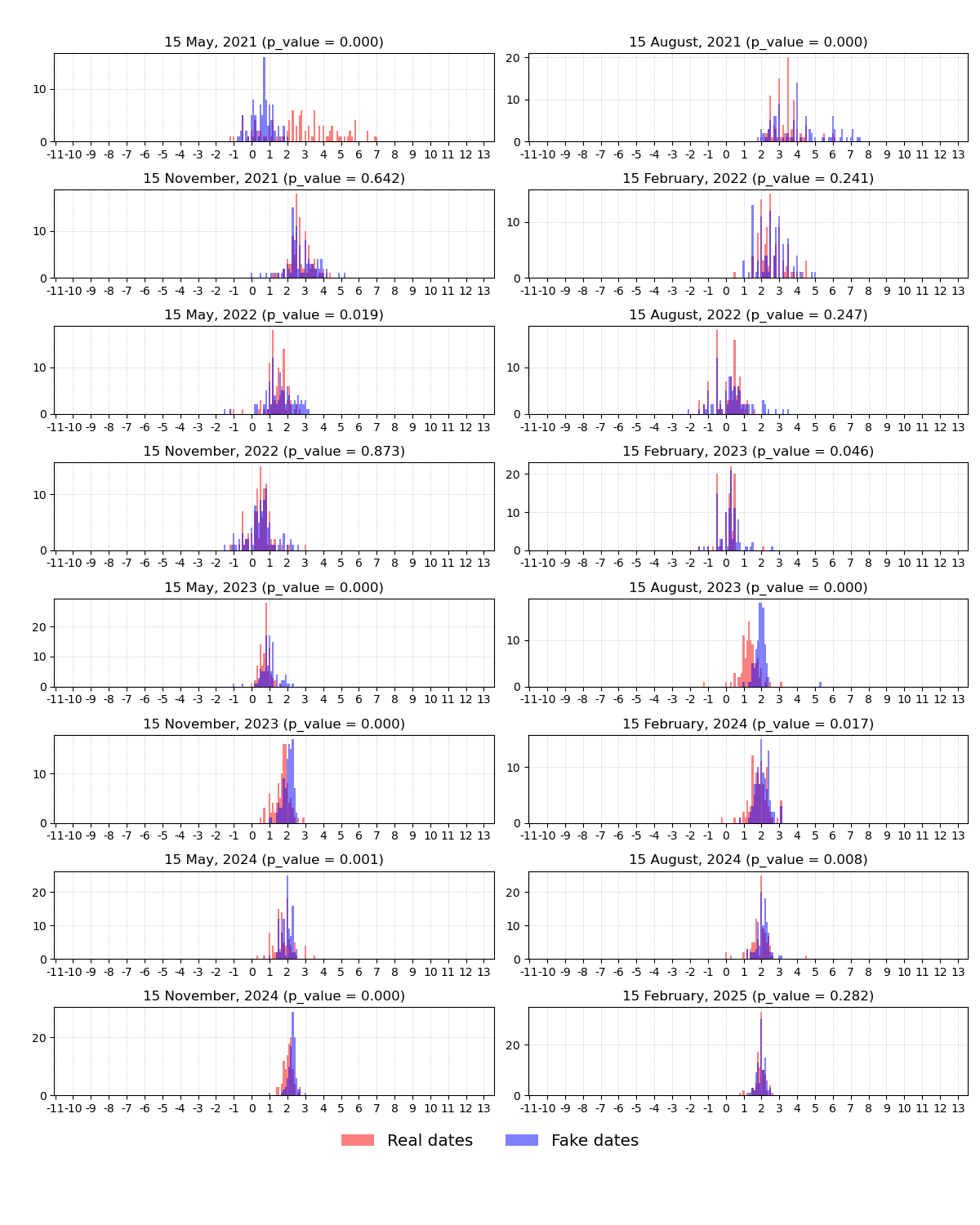}
	\caption{\textbf{Distributions of 1-year forecasts of real GDP growth (YoY) produced by the Kimi-K2 Instruct model.} The header shows the forecast date and the p-value for the Kolmogorov-Smirnov permutation test.}
    \label{fig: figure_B.1.3.5}
\end{figure}

\clearpage

\newgeometry{top=0.25cm,bottom=1.75cm,left=2cm,right=2cm,marginparwidth=2cm}

\subsection{Visualization of the distributions of Qwen3 Instruct forecasts}

\subsubsection{Fed interest rate}

\renewcommand{\thefigure}{B.2.1.\arabic{figure}}
\setcounter{figure}{0}

\begin{figure}[h!!]
	\centering \hspace*{-0.5cm}
	\includegraphics[scale=0.55]{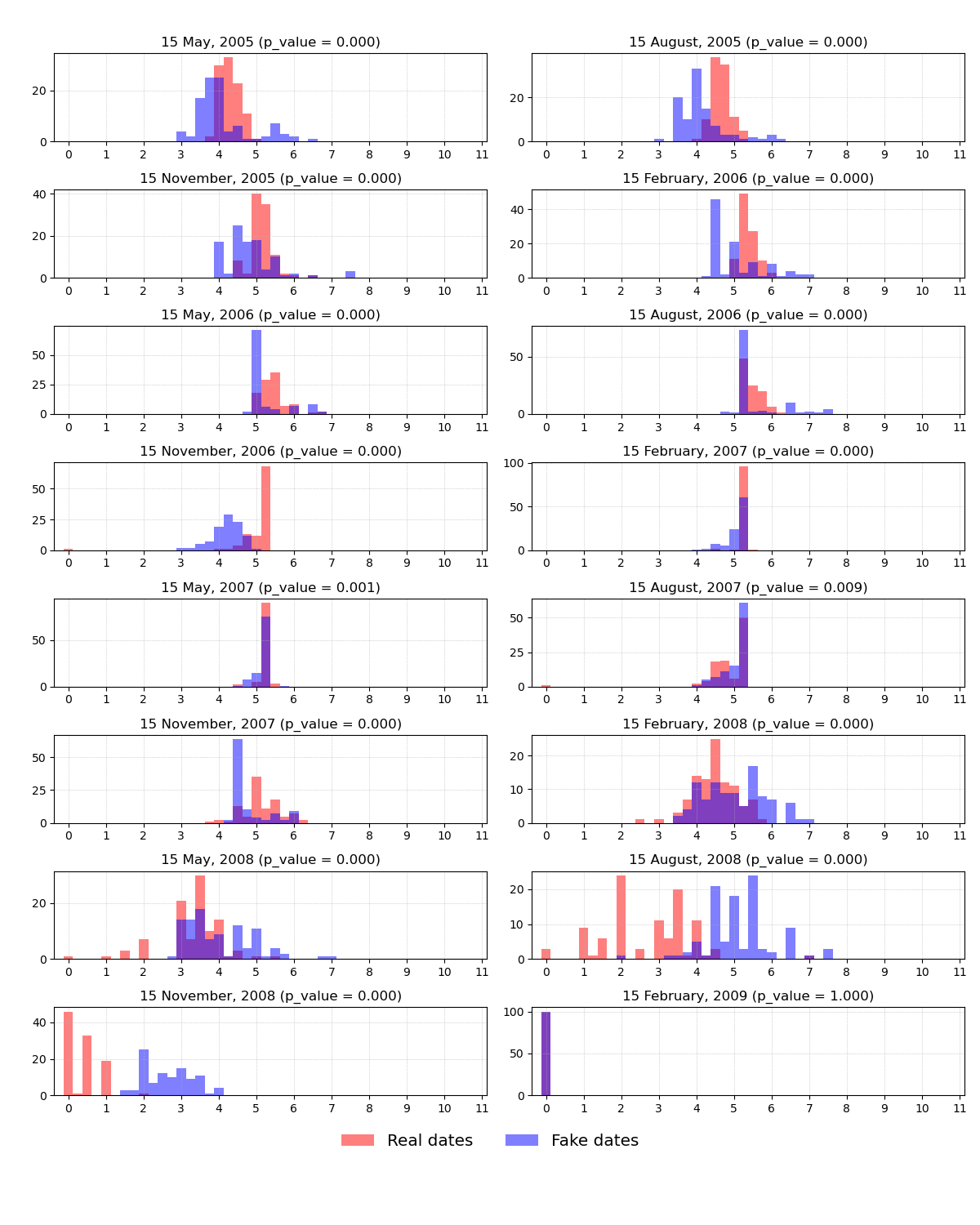}
	\caption{\textbf{Distributions of 1-year forecasts of the lower bound of the Fed interest rate produced by the Qwen3 Instruct model.} The header shows the forecast date and the p-value for the Kolmogorov-Smirnov permutation test.}
    \label{fig: figure_B.2.1.1}
\end{figure}

\clearpage

\restoregeometry

\begin{figure}[ht]
	\centering \hspace*{-0.5cm}
	\includegraphics[scale=0.55]{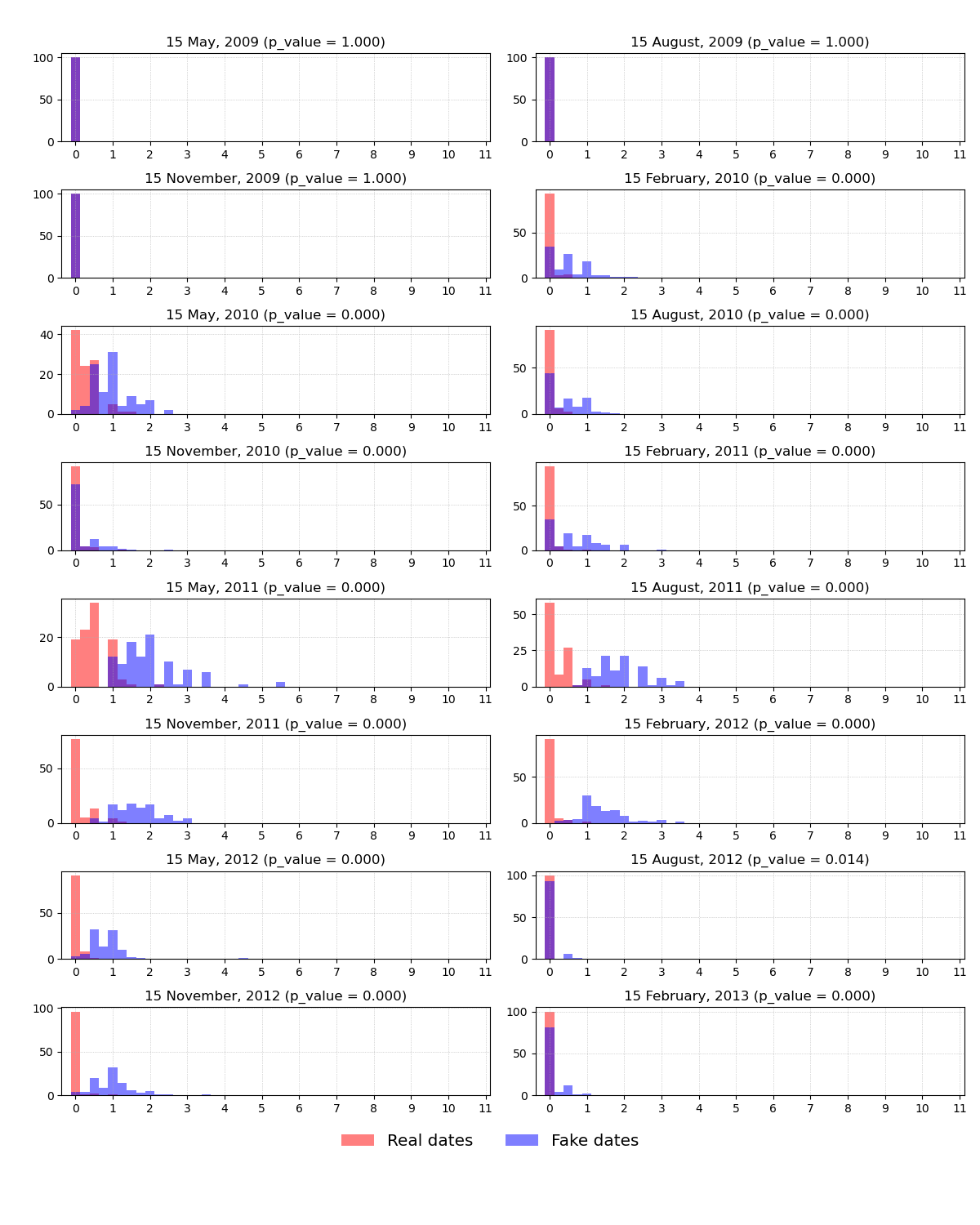}
	\caption{\textbf{Distributions of 1-year forecasts of the lower bound of the Fed interest rate produced by the Qwen3 Instruct model.} The header shows the forecast date and the p-value for the Kolmogorov-Smirnov permutation test.}
    \label{fig: figure_B.2.1.2}
\end{figure}

\begin{figure}[h!]
	\centering \hspace*{-0.5cm}
	\includegraphics[scale=0.55]{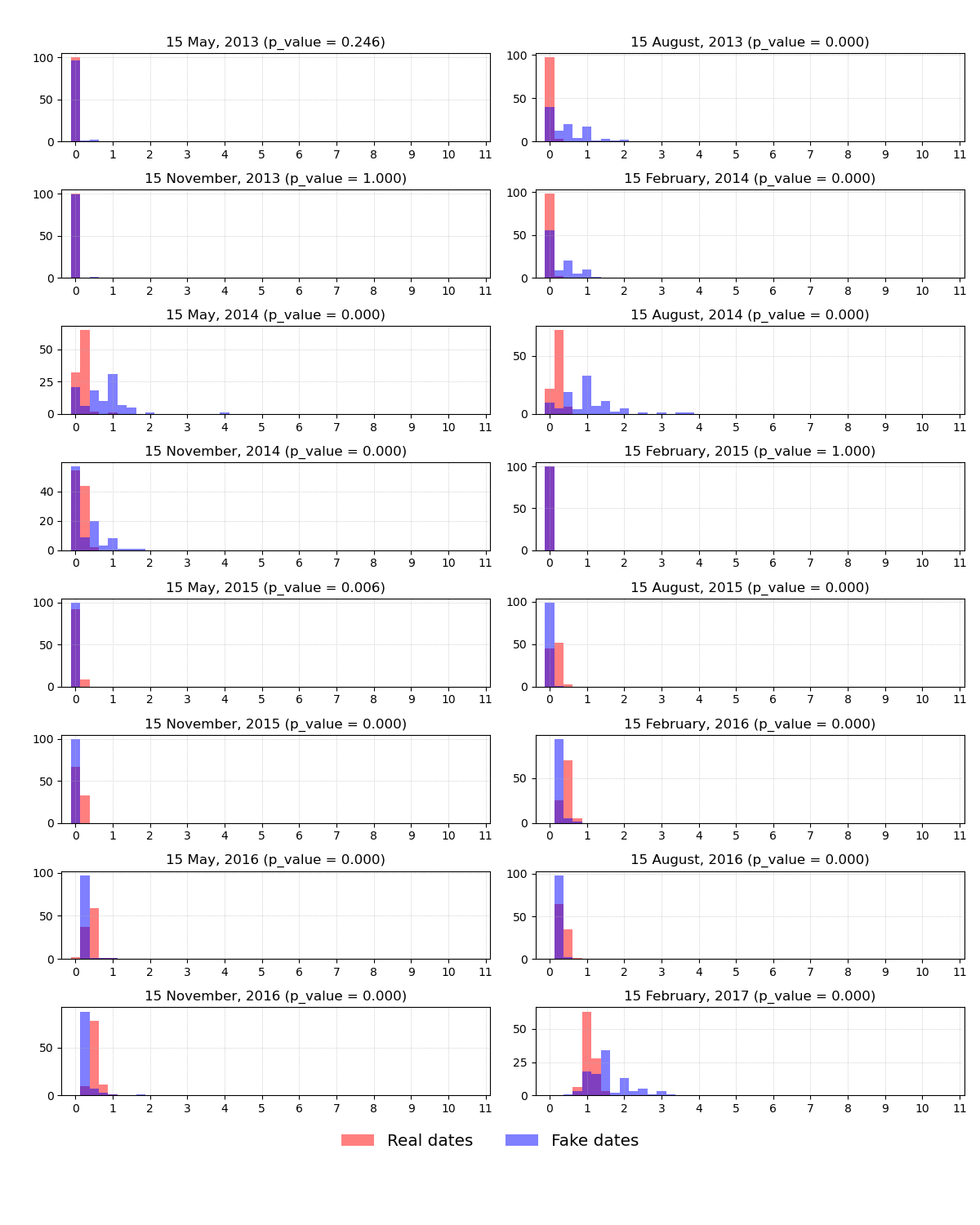}
	\caption{\textbf{Distributions of 1-year forecasts of the lower bound of the Fed interest rate produced by the Qwen3 Instruct model.} The header shows the forecast date and the p-value for the Kolmogorov-Smirnov permutation test.}
    \label{fig: figure_B.2.1.3}
\end{figure}

\begin{figure}[h!]
	\centering \hspace*{-0.5cm}
	\includegraphics[scale=0.55]{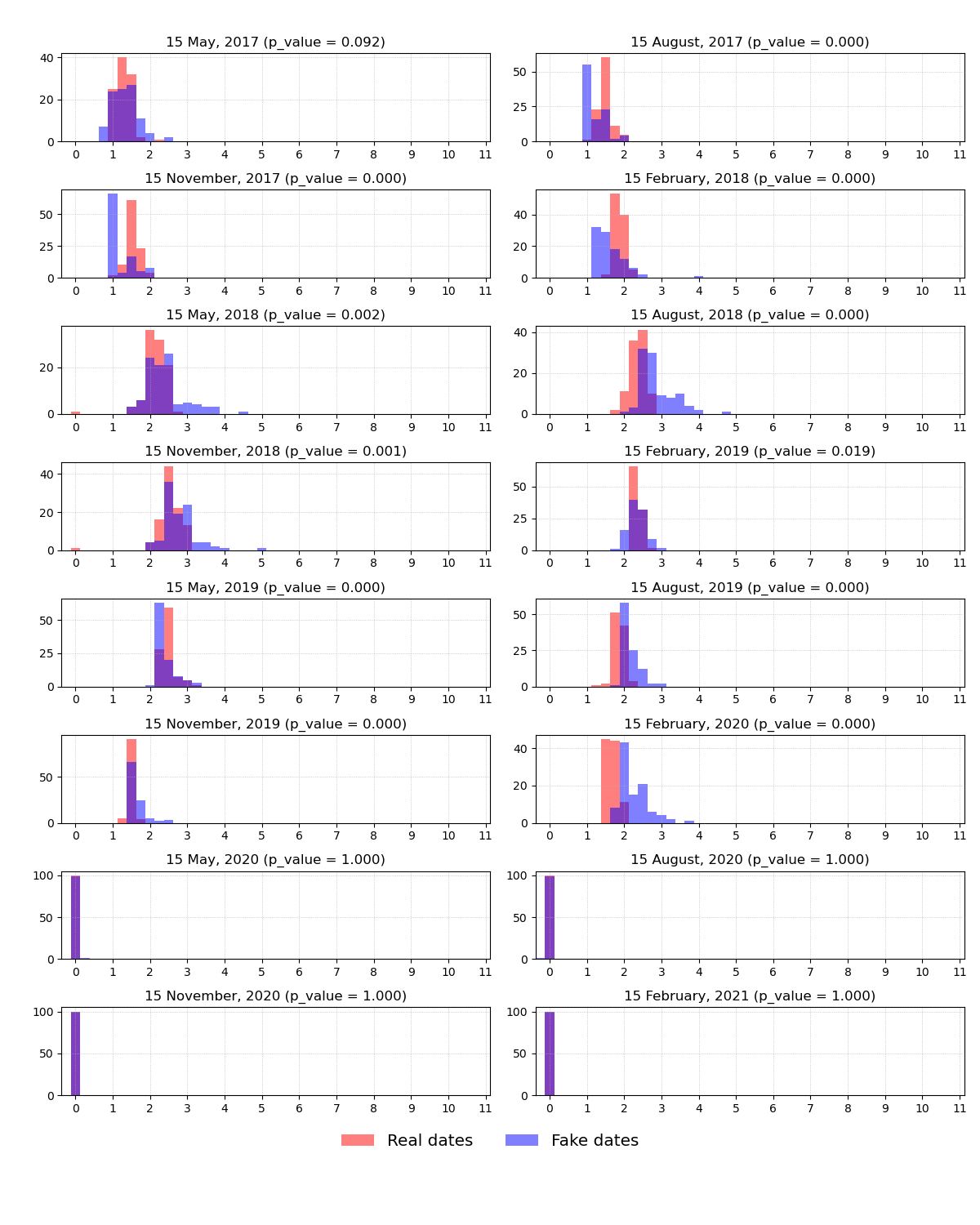}
	\caption{\textbf{Distributions of 1-year forecasts of the lower bound of the Fed interest rate produced by the Qwen3 Instruct model.} The header shows the forecast date and the p-value for the Kolmogorov-Smirnov permutation test.}
    \label{fig: figure_B.2.1.4}
\end{figure}

\begin{figure}[h!]
	\centering \hspace*{-0.5cm}
	\includegraphics[scale=0.55]{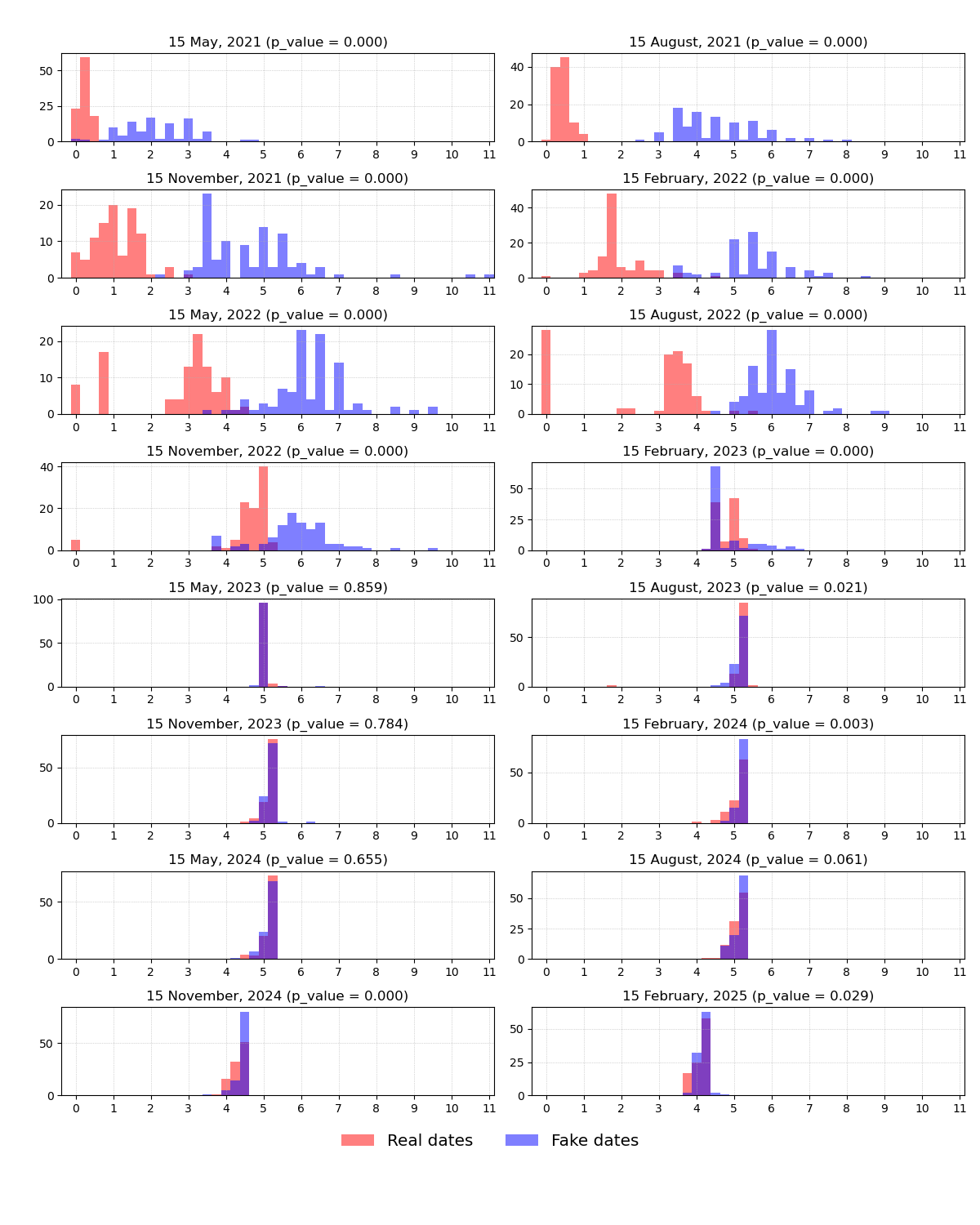}
	\caption{\textbf{Distributions of 1-year forecasts of the lower bound of the Fed interest rate produced by the Qwen3 Instruct model.} The header shows the forecast date and the p-value for the Kolmogorov-Smirnov permutation test.}
    \label{fig: figure_B.2.1.5}
\end{figure}

\clearpage

\newgeometry{top=1.85cm,bottom=2cm,left=2cm,right=2cm,marginparwidth=2cm}

\subsubsection{CPI growth (YoY)}

\renewcommand{\thefigure}{B.2.2.\arabic{figure}}
\setcounter{figure}{0}

\begin{figure}[h!]
	\centering \hspace*{-0.5cm}
	\includegraphics[scale=0.55]{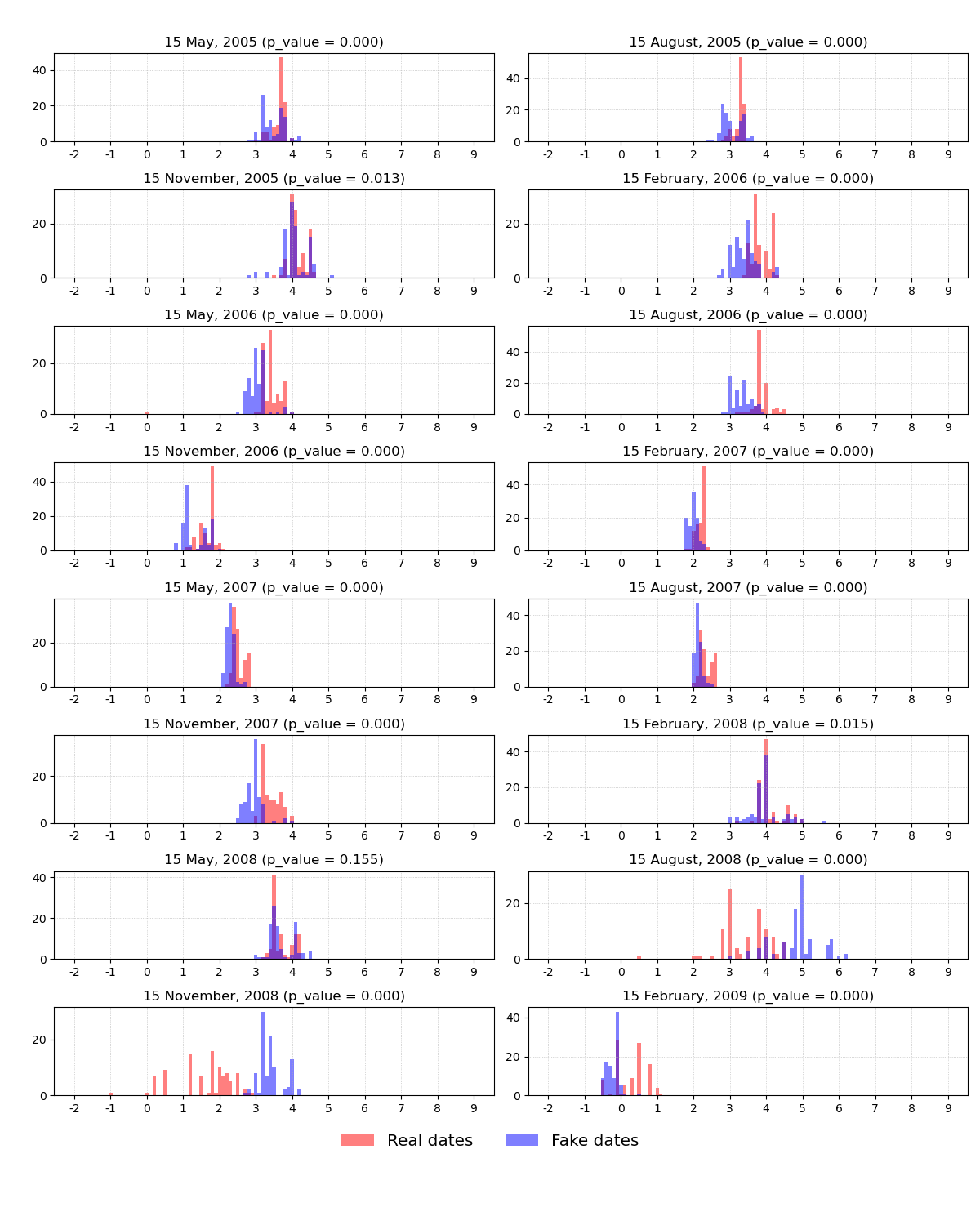}
	\caption{\textbf{Distributions of 1-year forecasts of CPI growth (YoY) produced by the Qwen3 Instruct model.} The header shows the forecast date and the p-value for the Kolmogorov-Smirnov permutation test.}
    \label{fig: figure_B.2.2.1}
\end{figure}

\clearpage

\restoregeometry

\begin{figure}[ht]
	\centering \hspace*{-0.5cm}
	\includegraphics[scale=0.55]{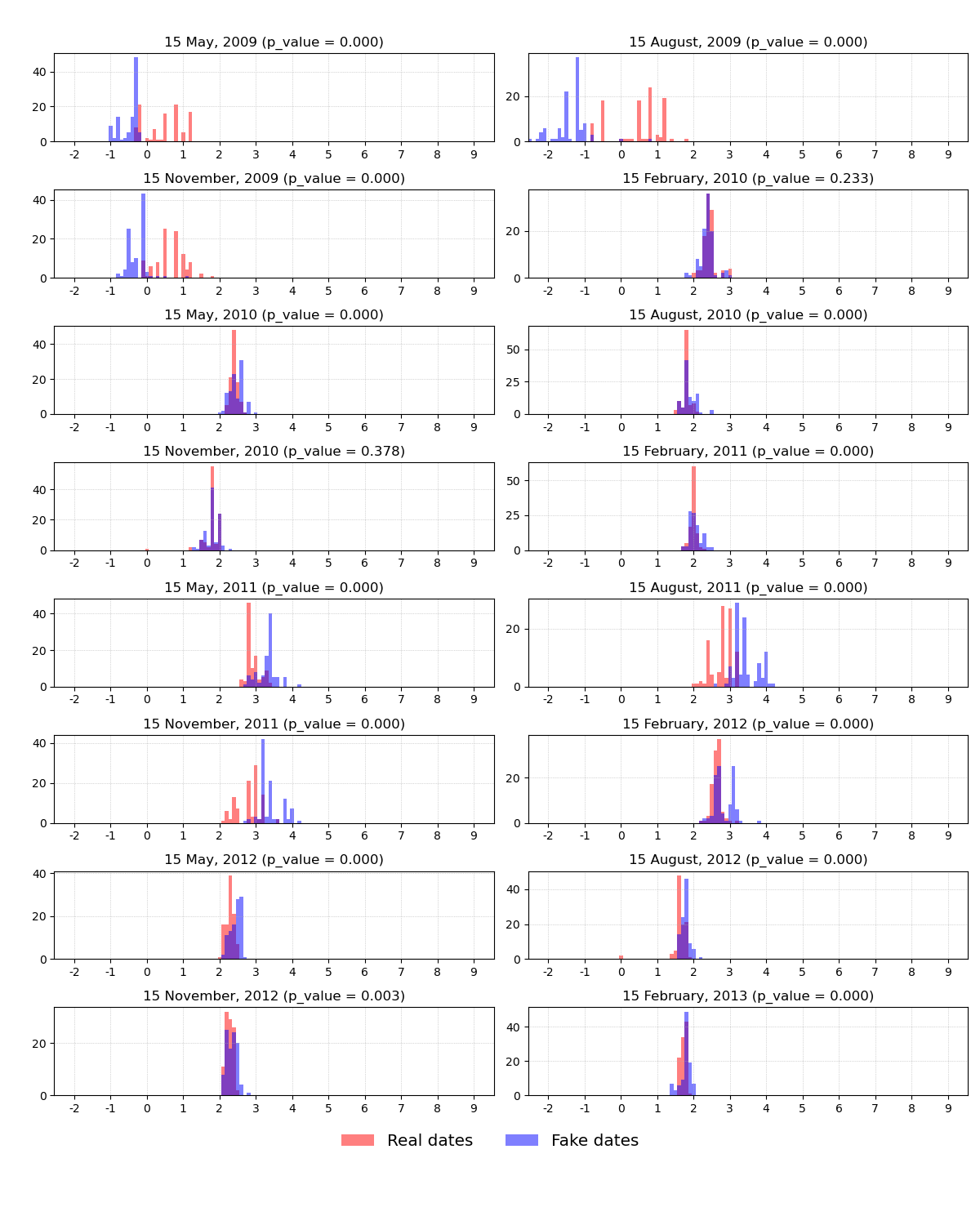}
	\caption{\textbf{Distributions of 1-year forecasts of CPI growth (YoY) produced by the Qwen3 Instruct model.} The header shows the forecast date and the p-value for the Kolmogorov-Smirnov permutation test.}
    \label{fig: figure_B.2.2.2}
\end{figure}

\begin{figure}[h!]
	\centering \hspace*{-0.5cm}
	\includegraphics[scale=0.55]{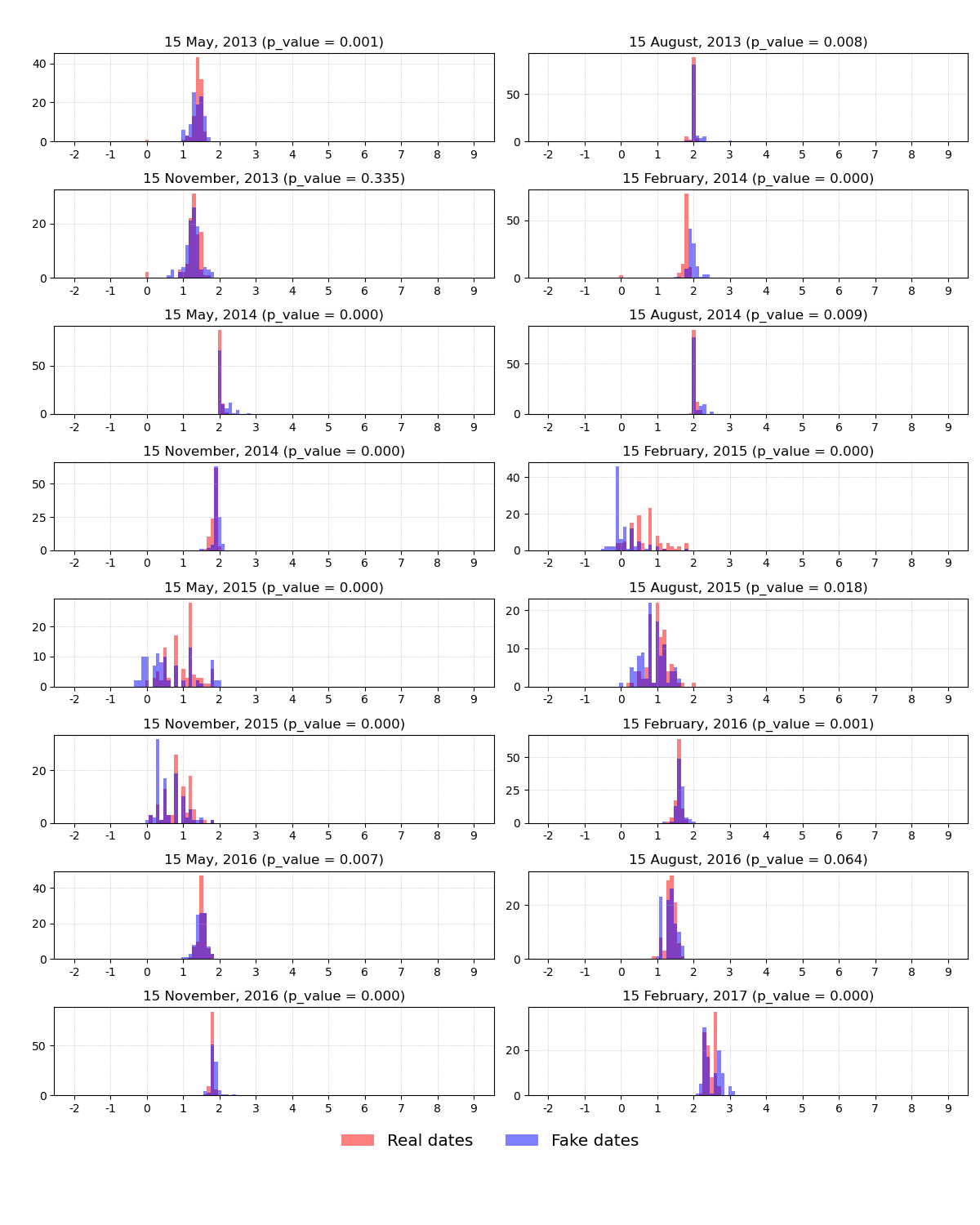}
	\caption{\textbf{Distributions of 1-year forecasts of CPI growth (YoY) produced by the Qwen3 Instruct model.} The header shows the forecast date and the p-value for the Kolmogorov-Smirnov permutation test.}
    \label{fig: figure_B.2.2.3}
\end{figure}

\begin{figure}[h!]
	\centering \hspace*{-0.5cm}
	\includegraphics[scale=0.55]{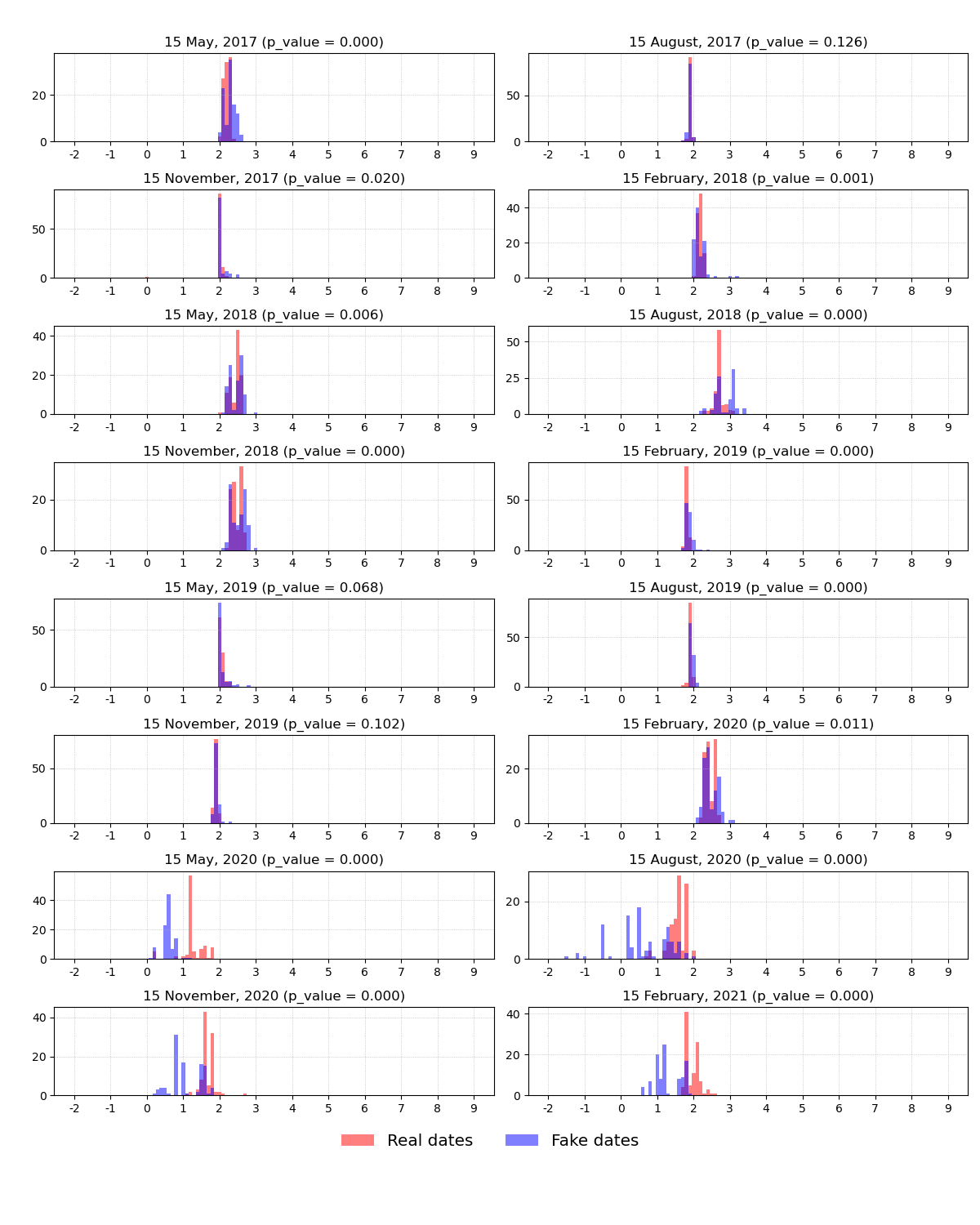}
	\caption{\textbf{Distributions of 1-year forecasts of CPI growth (YoY) produced by the Qwen3 Instruct model.} The header shows the forecast date and the p-value for the Kolmogorov-Smirnov permutation test.}
    \label{fig: figure_B.2.2.4}
\end{figure}

\begin{figure}[h!]
	\centering \hspace*{-0.5cm}
	\includegraphics[scale=0.55]{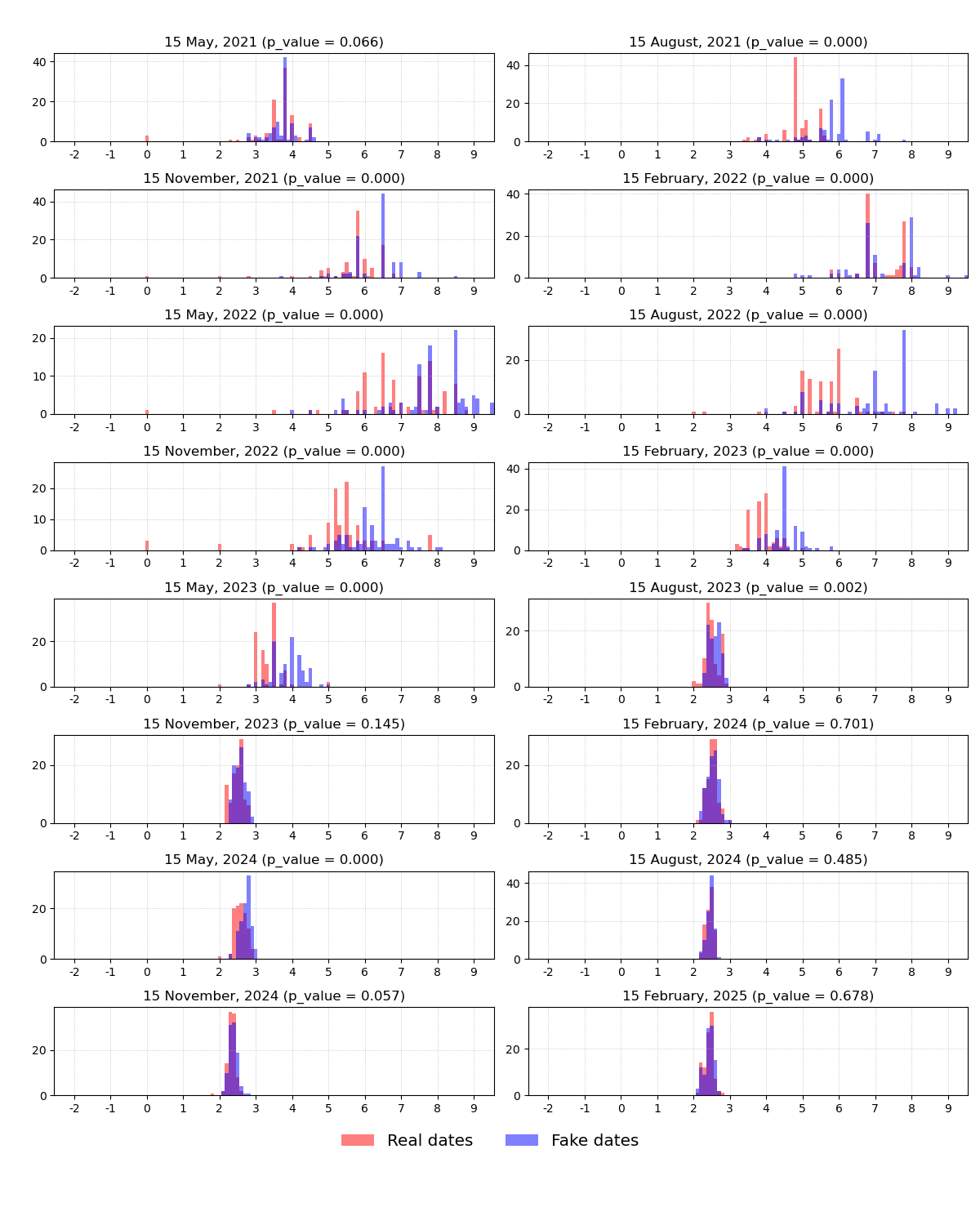}
	\caption{\textbf{Distributions of 1-year forecasts of CPI growth (YoY) produced by the Qwen3 Instruct model.} The header shows the forecast date and the p-value for the Kolmogorov-Smirnov permutation test.}
    \label{fig: figure_B.2.2.5}
\end{figure}

\clearpage

\newgeometry{top=1.85cm,bottom=2cm,left=2cm,right=2cm,marginparwidth=2cm}

\subsubsection{Real GDP growth (YoY)}

\renewcommand{\thefigure}{B.2.3.\arabic{figure}}
\setcounter{figure}{0}

\begin{figure}[h!]
	\centering \hspace*{-0.5cm}
	\includegraphics[scale=0.55]{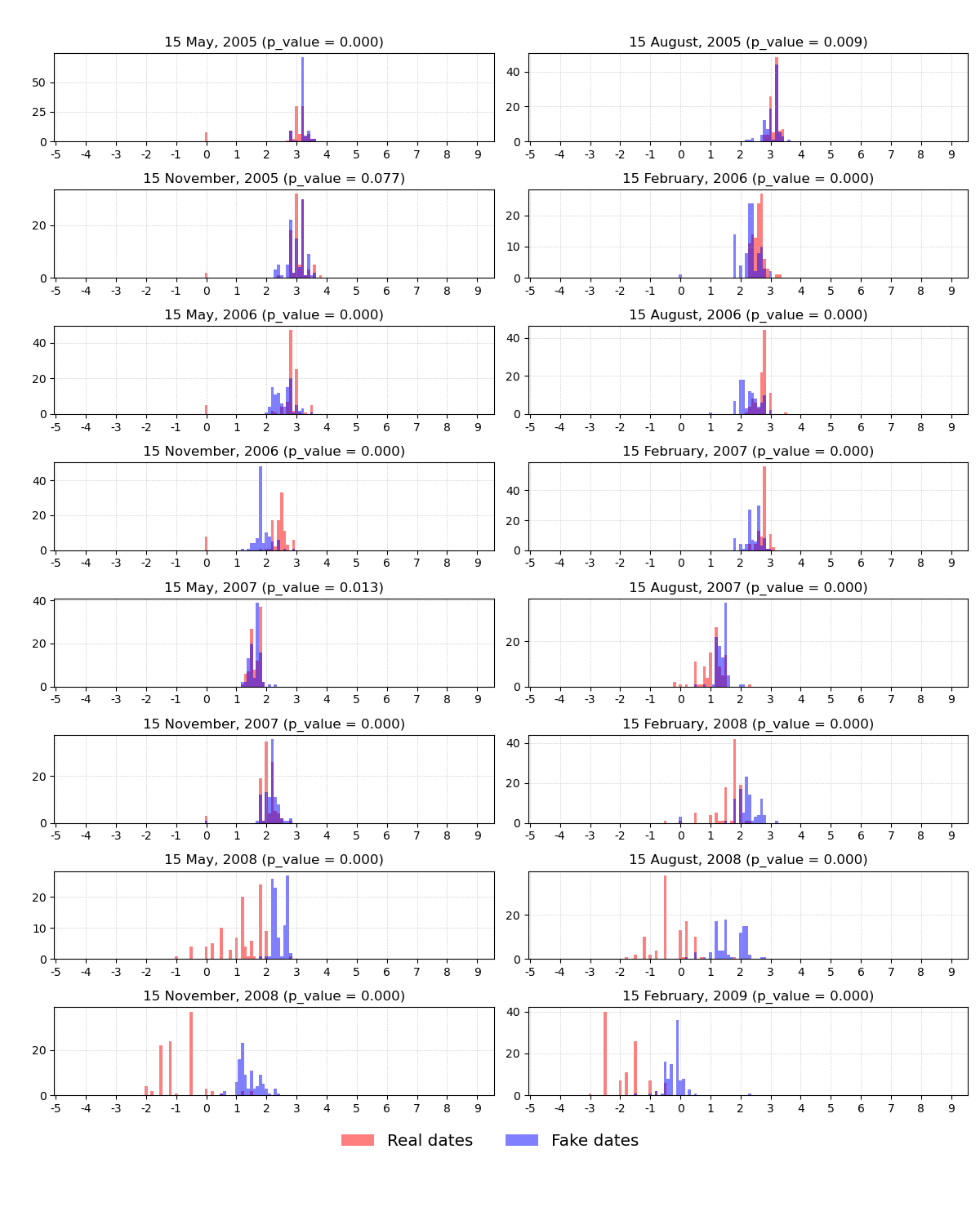}
	\caption{\textbf{Distributions of 1-year forecasts of real GDP growth (YoY) produced by the Qwen3 Instruct model.} The header shows the forecast date and the p-value for the Kolmogorov-Smirnov permutation test.}
    \label{fig: figure_B.2.3.1}
\end{figure}

\clearpage

\restoregeometry

\begin{figure}[ht]
	\centering \hspace*{-0.5cm}
	\includegraphics[scale=0.55]{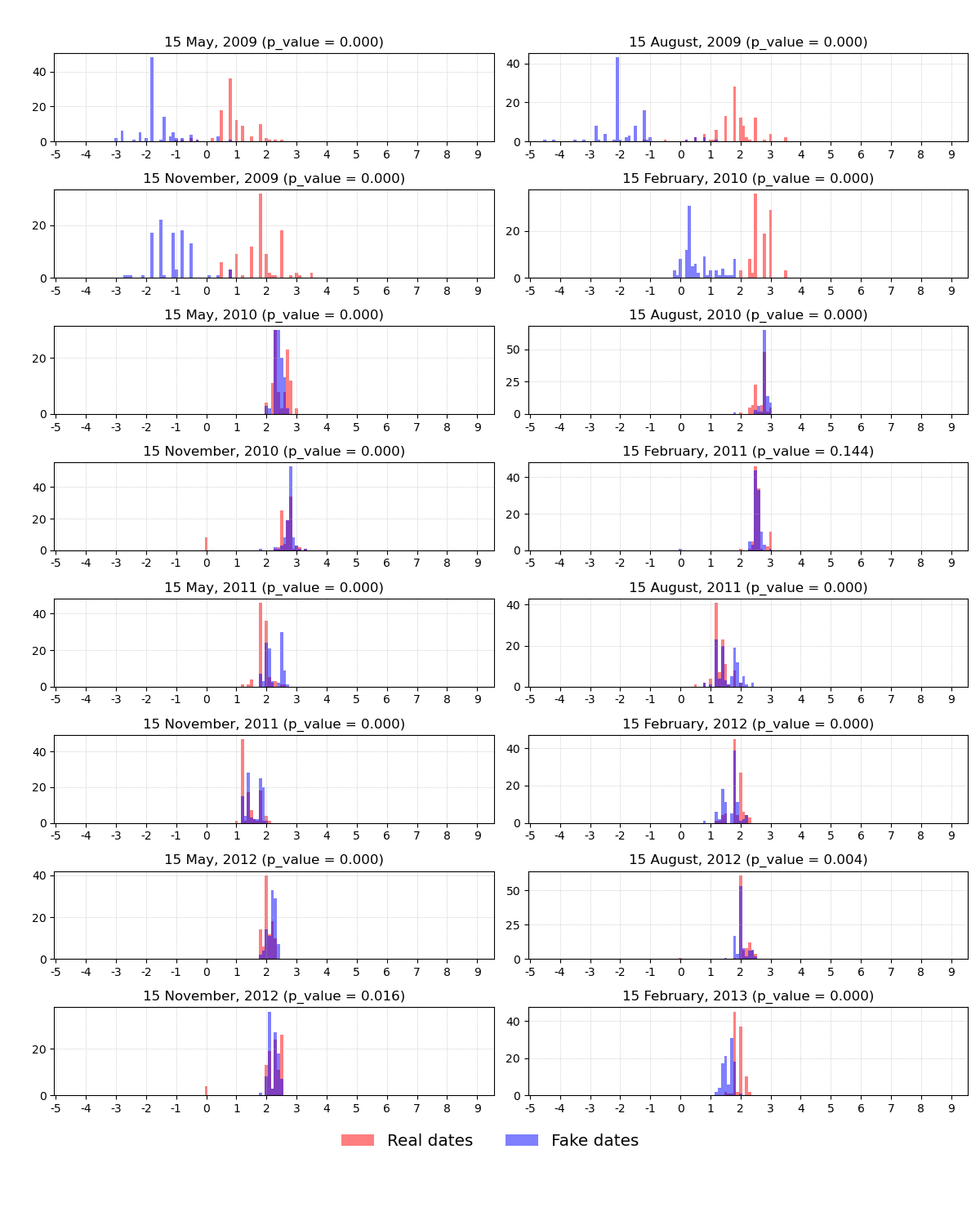}
	\caption{\textbf{Distributions of 1-year forecasts of real GDP growth (YoY) produced by the Qwen3 Instruct model.} The header shows the forecast date and the p-value for the Kolmogorov-Smirnov permutation test.}
    \label{fig: figure_B.2.3.2}
\end{figure}

\begin{figure}[h!]
	\centering \hspace*{-0.5cm}
	\includegraphics[scale=0.55]{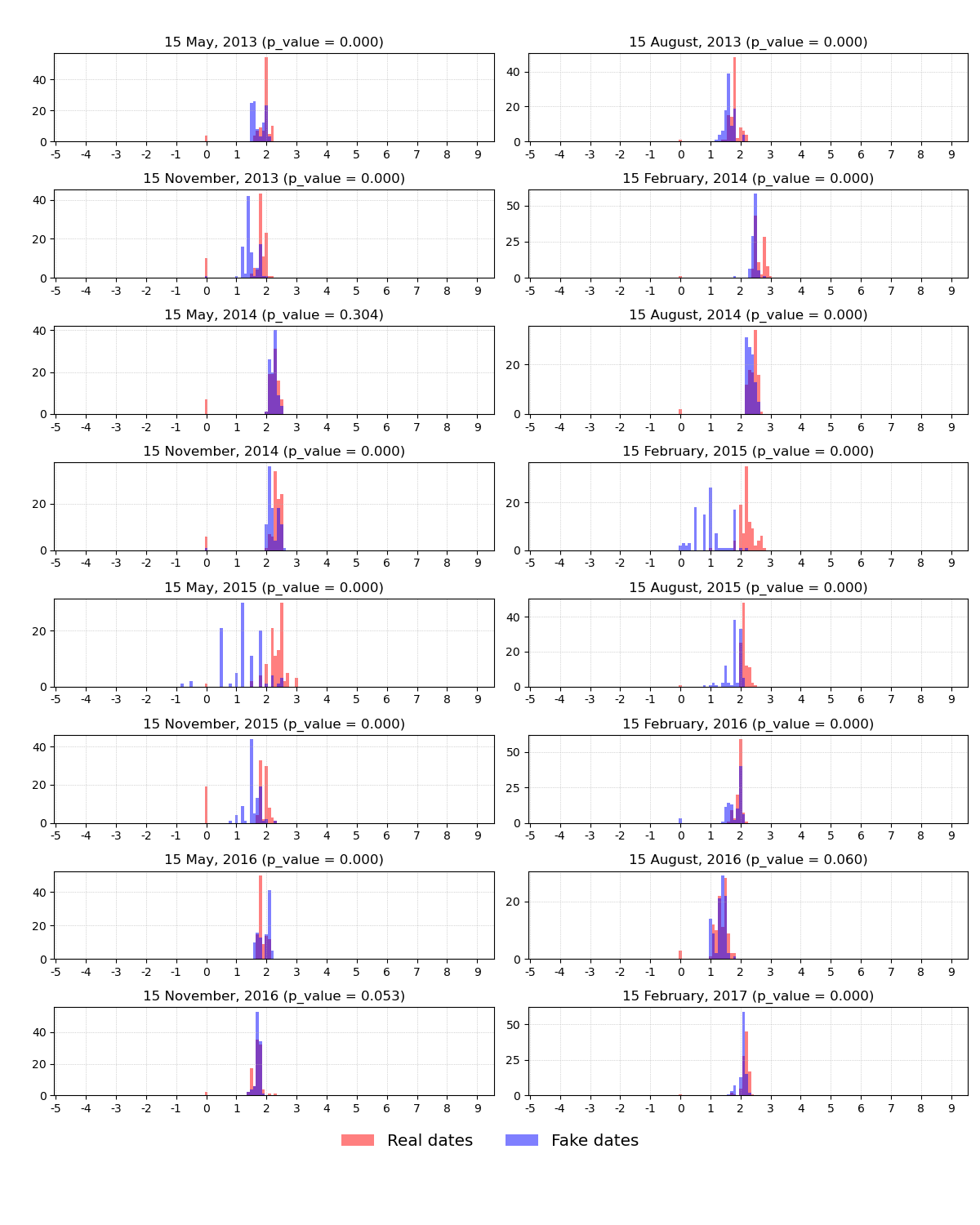}
	\caption{\textbf{Distributions of 1-year forecasts of real GDP growth (YoY) produced by the Qwen3 Instruct model.} The header shows the forecast date and the p-value for the Kolmogorov-Smirnov permutation test.}
    \label{fig: figure_B.2.3.3}
\end{figure}

\begin{figure}[h!]
	\centering \hspace*{-0.5cm}
	\includegraphics[scale=0.55]{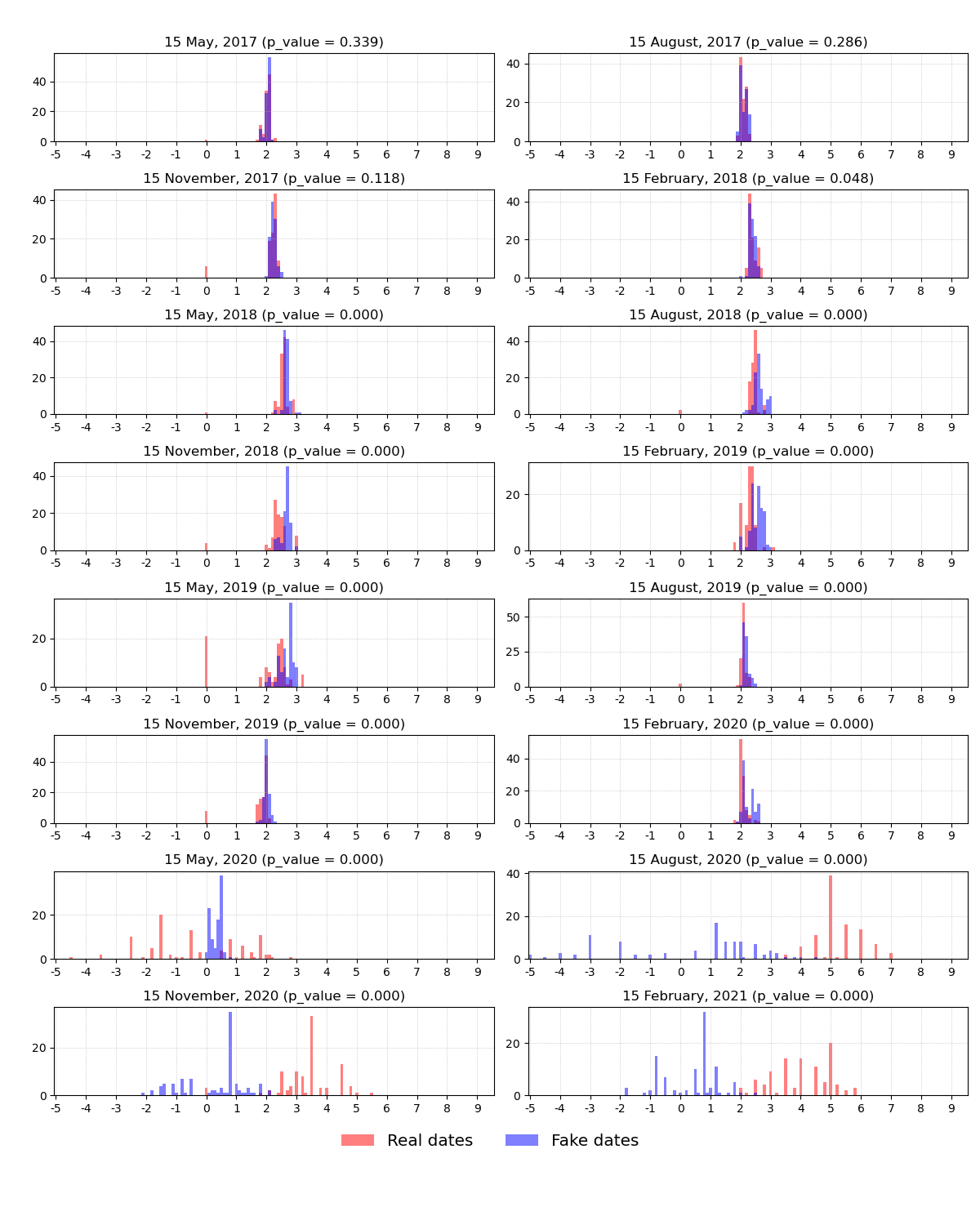}
	\caption{\textbf{Distributions of 1-year forecasts of real GDP growth (YoY) produced by the Qwen3 Instruct model.} The header shows the forecast date and the p-value for the Kolmogorov-Smirnov permutation test.}
    \label{fig: figure_B.2.3.4}
\end{figure}

\begin{figure}[h!]
	\centering \hspace*{-0.5cm}
	\includegraphics[scale=0.55]{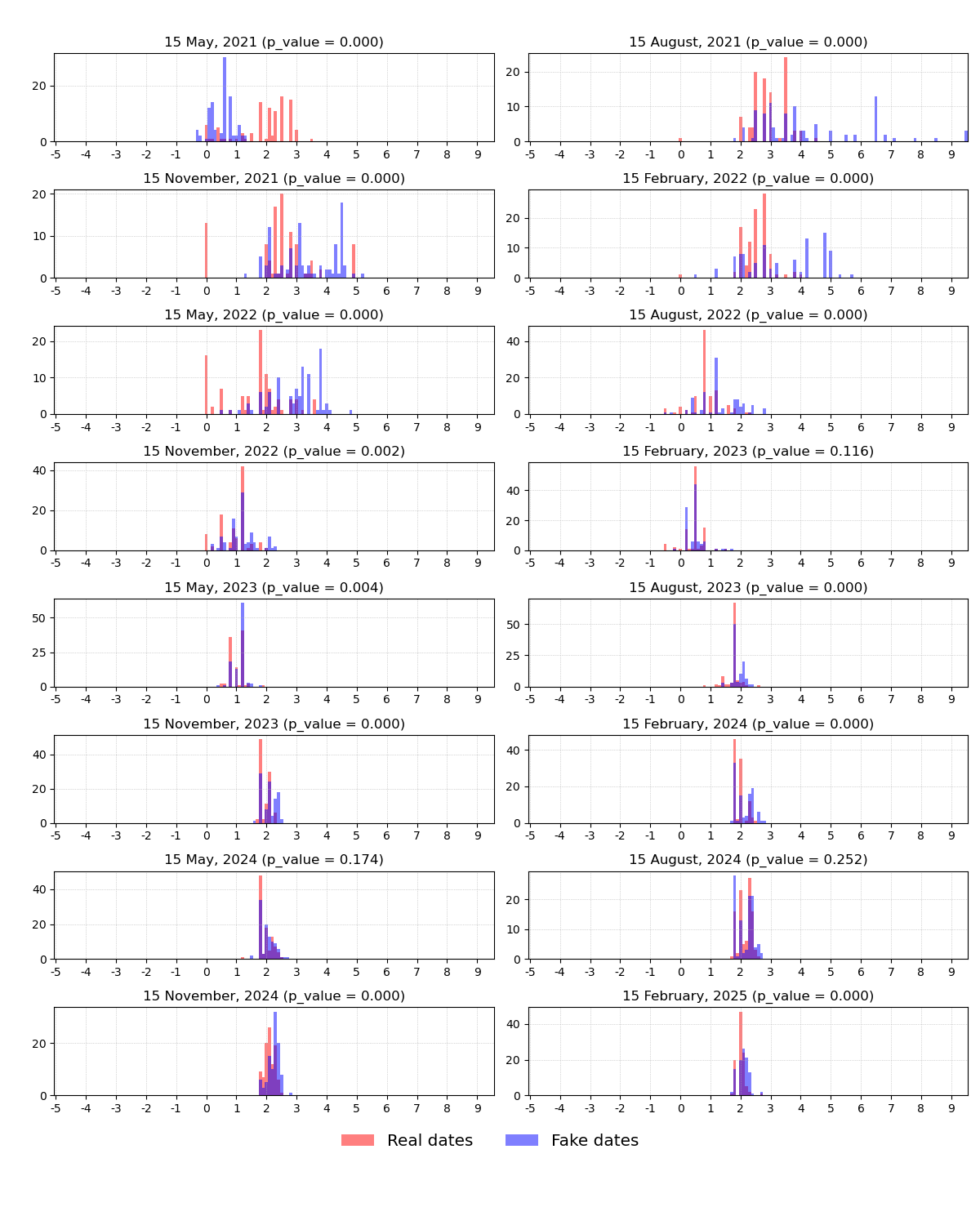}
	\caption{\textbf{Distributions of 1-year forecasts of real GDP growth (YoY) produced by the Qwen3 Instruct model.} The header shows the forecast date and the p-value for the Kolmogorov-Smirnov permutation test.}
    \label{fig: figure_B.2.3.5}
\end{figure}

\clearpage

\newgeometry{top=0.25cm,bottom=1.75cm,left=2cm,right=2cm,marginparwidth=2cm}

\subsection{Visualization of the distributions of DeepSeek-V3.1 forecasts}

\subsubsection{Fed interest rate}

\renewcommand{\thefigure}{B.3.1.\arabic{figure}}
\setcounter{figure}{0}

\begin{figure}[h!!]
	\centering \hspace*{-0.55cm}
	\includegraphics[scale=0.55]{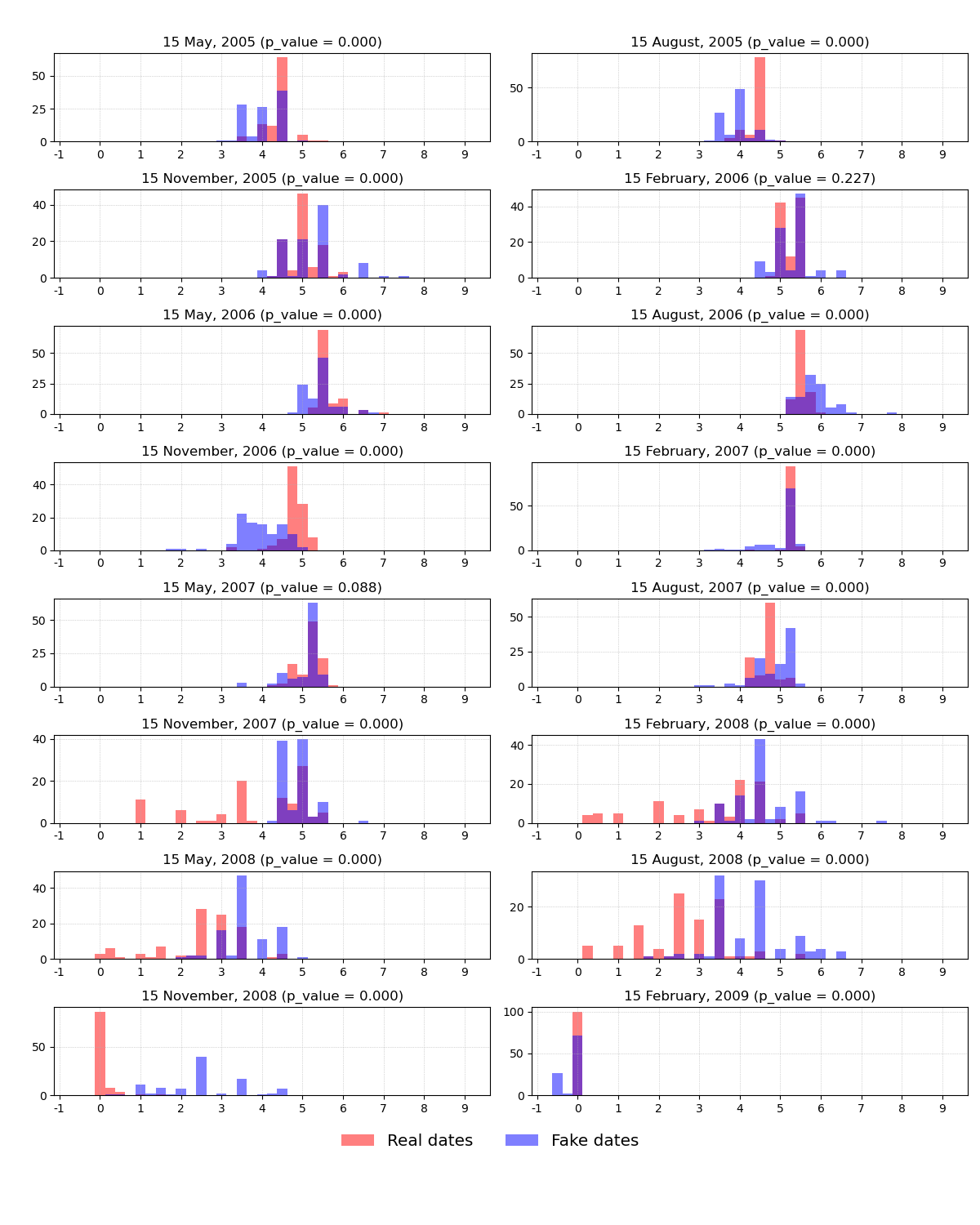}
	\caption{\textbf{Distributions of 1-year forecasts of the lower bound of the Fed interest rate produced by the DeepSeek-V3.1 model.} The header shows the forecast date and the p-value for the Kolmogorov-Smirnov permutation test.}
    \label{fig: figure_B.3.1.1}
\end{figure}

\clearpage

\restoregeometry

\begin{figure}[ht]
	\centering \hspace*{-0.5cm}
	\includegraphics[scale=0.55]{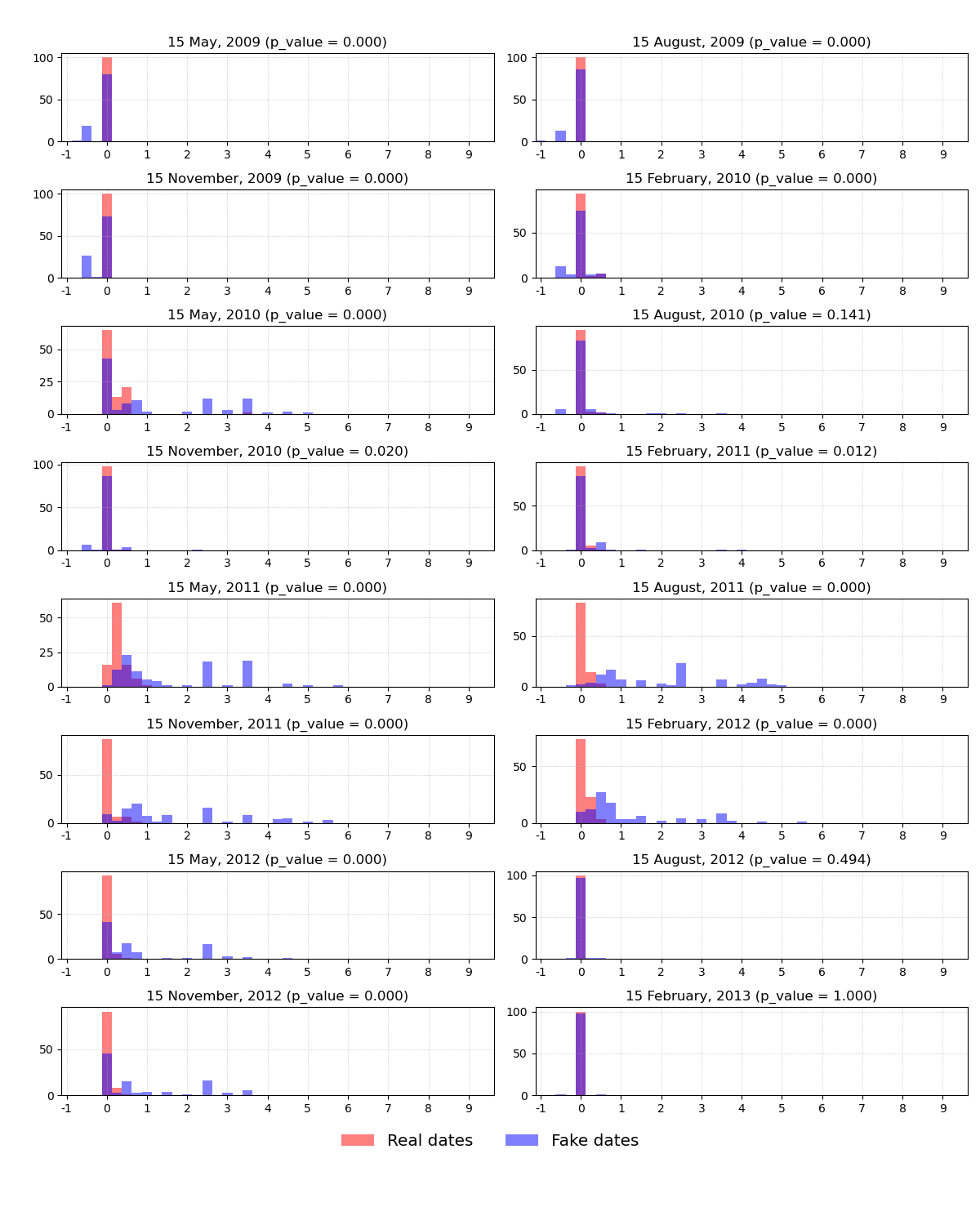}
	\caption{\textbf{Distributions of 1-year forecasts of the lower bound of the Fed interest rate produced by the DeepSeek-V3.1 model.} The header shows the forecast date and the p-value for the Kolmogorov-Smirnov permutation test.}
    \label{fig: figure_B.3.1.2}
\end{figure}

\begin{figure}[h!]
	\centering \hspace*{-0.5cm}
	\includegraphics[scale=0.55]{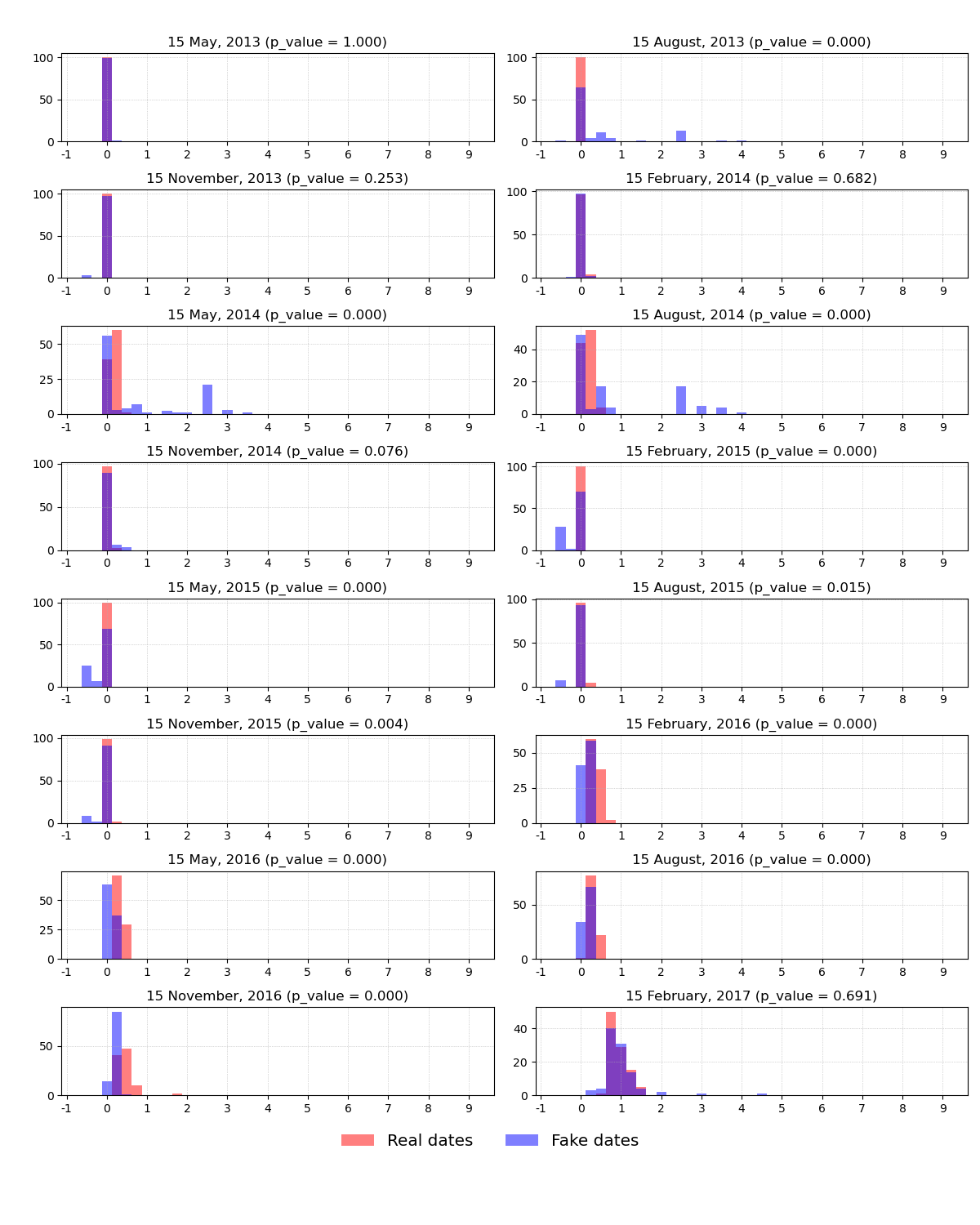}
	\caption{\textbf{Distributions of 1-year forecasts of the lower bound of the Fed interest rate produced by the DeepSeek-V3.1 model.} The header shows the forecast date and the p-value for the Kolmogorov-Smirnov permutation test.}
    \label{fig: figure_B.3.1.3}
\end{figure}

\begin{figure}[h!]
	\centering \hspace*{-0.5cm}
	\includegraphics[scale=0.55]{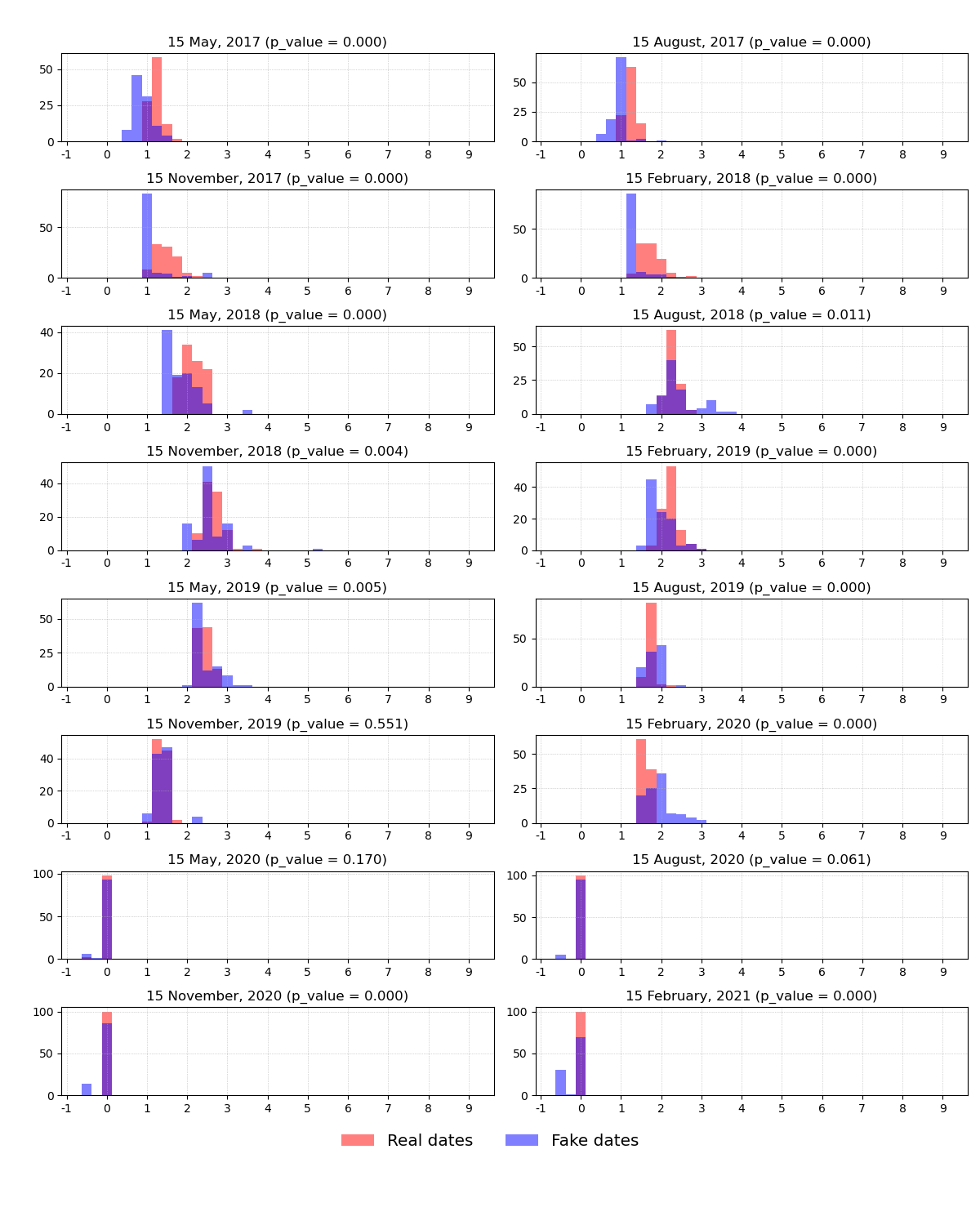}
	\caption{\textbf{Distributions of 1-year forecasts of the lower bound of the Fed interest rate produced by the DeepSeek-V3.1 model.} The header shows the forecast date and the p-value for the Kolmogorov-Smirnov permutation test.}
    \label{fig: figure_B.3.1.4}
\end{figure}

\begin{figure}[h!]
	\centering \hspace*{-0.5cm}
	\includegraphics[scale=0.55]{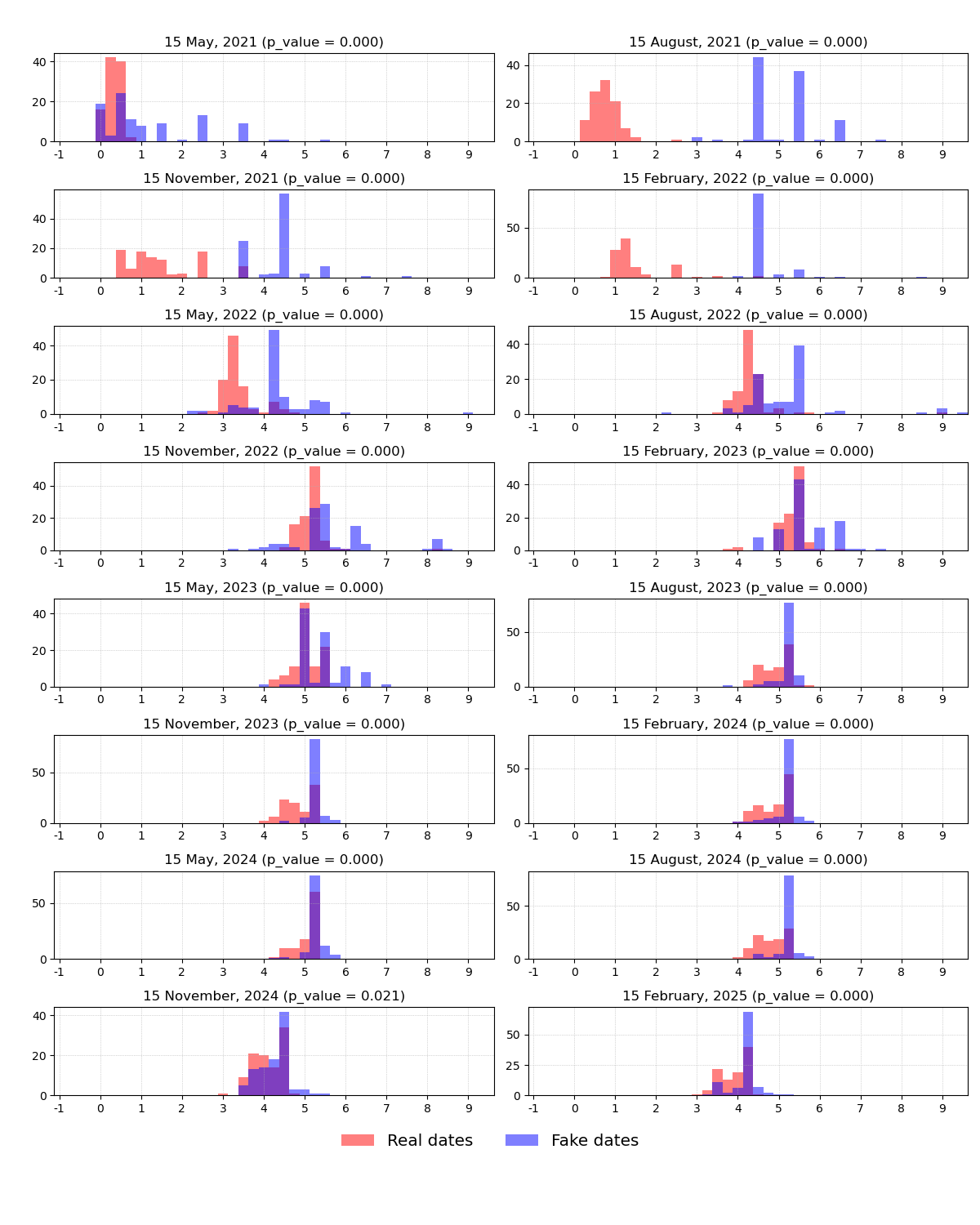}
	\caption{\textbf{Distributions of 1-year forecasts of the lower bound of the Fed interest rate produced by the DeepSeek-V3.1 model.} The header shows the forecast date and the p-value for the Kolmogorov-Smirnov permutation test.}
    \label{fig: figure_B.3.1.5}
\end{figure}

\clearpage

\newgeometry{top=1.85cm,bottom=2cm,left=2cm,right=2cm,marginparwidth=2cm}

\subsubsection{CPI growth (YoY)}

\renewcommand{\thefigure}{B.3.2.\arabic{figure}}
\setcounter{figure}{0}

\begin{figure}[h!]
	\centering \hspace*{-0.5cm}
	\includegraphics[scale=0.55]{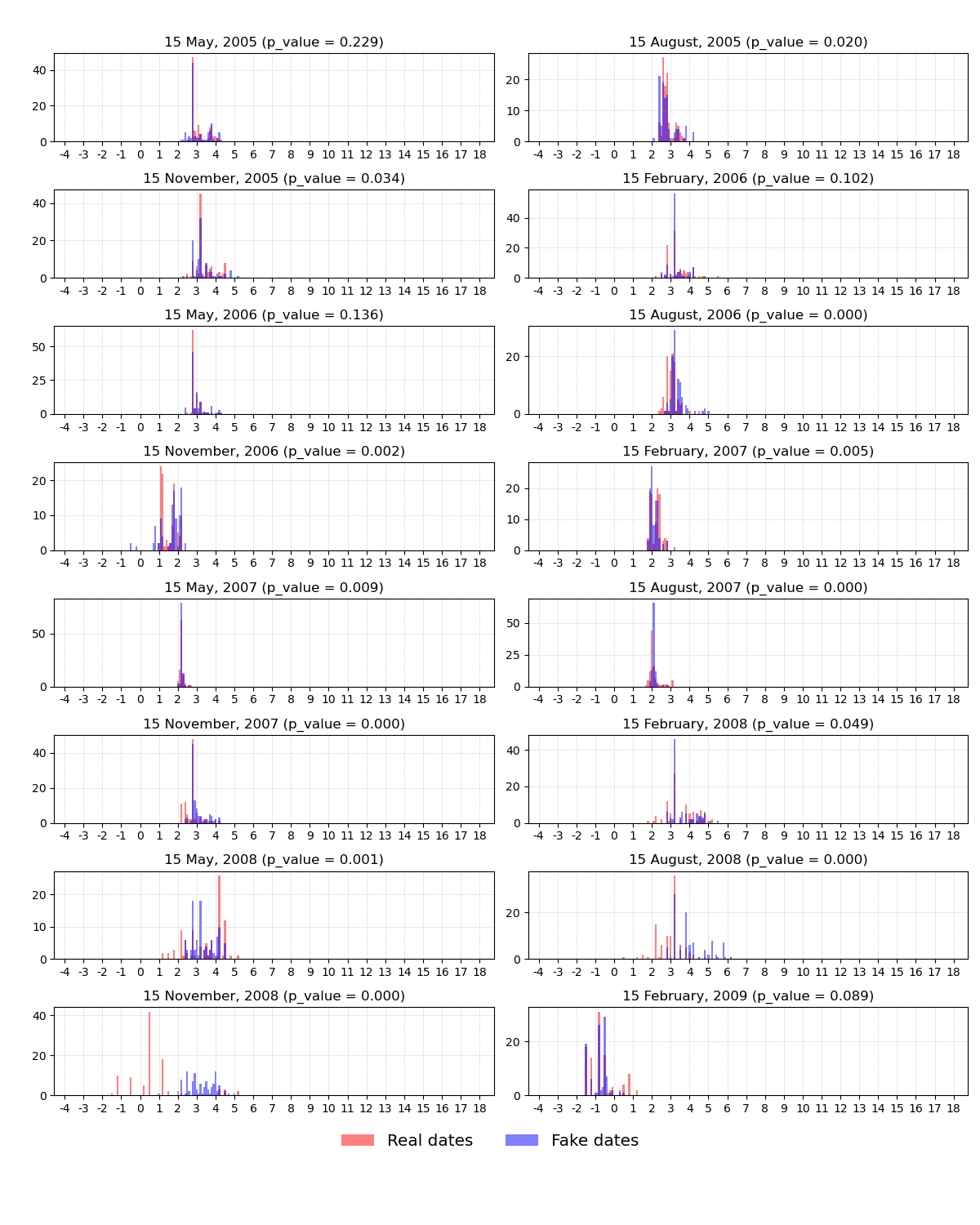}
	\caption{\textbf{Distributions of 1-year forecasts of CPI growth (YoY) produced by the DeepSeek-V3.1 model.} The header shows the forecast date and the p-value for the Kolmogorov-Smirnov permutation test.}
    \label{fig: figure_B.3.2.1}
\end{figure}

\clearpage

\restoregeometry

\begin{figure}[ht]
	\centering \hspace*{-0.5cm}
	\includegraphics[scale=0.55]{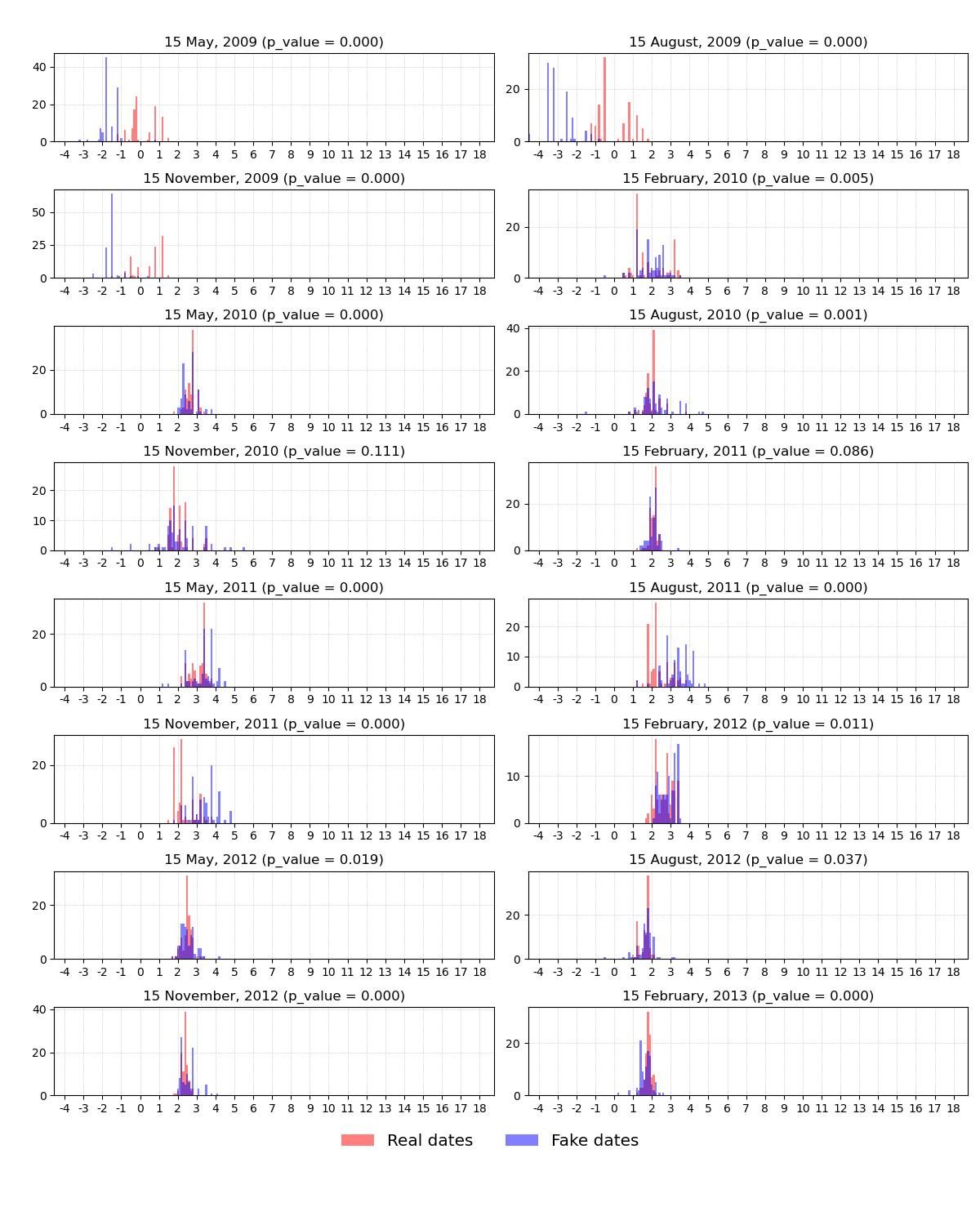}
	\caption{\textbf{Distributions of 1-year forecasts of CPI growth (YoY) produced by the DeepSeek-V3.1 model.} The header shows the forecast date and the p-value for the Kolmogorov-Smirnov permutation test.}
    \label{fig: figure_B.3.2.2}
\end{figure}

\begin{figure}[h!]
	\centering \hspace*{-0.5cm}
	\includegraphics[scale=0.55]{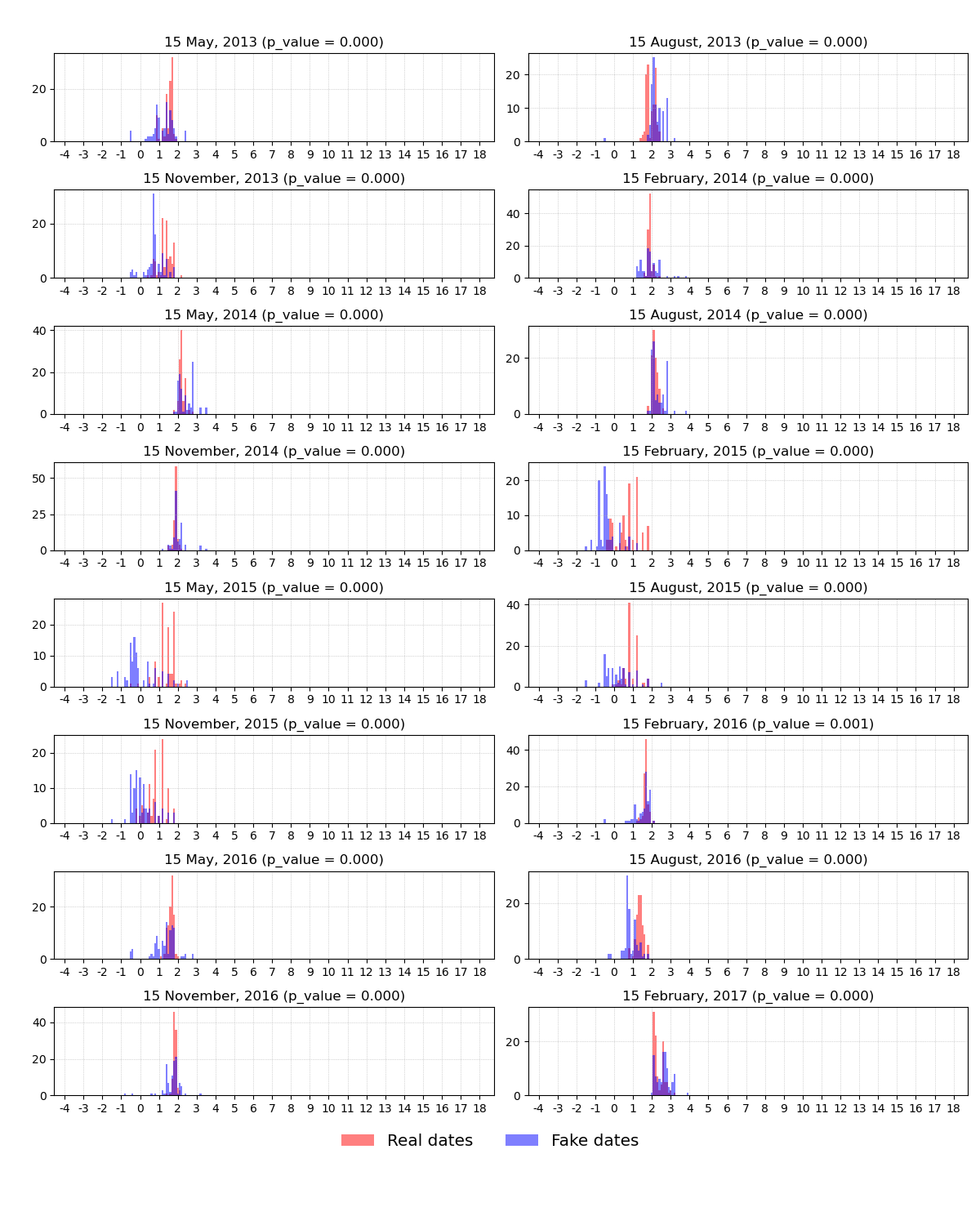}
	\caption{\textbf{Distributions of 1-year forecasts of CPI growth (YoY) produced by the DeepSeek-V3.1 model.} The header shows the forecast date and the p-value for the Kolmogorov-Smirnov permutation test.}
    \label{fig: figure_B.3.2.3}
\end{figure}

\begin{figure}[h!]
	\centering \hspace*{-0.5cm}
	\includegraphics[scale=0.55]{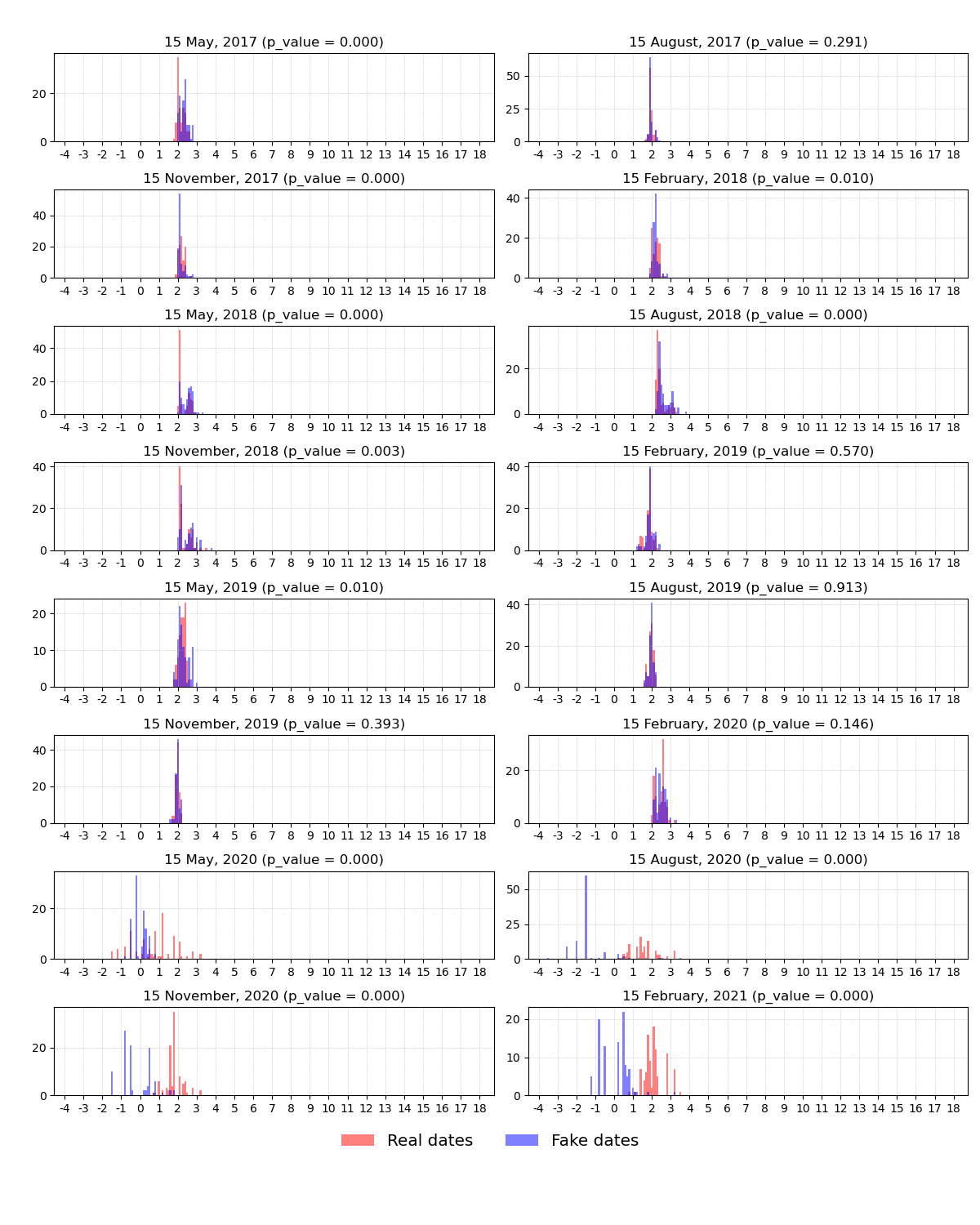}
	\caption{\textbf{Distributions of 1-year forecasts of CPI growth (YoY) produced by the DeepSeek-V3.1 model.} The header shows the forecast date and the p-value for the Kolmogorov-Smirnov permutation test.}
    \label{fig: figure_B.3.2.4}
\end{figure}

\begin{figure}[h!]
	\centering \hspace*{-0.5cm}
	\includegraphics[scale=0.55]{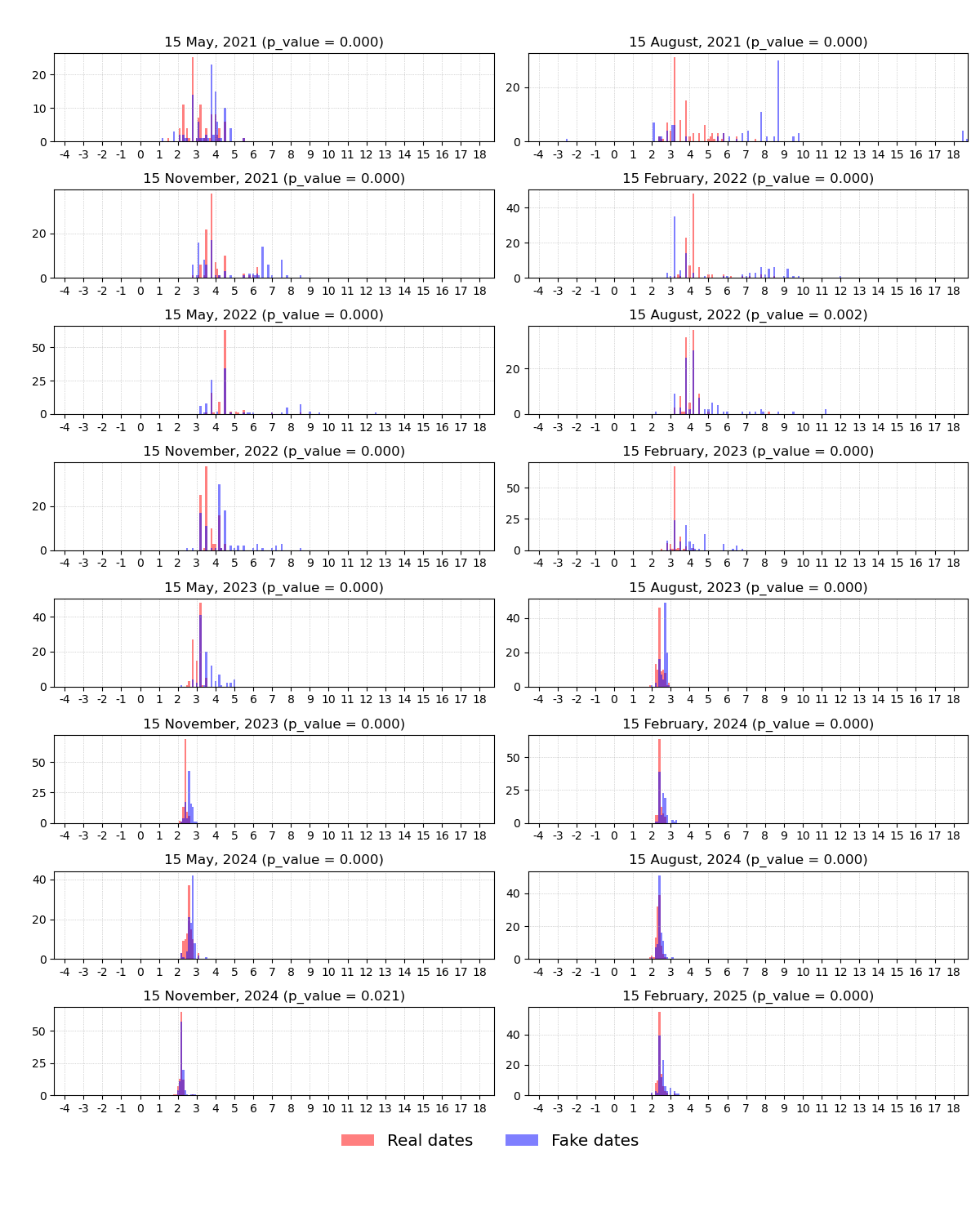}
	\caption{\textbf{Distributions of 1-year forecasts of CPI growth (YoY) produced by the DeepSeek-V3.1 model.} The header shows the forecast date and the p-value for the Kolmogorov-Smirnov permutation test.}
    \label{fig: figure_B.3.2.5}
\end{figure}

\clearpage

\newgeometry{top=1.85cm,bottom=2cm,left=2cm,right=2cm,marginparwidth=2cm}

\subsubsection{Real GDP growth (YoY)}

\renewcommand{\thefigure}{B.3.3.\arabic{figure}}
\setcounter{figure}{0}

\begin{figure}[h!]
	\centering \hspace*{-0.5cm}
	\includegraphics[scale=0.55]{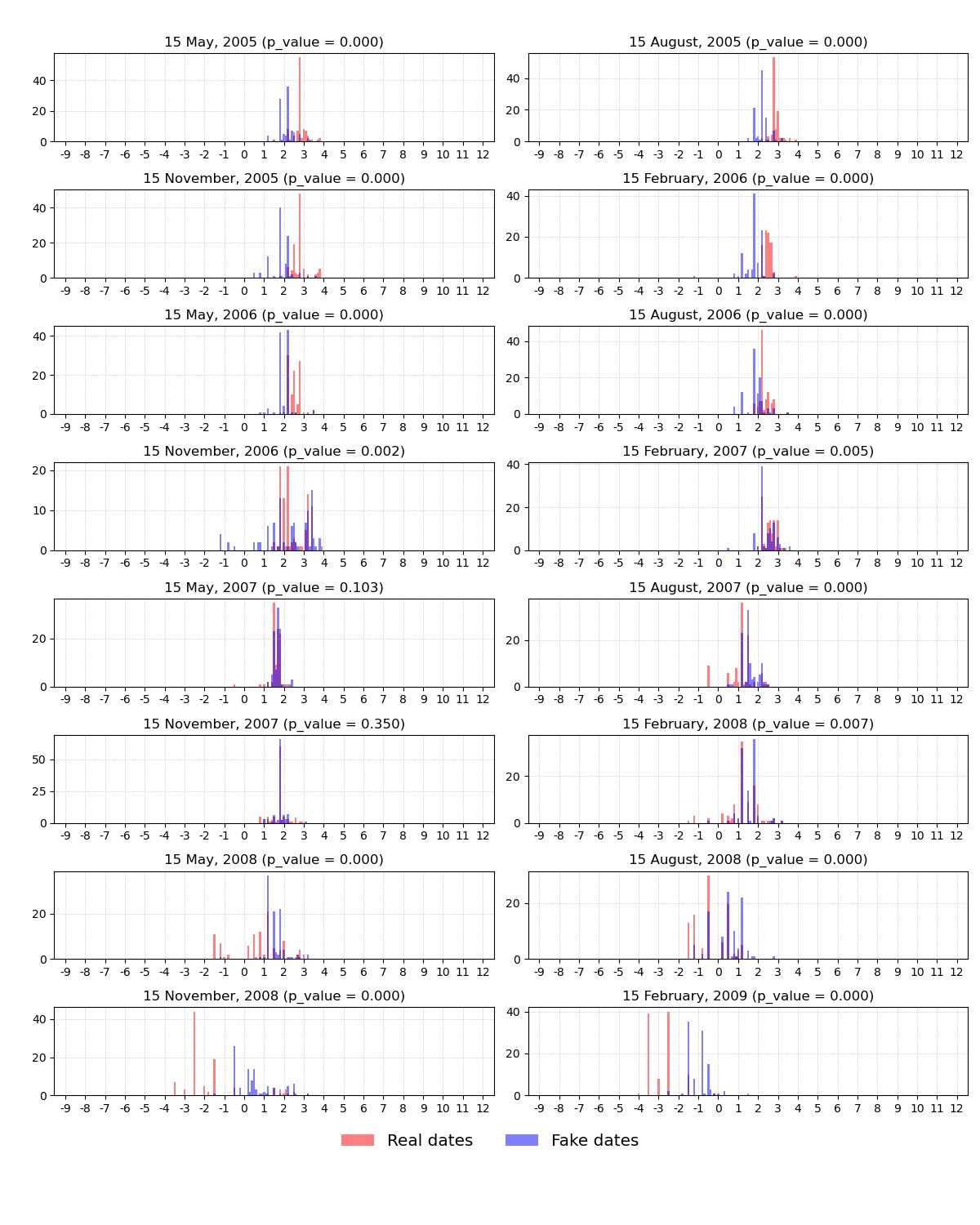}
	\caption{\textbf{Distributions of 1-year forecasts of real GDP growth (YoY) produced by the DeepSeek-V3.1 model.} The header shows the forecast date and the p-value for the Kolmogorov-Smirnov permutation test.}
    \label{fig: figure_B.3.3.1}
\end{figure}

\clearpage

\restoregeometry

\begin{figure}[ht]
	\centering \hspace*{-0.5cm}
	\includegraphics[scale=0.55]{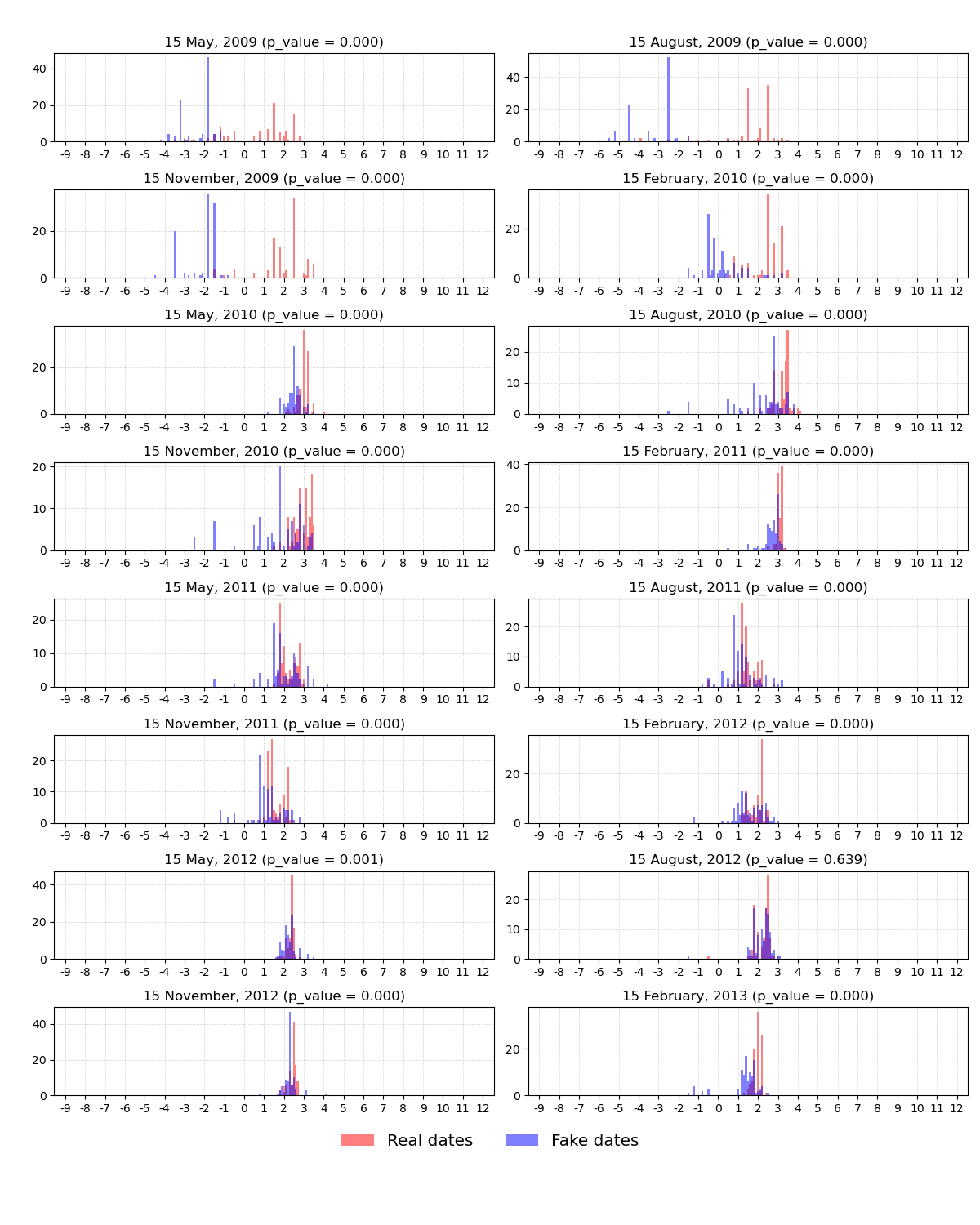}
	\caption{\textbf{Distributions of 1-year forecasts of real GDP growth (YoY) produced by the DeepSeek-V3.1 model.} The header shows the forecast date and the p-value for the Kolmogorov-Smirnov permutation test.}
    \label{fig: figure_B.3.3.2}
\end{figure}

\begin{figure}[h!]
	\centering \hspace*{-0.5cm}
	\includegraphics[scale=0.55]{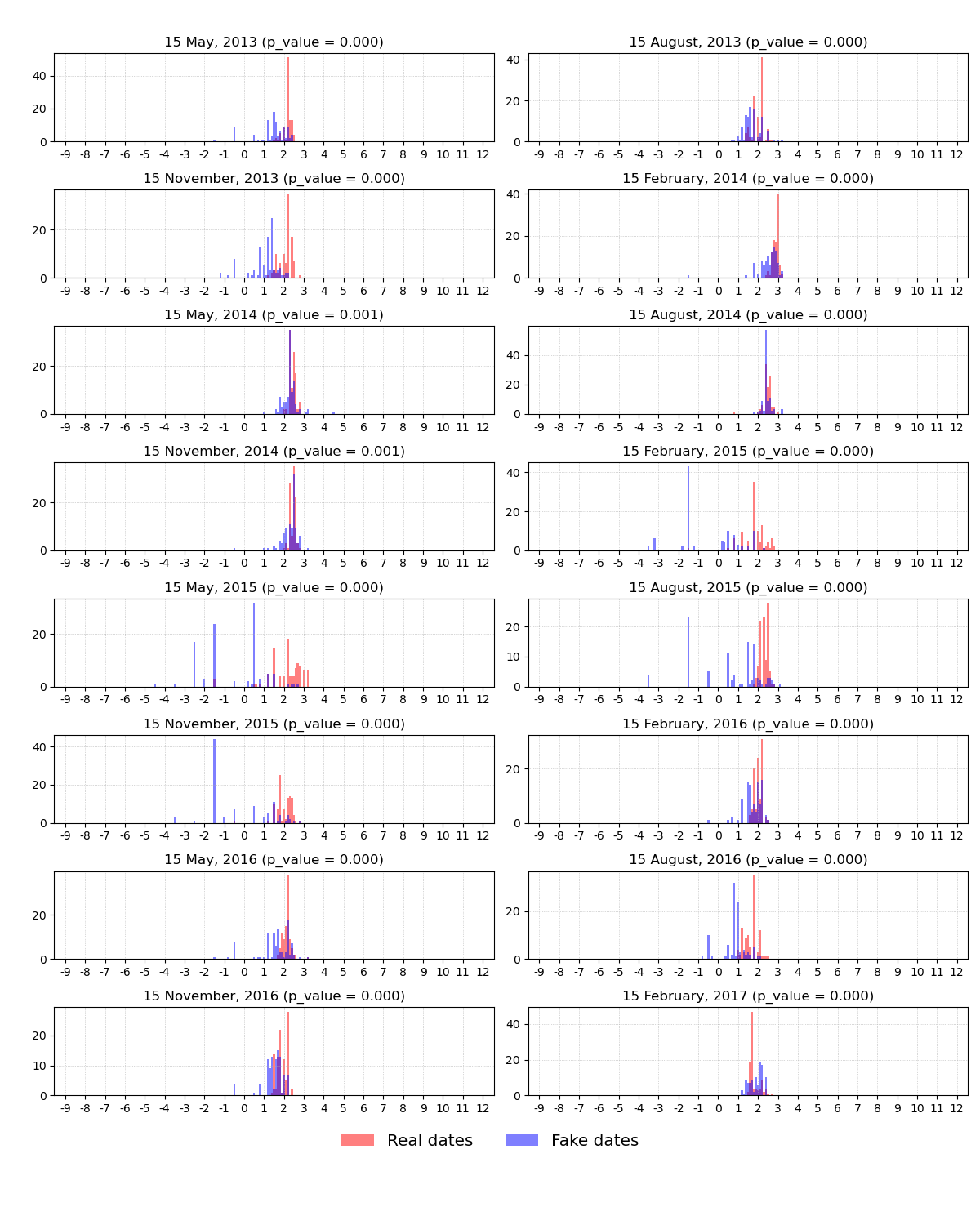}
	\caption{\textbf{Distributions of 1-year forecasts of real GDP growth (YoY) produced by the DeepSeek-V3.1 model.} The header shows the forecast date and the p-value for the Kolmogorov-Smirnov permutation test.}
    \label{fig: figure_B.3.3.3}
\end{figure}

\begin{figure}[h!]
	\centering \hspace*{-0.5cm}
	\includegraphics[scale=0.55]{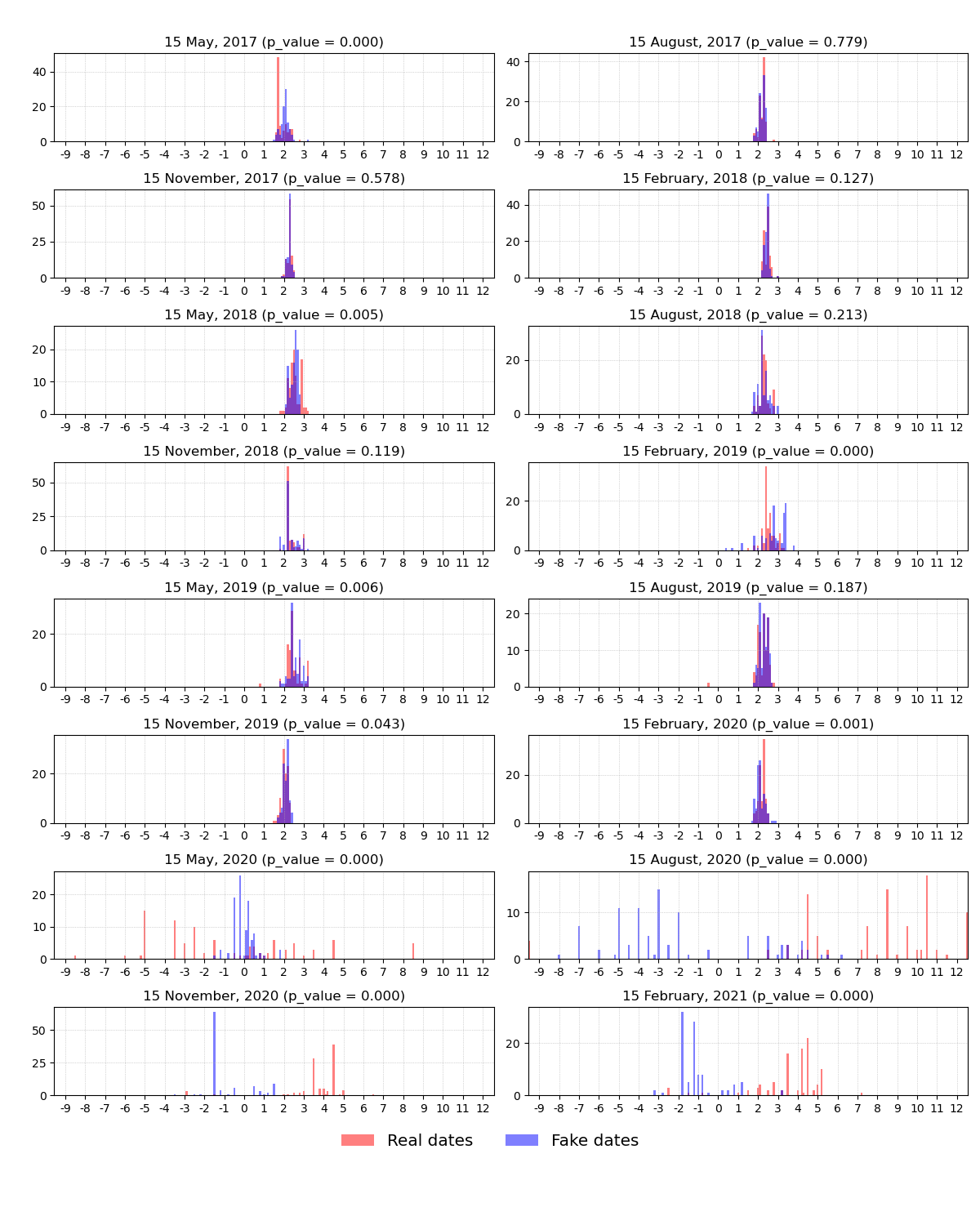}
	\caption{\textbf{Distributions of 1-year forecasts of real GDP growth (YoY) produced by the DeepSeek-V3.1 model.} The header shows the forecast date and the p-value for the Kolmogorov-Smirnov permutation test.}
    \label{fig: figure_B.3.3.4}
\end{figure}

\begin{figure}[h!]
	\centering \hspace*{-0.5cm}
	\includegraphics[scale=0.55]{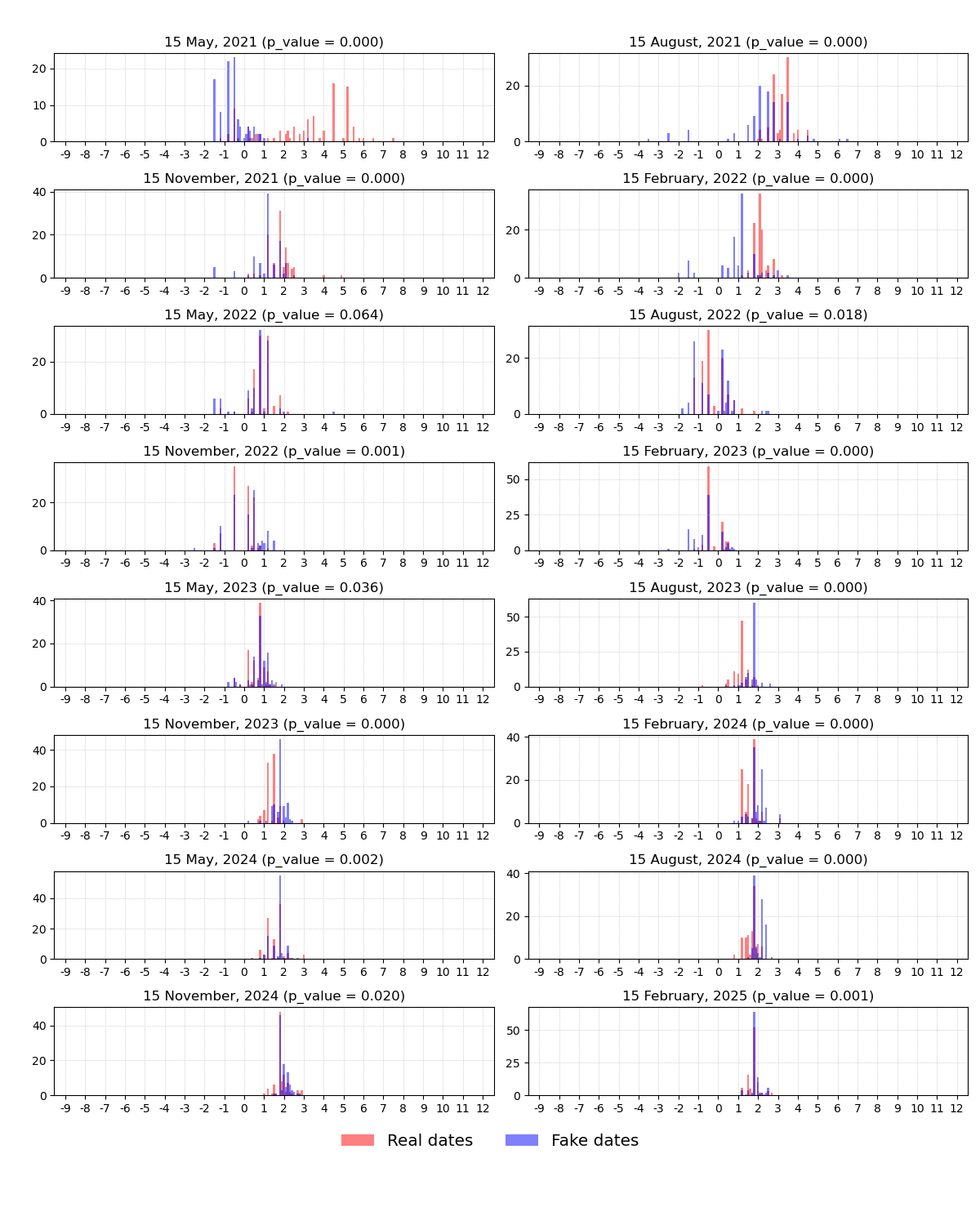}
	\caption{\textbf{Distributions of 1-year forecasts of real GDP growth (YoY) produced by the DeepSeek-V3.1 model.} The header shows the forecast date and the p-value for the Kolmogorov-Smirnov permutation test.}
    \label{fig: figure_B.3.3.5}
\end{figure}

\clearpage

\newgeometry{top=0.25cm,bottom=1.75cm,left=2cm,right=2cm,marginparwidth=2cm}

\appsection[app:appendix_c]{Results of checking Assumption 2}

\subsection{Distributions of GDP forecasts for a fake date with seasonality}

\renewcommand{\thefigure}{C.1.\arabic{figure}}
\setcounter{figure}{0}

\begin{figure}[h!!]
	\centering \hspace*{-0.5cm}
	\includegraphics[scale=0.55]{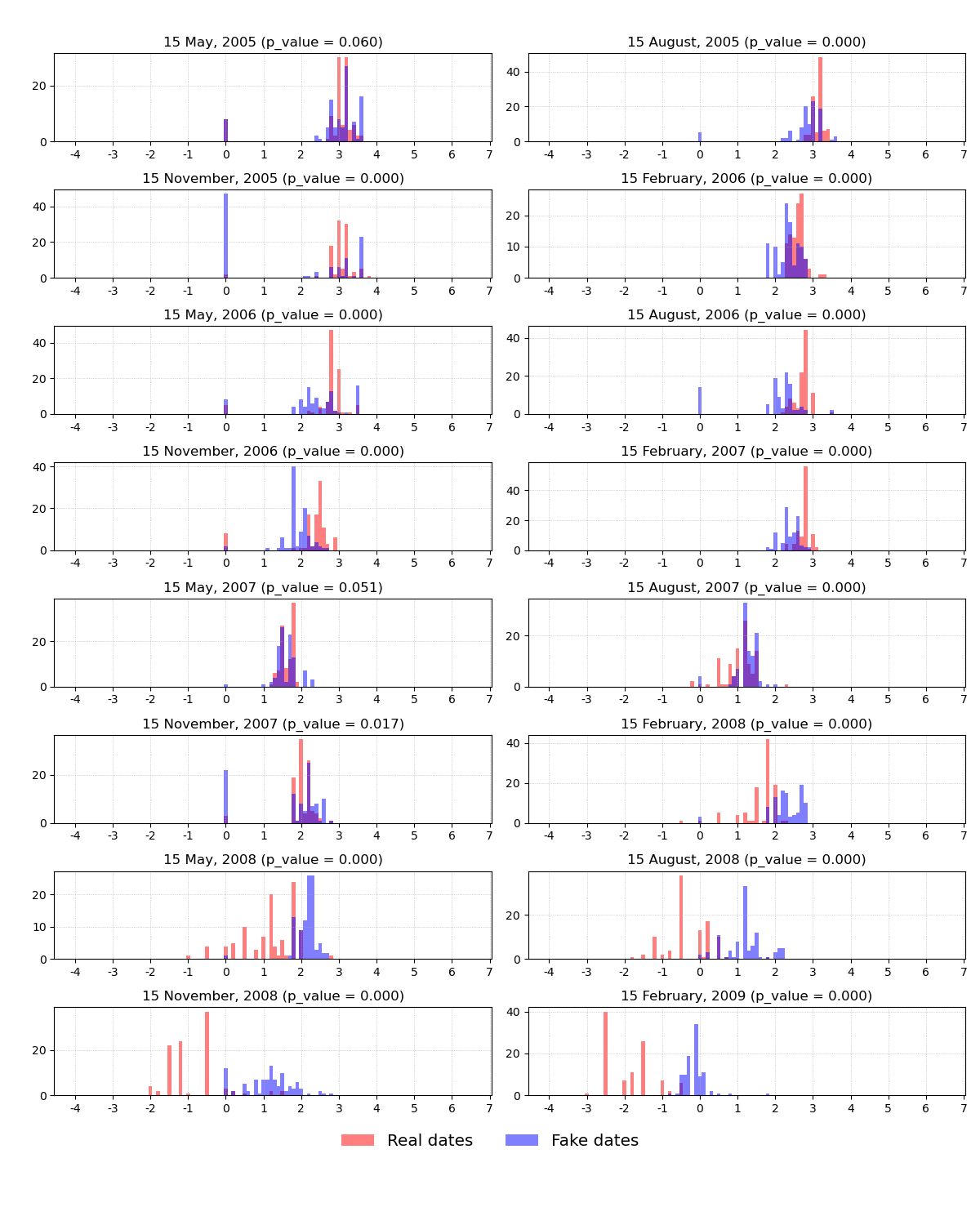}
	\caption{\textbf{Distributions of 1-year forecasts of real GDP growth (YoY) for a seasonal fake date produced by the Qwen3 Instruct model.} The header shows the forecast date and the p-value for the Kolmogorov-Smirnov permutation test.}
    \label{fig: figure_C.1.1}
\end{figure}

\clearpage

\restoregeometry

\begin{figure}[ht]
	\centering \hspace*{-0.5cm}
	\includegraphics[scale=0.55]{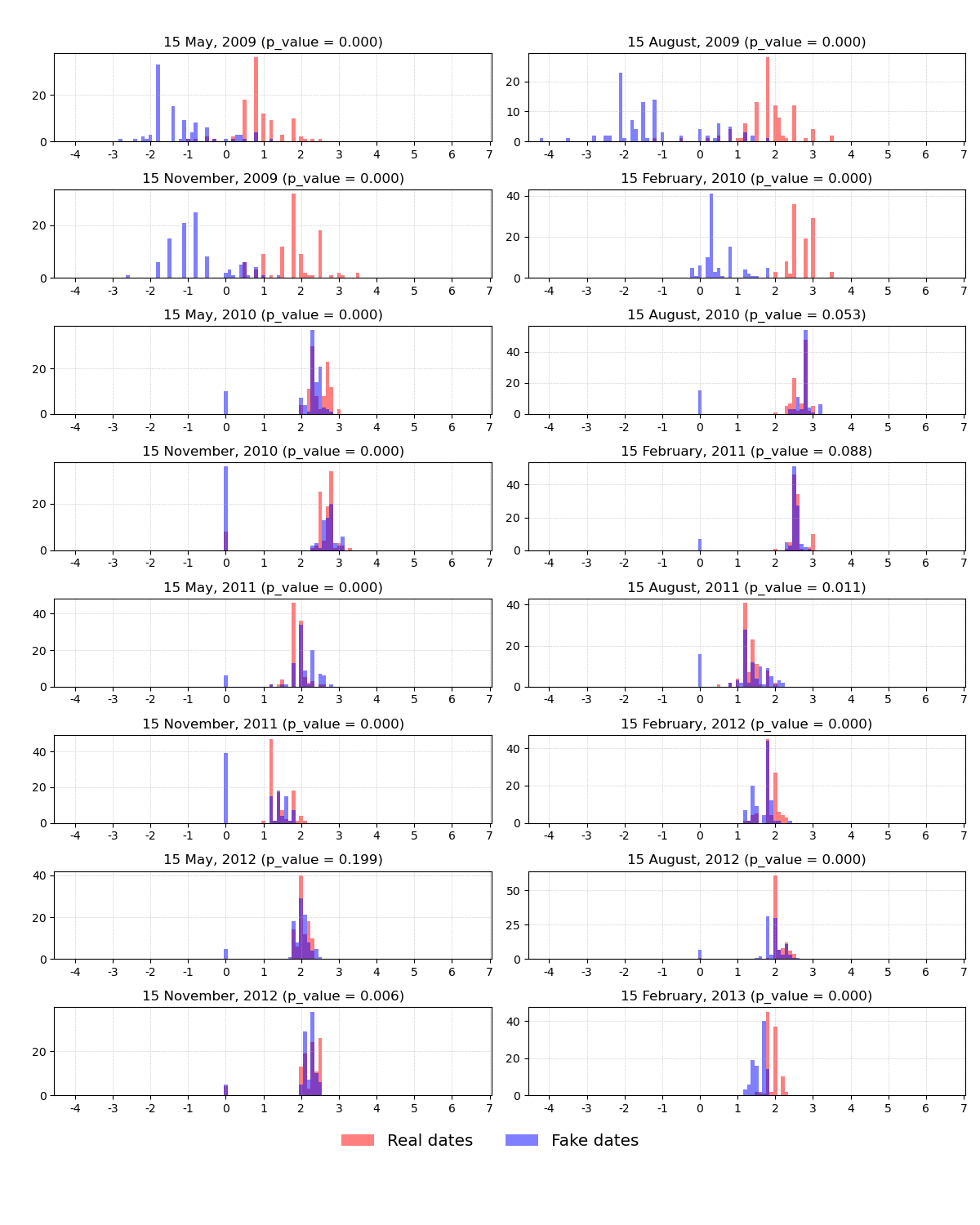}
	\caption{\textbf{Distributions of 1-year forecasts of real GDP growth (YoY) for a seasonal fake date produced by the Qwen3 Instruct model.} The header shows the forecast date and the p-value for the Kolmogorov-Smirnov permutation test.}
    \label{fig: figure_C.1.2}
\end{figure}

\begin{figure}[h!]
	\centering \hspace*{-0.5cm}
	\includegraphics[scale=0.55]{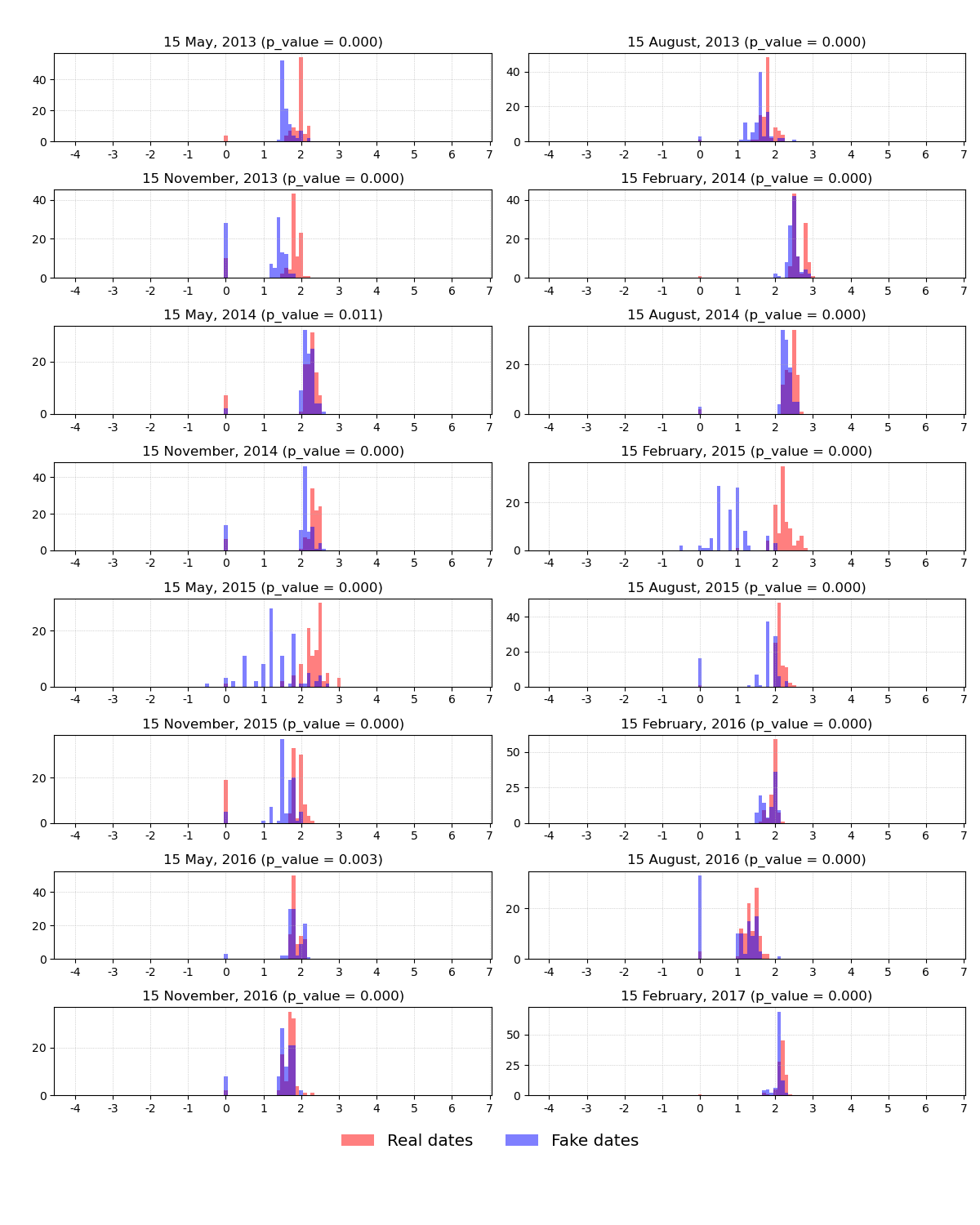}
	\caption{\textbf{Distributions of 1-year forecasts of real GDP growth (YoY) for a seasonal fake date produced by the Qwen3 Instruct model.} The header shows the forecast date and the p-value for the Kolmogorov-Smirnov permutation test.}
    \label{fig: figure_C.1.3}
\end{figure}

\begin{figure}[h!]
	\centering \hspace*{-0.5cm}
	\includegraphics[scale=0.55]{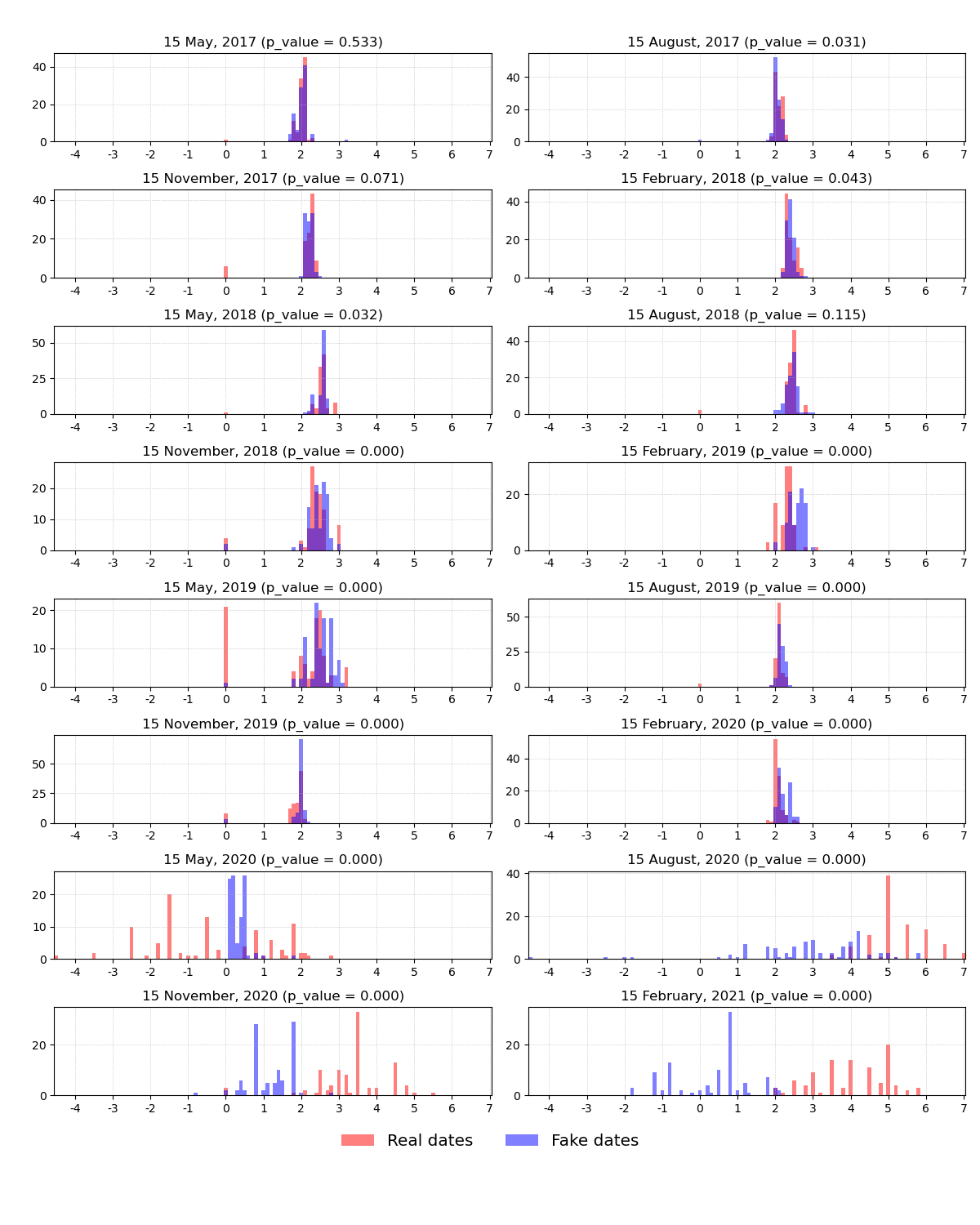}
	\caption{\textbf{Distributions of 1-year forecasts of real GDP growth (YoY) for a seasonal fake date produced by the Qwen3 Instruct model.} The header shows the forecast date and the p-value for the Kolmogorov-Smirnov permutation test.}
    \label{fig: figure_C.1.4}
\end{figure}

\begin{figure}[h!]
	\centering \hspace*{-0.5cm}
	\includegraphics[scale=0.55]{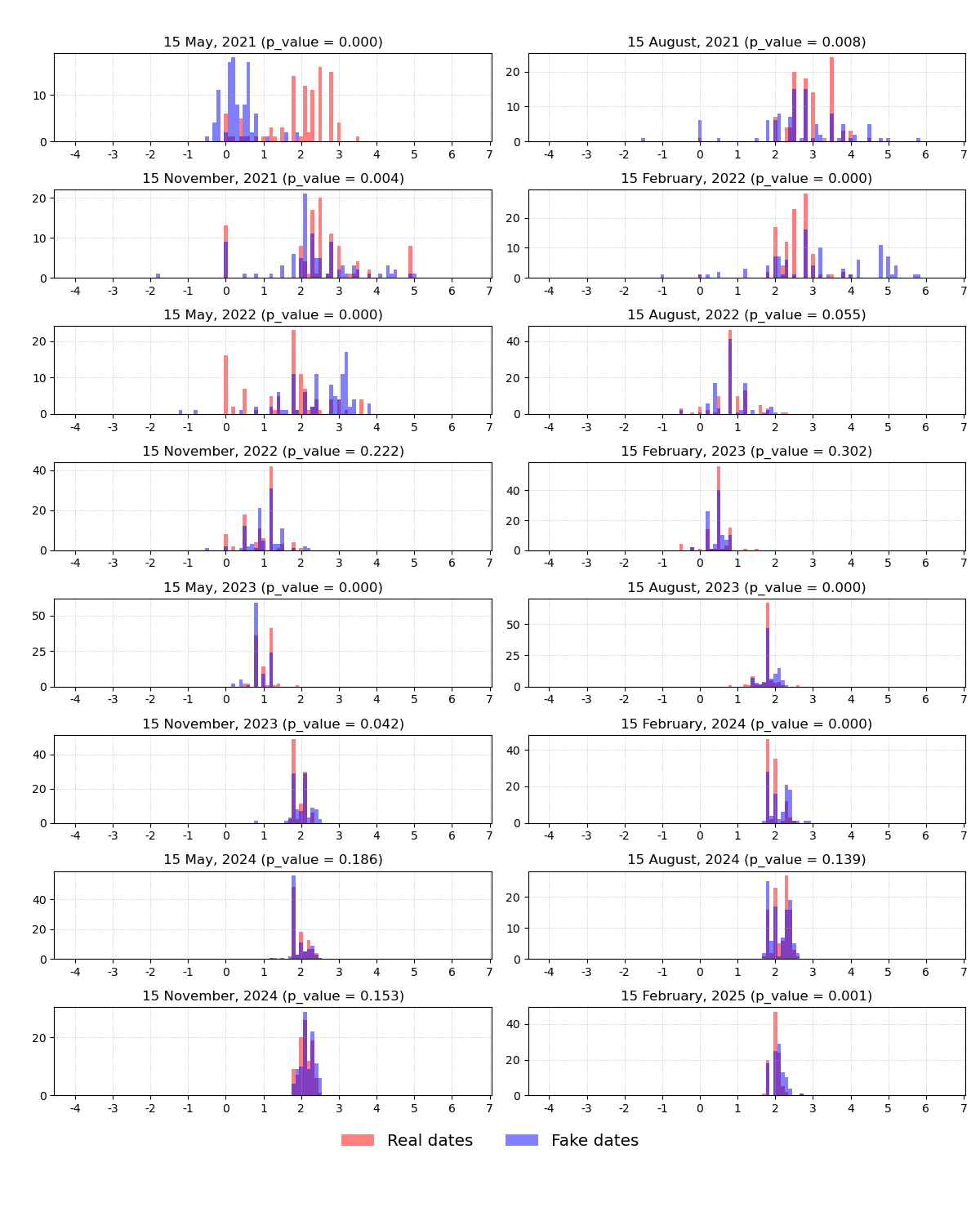}
	\caption{\textbf{Distributions of 1-year forecasts of real GDP growth (YoY) for a seasonal fake date produced by the Qwen3 Instruct model.} The header shows the forecast date and the p-value for the Kolmogorov-Smirnov permutation test.}
    \label{fig: figure_C.1.5}
\end{figure}

\clearpage

\begin{landscape}

\subsection{Tables comparing the p-values of forecast distributions}

\subsubsection{Kimi-K2 Instruct}

\renewcommand{\thetable}{C.2.1.\arabic{table}}
\setcounter{table}{0}

\begin{table}[h!]
\centering
\small
\setlength{\tabcolsep}{3.0pt}
\caption{\textbf{p-values ​​for comparing forecasts in the Assumption 2 checking for the Kimi-K2 Instruct model.} p-values for the Kolmogorov-Smirnov permutation test when comparing forecasts for the date specified in the row against February 15, 2030. The cutoff date is August 15, 2025. The date in the column is the date for which the macroeconomic statistics were taken.}
\begin{tabular}{l*{15}{c}}
\hline
& \multicolumn{5}{c}{Fed interest rate} 
& \multicolumn{5}{c}{CPI growth} 
& \multicolumn{5}{c}{Real GDP growth} \\
\cline{2-6}\cline{7-11}\cline{12-16}
& \shortstack{Nov.~15,\\2008}
& \shortstack{Aug.~15,\\2013}
& \shortstack{May~15,\\2020}
& \shortstack{Aug.~15,\\2021}
& \shortstack{Feb.~15,\\2025}
& \shortstack{Nov.~15,\\2008}
& \shortstack{Aug.~15,\\2013}
& \shortstack{May~15,\\2020}
& \shortstack{Aug.~15,\\2021}
& \shortstack{Feb.~15,\\2025}
& \shortstack{Nov.~15,\\2008}
& \shortstack{Aug.~15,\\2013}
& \shortstack{May~15,\\2020}
& \shortstack{Aug.~15,\\2021}
& \shortstack{Feb.~15,\\2025} \\
\hline
Aug.~15,~2025   & 0,750 & 0,412 & 1,000 & 0,402 & 0,086 & 0,020 & 0,157 & 0,075 & 0,033 & 0,805 & 0,064 & 0,525 & 0,867 & 0,033 & 0,002 \\
Nov.~15,~2025 & 0,200 & 0,694 & 1,000 & 0,039 & 0,041 & 0,295 & 0,719 & 0,015 & 0,895 & 0,014 & 0,509 & 0,017 & 0,028 & 0,349 & 0,004 \\
Feb.~15,~2026 & 0,027 & 0,654 & 1,000 & 0,391 & 0,727 & 0,067 & 0,108 & 0,641 & 0,958 & 0,375 & 0,141 & 0,672 & 0,752 & 0,970 & 0,037 \\
May~15,~2026      & 0,469 & 0,245 & 1,000 & 0,753 & 0,641 & 0,138 & 0,213 & 0,242 & 0,670 & 0,285 & 0,250 & 0,447 & 0,712 & 0,813 & 0,000 \\
Aug.~15,~2026   & 0,940 & 0,743 & 0,743 & 0,321 & 0,507 & 0,556 & 0,025 & 0,137 & 0,094 & 0,456 & 0,324 & 0,563 & 0,503 & 0,051 & 0,362 \\
Nov.~15,~2026 & 0,984 & 0,181 & 1,000 & 0,944 & 0,216 & 0,971 & 0,098 & 0,155 & 0,445 & 0,559 & 0,655 & 0,425 & 0,169 & 0,685 & 0,001 \\
Feb.~15,~2027 & 0,106 & 0,690 & 1,000 & 0,251 & 0,230 & 0,904 & 0,129 & 0,207 & 0,258 & 0,693 & 0,012 & 0,330 & 0,206 & 0,035 & 0,048 \\
May~15,~2027      & 0,114 & 0,054 & 1,000 & 0,757 & 0,424 & 0,802 & 0,237 & 0,804 & 0,336 & 0,261 & 0,626 & 0,814 & 0,284 & 0,283 & 0,003 \\
Aug.~15,~2027   & 0,011 & 0,848 & 1,000 & 0,636 & 0,780 & 0,058 & 0,011 & 0,789 & 0,451 & 0,132 & 0,500 & 0,633 & 0,273 & 0,578 & 0,850 \\
Nov.~15,~2027 & 0,151 & 0,987 & 0,496 & 0,948 & 0,755 & 0,995 & 0,247 & 0,351 & 0,132 & 0,689 & 0,849 & 0,654 & 0,096 & 0,710 & 0,622 \\
Feb.~15,~2028 & 0,383 & 0,324 & 1,000 & 0,211 & 0,638 & 0,278 & 0,017 & 0,155 & 0,806 & 0,352 & 0,115 & 0,427 & 0,227 & 0,225 & 0,820 \\
May~15,~2028      & 0,479 & 0,682 & 1,000 & 0,251 & 0,474 & 0,903 & 0,078 & 0,373 & 0,000 & 0,819 & 0,782 & 0,945 & 0,593 & 0,276 & 0,128 \\
Aug.~15,~2028   & 0,164 & 0,681 & 1,000 & 0,749 & 0,949 & 0,125 & 0,293 & 1,000 & 0,443 & 0,434 & 0,235 & 0,456 & 0,582 & 0,072 & 0,981 \\
Nov.~15,~2028 & 0,301 & 0,946 & 1,000 & 0,536 & 0,503 & 0,479 & 0,301 & 0,963 & 0,810 & 0,560 & 0,500 & 0,959 & 0,733 & 0,477 & 0,149 \\
Feb.~15,~2029 & 0,227 & 0,046 & 1,000 & 0,848 & 0,724 & 0,449 & 0,083 & 0,962 & 0,138 & 0,800 & 0,287 & 0,961 & 0,371 & 0,115 & 0,051 \\
May~15,~2029      & 0,052 & 0,334 & 1,000 & 0,397 & 0,760 & 0,711 & 0,017 & 0,879 & 0,797 & 0,065 & 0,400 & 0,959 & 0,615 & 0,071 & 0,091 \\
Aug.~15,~2029   & 0,272 & 0,261 & 1,000 & 0,865 & 0,747 & 0,611 & 0,027 & 0,993 & 0,030 & 0,016 & 0,088 & 0,644 & 0,204 & 0,039 & 0,618 \\
Nov.~15,~2029 & 0,102 & 0,980 & 1,000 & 0,523 & 0,336 & 0,320 & 0,135 & 0,800 & 0,066 & 0,359 & 0,480 & 0,343 & 0,733 & 0,353 & 0,409 \\
Feb.~15,~2030 & 1,000 & 1,000 & 1,000 & 1,000 & 1,000 & 1,000 & 1,000 & 1,000 & 1,000 & 1,000 & 1,000 & 1,000 & 1,000 & 1,000 & 1,000 \\
May~15,~2030      & 0,710 & 0,827 & 1,000 & 0,404 & 0,541 & 0,661 & 0,474 & 0,045 & 0,240 & 0,103 & 0,511 & 0,650 & 0,109 & 0,583 & 0,003 \\
\hline
\end{tabular}
\label{tab:table_C.2.1.1}
\end{table}

\begin{table}[h!]
\centering
\small
\setlength{\tabcolsep}{3.0pt}
\caption{\textbf{p-values ​​for comparing forecasts in the Assumption 2 checking for the Kimi-K2 Instruct model.} p-values for the Kolmogorov-Smirnov permutation test when comparing forecasts for the date specified in the row against February 15, 2030. The cutoff date is August 15, 2025. The date in the column is the date for which the macroeconomic statistics were taken.}
\begin{tabular}{l*{15}{c}}
\hline
& \multicolumn{5}{c}{Fed interesrt rate} 
& \multicolumn{5}{c}{CPI growth} 
& \multicolumn{5}{c}{Real GDP growth} \\
\cline{2-6}\cline{7-11}\cline{12-16}
& \shortstack{Nov.~15,\\2008}
& \shortstack{Aug.~15,\\2013}
& \shortstack{May~15,\\2020}
& \shortstack{Aug.~15,\\2021}
& \shortstack{Feb.~15,\\2025}
& \shortstack{Nov.~15,\\2008}
& \shortstack{Aug.~15,\\2013}
& \shortstack{May~15,\\2020}
& \shortstack{Aug.~15,\\2021}
& \shortstack{Feb.~15,\\2025}
& \shortstack{Nov.~15,\\2008}
& \shortstack{Aug.~15,\\2013}
& \shortstack{May~15,\\2020}
& \shortstack{Aug.~15,\\2021}
& \shortstack{Feb.~15,\\2025} \\
\hline
Aug.~15,~2030   & 0,601 & 0,810 & 1,000 & 0,528 & 0,238 & 0,908 & 0,048 & 0,665 & 0,460 & 0,032 & 0,001 & 0,008 & 0,002 & 0,286 & 0,264 \\
Nov.~15,~2030 & 0,307 & 0,392 & 1,000 & 0,097 & 0,502 & 0,779 & 0,004 & 0,212 & 0,452 & 0,853 & 0,041 & 0,028 & 0,000 & 0,461 & 0,496 \\
Feb.~15,~2031 & 0,865 & 0,922 & 0,491 & 0,639 & 0,119 & 0,680 & 0,084 & 0,991 & 0,268 & 0,982 & 0,870 & 0,002 & 0,256 & 0,144 & 0,307 \\
May~15,~2031      & 0,835 & 0,705 & 1,000 & 0,088 & 0,338 & 0,894 & 0,004 & 0,656 & 0,068 & 0,365 & 0,508 & 0,703 & 0,292 & 0,588 & 0,054 \\
Aug.~15,~2031   & 0,921 & 0,916 & 1,000 & 0,071 & 0,296 & 0,569 & 0,001 & 0,752 & 0,091 & 0,004 & 0,770 & 0,067 & 0,068 & 0,583 & 0,025 \\
Nov.~15,~2031 & 0,944 & 0,501 & 1,000 & 0,414 & 0,246 & 0,691 & 0,024 & 0,786 & 0,102 & 0,275 & 0,987 & 0,128 & 0,008 & 0,600 & 0,838 \\
Feb.~15,~2032 & 0,747 & 0,169 & 1,000 & 0,009 & 0,006 & 0,969 & 0,887 & 0,964 & 0,453 & 0,521 & 0,952 & 0,991 & 0,367 & 0,076 & 0,378 \\
May~15,~2032      & 0,837 & 0,930 & 1,000 & 0,239 & 0,096 & 0,074 & 0,206 & 0,242 & 0,689 & 0,483 & 0,319 & 0,526 & 0,848 & 0,148 & 0,693 \\
Aug.~15,~2032   & 0,470 & 0,749 & 1,000 & 0,093 & 0,385 & 0,001 & 0,002 & 0,049 & 0,906 & 0,048 & 0,133 & 0,042 & 0,166 & 0,158 & 0,284 \\
Nov.~15,~2032 & 0,930 & 0,839 & 0,501 & 0,002 & 0,713 & 0,423 & 0,973 & 0,695 & 0,903 & 0,266 & 0,003 & 0,026 & 0,034 & 0,024 & 0,971 \\
Feb.~15,~2033 & 0,937 & 0,033 & 0,499 & 0,126 & 0,329 & 0,102 & 0,353 & 0,662 & 0,957 & 0,565 & 0,990 & 0,090 & 0,006 & 0,199 & 0,737 \\
May~15,~2033      & 0,731 & 0,110 & 1,000 & 0,007 & 0,230 & 0,150 & 0,441 & 0,188 & 0,993 & 0,022 & 0,532 & 0,958 & 0,470 & 0,202 & 0,480 \\
Aug.~15,~2033   & 0,492 & 0,568 & 1,000 & 0,315 & 0,537 & 0,234 & 0,024 & 0,661 & 0,895 & 0,012 & 0,947 & 0,417 & 0,605 & 0,078 & 0,828 \\
Nov.~15,~2033 & 0,794 & 0,930 & 1,000 & 0,001 & 0,752 & 0,193 & 0,156 & 0,253 & 0,672 & 0,817 & 0,616 & 0,652 & 0,125 & 0,006 & 0,729 \\
Feb.~15,~2034 & 0,601 & 0,046 & 1,000 & 0,064 & 0,052 & 0,708 & 0,016 & 0,639 & 0,898 & 0,428 & 0,323 & 0,669 & 0,134 & 0,602 & 0,525 \\
May~15,~2034      & 0,994 & 0,710 & 1,000 & 0,005 & 0,766 & 0,047 & 0,023 & 0,581 & 0,892 & 0,045 & 0,409 & 0,658 & 0,454 & 0,051 & 0,605 \\
Aug.~15,~2034   & 0,385 & 0,930 & 1,000 & 0,000 & 0,999 & 0,006 & 0,004 & 0,779 & 0,992 & 0,003 & 0,748 & 0,655 & 0,115 & 0,025 & 0,000 \\
Nov.~15,~2034 & 0,735 & 0,319 & 1,000 & 0,013 & 0,941 & 0,066 & 0,016 & 0,384 & 1,000 & 0,223 & 0,761 & 0,873 & 0,014 & 0,016 & 0,714 \\
Feb.~15,~2035 & 0,976 & 0,105 & 0,496 & 0,030 & 0,327 & 0,568 & 0,641 & 0,152 & 0,898 & 0,181 & 0,041 & 0,882 & 0,049 & 0,053 & 0,779 \\
May~15,~2035      & 0,472 & 0,588 & 0,500 & 0,618 & 0,172 & 0,041 & 0,368 & 0,786 & 0,889 & 0,000 & 0,935 & 0,888 & 0,189 & 0,113 & 0,310 \\
Aug.~15,~2035      & 0,264 & 0,944 & 0,617 & 0,238 & 0,255 & 0,065 & 0,011 & 0,794 & 0,894 & 0,001 & 0,985 & 0,772 & 0,021 & 0,015 & 0,004 \\
Nov.~15,~2035      & 0,347 & 0,745 & 1,000 & 0,034 & 0,870 & 0,082 & 0,112 & 0,814 & 0,964 & 0,000 & 0,060 & 0,326 & 0,162 & 0,001 & 0,170 \\
\hline
\end{tabular}
\label{tab: table_C.2.1.2}
\end{table}

\clearpage
\subsubsection{Qwen3 Instruct}

\renewcommand{\thetable}{C.2.2.\arabic{table}}
\setcounter{table}{0}

\begin{table}[h!]
\centering
\small
\setlength{\tabcolsep}{3.0pt}
\caption{\textbf{p-values ​​for comparing forecasts in the Assumption 2 checking for the Qwen3 Instruct model.} p-values for the Kolmogorov-Smirnov permutation test when comparing forecasts for the date specified in the row against February 15, 2030. The cutoff date is August 15, 2025. The date in the column is the date for which the macroeconomic statistics were taken.}
\begin{tabular}{l*{15}{c}}
\hline
& \multicolumn{5}{c}{Fed interest rate} 
& \multicolumn{5}{c}{CPI growth} 
& \multicolumn{5}{c}{Real GDP growth} \\
\cline{2-6}\cline{7-11}\cline{12-16}
& \shortstack{Nov.~15,\\2008}
& \shortstack{Aug.~15,\\2013}
& \shortstack{May~15,\\2020}
& \shortstack{Aug.~15,\\2021}
& \shortstack{Feb.~15,\\2025}
& \shortstack{Nov.~15,\\2008}
& \shortstack{Aug.~15,\\2013}
& \shortstack{May~15,\\2020}
& \shortstack{Aug.~15,\\2021}
& \shortstack{Feb.~15,\\2025}
& \shortstack{Nov.~15,\\2008}
& \shortstack{Aug.~15,\\2013}
& \shortstack{May~15,\\2020}
& \shortstack{Aug.~15,\\2021}
& \shortstack{Feb.~15,\\2025} \\
\hline
Aug.~15,~2025   & 0,036 & 0,001 & 1,000 & 0,000 & 0,148 & 0,003 & 0,007 & 0,000 & 0,192 & 0,070 & 0,001 & 0,001 & 0,000 & 0,000 & 0,004 \\
Nov.~15,~2025 & 0,067 & 0,213 & 1,000 & 0,457 & 0,307 & 0,869 & 0,702 & 0,000 & 0,000 & 0,566 & 0,518 & 0,000 & 0,000 & 0,000 & 0,151 \\
Feb.~15,~2026 & 0,186 & 0,041 & 1,000 & 0,447 & 0,931 & 0,764 & 0,001 & 0,005 & 0,000 & 0,083 & 0,013 & 0,113 & 0,000 & 0,003 & 0,030 \\
May~15,~2026      & 0,185 & 0,557 & 1,000 & 0,022 & 0,872 & 0,990 & 0,200 & 0,000 & 0,001 & 0,434 & 0,007 & 0,000 & 0,011 & 0,000 & 0,000 \\
Aug.~15,~2026   & 0,134 & 0,445 & 1,000 & 0,019 & 0,039 & 0,034 & 0,075 & 0,206 & 0,000 & 0,247 & 0,004 & 0,000 & 0,214 & 0,000 & 0,249 \\
Nov.~15,~2026 & 0,021 & 0,786 & 1,000 & 0,901 & 0,140 & 0,941 & 0,419 & 0,551 & 0,000 & 0,030 & 0,228 & 0,000 & 0,965 & 0,000 & 0,455 \\
Feb.~15,~2027 & 0,782 & 0,474 & 1,000 & 0,916 & 0,041 & 0,007 & 0,064 & 0,055 & 0,000 & 0,695 & 0,134 & 0,227 & 0,005 & 0,003 & 0,312 \\
May~15,~2027      & 0,112 & 0,346 & 1,000 & 0,075 & 0,006 & 0,787 & 0,200 & 0,121 & 0,000 & 0,000 & 0,000 & 0,000 & 0,087 & 0,000 & 0,014 \\
Aug.~15,~2027   & 0,254 & 0,113 & 1,000 & 0,821 & 0,013 & 0,164 & 0,003 & 0,842 & 0,005 & 0,084 & 0,000 & 0,000 & 0,307 & 0,000 & 0,319 \\
Nov.~15,~2027 & 0,888 & 0,584 & 1,000 & 0,596 & 1,000 & 0,749 & 0,127 & 0,817 & 0,002 & 0,196 & 0,007 & 0,000 & 0,551 & 0,000 & 0,030 \\
Feb.~15,~2028 & 0,433 & 0,092 & 1,000 & 0,008 & 0,003 & 0,027 & 0,539 & 0,715 & 0,105 & 0,116 & 0,000 & 0,161 & 0,000 & 0,009 & 0,346 \\
May~15,~2028      & 0,100 & 0,276 & 1,000 & 0,971 & 0,506 & 0,349 & 0,958 & 0,053 & 0,001 & 0,331 & 0,635 & 0,000 & 0,157 & 0,000 & 0,149 \\
Aug.~15,~2028   & 0,777 & 0,127 & 1,000 & 0,458 & 0,468 & 0,392 & 0,048 & 0,817 & 0,027 & 0,030 & 0,389 & 0,000 & 0,011 & 0,000 & 0,536 \\
Nov.~15,~2028 & 0,417 & 0,443 & 1,000 & 0,157 & 0,212 & 0,955 & 0,385 & 0,104 & 0,054 & 0,552 & 0,242 & 0,000 & 0,333 & 0,000 & 0,804 \\
Feb.~15,~2029 & 0,181 & 0,002 & 1,000 & 0,675 & 0,344 & 0,751 & 0,107 & 0,208 & 0,000 & 0,016 & 0,017 & 0,002 & 0,086 & 0,000 & 0,370 \\
May~15,~2029      & 0,879 & 0,463 & 1,000 & 0,720 & 0,555 & 0,045 & 0,031 & 0,704 & 0,001 & 0,000 & 0,000 & 0,000 & 0,002 & 0,000 & 0,139 \\
Aug.~15,~2029   & 0,889 & 0,031 & 1,000 & 0,051 & 1,000 & 0,631 & 0,005 & 0,845 & 0,107 & 0,001 & 0,007 & 0,000 & 0,000 & 0,000 & 0,001 \\
Nov.~15,~2029 & 0,101 & 0,060 & 1,000 & 0,048 & 0,431 & 0,071 & 0,688 & 0,369 & 0,000 & 0,000 & 0,000 & 0,000 & 0,037 & 0,000 & 0,019 \\
Feb.~15,~2030 & 1,000 & 1,000 & 1,000 & 1,000 & 1,000 & 1,000 & 1,000 & 1,000 & 1,000 & 1,000 & 1,000 & 1,000 & 1,000 & 1,000 & 1,000 \\
May~15,~2030      & 0,087 & 0,114 & 1,000 & 0,010 & 0,318 & 0,125 & 0,314 & 0,680 & 0,077 & 0,060 & 0,016 & 0,000 & 0,000 & 0,000 & 0,000 \\
\hline
\end{tabular}
\label{tab: table_C.2.2.1}
\end{table}

\begin{table}[h!]
\centering
\small
\setlength{\tabcolsep}{3.0pt}
\caption{\textbf{p-values ​​for comparing forecasts in the Assumption 2 checking for the Qwen3 Instruct model.} p-values for the Kolmogorov-Smirnov permutation test when comparing forecasts for the date specified in the row against February 15, 2030. The cutoff date is August 15, 2025. The date in the column is the date for which the macroeconomic statistics were taken.}
\begin{tabular}{l*{15}{c}}
\hline
& \multicolumn{5}{c}{Fed interest rate} 
& \multicolumn{5}{c}{CPI growth} 
& \multicolumn{5}{c}{Real GDP growth} \\
\cline{2-6}\cline{7-11}\cline{12-16}
& \shortstack{Nov.~15,\\2008}
& \shortstack{Aug.~15,\\2013}
& \shortstack{May~15,\\2020}
& \shortstack{Aug.~15,\\2021}
& \shortstack{Feb.~15,\\2025}
& \shortstack{Nov.~15,\\2008}
& \shortstack{Aug.~15,\\2013}
& \shortstack{May~15,\\2020}
& \shortstack{Aug.~15,\\2021}
& \shortstack{Feb.~15,\\2025}
& \shortstack{Nov.~15,\\2008}
& \shortstack{Aug.~15,\\2013}
& \shortstack{May~15,\\2020}
& \shortstack{Aug.~15,\\2021}
& \shortstack{Feb.~15,\\2025} \\
\hline
Aug.~15,~2030   & 0,008 & 0,370 & 1,000 & 0,819 & 0,930 & 0,417 & 0,001 & 0,837 & 0,027 & 0,039 & 0,079 & 0,000 & 0,018 & 0,000 & 0,030 \\
Nov.~15,~2030 & 0,051 & 0,777 & 1,000 & 0,906 & 0,874 & 0,410 & 0,753 & 0,272 & 0,057 & 0,020 & 0,026 & 0,000 & 0,365 & 0,000 & 0,243 \\
Feb.~15,~2031 & 0,012 & 0,113 & 1,000 & 0,193 & 0,619 & 0,617 & 0,060 & 0,248 & 0,001 & 0,616 & 0,000 & 0,085 & 0,057 & 0,098 & 0,908 \\
May~15,~2031      & 0,011 & 0,123 & 1,000 & 0,003 & 0,842 & 0,728 & 0,094 & 0,278 & 0,007 & 0,138 & 0,000 & 0,000 & 0,002 & 0,008 & 0,089 \\
Aug.~15,~2031   & 0,991 & 0,031 & 1,000 & 0,700 & 0,654 & 0,421 & 0,033 & 0,580 & 0,001 & 0,000 & 0,000 & 0,000 & 0,043 & 0,000 & 0,975 \\
Nov.~15,~2031 & 0,450 & 0,691 & 1,000 & 0,353 & 0,916 & 0,951 & 0,549 & 0,977 & 0,005 & 0,558 & 0,066 & 0,000 & 0,354 & 0,142 & 0,972 \\
Feb.~15,~2032 & 0,765 & 0,657 & 1,000 & 0,220 & 0,311 & 0,668 & 0,702 & 0,924 & 0,028 & 0,012 & 0,147 & 0,134 & 0,003 & 0,048 & 0,094 \\
May~15,~2032      & 0,537 & 0,981 & 1,000 & 0,346 & 0,686 & 0,501 & 0,823 & 0,451 & 0,001 & 0,977 & 0,040 & 0,003 & 0,136 & 0,000 & 0,002 \\
Aug.~15,~2032   & 0,516 & 0,001 & 1,000 & 0,268 & 1,000 & 0,099 & 0,026 & 0,039 & 0,317 & 0,018 & 0,000 & 0,000 & 0,002 & 0,000 & 0,727 \\
Nov.~15,~2032 & 0,330 & 0,033 & 1,000 & 0,009 & 0,346 & 0,619 & 0,520 & 0,254 & 0,144 & 0,309 & 0,231 & 0,018 & 0,908 & 0,001 & 0,034 \\
Feb.~15,~2033 & 0,437 & 0,883 & 1,000 & 0,112 & 1,000 & 0,102 & 0,397 & 0,061 & 0,510 & 0,680 & 0,008 & 0,798 & 0,058 & 0,034 & 0,697 \\
May~15,~2033      & 0,342 & 0,412 & 1,000 & 0,474 & 0,768 & 0,226 & 0,484 & 0,061 & 0,384 & 0,346 & 0,018 & 0,516 & 0,120 & 0,000 & 0,001 \\
Aug.~15,~2033   & 0,075 & 0,000 & 1,000 & 0,295 & 0,091 & 0,106 & 0,004 & 0,157 & 0,000 & 0,330 & 0,000 & 0,000 & 0,000 & 0,000 & 0,015 \\
Nov.~15,~2033 & 0,428 & 0,007 & 1,000 & 0,356 & 0,004 & 0,464 & 0,567 & 0,142 & 0,000 & 0,806 & 0,019 & 0,002 & 0,618 & 0,000 & 0,334 \\
Feb.~15,~2034 & 0,552 & 0,688 & 1,000 & 0,575 & 0,064 & 0,989 & 0,878 & 0,044 & 0,021 & 0,406 & 0,000 & 0,622 & 0,005 & 0,458 & 0,000 \\
May~15,~2034      & 0,029 & 0,003 & 1,000 & 0,585 & 0,546 & 0,331 & 0,821 & 0,024 & 0,003 & 0,535 & 0,027 & 0,002 & 0,946 & 0,000 & 0,776 \\
Aug.~15,~2034   & 0,045 & 0,013 & 0,493 & 0,050 & 1,000 & 0,096 & 0,864 & 0,260 & 0,083 & 0,049 & 0,004 & 0,000 & 0,118 & 0,000 & 0,499 \\
Nov.~15,~2034 & 0,778 & 0,067 & 1,000 & 0,364 & 0,551 & 0,304 & 0,535 & 0,071 & 0,377 & 0,025 & 0,001 & 0,042 & 0,294 & 0,000 & 0,002 \\
Feb.~15,~2035 & 0,873 & 0,267 & 1,000 & 0,690 & 0,782 & 0,238 & 0,174 & 0,251 & 0,000 & 0,038 & 0,180 & 0,006 & 0,897 & 0,098 & 0,000 \\
May~15,~2035      & 0,669 & 0,283 & 1,000 & 0,154 & 0,601 & 0,844 & 0,043 & 0,271 & 0,000 & 0,995 & 0,000 & 0,000 & 0,371 & 0,000 & 0,912 \\
Aug.~15,~2035      & 0,251 & 0,426 & 1,000 & 0,525 & 0,143 & 0,943 & 0,002 & 0,391 & 0,007 & 0,013 & 0,000 & 0,000 & 0,011 & 0,000 & 0,972 \\
Nov.~15,~2035      & 0,189 & 0,295 & 1,000 & 0,372 & 0,434 & 0,252 & 0,129 & 0,097 & 0,363 & 0,120 & 0,001 & 0,000 & 0,014 & 0,015 & 0,000 \\
\hline
\end{tabular}
\label{tab: table_C.2.2.2}
\end{table}

\clearpage
\subsubsection{DeepSeek-V3.1}

\renewcommand{\thetable}{C.2.3.\arabic{table}}
\setcounter{table}{0}

\begin{table}[h!]
\centering
\small
\setlength{\tabcolsep}{3.0pt}
\caption{\textbf{p-values ​​for comparing forecasts in the Assumption 2 checking for the DeepSeek-V3.1 model.} p-values for the Kolmogorov-Smirnov permutation test when comparing forecasts for the date specified in the row against February 15, 2030. The cutoff date is August 15, 2025. The date in the column is the date for which the macroeconomic statistics were taken.}
\begin{tabular}{l*{15}{c}}
\hline
& \multicolumn{5}{c}{Fed interest rate} 
& \multicolumn{5}{c}{CPI growth} 
& \multicolumn{5}{c}{Real GDP growth} \\
\cline{2-6}\cline{7-11}\cline{12-16}
& \shortstack{Nov.~15,\\2008}
& \shortstack{Aug.~15,\\2013}
& \shortstack{May~15,\\2020}
& \shortstack{Aug.~15,\\2021}
& \shortstack{Feb.~15,\\2025}
& \shortstack{Nov.~15,\\2008}
& \shortstack{Aug.~15,\\2013}
& \shortstack{May~15,\\2020}
& \shortstack{Aug.~15,\\2021}
& \shortstack{Feb.~15,\\2025}
& \shortstack{Nov.~15,\\2008}
& \shortstack{Aug.~15,\\2013}
& \shortstack{May~15,\\2020}
& \shortstack{Aug.~15,\\2021}
& \shortstack{Feb.~15,\\2025} \\
\hline
Aug.~15,~2025   & 0,000 & 0,030 & 0,000 & 0,000 & 0,448 & 0,000 & 0,093 & 0,000 & 0,000 & 0,000 & 0,055 & 0,070 & 0,022 & 0,002 & 0,000 \\
Nov.~15,~2025 & 0,035 & 0,404 & 0,388 & 0,000 & 0,000 & 0,000 & 0,789 & 0,000 & 0,000 & 0,000 & 0,088 & 0,078 & 0,027 & 0,187 & 0,000 \\
Feb.~15,~2026 & 0,043 & 0,002 & 0,284 & 0,278 & 0,000 & 0,020 & 0,275 & 0,003 & 0,277 & 0,000 & 0,001 & 0,544 & 0,033 & 0,080 & 0,056 \\
May~15,~2026      & 0,008 & 0,018 & 0,163 & 0,000 & 0,035 & 0,001 & 0,029 & 0,002 & 0,796 & 0,000 & 0,000 & 0,274 & 0,039 & 0,385 & 0,002 \\
Aug.~15,~2026   & 0,010 & 0,016 & 0,018 & 0,000 & 0,001 & 0,046 & 0,063 & 0,295 & 0,270 & 0,000 & 0,081 & 0,252 & 0,030 & 0,003 & 0,336 \\
Nov.~15,~2026 & 0,000 & 0,050 & 0,103 & 0,001 & 0,051 & 0,095 & 0,208 & 0,002 & 0,212 & 0,000 & 0,005 & 0,269 & 0,034 & 0,089 & 0,250 \\
Feb.~15,~2027 & 0,324 & 0,463 & 1,000 & 0,718 & 0,000 & 0,046 & 0,108 & 0,000 & 0,038 & 0,000 & 0,076 & 0,211 & 0,071 & 0,624 & 0,691 \\
May~15,~2027      & 0,241 & 0,107 & 0,071 & 0,000 & 0,704 & 0,509 & 0,077 & 0,000 & 0,110 & 0,005 & 0,012 & 0,669 & 0,032 & 0,007 & 0,861 \\
Aug.~15,~2027   & 0,002 & 0,069 & 0,051 & 0,004 & 0,538 & 0,046 & 0,051 & 0,000 & 0,557 & 0,011 & 0,000 & 0,134 & 0,010 & 0,009 & 0,518 \\
Nov.~15,~2027 & 0,021 & 0,019 & 0,146 & 0,255 & 0,007 & 0,001 & 0,085 & 0,159 & 0,220 & 0,009 & 0,000 & 0,207 & 0,014 & 0,017 & 0,445 \\
Feb.~15,~2028 & 0,226 & 0,004 & 0,073 & 0,365 & 0,380 & 0,410 & 0,449 & 0,012 & 0,051 & 0,000 & 0,001 & 0,058 & 0,048 & 0,512 & 0,448 \\
May~15,~2028      & 0,091 & 0,011 & 0,263 & 0,076 & 0,291 & 0,654 & 0,267 & 0,152 & 0,149 & 0,992 & 0,000 & 0,108 & 0,026 & 0,740 & 0,251 \\
Aug.~15,~2028   & 0,000 & 0,025 & 0,095 & 0,074 & 0,155 & 0,647 & 0,343 & 0,110 & 0,283 & 0,001 & 0,000 & 0,583 & 0,015 & 0,610 & 0,738 \\
Nov.~15,~2028 & 0,000 & 0,133 & 0,227 & 0,009 & 0,000 & 0,020 & 0,589 & 0,350 & 0,786 & 0,003 & 0,000 & 0,145 & 0,192 & 0,014 & 0,703 \\
Feb.~15,~2029 & 0,017 & 0,074 & 1,000 & 0,004 & 0,199 & 0,031 & 0,017 & 0,108 & 0,337 & 0,495 & 0,000 & 0,023 & 0,101 & 0,482 & 0,933 \\
May~15,~2029      & 0,108 & 0,019 & 0,199 & 0,074 & 0,097 & 0,340 & 0,363 & 0,980 & 0,365 & 0,884 & 0,000 & 0,347 & 0,101 & 0,866 & 0,710 \\
Aug.~15,~2029   & 0,010 & 0,009 & 0,010 & 0,100 & 0,109 & 0,329 & 0,149 & 0,820 & 0,217 & 0,642 & 0,000 & 0,187 & 0,081 & 0,001 & 0,585 \\
Nov.~15,~2029 & 0,065 & 0,034 & 1,000 & 0,000 & 0,057 & 0,017 & 0,607 & 0,047 & 0,461 & 0,164 & 0,001 & 0,680 & 0,356 & 0,000 & 0,009 \\
Feb.~15,~2030 & 1,000 & 1,000 & 1,000 & 1,000 & 1,000 & 1,000 & 1,000 & 1,000 & 1,000 & 1,000 & 1,000 & 1,000 & 1,000 & 1,000 & 1,000 \\
May~15,~2030      & 0,047 & 0,843 & 0,548 & 0,463 & 0,505 & 0,528 & 0,017 & 0,489 & 0,353 & 0,135 & 0,000 & 0,362 & 0,065 & 0,827 & 0,200 \\
\hline
\end{tabular}
\label{tab: table_C.2.3.1}
\end{table}

\begin{table}[h!]
\centering
\small
\setlength{\tabcolsep}{3.0pt}
\caption{\textbf{p-values ​​for comparing forecasts in the Assumption 2 checking for the DeepSeek-V3.1 model.} p-values for the Kolmogorov-Smirnov permutation test when comparing forecasts for the date specified in the row against February 15, 2030. The cutoff date is August 15, 2025. The date in the column is the date for which the macroeconomic statistics were taken.}
\begin{tabular}{l*{15}{c}}
\hline
& \multicolumn{5}{c}{Fed interest rate} 
& \multicolumn{5}{c}{CPI growth} 
& \multicolumn{5}{c}{Real GDP growth} \\
\cline{2-6}\cline{7-11}\cline{12-16}
& \shortstack{Nov.~15,\\2008}
& \shortstack{Aug.~15,\\2013}
& \shortstack{May~15,\\2020}
& \shortstack{Aug.~15,\\2021}
& \shortstack{Feb.~15,\\2025}
& \shortstack{Nov.~15,\\2008}
& \shortstack{Aug.~15,\\2013}
& \shortstack{May~15,\\2020}
& \shortstack{Aug.~15,\\2021}
& \shortstack{Feb.~15,\\2025}
& \shortstack{Nov.~15,\\2008}
& \shortstack{Aug.~15,\\2013}
& \shortstack{May~15,\\2020}
& \shortstack{Aug.~15,\\2021}
& \shortstack{Feb.~15,\\2025} \\
\hline
Aug.~15,~2030   & 0,000 & 0,217 & 0,335 & 0,054 & 0,060 & 0,014 & 0,107 & 0,413 & 0,677 & 0,175 & 0,000 & 0,781 & 0,086 & 0,165 & 0,177 \\
Nov.~15,~2030 & 0,506 & 0,803 & 0,668 & 0,613 & 0,000 & 0,020 & 0,384 & 0,058 & 0,965 & 0,025 & 0,013 & 0,267 & 0,125 & 0,337 & 0,054 \\
Feb.~15,~2031 & 0,136 & 0,838 & 0,543 & 0,061 & 0,896 & 0,312 & 0,574 & 0,020 & 0,002 & 0,051 & 0,011 & 0,001 & 0,053 & 0,113 & 0,925 \\
May~15,~2031      & 0,863 & 0,999 & 0,498 & 0,419 & 0,003 & 0,638 & 0,006 & 0,012 & 0,792 & 0,200 & 0,001 & 0,903 & 0,001 & 0,459 & 0,222 \\
Aug.~15,~2031   & 0,000 & 0,722 & 1,000 & 0,096 & 0,072 & 0,243 & 0,718 & 0,052 & 0,077 & 0,068 & 0,001 & 0,049 & 0,030 & 0,395 & 0,002 \\
Nov.~15,~2031 & 0,001 & 0,587 & 0,427 & 0,067 & 0,489 & 0,097 & 0,186 & 0,272 & 0,056 & 0,183 & 0,090 & 0,687 & 0,331 & 0,654 & 0,020 \\
Feb.~15,~2032 & 0,008 & 0,683 & 0,828 & 0,237 & 0,137 & 0,645 & 0,834 & 0,395 & 0,082 & 0,174 & 0,124 & 0,263 & 0,016 & 0,159 & 0,667 \\
May~15,~2032      & 0,783 & 0,495 & 0,869 & 0,256 & 0,087 & 0,256 & 0,373 & 0,373 & 0,354 & 0,133 & 0,000 & 0,450 & 0,043 & 0,720 & 0,054 \\
Aug.~15,~2032   & 0,027 & 0,243 & 0,335 & 0,335 & 0,749 & 0,323 & 0,160 & 0,833 & 0,439 & 0,227 & 0,000 & 0,353 & 0,062 & 0,210 & 0,356 \\
Nov.~15,~2032 & 0,035 & 0,899 & 0,908 & 0,073 & 0,317 & 0,000 & 0,705 & 0,283 & 0,788 & 0,650 & 0,000 & 0,971 & 0,063 & 0,756 & 0,034 \\
Feb.~15,~2033 & 0,307 & 0,081 & 0,383 & 0,000 & 0,263 & 0,089 & 0,203 & 0,598 & 0,358 & 0,348 & 0,001 & 0,128 & 0,020 & 0,000 & 0,682 \\
May~15,~2033      & 0,302 & 0,004 & 0,646 & 0,151 & 0,422 & 0,086 & 0,618 & 0,315 & 0,456 & 0,795 & 0,000 & 0,221 & 0,069 & 0,854 & 0,094 \\
Aug.~15,~2033   & 0,000 & 0,132 & 0,104 & 0,001 & 0,602 & 0,321 & 0,199 & 0,830 & 0,449 & 0,496 & 0,000 & 0,899 & 0,088 & 0,985 & 0,030 \\
Nov.~15,~2033 & 0,007 & 0,353 & 1,000 & 0,000 & 0,009 & 0,026 & 0,374 & 0,100 & 0,111 & 0,106 & 0,000 & 0,909 & 0,132 & 0,232 & 0,230 \\
Feb.~15,~2034 & 0,050 & 0,400 & 0,075 & 0,000 & 0,835 & 0,335 & 0,824 & 0,455 & 0,109 & 0,186 & 0,000 & 0,030 & 0,021 & 0,629 & 0,703 \\
May~15,~2034      & 0,141 & 0,675 & 1,000 & 0,000 & 0,002 & 0,328 & 0,271 & 0,732 & 0,118 & 0,115 & 0,000 & 0,821 & 0,043 & 0,948 & 0,137 \\
Aug.~15,~2034   & 0,009 & 0,102 & 0,930 & 0,013 & 0,199 & 0,084 & 0,591 & 0,603 & 0,206 & 0,419 & 0,000 & 0,710 & 0,018 & 0,169 & 0,147 \\
Nov.~15,~2034 & 0,067 & 0,251 & 1,000 & 0,000 & 0,760 & 0,016 & 0,833 & 0,169 & 0,445 & 0,025 & 0,000 & 0,240 & 0,002 & 0,740 & 0,378 \\
Feb.~15,~2035 & 0,138 & 0,189 & 0,247 & 0,000 & 0,725 & 0,094 & 0,475 & 0,000 & 0,054 & 0,565 & 0,011 & 0,793 & 0,022 & 0,059 & 0,531 \\
May~15,~2035      & 0,500 & 0,105 & 1,000 & 0,059 & 0,028 & 0,414 & 0,038 & 0,176 & 0,077 & 0,030 & 0,002 & 0,675 & 0,109 & 0,852 & 0,043 \\
Aug.~15,~2035      & 0,025 & 0,007 & 1,000 & 0,004 & 0,436 & 0,342 & 0,001 & 0,755 & 0,435 & 0,091 & 0,000 & 0,894 & 0,028 & 0,307 & 0,002 \\
Nov.~15,~2035      & 0,050 & 0,096 & 1,000 & 0,020 & 0,582 & 0,095 & 0,166 & 0,023 & 0,808 & 0,022 & 0,000 & 0,774 & 0,064 & 0,300 & 0,185 \\
\hline
\end{tabular}
\label{tab: table_C.2.3.2}
\end{table}

\end{landscape}

\clearpage

\appsection[app:appendix_d]{Examples of LLMs' responses for Fake date test I}
\subsection{Examples of LLMs' responses with the forecast date of November 15, 2008}

\subsubsection{Kimi-K2 Instruct}

\renewcommand{\thefigure}{D.1.1.\arabic{figure}}
\setcounter{figure}{0}

\begin{figure}[ht]
\begin{tcolorbox}[promptbox]
Given the facts available as of November 15, 2008:

- The effective lower bound of the federal-funds target range is 1.0 \%.

- Real GDP grew only 0.8 \% year-over-year in Q3 2008, indicating a sharp slowdown from the growth rates of 2006-2007.

- Headline CPI inflation is 3.7 \% YoY, but this is largely driven by the oil-price spike that has already reversed; market-based inflation expectations and survey measures are falling rapidly.

- Credit markets remain severely impaired, equity prices have fallen ~40 \% from their 2007 peaks, and unemployment is rising.

- Historical FOMC reaction patterns (1990-91, 2001) show that when growth slows sharply and inflation expectations retreat, the Committee cuts aggressively until the real funds rate is clearly negative to cushion the downturn.

With growth near zero and financial stress intense, the Fed is expected to cut further. The likely endpoint implied by both the Taylor-rule logic (with a 0.8 \% output gap and falling inflation) and past easing cycles is a target range whose lower bound reaches the practical floor. Therefore, the lower bound of the target range is expected to bottom out at 0 \% by mid-2009 and remain there through November 15, 2009.

Interest rate forecast: 0.0\%
\end{tcolorbox}
\caption{\textbf{Example of a response for 1-year forecast of the lower bound of the Fed interest rate for a real date by the Kimi-K2 Instruct model.} Formatting was slightly adjusted for readability.}
\label{fig: figure_D.1.1.1}
\end{figure}

\clearpage

\begin{figure}[ht]
\begin{tcolorbox}[promptbox]
Based solely on the data provided and without drawing on historical patterns or post-2005 knowledge:

- Real GDP growth (0.8\% YoY) is low, suggesting the economy is operating below potential.

- Inflation (3.7\% YoY) is above the Fed's presumed long-run target (which was 2\% in 2005), indicating persistent price pressures.

- The current policy rate (1.0\%) is accommodative relative to both inflation and any implicit neutral rate.

Given high inflation despite weak growth, the Fed is likely to prioritize inflation control. To bring inflation down toward target, it will need to tighten policy. Over the next year, the Fed is expected to raise the lower bound gradually. Applying a simple Taylor-type adjustment (without coefficients), the rate would rise toward a level at least matching inflation minus a small real-rate buffer - say roughly inflation minus 1\%. A 3.7\% CPI minus ~1\% implies around 2.7-3.0\%. Allowing for gradual implementation and smoothing, a 2.5\% lower bound is a plausible midpoint for February 2031.

Interest rate forecast: 2.5\%
\end{tcolorbox}
\caption{\textbf{Example of a response for 1-year forecast of the lower bound of the Fed interest rate for a fake date by the Kimi-K2 Instruct model.} Formatting was slightly adjusted for readability.}
\label{fig: figure_D.1.1.2}
\end{figure}

\clearpage

\subsubsection{Qwen3 Instruct}

\renewcommand{\thefigure}{D.1.2.\arabic{figure}}
\setcounter{figure}{0}

\begin{figure}[h!!]
\begin{tcolorbox}[promptbox, fontupper=\scriptsize\ttfamily]
To forecast the lower bound of the Fed interest rate on November 15, 2009, we must rely only on information available up to November 15, 2005, and the specific statistics provided in the prompt: as of today (November 15, 2008), the current lower bound of the Fed interest rate is 1.0\%, US real GDP growth in Q3 2008 was 0.8\% (YoY), and US inflation (CPI) in October 2008 was 3.7\% (YoY).

Given that we cannot use any real-world knowledge beyond November 15, 2005 (such as the actual onset of the financial crisis, government interventions, or subsequent policy responses), our forecast must be based solely on the macroeconomic data provided and the behavior of monetary policy up to 2005.

Prior to 2005, the Federal Reserve typically responded to economic conditions by adjusting the federal funds rate to balance inflation and output. During periods of slowing growth and rising inflation (a stagflationary signal), the Fed often maintained or slightly raised rates to combat inflation, unless growth fell sharply into recession.

However, in this scenario, real GDP growth is still positive at 0.8\% in Q3 2008, though low, and inflation is relatively high at 3.7\%. Based on pre-2005 policy patterns, such a combination might prompt the Fed to hold rates steady or possibly raise them if inflation were accelerating. But since the current rate is already at 1.0\%, which is relatively low, and given that growth is weakening, the Fed might consider further rate cuts if it judged that the downside growth risks outweighed inflation concerns.

Still, based strictly on pre-2005 precedent, the Fed was generally reluctant to cut rates aggressively in the face of above-target inflation, especially when growth was still positive. For example, during the 1970s stagflation, rates were not cut despite slow growth due to high inflation. In contrast, in 2001, the Fed cut rates aggressively because inflation was low and growth was weakening.

In this case, inflation is relatively high (3.7\%), which would have been seen as a constraint on rate cuts under pre-2005 policy logic. However, if the Fed perceived the slowdown as the beginning of a deeper downturn, some easing might be expected. But without knowledge of events after 2005 - such as the collapse of Lehman Brothers or the severity of the financial crisis—we cannot assume an extraordinary policy response.

Given the data: low but positive growth and high inflation, and assuming the Fed follows a Taylor-rule-like policy based on pre-2005 behavior, the Fed might hold rates steady or cut only modestly if at all. However, since the rate is already at 1.0\%, and considering the inflationary pressure, the Fed might be hesitant to go significantly lower.

But the prompt gives a current rate of 1.0\% on November 15, 2008, which is lower than what would have been typical under high inflation in earlier decades. This suggests that by 2008, the Fed may already be responding to exceptional conditions. However, we are restricted from using knowledge after November 15, 2005. Therefore, we must assume that such a low rate in the face of 3.7\% inflation would be highly unusual and possibly unsustainable under prior norms.

Nonetheless, the provided current rate is 1.0\%, so we accept that as the starting point. Given the weak growth (0.8\%) and elevated inflation, the Fed might pause or cut slightly further if growth continues to weaken. But without knowledge of systemic financial collapse, a cautious approach would limit the extent of rate cuts.

Assuming moderate further weakening in growth over the next year, but inflation moderating somewhat due to global demand slowdown (which could be inferred from commodity price trends evident by late 2008, though we cannot use post-2005 knowledge of their causes), the Fed might lower rates slightly below 1.0\% but not to zero.

However, based on the actual policy path up to 2005, and the fact that the rate is already low with high inflation, the probability of deep cuts is limited in this constrained knowledge framework.

But the current rate is already 1.0\% - lower than pre-2005 patterns would likely allow under 3.7\% inflation - suggesting that the Fed is already in an easing cycle. If the economy continues to slow, and inflation eventually falls (possibly due to reduced demand), the Fed might continue cutting into 2009.

Given the low growth and the momentum of easing, a reasonable forecast under a cautious extrapolation would be a further cut to near zero, but only if inflation falls. However, without knowing post-2005 events, we cannot anticipate the full extent of the crisis.

Still, the combination of sub-1\% growth and high inflation presents a policy dilemma. In prior decades, the Fed would likely not cut much further. Therefore, the most plausible forecast under pre-2005 logic is that the Fed holds rates steady or cuts only modestly.

But given the provided current rate of 1.0\% already reflects a significant easing (unusual for 3.7\% inflation), and assuming some continued weakness, a small cut to 0.5\% might occur.

Interest rate forecast: 0.5\%
\end{tcolorbox}
\caption{\textbf{Example of a response for 1-year forecast of the lower bound of the Fed interest rate for a real date by the Qwen3 Instruct model.} Formatting was slightly adjusted for readability.}
\label{fig: figure_D.1.2.1}
\end{figure}

\clearpage

\begin{figure}[ht]
\begin{tcolorbox}[promptbox]
To forecast the lower bound of the Fed interest rate on February 15, 2031, we rely solely on the data provided and the Federal Reserve's typical policy considerations around inflation and economic growth, using only knowledge available up to November 15, 2005.

As of February 15, 2030, the lower bound of the Fed interest rate is 1.0\%. Inflation in January 2030, measured by CPI, is 3.7\% year-over-year, which exceeds the Federal Reserve's commonly understood inflation target (historically around 2\%, based on knowledge up to 2005). Real GDP growth in Q4 2029 was 0.8\% year-over-year, indicating modest but low economic growth.

Based on pre-2005 understanding of monetary policy, central banks typically respond to inflation above target by raising interest rates to cool demand and bring inflation down. Even with moderate GDP growth, persistently high inflation (3.7\%) relative to the target would likely prompt the Fed to tighten monetary policy.

Given that the current rate is 1.0\% and inflation is 1.7 percentage points above a 2\% target, there would be strong pressure to increase rates over the coming year to restore price stability. A gradual hiking path might be expected to avoid disrupting growth, but with inflation still elevated, the Fed would likely move rates into a more restrictive territory.

Assuming a policy response consistent with pre-2005 understanding of the Taylor rule, which suggests adjusting rates in response to inflation and output gaps, the Fed would likely raise rates by several increments over the year. A typical response might involve increasing the federal funds rate by more than the inflation deviation from target, especially if inflation expectations were perceived to be rising.

Considering the starting rate of 1.0\%, inflation of 3.7\%, and sluggish growth, a reasonable forecast - based on historical policy reactions known before November 15, 2005 - would imply a tightening cycle that raises the lower bound of the Fed interest rate to a level closer to or slightly above the inflation rate to achieve a positive real interest rate.

Therefore, a plausible forecast for the lower bound of the Fed interest rate on February 15, 2031, is approximately 3.0\%, reflecting a balance between controlling inflation and not overly constraining low growth.

Interest rate forecast: 3.0\%
\end{tcolorbox}
\caption{\textbf{Example of a response for 1-year forecast of the lower bound of the Fed interest rate for a fake date by the Qwen3 Instruct model.} Formatting was slightly adjusted for readability.}
\label{fig: figure_D.1.2.2}
\end{figure}

\clearpage

\subsubsection{DeepSeek-V3.1}

\renewcommand{\thefigure}{D.1.3.\arabic{figure}}
\setcounter{figure}{0}

\begin{figure}[ht]
\begin{tcolorbox}[promptbox]
Based solely on the data provided, which reflects the economic situation as of November 15, 2008, the outlook for the Fed interest rate over the next year is for significant further cuts.

The key indicator is the sharp slowdown in real GDP growth to 0.8\% in Q3 2008. This is a very low rate of growth and suggests the economy is stagnating or potentially entering a recession. When economic growth slows dramatically, the typical central bank response is to lower interest rates to stimulate borrowing, spending, and investment.

Although the inflation rate of 3.7\% is above what is typically considered comfortable (often around 2\%), the risk of a severe economic contraction appears to be the more immediate and pressing concern for policymakers. In such a situation, the Fed would likely prioritize supporting growth and employment over immediately bringing down inflation. The current rate of 1.0\% is low, but given the severity of the growth slowdown, there is room and reason to lower it further, even to unprecedented levels.

Therefore, the forecast is for the Fed to continue cutting rates aggressively to combat the economic downturn. The most likely scenario is that the rate will be lowered to its effective lower bound, a level near zero.

Interest rate forecast: 0.0\%
\end{tcolorbox}
\caption{\textbf{Example of a response for 1-year forecast of the lower bound of the Fed interest rate for a real date by the DeepSeek-V3.1 model.} Formatting was slightly adjusted for readability.}
\label{fig: figure_D.1.3.1}
\end{figure}

\begin{figure}[h!!]
\begin{tcolorbox}[promptbox]
Based solely on the data provided (GDP growth of 0.8\% and inflation of 3.7\%), the economic situation appears to be one of stagflation-lite: economic growth is very slow, but inflation remains significantly above the typical Fed target of 2\%. The Fed's primary mandates are to promote maximum employment and price stability. With inflation at 3.7\%, the priority would be to combat high prices, even though growth is anemic. A very low interest rate of 1.0\% is likely considered highly stimulative and inappropriate for an economy with inflation this high. Therefore, the logical policy action would be to raise the interest rate to a level that is more restrictive to cool down inflation, even if it risks slowing the already weak growth further. A moderate increase would be a reasonable expectation.

Interest rate forecast: 2.5\%
\end{tcolorbox}
\caption{\textbf{Example of a response for 1-year forecast of the lower bound of the Fed interest rate for a fake date by the DeepSeek-V3.1 model.} Formatting was slightly adjusted for readability.}
\label{fig: figure_D.1.3.2}
\end{figure}

\clearpage

\subsection{Examples of LLMs' responses with the forecast date of May 15, 2021}

\subsubsection{Kimi-K2 Instruct}

\renewcommand{\thefigure}{D.2.1.\arabic{figure}}
\setcounter{figure}{0}

\begin{figure}[ht]
\begin{tcolorbox}[promptbox]
Given the data as of May 15, 2021, the US economy is in a peculiar state:

- The Fed has set the lower bound of the interest rate at 0.0\%, indicating an extremely accommodative monetary stance aimed at stimulating economic activity.

- Real GDP growth in Q1 2021 was merely 0.4\% YoY, which is quite low, suggesting the economy is still recovering or facing significant headwinds.

- CPI inflation in April 2021 was 4.2\% YoY, which is notably high compared to the Fed's typical 2\% target, indicating potential overheating or supply-side constraints.

Given this combination of low real GDP growth and high inflation, several factors might influence the trajectory to Q1 2022:

1. **Monetary Policy**: With rates at the zero lower bound, the Fed's ability to further stimulate demand through rate cuts is exhausted. However, the high inflation might pressure the Fed to consider tightening, though as of May 2021, no hikes have occurred. The lagged effects of ultra-low rates could continue to support growth into 2022, albeit modestly.

2. **Inflation Dynamics**: The 4.2\% inflation suggests demand may be running ahead of supply, which could either:

   - Lead to a growth slowdown if the Fed tightens or if higher prices erode real incomes, or
   
   - Reflect a robust rebound in demand that could sustain higher growth if supply catches up.
   
3. **Base Effects**: Q1 2021 growth was weak, creating a low base for YoY comparisons in Q1 2022. Even a modest sequential improvement in economic activity could translate into higher YoY growth.

4. **Momentum**: The combination of fiscal support (assuming it continues), pent-up demand (from pandemic-related restrictions), and the low base effect could push growth higher.

Balancing these factors, the most likely outcome is a moderate acceleration in real GDP growth from the Q1 2021 level, driven by base effects and continued policy support, despite inflationary pressures. A growth rate in the 3-4\% range seems plausible, accounting for some normalization without implying a full boom.

GDP growth forecast: 3.5\%
\end{tcolorbox}
\caption{\textbf{Example of a response for 1-year forecast of real GDP growth (YoY) for a real date by the Kimi-K2 Instruct model.} Formatting was slightly adjusted for readability.}
\label{fig: figure_D.2.1.1}
\end{figure}

\clearpage

\begin{figure}[ht]
\begin{tcolorbox}[promptbox]
Given the data provided as of February 15, 2030, we have the following key macroeconomic indicators:

- Fed funds rate lower bound: 0.0\%

- US real GDP growth (Q4 2029, YoY): 0.4\%

- US CPI inflation (January 2030, YoY): 4.2\%

These indicators suggest a situation of **stagflationary pressure**: growth is near zero while inflation is elevated (above 2\% target). The Fed's policy rate is at the zero lower bound, limiting conventional monetary easing tools.
Using only pre-May 2018 economic theory and patterns:

1. **Inflation at 4.2\%** signals demand-side or cost-push pressures. With interest rates already at zero, the Fed has little room to stimulate demand without risking further inflation. This constrains the ability to boost growth via lower rates.

2. **Real GDP growth at 0.4\%** indicates near-stagnation. With inflation high and policy rates low, the Fed is likely constrained from easing further, and may even face pressure to tighten if inflation persists. Tightening would further weigh on growth.

3. **Historical precedent (pre-May 2018)** shows that when inflation is above target and growth is weak, growth tends to remain subdued or decline slightly unless inflation is brought under control. The limited policy space at the zero lower bound historically exacerbates this dynamic.

Given these constraints and the lack of room for policy stimulus, real GDP growth is unlikely to accelerate. Instead, it is more likely to remain low or decline modestly over the year as inflation and policy constraints persist.

GDP growth forecast: 0.1\%
\end{tcolorbox}
\caption{\textbf{Example of a response for 1-year forecast of real GDP growth (YoY) for a fake date by the Kimi-K2 Instruct model.} Formatting was slightly adjusted for readability.}
\label{fig: figure_D.2.1.2}
\end{figure}

\clearpage

\subsubsection{Qwen3 Instruct}

\renewcommand{\thefigure}{D.2.2.\arabic{figure}}
\setcounter{figure}{0}

\begin{figure}[h!!]
\begin{tcolorbox}[promptbox]
To forecast U.S. real GDP growth for Q1 2022, we must rely solely on data available up to May 15, 2018, and the statistics provided in the prompt, which include the current (as of May 15, 2021) Fed interest rate lower bound of 0.0\%, U.S. real GDP growth of 0.4\% (YoY) in Q1 2021, and U.S. inflation (CPI) of 4.2\% (YoY) in April 2021. However, the instruction prohibits using any real-world knowledge after May 15, 2018, including events such as the full impact of monetary policy responses to the 2020 pandemic, fiscal stimulus packages, or post-2018 economic trends.

Given that, we must interpret the provided Q1 2021 GDP growth (0.4\% YoY) and April 2021 inflation (4.2\%) as inputs, but cannot use post-2018 understanding of how such inflation readings or low interest rates affected growth. Based only on pre-May 2018 economic relationships, a 4.2\% inflation rate would typically be associated with tightening monetary policy to prevent overheating. However, the Fed rate is at 0.0\%, which contradicts historical precedents before 2018 where such low rates were typically seen during recessions or deep economic slack (e.g., post-2008 crisis).

Prior to 2018, sustained inflation above 4\% with near-zero interest rates was not observed; normally, the Fed would have raised rates in response. Therefore, under pre-2018 logic, the combination of high inflation and zero rates would be inconsistent and suggest either extreme policy accommodation (unseen before 2018 outside of crises) or data anomalies.

Assuming the provided data is valid and considering that real GDP growth was only 0.4\% YoY in Q1 2021 despite 4.2\% inflation and zero interest rates - conditions that, before 2018, might have spurred stronger growth due to stimulative policy - we infer significant underlying economic weakness. The low growth despite high inflation suggests stagflationary pressures. In pre-2018 economic models, such stagflation (e.g., 1970s) often led to weak subsequent growth as the central bank would eventually have to raise rates to combat inflation, dampening activity.

However, since the Fed rate remains at 0.0\%, policy remains highly accommodative. In historical models before 2018, prolonged accommodation during high inflation was rare, but when rates were kept low during recoveries (e.g., post-2009), GDP growth was modest, averaging around 2\% in the medium term.

Therefore, adjusting for the unusual combination of high inflation and low rates with minimal growth, and applying pre-2018 cyclical logic, we project a moderate rebound in real GDP growth for Q1 2022, assuming some lagged effect of monetary stimulus and a base effect from weak prior growth, but constrained by inflationary drag.

GDP growth forecast: 2.1\%
\end{tcolorbox}
\caption{\textbf{Example of a response for 1-year forecast of real GDP growth (YoY) for a real date by the Qwen3 Instruct model.} Formatting was slightly adjusted for readability.}
\label{fig: figure_D.2.2.1}
\end{figure}

\clearpage

\begin{figure}[ht]
\begin{tcolorbox}[promptbox]
To forecast U.S. real GDP growth for Q4 2030, we rely only on information available up to May 15, 2018, and the data provided in the prompt: as of February 15, 2030, the lower bound of the Fed interest rate is 0.0\%, Q4 2029 real GDP growth was 0.4\% year-over-year (YoY), and January 2030 CPI inflation was 4.2\% YoY.

Given the Federal Reserve's dual mandate of price stability and maximum employment, and based on pre-May 15, 2018 understanding of monetary policy, a central bank typically responds to elevated inflation by tightening monetary policy - raising interest rates to reduce demand and bring inflation under control. However, in this scenario, the Fed funds rate lower bound is at 0.0\%, which suggests the central bank has limited ability to raise rates, possibly due to constraints perceived before May 2018, such as the zero lower bound (ZLB) problem.

With inflation at 4.2\% YoY - well above the Fed's then-understood unofficial target of around 2\% - and policy rates at 0.0\%, this implies a significant deviation from conventional policy responses. According to economic principles known before May 2018, such a combination of high inflation and zero policy rates would be unusual and potentially destabilizing, as it suggests a lack of policy tightening in the face of inflationary pressures. This could lead to inflation expectations becoming unanchored, further driving inflation.

However, real GDP growth in Q4 2029 was only 0.4\% YoY, indicating weak economic activity. Typically, such low growth, especially near or below potential, would reduce inflationary pressures. The combination of high inflation and low growth resembles stagflation, a phenomenon observed in the 1970s and understood by May 2018 as particularly challenging for policymakers.

With interest rates at the zero lower bound, the Fed would likely be constrained in using conventional tools to combat inflation. According to pre-May 2018 knowledge, the Fed might rely on forward guidance or quantitative tightening, but the prompt does not provide information on balance sheet policy or communication strategy.

In a stagflationary environment with limited monetary policy space, economic growth typically remains subdued. High inflation erodes real incomes and creates uncertainty, which can reduce consumption and investment. Even though monetary policy is nominally accommodative (rate at 0.0\%), its real stance is tightening due to high inflation, which effectively raises real interest rates. Real interest rate = nominal rate - inflation. At 0.0\% nominal and 4.2\% inflation, the real rate is -4.2\%, which might seem stimulative, but if inflation expectations are high and persistent, businesses and consumers may act cautiously, limiting spending.

Moreover, if inflation remains high and the Fed is unable or unwilling to raise rates (possibly due to concerns about growth), confidence could deteriorate, leading to reduced investment and hiring. Given the weak starting growth rate (0.4\%) and persistent inflation, it is unlikely that growth accelerates significantly over the next year without a strong positive shock - none of which can be assumed post-May 2018.

Considering the weak prior growth, high inflation, and constrained monetary policy, the outlook for real GDP growth in Q4 2030 is likely stagnant or slightly negative. Based on historical patterns of stagflation known before May 2018 (e.g., mid-1970s), growth tends to remain low or contract slightly.

Therefore, the forecast for Q4 2030 real GDP growth is slightly below the Q4 2029 level, reflecting continued economic weakness amid high inflation and limited policy options.

GDP growth forecast: 0.1\%
\end{tcolorbox}
\caption{\textbf{Example of a response for 1-year forecast of real GDP growth (YoY) for a fake date by the Qwen3 Instruct model.} Formatting was slightly adjusted for readability.}
\label{fig: figure_D.2.2.2}
\end{figure}

\clearpage

\subsubsection{DeepSeek-V3.1}

\renewcommand{\thefigure}{D.2.3.\arabic{figure}}
\setcounter{figure}{0}

\begin{figure}[ht]
\begin{tcolorbox}[promptbox]
Based on the provided statistics as of May 15, 2021, the forecast for US YoY real GDP growth for Q1 2022 is derived from the following reasoning:

The Fed interest rate is at its effective lower bound (0.0\%), indicating an accommodative monetary policy aimed at stimulating economic activity. The real GDP growth for Q1 2021 was only 0.4\% (YoY), which is low, but this period likely reflects the ongoing recovery from the economic disruptions caused by the COVID-19 pandemic. The high inflation rate of 4.2\% (YoY) in April 2021 suggests strong demand pressures or supply chain constraints, which could be indicative of an overheating economy or rapid rebound.

Given the low base of GDP growth in Q1 2021 (0.4\%), the accommodative monetary policy, and high inflationary pressures, it is reasonable to expect a significant acceleration in economic activity over the next year. The low interest rates should encourage borrowing and investment, while the high inflation may reflect robust consumer demand. Therefore, YoY real GDP growth for Q1 2022 is forecast to be substantially higher, reflecting a strong recovery from the depressed level a year earlier.

GDP growth forecast: 5.2\%
\end{tcolorbox}
\caption{\textbf{Example of a response for 1-year forecast of real GDP growth (YoY) for a real date by the DeepSeek-V3.1 model.} Formatting was slightly adjusted for readability.}
\label{fig: figure_D.2.3.1}
\end{figure}

\begin{figure}[h!!]
\begin{tcolorbox}[promptbox]
Based solely on the provided statistics and the instruction to not use any external knowledge, the forecast is derived from the relationship between inflation, interest rates, and economic growth.

The current inflation rate of 4.2\% is significantly above the typical 2\% target, while the interest rate is at its effective lower bound of 0.0\%. With inflation high and the Fed unable to lower rates further to stimulate the economy, the most probable action for the Fed would be to raise interest rates to combat inflation. Raising interest rates makes borrowing more expensive for businesses and consumers, which typically slows down economic activity, investment, and spending. This action is taken to reduce demand in the economy and bring inflation down to a target level. Consequently, a tightening of monetary policy through rate hikes would be expected to lead to lower real GDP growth in the near future. Given the current high inflation, the Fed's response would likely be assertive, leading to a noticeable slowdown in growth by Q4 2030. The most direct outcome based on this dynamic is a forecast of negative growth.

GDP growth forecast: -0.8\%
\end{tcolorbox}
\caption{\textbf{Example of a response for 1-year forecast of real GDP growth (YoY) for a fake date by the DeepSeek-V3.1 model.} Formatting was slightly adjusted for readability.}
\label{fig: figure_D.2.3.2}
\end{figure}

\clearpage

\begin{landscape}

\appsection[app:appendix_e]{p-values for comparing forecasts in Fake date test II}

\renewcommand{\thetable}{E.\arabic{table}}
\setcounter{table}{0}

\begin{table}[h!]
\centering
\caption{\textbf{p-values for comparing forecasts in Fake date test II.} p-values for the Kolmogorov-Smirnov permutation test when comparing forecasts for real and fake dates.}
\small
\setlength{\tabcolsep}{3.0pt}
\begin{tabular}{l*{9}{c}}
\hline
& \multicolumn{3}{c}{Kimi-K2 Instruct} 
& \multicolumn{3}{c}{Qwen3 Instruct} 
& \multicolumn{3}{c}{DeepSeek-V3.1} \\
\cline{2-4}\cline{5-7}\cline{8-10}
& \shortstack{Fed int. rate}
& \shortstack{CPI growth}
& \shortstack{Real GDP growth}
& \shortstack{Fed int. rate}
& \shortstack{CPI growth}
& \shortstack{Real GDP growth}
& \shortstack{Fed int. rate}
& \shortstack{CPI growth}
& \shortstack{Real GDP growth} \\
\hline
May 15, 2005 & 0,000 & 0,791 & 0,000 & 0,000 & 0,099 & 0,005 & 0,000 & 0,086 & 0,000 \\
Aug. 15, 2005 & 0,000 & 0,005 & 0,000 & 0,000 & 0,000 & 0,000 & 0,000 & 0,009 & 0,000 \\
Nov. 15, 2005 & 0,301 & 0,000 & 0,000 & 0,000 & 0,000 & 0,000 & 0,000 & 0,020 & 0,000 \\
Feb. 15, 2006 & 0,000 & 0,021 & 0,000 & 0,000 & 0,000 & 0,000 & 0,005 & 0,008 & 0,000 \\
May 15, 2006 & 0,001 & 0,380 & 0,000 & 0,000 & 0,000 & 0,000 & 0,000 & 0,012 & 0,000 \\
Aug. 15, 2006 & 0,000 & 0,038 & 0,000 & 0,000 & 0,000 & 0,000 & 0,001 & 0,000 & 0,000 \\
Nov. 15, 2006 & 0,000 & 0,000 & 0,087 & 0,000 & 0,000 & 0,000 & 0,000 & 0,226 & 0,000 \\
Feb. 15, 2007 & 0,000 & 0,002 & 0,000 & 0,000 & 0,000 & 0,000 & 0,000 & 0,005 & 0,000 \\
May 15, 2007 & 0,000 & 0,000 & 0,005 & 0,003 & 0,000 & 0,000 & 0,125 & 0,178 & 0,000 \\
Aug. 15, 2007 & 0,000 & 0,375 & 0,001 & 0,012 & 0,000 & 0,000 & 0,000 & 0,009 & 0,026 \\
Nov. 15, 2007 & 0,000 & 0,003 & 0,000 & 0,000 & 0,000 & 0,000 & 0,000 & 0,008 & 0,070 \\
Feb. 15, 2008 & 0,000 & 0,000 & 0,000 & 0,000 & 0,000 & 0,000 & 0,000 & 0,024 & 0,009 \\
May 15, 2008 & 0,000 & 0,000 & 0,000 & 0,000 & 0,349 & 0,000 & 0,000 & 0,000 & 0,000 \\
Aug. 15, 2008 & 0,000 & 0,000 & 0,000 & 0,000 & 0,000 & 0,000 & 0,000 & 0,000 & 0,004 \\
Nov. 15, 2008 & 0,000 & 0,000 & 0,000 & 0,000 & 0,000 & 0,000 & 0,000 & 0,000 & 0,000 \\
Feb. 15, 2009 & 1,000 & 0,000 & 0,000 & 1,000 & 0,000 & 0,000 & 0,000 & 0,000 & 0,000 \\
May 15, 2009 & 1,000 & 0,000 & 0,318 & 1,000 & 0,000 & 0,000 & 0,000 & 0,000 & 0,000 \\
Aug. 15, 2009 & 1,000 & 0,000 & 0,000 & 1,000 & 0,000 & 0,000 & 0,000 & 0,000 & 0,000 \\
Nov. 15, 2009 & 1,000 & 0,000 & 0,000 & 1,000 & 0,000 & 0,000 & 0,000 & 0,000 & 0,000 \\
Feb. 15, 2010 & 0,065 & 0,000 & 0,000 & 0,000 & 0,199 & 0,000 & 0,000 & 0,000 & 0,000 \\
\hline
\end{tabular}
\label{tab: table_E.1}
\end{table}

\begin{table}[h!]
\centering
\caption{\textbf{p-values for comparing forecasts in Fake date test II.} p-values for the Kolmogorov-Smirnov permutation test when comparing forecasts for real and fake dates.}
\small
\setlength{\tabcolsep}{3.0pt}
\begin{tabular}{l*{9}{c}}
\hline
& \multicolumn{3}{c}{Kimi-K2 Instruct} 
& \multicolumn{3}{c}{Qwen3 Instruct} 
& \multicolumn{3}{c}{DeepSeek-V3.1} \\
\cline{2-4}\cline{5-7}\cline{8-10}
& \shortstack{Fed int. rate}
& \shortstack{CPI growth}
& \shortstack{Real GDP growth}
& \shortstack{Fed int. rate}
& \shortstack{CPI growth}
& \shortstack{Real GDP growth}
& \shortstack{Fed int. rate}
& \shortstack{CPI growth}
& \shortstack{Real GDP growth} \\
\hline
May 15, 2010 & 0,000 & 0,000 & 0,000 & 0,000 & 0,001 & 0,000 & 0,000 & 0,261 & 0,000 \\
Aug. 15, 2010 & 0,000 & 0,196 & 0,824 & 0,000 & 0,147 & 0,000 & 0,053 & 0,006 & 0,000 \\
Nov. 15, 2010 & 0,000 & 0,000 & 0,520 & 0,000 & 0,000 & 0,006 & 0,218 & 0,017 & 0,000 \\
Feb. 15, 2011 & 0,000 & 0,000 & 0,000 & 0,000 & 0,000 & 0,000 & 0,010 & 0,006 & 0,000 \\
May 15, 2011 & 0,000 & 0,000 & 0,015 & 0,000 & 0,000 & 0,000 & 0,000 & 0,000 & 0,000 \\
Aug. 15, 2011 & 0,000 & 0,000 & 0,006 & 0,000 & 0,000 & 0,103 & 0,000 & 0,000 & 0,000 \\
Nov. 15, 2011 & 0,000 & 0,000 & 0,060 & 0,000 & 0,000 & 0,000 & 0,000 & 0,000 & 0,000 \\
Feb. 15, 2012 & 0,000 & 0,001 & 0,000 & 0,000 & 0,000 & 0,000 & 0,000 & 0,000 & 0,000 \\
May 15, 2012 & 0,000 & 0,043 & 0,002 & 0,000 & 0,000 & 0,177 & 0,000 & 0,000 & 0,146 \\
Aug. 15, 2012 & 0,000 & 0,043 & 0,087 & 0,000 & 0,550 & 0,000 & 0,029 & 0,001 & 0,023 \\
Nov. 15, 2012 & 0,000 & 0,005 & 0,009 & 0,000 & 0,001 & 0,000 & 0,000 & 0,000 & 0,001 \\
Feb. 15, 2013 & 0,001 & 0,173 & 0,000 & 0,000 & 0,000 & 0,000 & 0,497 & 0,028 & 0,000 \\
May 15, 2013 & 0,035 & 0,026 & 0,000 & 0,007 & 0,000 & 0,000 & 0,241 & 0,000 & 0,000 \\
Aug. 15, 2013 & 0,000 & 0,048 & 0,000 & 0,000 & 0,029 & 0,000 & 0,000 & 0,000 & 0,000 \\
Nov. 15, 2013 & 0,747 & 0,000 & 0,000 & 0,258 & 0,000 & 0,000 & 0,117 & 0,000 & 0,000 \\
Feb. 15, 2014 & 0,001 & 0,001 & 0,000 & 0,000 & 0,000 & 0,000 & 0,002 & 0,000 & 0,000 \\
May 15, 2014 & 0,000 & 0,226 & 0,000 & 0,000 & 0,001 & 0,000 & 0,000 & 0,000 & 0,000 \\
Aug. 15, 2014 & 0,000 & 0,178 & 0,000 & 0,000 & 0,000 & 0,000 & 0,000 & 0,000 & 0,000 \\
Nov. 15, 2014 & 0,000 & 0,035 & 0,000 & 0,000 & 0,000 & 0,000 & 0,220 & 0,000 & 0,001 \\
Feb. 15, 2015 & 0,000 & 0,000 & 0,000 & 0,006 & 0,000 & 0,000 & 0,000 & 0,000 & 0,000 \\
\hline
\end{tabular}
\label{tab: table_E.2}
\end{table}

\begin{table}[h!]
\centering
\caption{\textbf{p-values for comparing forecasts in Fake date test II.} p-values for the Kolmogorov-Smirnov permutation test when comparing forecasts for real and fake dates.}
\small
\setlength{\tabcolsep}{3.0pt}
\begin{tabular}{l*{9}{c}}
\hline
& \multicolumn{3}{c}{Kimi-K2 Instruct} 
& \multicolumn{3}{c}{Qwen3 Instruct} 
& \multicolumn{3}{c}{DeepSeek-V3.1} \\
\cline{2-4}\cline{5-7}\cline{8-10}
& \shortstack{Fed int. rate}
& \shortstack{CPI growth}
& \shortstack{Real GDP growth}
& \shortstack{Fed int. rate}
& \shortstack{CPI growth}
& \shortstack{Real GDP growth}
& \shortstack{Fed int. rate}
& \shortstack{CPI growth}
& \shortstack{Real GDP growth} \\
\hline
May 15, 2015 & 0,000 & 0,005 & 0,000 & 0,000 & 0,000 & 0,000 & 0,000 & 0,000 & 0,000 \\
Aug. 15, 2015 & 0,000 & 0,000 & 0,000 & 0,000 & 0,000 & 0,000 & 0,000 & 0,000 & 0,000 \\
Nov. 15, 2015 & 0,000 & 0,000 & 0,000 & 0,000 & 0,000 & 0,000 & 0,004 & 0,000 & 0,000 \\
Feb. 15, 2016 & 0,000 & 0,954 & 0,001 & 0,000 & 0,000 & 0,000 & 0,000 & 0,021 & 0,000 \\
May 15, 2016 & 0,000 & 0,056 & 0,000 & 0,000 & 0,000 & 0,000 & 0,000 & 0,003 & 0,000 \\
Aug. 15, 2016 & 0,000 & 0,000 & 0,000 & 0,000 & 0,000 & 0,000 & 0,000 & 0,001 & 0,000 \\
Nov. 15, 2016 & 0,000 & 0,020 & 0,000 & 0,000 & 0,154 & 0,000 & 0,000 & 0,007 & 0,001 \\
Feb. 15, 2017 & 0,002 & 0,011 & 0,702 & 0,000 & 0,000 & 0,000 & 0,036 & 0,000 & 0,000 \\
May 15, 2017 & 0,000 & 0,024 & 0,155 & 0,000 & 0,000 & 0,000 & 0,000 & 0,000 & 0,005 \\
Aug. 15, 2017 & 0,000 & 0,396 & 0,063 & 0,000 & 0,282 & 0,000 & 0,000 & 0,004 & 0,599 \\
Nov. 15, 2017 & 0,000 & 0,232 & 0,003 & 0,000 & 0,000 & 0,000 & 0,000 & 0,017 & 0,055 \\
Feb. 15, 2018 & 0,000 & 0,806 & 0,005 & 0,000 & 0,002 & 0,000 & 0,000 & 0,015 & 0,766 \\
May 15, 2018 & 0,000 & 0,037 & 0,001 & 0,000 & 0,001 & 0,000 & 0,000 & 0,000 & 0,423 \\
Aug. 15, 2018 & 0,000 & 0,000 & 0,000 & 0,001 & 0,000 & 0,000 & 0,245 & 0,000 & 0,001 \\
Nov. 15, 2018 & 0,000 & 0,000 & 0,001 & 0,000 & 0,000 & 0,001 & 0,018 & 0,000 & 0,000 \\
Feb. 15, 2019 & 0,000 & 0,589 & 0,000 & 0,000 & 0,000 & 0,000 & 0,000 & 0,000 & 0,000 \\
May 15, 2019 & 0,000 & 0,001 & 0,000 & 0,000 & 0,003 & 0,000 & 0,000 & 0,000 & 0,000 \\
Aug. 15, 2019 & 0,000 & 0,012 & 0,000 & 0,000 & 0,000 & 0,000 & 0,000 & 0,079 & 0,216 \\
Nov. 15, 2019 & 0,000 & 0,137 & 0,000 & 0,000 & 0,191 & 0,000 & 0,080 & 0,002 & 0,000 \\
Feb. 15, 2020 & 0,000 & 0,000 & 0,000 & 0,000 & 0,004 & 0,000 & 0,000 & 0,002 & 0,183 \\
\hline
\end{tabular}
\label{tab: table_E.3}
\end{table}

\begin{table}[h!]
\centering
\caption{\textbf{p-values for comparing forecasts in Fake date test II.} p-values for the Kolmogorov-Smirnov permutation test when comparing forecasts for real and fake dates.}
\small
\setlength{\tabcolsep}{3.0pt}
\begin{tabular}{l*{9}{c}}
\hline
& \multicolumn{3}{c}{Kimi-K2 Instruct} 
& \multicolumn{3}{c}{Qwen3 Instruct} 
& \multicolumn{3}{c}{DeepSeek-V3.1} \\
\cline{2-4}\cline{5-7}\cline{8-10}
& \shortstack{Fed int. rate}
& \shortstack{CPI growth}
& \shortstack{Real GDP growth}
& \shortstack{Fed int. rate}
& \shortstack{CPI growth}
& \shortstack{Real GDP growth}
& \shortstack{Fed int. rate}
& \shortstack{CPI growth}
& \shortstack{Real GDP growth} \\
\hline
May 15, 2020 & 1,000 & 0,000 & 0,000 & 1,000 & 0,000 & 0,000 & 0,252 & 0,000 & 0,000 \\
Aug. 15, 2020 & 1,000 & 0,000 & 0,000 & 1,000 & 0,000 & 0,000 & 0,063 & 0,000 & 0,000 \\
Nov. 15, 2020 & 1,000 & 0,000 & 0,000 & 1,000 & 0,000 & 0,000 & 0,000 & 0,000 & 0,000 \\
Feb. 15, 2021 & 1,000 & 0,000 & 0,000 & 1,000 & 0,000 & 0,000 & 0,002 & 0,000 & 0,000 \\
May 15, 2021 & 0,000 & 0,398 & 0,000 & 0,000 & 0,000 & 0,000 & 0,000 & 0,000 & 0,000 \\
Aug. 15, 2021 & 0,000 & 0,000 & 0,000 & 0,000 & 0,000 & 0,000 & 0,000 & 0,000 & 0,003 \\
Nov. 15, 2021 & 0,000 & 0,000 & 0,068 & 0,000 & 0,000 & 0,000 & 0,000 & 0,000 & 0,000 \\
Feb. 15, 2022 & 0,000 & 0,000 & 0,001 & 0,000 & 0,000 & 0,000 & 0,000 & 0,000 & 0,000 \\
May 15, 2022 & 0,000 & 0,000 & 0,000 & 0,000 & 0,000 & 0,000 & 0,000 & 0,000 & 0,000 \\
Aug. 15, 2022 & 0,000 & 0,000 & 0,295 & 0,000 & 0,000 & 0,000 & 0,000 & 0,000 & 0,000 \\
Nov. 15, 2022 & 0,000 & 0,000 & 0,015 & 0,000 & 0,000 & 0,000 & 0,000 & 0,000 & 0,152 \\
Feb. 15, 2023 & 0,000 & 0,000 & 0,066 & 0,000 & 0,000 & 0,000 & 0,001 & 0,000 & 0,018 \\
May 15, 2023 & 0,000 & 0,000 & 0,000 & 0,152 & 0,000 & 0,231 & 0,005 & 0,000 & 0,086 \\
Aug. 15, 2023 & 0,209 & 0,005 & 0,000 & 0,000 & 0,000 & 0,000 & 0,000 & 0,000 & 0,000 \\
Nov. 15, 2023 & 0,218 & 0,000 & 0,000 & 0,003 & 0,000 & 0,000 & 0,319 & 0,000 & 0,000 \\
Feb. 15, 2024 & 0,000 & 0,000 & 0,000 & 0,000 & 0,633 & 0,001 & 0,000 & 0,000 & 0,013 \\
May 15, 2024 & 0,400 & 0,001 & 0,002 & 0,009 & 0,005 & 0,172 & 0,017 & 0,000 & 0,739 \\
Aug. 15, 2024 & 0,001 & 0,030 & 0,000 & 0,179 & 0,038 & 0,000 & 0,000 & 0,000 & 0,000 \\
Nov. 15, 2024 & 0,000 & 0,001 & 0,000 & 0,000 & 0,000 & 0,003 & 0,000 & 0,001 & 0,000 \\
Feb. 15, 2025 & 0,001 & 0,074 & 0,005 & 0,000 & 0,000 & 0,109 & 0,002 & 0,000 & 0,005 \\
\hline
\end{tabular}
\label{tab: table_E.4}
\end{table}

\end{landscape}

\clearpage

\appsection[app:appendix_f]{Examples of LLMs' responses for Fake date test II with the forecast date of May 15, 2021}

\subsection{Kimi-K2 Instruct}

\subsubsection{Fed interest rate}

\renewcommand{\thefigure}{F.1.1.\arabic{figure}}
\setcounter{figure}{0}

\begin{figure}[ht]
\begin{tcolorbox}[promptbox]
To forecast the lower bound of the Fed interest rate on May 15, 2022, we need to assess how the Federal Reserve might respond to the current macroeconomic conditions as of May 15, 2021.

**Key data (as of May 15, 2021):**

- Fed funds rate lower bound: **0.0\%**

- Real GDP growth (Q1 2021 YoY): **0.4\%**

- CPI inflation (April 2021 YoY): **4.2\%**

**Analysis:**

1. **Inflation is elevated**: A 4.2\% CPI inflation rate is significantly above the Fed's 2\% target. This suggests inflationary pressures are building, possibly due to supply bottlenecks or fiscal stimulus effects.

2. **Growth is subdued**: Real GDP growth at 0.4\% YoY is quite low, indicating the economy is still far from full recovery, especially considering the pandemic context.

3. **Policy stance**: The Fed has kept rates at the zero lower bound since March 2020, signaling a commitment to accommodative policy to support recovery. However, if inflation remains persistently above target and growth improves, the Fed may begin to consider tightening.

4. **Forward guidance (as of May 2021)**: The Fed has indicated it will not raise rates until it sees ``substantial further progress'' toward its dual mandate goals. The current data show inflation above target but growth still weak, so the Fed may tolerate higher inflation temporarily.

Given the balance between rising inflation and still-weak growth, the Fed is unlikely to raise rates aggressively in the near term. However, if inflation remains elevated and growth picks up over the next year, a modest increase in the lower bound could occur by May 2022.

A reasonable forecast is that the Fed begins a cautious tightening cycle, raising the lower bound slightly to signal a shift in stance, while still maintaining an overall accommodative posture.

Interest rate forecast: 0.25\%
\end{tcolorbox}
\caption{\textbf{Example of a response for 1-year forecast of the lower bound of the Fed interest rate for a real date by the Kimi-K2 Instruct model.} Formatting was slightly adjusted for readability.}
\label{fig: figure_F.1.1.1}
\end{figure}

\clearpage

\begin{figure}[ht]
\begin{tcolorbox}[promptbox]
Given the current data:

- Fed lower bound: 0.0\%

- Q4 2029 real GDP growth: 0.4\% YoY

- January 2030 CPI inflation: 4.2\% YoY

At the start of 2030 the Fed faces a clear imbalance: inflation is running well above its 2\% target while the economy is barely growing. With the policy rate already at the zero lower bound, the Fed has little room to ease further and strong reason to tighten if it wishes to keep the 2\% inflation target credible. A growth rate of 0.4\% is weak, but it is not yet an outright contraction; meanwhile, 4.2\% CPI indicates that price pressures are persistent.

The Fed's dual mandate calls for maximum employment and price stability. At this juncture, the inflation leg is far off target and the growth leg is only marginally weak. Historical tightening cycles (as of May 2021) show the FOMC willing to raise rates when inflation exceeds target materially, even if growth is sluggish, provided unemployment remains low (and the labor market, while not explicitly given, is unlikely to be in deep distress if GDP is still positive). With the policy rate at zero, the Fed has ample space to lift rates without risking an immediate hard landing.

Taking a standard quarterly-tightening cadence of 25 bp increments beginning as early as March 2030, the Committee could deliver eight hikes by February 2031. That would raise the lower bound from 0.0\% to 2.0\%. Given the inflation gap and the still-positive GDP print, such a cumulative move is plausible; larger or faster increases would risk tipping an already fragile expansion into recession, while fewer hikes would leave inflation expectations unanchored. Therefore the most reasonable forecast is that the lower bound will reach 2\%.

Interest rate forecast: 2.0\%
\end{tcolorbox}
\caption{\textbf{Example of a response for 1-year forecast of the lower bound of the Fed interest rate for a fake date by the Kimi-K2 Instruct model.} Formatting was slightly adjusted for readability.}
\label{fig: figure_F.1.1.2}
\end{figure}

\clearpage

\subsubsection{CPI growth (YoY)}

\renewcommand{\thefigure}{F.1.2.\arabic{figure}}
\setcounter{figure}{0}

\begin{figure}[ht]
\begin{tcolorbox}[promptbox]
Given that we are on May 15, 2021, the Fed funds rate is at the zero lower bound (0.0\%), Q1 2021 real GDP grew only 0.4 \% YoY (indicating still-sluggish demand), and headline CPI inflation has already jumped to 4.2\% YoY as of April 2021, the key question is how quickly this elevated inflation will fade or persist during the coming twelve months.

Mechanical base effects

The April 2020 CPI level was depressed by the initial pandemic shock, so the April 2021 print contains a sizeable ``base-effect'' boost. Once that low base rolls out of the 12-month window, the same mechanical effect will push the April 2022 YoY reading lower, all else equal.

Momentum in prices, wages, and reopening  

On the other hand, strong sequential momentum is visible in reopening-sensitive categories (used cars, airfares, lodging, food services). Supply bottlenecks and shipping delays are still acute. Surveys (ISM, NFIB) point to widespread input-cost pressures and planned price increases. The labor market is tightening quickly - April payrolls surprised to the downside, but average hourly earnings are already rising at a 4-5\% annualized pace in the latest three-month change. These forces argue for persistent core inflation.

Policy backdrop  

With the policy rate stuck at zero and the Fed emphasizing that it will look through ``transitory'' overshoots, there is no immediate monetary brake on demand. Fiscal impulses from ARPA remain in the pipeline through mid-2022. Hence, demand-side pressures are unlikely to cool abruptly.

Putting it together  

The mechanical base effect alone would drag headline CPI down roughly 1½-2 pp by April 2022. Offsetting that, reopening and cost-push pressures could add 1-1¼ pp. Netting these forces yields an inflation rate that is lower than the current 4.2\% but still clearly above the pre-pandemic norm.

Inflation forecast: 3.1\%
\end{tcolorbox}
\caption{\textbf{Example of a response for 1-year forecast of CPI growth (YoY) for a real date by the Kimi-K2 Instruct model.} Formatting was slightly adjusted for readability.}
\label{fig: figure_F.1.2.1}
\end{figure}

\clearpage

\begin{figure}[ht]
\begin{tcolorbox}[promptbox]
Given the data provided as of February 15, 2030:

- Fed policy rate floor: 0.0\%

- Latest real GDP growth (Q4 2029, YoY): 0.4\%

- Latest CPI inflation (January 2030, YoY): 4.2\%

The economy is in a low-growth, moderately-high-inflation environment.

With the policy rate already at its effective lower bound, the Fed has very little room to tighten monetary conditions; therefore the current stance is highly accommodative. Under such conditions, inflation tends to remain above target for some time because (1) real rates are negative and stimulating demand, (2) the output gap is not wide enough to create substantial disinflationary pressure, and (3) there is no evidence of imminent aggressive tightening. Historical episodes of similar starting points (pre-May-2021 data) suggest that inflation declines only gradually unless policy is tightened meaningfully, which is not possible here.

Taking into account the sluggish growth and the absence of tightening, inflation is expected to drift lower over the year but stay meaningfully above the pre-2021 norm of ~2\%. A plausible path would shave roughly one percentage point off the current 4.2\% figure by January 2021, reflecting mild base effects and some erosion of momentum.

Inflation forecast: 3.2\%
\end{tcolorbox}
\caption{\textbf{Example of a response for 1-year forecast of CPI growth (YoY) for a fake date by the Kimi-K2 Instruct model.} Formatting was slightly adjusted for readability.}
\label{fig: figure_F.1.2.2}
\end{figure}

\clearpage

\subsubsection{Real GDP growth (YoY)}

\renewcommand{\thefigure}{F.1.3.\arabic{figure}}
\setcounter{figure}{0}

\begin{figure}[ht]
\begin{tcolorbox}[promptbox]
Explanation:

Given the data available up to May 15, 2021, we observe:

- The Fed's policy rate is at the zero lower bound (0.0\%), indicating highly accommodative monetary policy.

- US real GDP growth in Q1 2021 was a modest 0.4\% year-over-year.

- Headline CPI inflation in April 2021 was already 4.2\% year-over-year, well above the Fed's 2\% target.

These conditions mirror the early stages of a reopening-driven rebound: extremely low base effects from the pandemic shock in 2020, unprecedented fiscal and monetary stimulus still in the pipeline (e.g., the American Rescue Plan enacted in March 2021), and the accelerating vaccination rollout that should allow service-sector activity to revive. With policy settings remaining ultra-loose and households holding elevated excess savings, sequential momentum is set to strengthen markedly through the second half of 2021. Therefore, by Q1 2022 the YoY comparison will be against the still-depressed levels of Q1 2021, magnifying growth. Historical patterns from past post-recession rebounds (such as after the 2008-09 recession) suggest that when policy support is large and the output gap wide, real GDP often prints in the 5\%-7\% YoY range for several quarters. Balancing the fading boost from fiscal stimulus as 2021 progresses against the drag from higher inflation and possible supply bottlenecks, a central-tendency estimate is around 6\% real GDP growth year-over-year in the first quarter of 2022.

GDP growth forecast: 6.0\%
\end{tcolorbox}
\caption{\textbf{Example of a response for 1-year forecast of real GDP growth (YoY) for a real date by the Kimi-K2 Instruct model.} Formatting was slightly adjusted for readability.}
\label{fig: figure_F.1.3.1}
\end{figure}

\clearpage

\begin{figure}[ht]
\begin{tcolorbox}[promptbox]
Given the data as of February 15, 2030, I must use only information available up to May 15, 2021. That means I cannot incorporate any post-May 2021 economic developments, models, or data. I will therefore rely solely on the provided statistics and what was known before May 2021.

The key inputs are:

- Fed policy rate lower bound: 0.0\%

- Q4 2029 real GDP growth (YoY): 0.4\%

- January 2030 CPI inflation (YoY): 4.2\%

From a pre-May 2021 perspective:
- A 0.0\% policy rate is consistent with monetary accommodation typical after periods of economic weakness or stagnation.  

- 0.4\% real GDP growth is extremely low—below potential and consistent with stagnation or near-recession conditions.

- 4.2\% CPI inflation is elevated relative to the Fed's historical 2\% target, suggesting some combination of supply constraints or demand pressures, but not hyperinflation.

Given these three facts alone and without any post-May 2021 context, the most straightforward inference is that the U.S. economy in late 2029 and early 2030 is experiencing stagflation-like conditions: very weak real growth accompanied by above-target inflation. Monetary policy is already at its lower bound, implying limited additional conventional stimulus. Under such conditions, and absent any information suggesting policy shifts or external shocks, the economy is unlikely to accelerate sharply. Historical patterns prior to May 2021 suggest that when growth is near zero and inflation is elevated, recovery tends to be slow and gradual. Therefore, I expect only modest improvement by Q4 2030.

GDP growth forecast: 1.2\%
\end{tcolorbox}
\caption{\textbf{Example of a response for 1-year forecast of real GDP growth (YoY) for a fake date by the Kimi-K2 Instruct model.} Formatting was slightly adjusted for readability.}
\label{fig: figure_F.1.3.2}
\end{figure}

\clearpage

\subsection{Qwen3 Instruct}

\subsubsection{Fed interest rate}

\renewcommand{\thefigure}{F.2.1.\arabic{figure}}
\setcounter{figure}{0}

\begin{figure}[ht]
\begin{tcolorbox}[promptbox]
As of May 15, 2021, the Federal Reserve has maintained a highly accommodative monetary policy stance, with the lower bound of the federal funds rate at 0.0\%, in response to the economic disruptions caused by the pandemic. The U.S. real GDP growth in Q1 2021 was 0.4\% year-over-year, indicating a still-recovering economy that has not yet returned to pre-pandemic growth trends. However, inflation in April 2021 reached 4.2\% year-over-year, a notable increase that exceeds the Fed's 2\% target and may raise concerns about price stability.

Despite the elevated inflation reading, the Federal Reserve has previously communicated that it expects transitory factors - such as base effects from low prices in 2020 and supply chain disruptions - to drive the near-term rise in inflation. Given the modest GDP growth and the Fed's commitment to a flexible average inflation targeting (FAIT) framework, which allows inflation to run moderately above 2\% for some time to compensate for prior undershoots, the Fed is likely to remain patient before tightening policy.

Moreover, as of May 2021, the labor market has not yet achieved ``maximum employment'', another key component of the Fed's dual mandate. Therefore, the Fed is expected to maintain low interest rates through 2021 and into 2022 to support the recovery. Market expectations and Fed dot plots as of mid-2021 suggest that liftoff - the first rate hike - is not anticipated until 2023. Therefore, even with elevated inflation, the Fed is unlikely to raise rates by May 15, 2022, due to its focus on sustainable, broad-based economic recovery.

Interest rate forecast: 0.0\%
\end{tcolorbox}
\caption{\textbf{Example of a response for 1-year forecast of the lower bound of the Fed interest rate for a real date by the Qwen3 Instruct model.} Formatting was slightly adjusted for readability.}
\label{fig: figure_F.2.1.1}
\end{figure}

\clearpage

\begin{figure}[ht]
\begin{tcolorbox}[promptbox]
To forecast the lower bound of the Fed interest rate on February 15, 2031, we analyze the most recently available economic indicators as of February 15, 2030: US real GDP growth in Q4 2029 was 0.4\% year-over-year, indicating very weak economic growth, and US inflation in January 2030 was 4.2\% year-over-year, which is significantly above the Federal Reserve's typical long-term inflation target (assumed to be around 2\%, based on pre-May 2021 historical context). The current lower bound of the Fed interest rate is 0.0\%, suggesting that monetary policy is already at its most accommodative level in terms of nominal rates.

Given the combination of high inflation (4.2\%) and very low growth (0.4\%), the economy exhibits a stagflationary environment. If the Federal Reserve prioritizes inflation control - as it has historically done during periods of persistently high inflation - it may seek to tighten policy. However, with the rate already at 0.0\%, the Fed cannot lower rates further, and any tightening would require raising rates. Yet, with growth barely positive, the Fed may be cautious about aggressive rate hikes that could tip the economy into contraction.

However, because the rate is at 0.0\%, which is unusually low given 4.2\% inflation (implying a deeply negative real interest rate), it is likely that the Fed would begin raising rates to regain policy space and anchor inflation expectations. The persistence of inflation at 4.2\% - well above target - would likely prompt a policy response, assuming the Fed follows a Taylor-rule-like approach where rates are adjusted based on deviations of inflation from target and output from potential.

Therefore, even with weak growth, sustained inflation above 4\% would likely force the Fed to increase the lower bound of the federal funds rate over the coming year to combat inflationary pressures, especially if inflation remains elevated through early 2030 and into 2031. A gradual normalization path would suggest moderate rate increases, moving the lower bound off zero.

Based on the need to address above-target inflation despite weak growth, a reasonable forecast is that the Fed will raise the lower bound to a level that begins to constrain inflation but avoids deepening economic stagnation.

Interest rate forecast: 2.0\%
\end{tcolorbox}
\caption{\textbf{Example of a response for 1-year forecast of the lower bound of the Fed interest rate for a fake date by the Qwen3 Instruct model.} Formatting was slightly adjusted for readability.}
\label{fig: figure_F.2.1.2}
\end{figure}

\clearpage

\subsubsection{CPI growth (YoY)}

\renewcommand{\thefigure}{F.2.2.\arabic{figure}}
\setcounter{figure}{0}

\begin{figure}[h!!]
\begin{tcolorbox}[promptbox]
To forecast U.S. year-over-year inflation for April 2022, we consider the available data as of May 15, 2021. At that time, the Federal Reserve has maintained the lower bound of the federal funds rate at 0.0\%, indicating highly accommodative monetary policy in response to the economic conditions following the pandemic. This low interest rate environment tends to support demand and can contribute to upward pressure on prices.

Additionally, U.S. real GDP growth in Q1 2021 was 0.4\% year-over-year, suggesting that the economy is still recovering and operating below its potential output. In such a scenario, slack in the economy might normally dampen inflationary pressures. However, the year-over-year CPI inflation in April 2021 was already elevated at 4.2\%, which is significantly above the Federal Reserve's long-run target of 2\%. This high inflation reading is influenced by base effects due to the sharp drop in prices during the early phase of the pandemic in spring 2020, as well as supply chain disruptions and surging demand in certain sectors as the economy reopens.

Given the base effects, the 4.2\% inflation rate in April 2021 will directly affect the year-over-year reading in April 2022, unless prices stabilize or decline. Because prices in April 2020 were exceptionally low due to the pandemic shock, the comparison for April 2022 will be against a relatively normal price level (April 2021), meaning the base effect will diminish but may still contribute to above-average inflation if current price pressures persist.

Moreover, with continued fiscal stimulus, accommodative monetary policy, and ongoing supply constraints, inflationary pressures are likely to persist into 2022. Although the economy has slack, sector-specific bottlenecks (e.g., in semiconductors, transportation, and energy) and strong consumer demand fueled by savings accumulated during lockdowns can keep prices elevated.

Considering these factors - persistent demand, supply constraints, ongoing monetary and fiscal stimulus, and diminishing but still relevant base effects - it is reasonable to expect that inflation will remain elevated in April 2022 compared to pre-pandemic norms, though potentially moderating slightly from the April 2021 peak if supply adjusts and base effects fade.

However, since inflation pressures appear broadening beyond just temporary factors by mid-2021 (based on data available up to that point), and with no indication of monetary tightening (given the 0.0\% rate), inflation is likely to remain high.

Thus, a plausible forecast incorporates continued high inflation, though slightly lower than the 4.2\% print if some normalization occurs, or potentially similar if pressures persist.

Inflation forecast: 3.8\%
\end{tcolorbox}
\caption{\textbf{Example of a response for 1-year forecast of CPI growth (YoY) for a real date by the Qwen3 Instruct model.} Formatting was slightly adjusted for readability.}
\label{fig: figure_F.2.2.1}
\end{figure}

\clearpage

\begin{figure}[ht]
\begin{tcolorbox}[promptbox]
To forecast U.S. year-over-year (YoY) inflation for January 2031, we analyze the most recent available data as of February 15, 2030. At that time, the Federal Reserve's lower bound for the interest rate is 0.0\%, indicating that monetary policy is likely accommodative or constrained by the zero lower bound. Historically, such a policy stance tends to support economic activity and can contribute to inflationary pressures if sustained, especially if output gaps are closing or expectations become unanchored.

U.S. real GDP growth in Q4 2029 was 0.4\% (YoY), which indicates a very sluggish economy. Low growth typically reduces inflationary pressures due to subdued demand. However, inflation in January 2030 stands at 4.2\% (YoY), which is significantly above the Fed's traditional 2\% target. This suggests that inflation is being driven more by supply-side factors, inflation expectations, or cost-push dynamics rather than robust demand.

With the Fed at the zero lower bound, it lacks conventional tools to cool demand, implying limited ability to combat inflation actively. If inflation expectations remain elevated due to prior high readings and policy constrained by the zero lower bound, inflation may persist above target even in the face of low growth - a scenario resembling stagflation.

Assuming inflation expectations remain sticky and there are no disinflationary shocks (such as a major collapse in demand or oil prices), and given no new information beyond May 2021 about technological, geopolitical, or policy breakthroughs that could alter the trajectory, inflation is likely to remain elevated but may trend slightly downward due to base effects or gradual adjustments in supply chains.

However, with no tightening capacity (interest rates at 0.0\%) and growth stagnant, disinflationary forces are weak. Therefore, inflation is expected to remain above target but may decline modestly from 4.2\% if prior high readings drop out of the 12-month comparison.

A reasonable assumption is that inflation declines gradually by 0.5 to 0.8 percentage points over the year due to base effects and mild slack in the economy, despite accommodative policy.

Inflation forecast: 3.6\%
\end{tcolorbox}
\caption{\textbf{Example of a response for 1-year forecast of CPI growth (YoY) for a fake date by the Qwen3 Instruct model.} Formatting was slightly adjusted for readability.}
\label{fig: figure_F.2.2.2}
\end{figure}

\clearpage

\subsubsection{Real GDP growth (YoY)}

\renewcommand{\thefigure}{F.2.3.\arabic{figure}}
\setcounter{figure}{0}

\begin{figure}[ht]
\begin{tcolorbox}[promptbox]
To forecast U.S. real GDP growth for Q1 2022 (year-over-year), we consider the economic conditions and policy environment as of May 15, 2021. As of that date, the Federal Reserve maintained a highly accommodative monetary policy stance, with the lower bound of the federal funds rate at 0.0\%, indicating continued support for economic recovery following the pandemic-induced recession of 2020. This low-rate environment is conducive to investment and consumption, which can boost GDP growth.

Real GDP growth in Q1 2021 was 0.4\% year-over-year, which is modest but reflects the base effects from the sharp contraction in Q1 2020 due to the initial wave of the pandemic. With widespread vaccination efforts underway in early 2021 and fiscal stimulus (including the \$1.9 trillion American Rescue Plan enacted in March 2021), strong demand-side support was expected to drive above-trend growth in 2021 and early 2022.

Additionally, inflation as measured by CPI reached 4.2\% YoY in April 2021, indicating rising price pressures, partly due to supply constraints and base effects from low prices in 2020. While inflation could prompt concerns about future tightening, as of May 15, 2021, the Fed signaled it would tolerate higher inflation temporarily and not raise rates immediately. Thus, monetary conditions were expected to remain stimulative through 2021.

Given the combination of fiscal stimulus, accommodative monetary policy, and the ongoing reopening of the economy, a significant rebound in economic activity was anticipated. The low base from Q1 2021 (0.4\% YoY) and continued recovery in consumption, housing, and business investment point to a strong year-over-year growth number for Q1 2022, even if quarter-over-quarter momentum moderated.

Therefore, factoring in the stimulus effects, reopening dynamics, and base-year comparisons, real GDP growth for Q1 2022 (YoY) is forecast to be elevated relative to trend.

GDP growth forecast: 4.8\%
\end{tcolorbox}
\caption{\textbf{Example of a response for 1-year forecast of real GDP growth (YoY) for a real date by the Qwen3 Instruct model.} Formatting was slightly adjusted for readability.}
\label{fig: figure_F.2.3.1}
\end{figure}

\clearpage

\begin{figure}[ht]
\begin{tcolorbox}[promptbox]
To forecast US real GDP growth for Q4 2030, we consider the available data as of early 2030. The US real GDP growth in Q4 2029 was 0.4\% year-over-year, indicating very sluggish economic activity. Inflation in January 2030 stood at 4.2\% year-over-year, which remains above the Federal Reserve's typical target, suggesting inflationary pressures are still present. However, the federal funds rate lower bound is at 0.0\%, which is unusually low given the inflation level. This implies that monetary policy is highly accommodative despite inflation, possibly in response to weak growth or external shocks.

With policy rates at the zero lower bound and inflation still elevated, the central bank has limited capacity to stimulate the economy further through conventional means. The combination of low growth and high inflation points to a period of stagflationary pressure. Given the weak starting point of 0.4\% growth in late 2029 and the constraints of both policy and inflation, there is limited room for strong acceleration in output. Without evidence of significant fiscal stimulus or external improvements (which we cannot assume beyond May 2021 knowledge), the momentum suggests another year of subpar growth.

Therefore, assuming persistence in weak trends and limited policy effectiveness, real GDP growth for Q4 2030 is expected to remain low, possibly improving only modestly due to base effects or stabilization efforts, but not enough to break into stronger expansion territory.

GDP growth forecast: 0.6\%
\end{tcolorbox}
\caption{\textbf{Example of a response for 1-year forecast of real GDP growth (YoY) for a fake date by the Qwen3 Instruct model.} Formatting was slightly adjusted for readability.}
\label{fig: figure_F.2.3.2}
\end{figure}

\clearpage

\subsection{DeepSeek-V3.1}

\subsubsection{Fed interest rate}

\renewcommand{\thefigure}{F.3.1.\arabic{figure}}
\setcounter{figure}{0}

\begin{figure}[ht]
\begin{tcolorbox}[promptbox]
Based on the statistics provided (real GDP growth at 0.4\% and CPI inflation at 4.2\%), the US economy is presenting a significant policy dilemma for the Fed.

On one hand, the GDP growth figure of 0.4\% is very low, suggesting the economic recovery from the pandemic is still fragile and may require continued accommodative monetary policy (i.e., low interest rates) to support it.

On the other hand, the inflation figure of 4.2\% is significantly above the Fed's typical 2\% target. This level of inflation could signal overheating and might require the Fed to tighten monetary policy (i.e., raise interest rates) to prevent inflation from becoming entrenched.

Given that the Fed has repeatedly stated it will be ``patient'' and allow inflation to run moderately above target for some time to achieve a broad and inclusive recovery, the immediate reaction might be to hold rates steady. However, an inflation rate more than double their target is a strong signal. The forecast must consider that the Fed will be watching incoming data over the next 12 months. If the high inflation persists for several more months, which is a plausible scenario given the data from April 2021, the Fed would likely feel compelled to act to maintain its credibility.

Therefore, the most probable action over the one-year period is a cautious move to tighten policy in response to the persistent high inflation, leading to at least one rate increase.

Interest rate forecast: 0.25\%
\end{tcolorbox}
\caption{\textbf{Example of a response for 1-year forecast of the lower bound of the Fed interest rate for a real date by the DeepSeek-V3.1 model.} Formatting was slightly adjusted for readability.}
\label{fig: figure_F.3.1.1}
\end{figure}

\clearpage

\begin{figure}[ht]
\begin{tcolorbox}[promptbox]
Based on the limited data provided and the instruction to ignore any real-world knowledge after May 2021, the forecast must rely solely on a simplified interpretation of basic monetary policy principles.

The provided statistics show an economy with moderate but positive real GDP growth (0.4\% YoY) and inflation that is significantly above a typical central bank target of 2\% (4.2\% YoY). The primary tool a central bank uses to combat high inflation is raising interest rates. By increasing the cost of borrowing, the Fed would aim to cool down economic activity, reduce demand, and bring inflation down toward its target.

Given that the current interest rate is at a very low level (0.0\%), there is considerable room for the Fed to raise rates. With inflation more than double the target, the logical policy action would be to implement a series of rate increases over the coming year to address this.

Therefore, the forecast is for a higher federal funds rate in one year's time.

Interest rate forecast: 2.5\%
\end{tcolorbox}
\caption{\textbf{Example of a response for 1-year forecast of the lower bound of the Fed interest rate for a fake date by the DeepSeek-V3.1 model.} Formatting was slightly adjusted for readability.}
\label{fig: figure_F.3.1.2}
\end{figure}

\clearpage

\subsubsection{CPI growth (YoY)}

\renewcommand{\thefigure}{F.3.2.\arabic{figure}}
\setcounter{figure}{0}

\begin{figure}[ht]
\begin{tcolorbox}[promptbox]
Let's break down the problem using only the given data and no external knowledge.

**Step 1: Understand the given data (as of May 15, 2021)**

- Fed interest rate lower bound: 0.0\%

- US real GDP growth Q1 2021 (YoY): 0.4\%

- US inflation (CPI) April 2021 (YoY): 4.2\%

Interpretation: GDP growth is low (just 0.4\% YoY in Q1), but inflation is relatively high (4.2\% YoY). This suggests possible temporary factors (e.g., base effects from pandemic shock in early 2020, supply chain issues) rather than overheating from strong growth.

**Step 2: Reason about inflation dynamics without external knowledge**  

The Fed is keeping rates at 0\%, suggesting they view inflation as transitory and growth as still needing support. If GDP growth remains modest, demand-pull inflation may be limited.

April 2021's 4.2\% is influenced by the low base from April 2020 (pandemic collapse in prices). By April 2022, the base will be April 2021 (already higher), so base effects will not boost YoY inflation further.

**Step 3: Projection**

If inflation normalizes after temporary factors fade, and given modest growth, inflation in April 2022 should be closer to the Fed's long-term target (often near 2\%). But since the data says nothing about long-term targets, we must infer:

- Without strong growth, inflation shouldn't stay extremely elevated.

- Without new shocks, it should drop from 4.2\% but remain positive.

A reasonable estimate with only this info: between 2.0\% and 3.0\%.

Given the low GDP growth figure (0.4\% YoY in Q1 2021), the economy isn’t booming, so inflation pressure should ease. Pick midpoint of likely range: **2.5\%**.  

Inflation forecast: 2.5\%
\end{tcolorbox}
\caption{\textbf{Example of a response for 1-year forecast of CPI growth (YoY) for a real date by the DeepSeek-V3.1 model.} Formatting was slightly adjusted for readability.}
\label{fig: figure_F.3.2.1}
\end{figure}

\clearpage

\begin{figure}[ht]
\begin{tcolorbox}[promptbox]
Based on the statistics provided:

- The Federal Reserve's interest rate is at its lower bound of 0.0\%. This is historically a very accommodative monetary policy stance, typically used to stimulate the economy during periods of weak growth or recession.

- The most recent real GDP growth figure (Q4 2029, YoY) is 0.4\%. This indicates the US economy is growing, but at a very slow pace, well below its long-term trend. This suggests economic weakness.

- The most recent inflation figure (January 2030, YoY) is 4.2\%. This is above the typical Federal Reserve target of 2\%, indicating persistent inflationary pressures.

To forecast inflation for January 2031, we must consider the interaction between these factors. The high inflation of 4.2\% would normally prompt the Federal Reserve to raise interest rates to cool down the economy and bring prices under control. However, with GDP growth being so anemic (0.4\%), the Fed is faced with a significant dilemma. Raising rates to fight inflation could easily push the very slow-growing economy into a recession.

Given that the interest rate is already at 0.0\%, it suggests the Fed has prioritized supporting economic growth over aggressively fighting inflation. The extremely low GDP growth figure of 0.4\% provides a strong reason for the Fed to maintain this accommodative stance. Continuing with such stimulative policy (0.0\% rates) in an environment where inflation is already high (4.2\%) is likely to sustain inflationary pressures. The weak GDP growth might moderate the inflation slightly, as lower economic activity can reduce demand, but the primary driver will be the ongoing highly accommodative monetary policy.

Therefore, it is likely that inflation will remain elevated, but may decrease somewhat from its current level due to the weak economic growth. A forecast slightly above the Fed's target, but below the current rate, seems most plausible.

Inflation forecast: 3.1\%
\end{tcolorbox}
\caption{\textbf{Example of a response for 1-year forecast of CPI growth (YoY) for a fake date by the DeepSeek-V3.1 model.} Formatting was slightly adjusted for readability.}
\label{fig: figure_F.3.2.2}
\end{figure}

\clearpage

\subsubsection{Real GDP growth (YoY)}

\renewcommand{\thefigure}{F.3.3.\arabic{figure}}
\setcounter{figure}{0}

\begin{figure}[ht]
\begin{tcolorbox}[promptbox]
Let's break this down step by step:

1. **Starting point** - On May 15, 2021, the Fed interest rate lower bound is 0.0\%, Q1 2021 real GDP growth was 0.4\% (YoY), and April 2021 CPI inflation was 4.2\% (YoY).

2. **Context from early 2021** - The US economy is recovering from the pandemic shock. Fiscal stimulus and vaccine rollout are boosting demand, but supply chain disruptions and labor market frictions are pushing inflation higher.

3. **Forward projection** - By Q1 2022 (compared to Q1 2021), the base effect will matter because Q1 2021 growth was low (0.4\%). If the economy sustains recovery through 2021, Q1 2022 should show strong year-over-year growth simply because Q1 2021 was still subdued.

4. **No knowledge after May 15, 2021** - I cannot use actual 2021-2022 data, but based on the typical recovery pattern from a downturn and the data given, real GDP growth could accelerate sharply over the next year.

5. **Estimation** - Many early 2021 forecasts projected 2021 annual real GDP growth around 6-7\%, with Q1 2022 YoY growth possibly in the 4-5\% range, reflecting a strong rebound but also some normalization from peak growth rates in mid-2021.

Given the low base of 0.4\% for Q1 2021, a reasonable estimate for Q1 2022 YoY would be around **4.5\%**.

GDP growth forecast: 4.5\%
\end{tcolorbox}
\caption{\textbf{Example of a response for 1-year forecast of real GDP growth (YoY) for a real date by the DeepSeek-V3.1 model.} Formatting was slightly adjusted for readability.}
\label{fig: figure_F.3.3.1}
\end{figure}

\clearpage

\begin{figure}[ht]
\begin{tcolorbox}[promptbox]
Based solely on the data provided and without using external knowledge, here is the forecast reasoning:

The provided data shows an economy with very low growth (0.4\% YoY real GDP growth in Q4 2029) but significantly above-target inflation (4.2\% CPI in January 2030). The Federal Funds Rate is at its lower bound of 0.0\%, which is an extremely accommodative monetary policy stance, typically used to stimulate a weak economy.

This combination presents a dilemma. The high inflation suggests the economy may be overheating and requires tighter monetary policy (higher interest rates) to cool down. However, the very low GDP growth figure suggests economic weakness, which would normally call for continued or even more stimulus.

Given that the Fed's primary mandate is often considered to be price stability, the persistently high inflation of 4.2\% is likely the dominant concern. Even with modest growth, such an inflation rate is unsustainable. Therefore, the logical monetary policy response, even from this limited data, would be for the Fed to begin raising interest rates away from the 0\% lower bound sometime in 2030 to combat inflation.

The effect of raising interest rates is to slow down economic activity by making borrowing more expensive. This would likely further suppress GDP growth. Starting from an already low base of 0.4\%, the act of tightening monetary policy is forecast to push real GDP growth even lower. It is highly probable that the rate hikes necessary to bring inflation under control would cause growth to contract.

GDP growth forecast: -0.5\%
\end{tcolorbox}
\caption{\textbf{Example of a response for 1-year forecast of real GDP growth (YoY) for a fake date by the DeepSeek-V3.1 model.} Formatting was slightly adjusted for readability.}
\label{fig: figure_F.3.3.2}
\end{figure}

\end{document}